%% file: MAIN.tex
\documentclass[aps,prx,reprint,groupedaddress,superscriptaddress,floatfix]{revtex4-2}

\usepackage[T1]{fontenc}
\usepackage{newtxtext}
\usepackage{newtxmath}
\usepackage{bm}

\usepackage{titlesec}

\usepackage{amsmath}
\usepackage{dsfont}
\usepackage{mathtools}
\usepackage{amsfonts}
\usepackage{bbold}

\usepackage{color}    
\usepackage{xcolor}
\usepackage[colorlinks,citecolor=blue,linkcolor=blue,urlcolor=blue]{hyperref}

\usepackage{graphicx}
\usepackage[all]{hypcap}
\usepackage[english]{babel}  
\usepackage{ragged2e}        

\usepackage{enumitem}

\usepackage{tabularx}
\usepackage{makecell}
\usepackage{multirow} 
\usepackage{booktabs}  
\usepackage{siunitx}   

\graphicspath{{./img/}}

\usepackage{hyperref}

\usepackage{array}                
\usepackage{ifpdf}				
\usepackage{bbm}					

\usepackage{amscd}				
\usepackage[mathscr]{eucal}      
\usepackage{float}				
 \usepackage{epsfig}				
\usepackage{dcolumn}				

\usepackage{subcaption}			
\usepackage{cleveref}
\usepackage{etoolbox}

\makeatletter
\let\addcontentslineOriginal\addcontentsline
\let\addcontentsline\@gobblethree
\makeatother

\begin{document}
\input{CONTENT.tex}

\bibliographystyle{apsrev4-2}
\bibliography{references}

\clearpage

\makeatletter
\let\addcontentsline\addcontentslineOriginal
\makeatother

\renewcommand{\thesection}{S\Roman{section}}
\renewcommand{\thesubsection}{S\Roman{section}.\Alph{subsection}}
\renewcommand{\theequation}{S\arabic{equation}}
\renewcommand{\thefigure}{S\arabic{figure}}
\renewcommand{\thetable}{S\arabic{table}}

\setcounter{section}{0}
\setcounter{subsection}{0}
\setcounter{equation}{0}
\setcounter{figure}{0}
\setcounter{table}{0}


\newcommand{\suppsection}[2][]{%
  \refstepcounter{section}%
  \phantomsection%
  \section*{\texorpdfstring{S\Roman{section}\quad #2}{S\Roman{section} #2}}%
  \addcontentsline{toc}{section}{S\Roman{section}\quad #2}%
  \ifstrempty{#1}{}{%
    \label{#1}%
  }%
}

\newcommand{\suppsubsection}[2][]{%
  \refstepcounter{subsection}%
  \phantomsection%
  \subsection*{#2}
  \addcontentsline{toc}{subsection}{\thesubsection\quad #2}%
  \ifstrempty{#1}{}{%
    \label{#1}%
  }%
}

\input{SUPPLEMENT.tex}
\end{document}

%% file: CONTENT.tex
\title{Higher-order spacings in the superposed spectra of random matrices with comparison to spacing ratios and application to complex systems}

\author{Sashmita Rout}
\email{sashmitaa111@gmail.com}
\affiliation{Department of Physics, Visvesvaraya  National Institute of Technology, Nagpur 440010, India}

\author{Udaysinh T. Bhosale}
\email{udaysinhbhosale@phy.vnit.ac.in}
\affiliation{Department of Physics, Visvesvaraya  National Institute of Technology, Nagpur 440010, India}

\date{\today}

\begin{abstract}
Higher-order spacing statistics in the $m$ superposed spectra of circular random matrices of the same class are studied numerically. We conjecture that for given $m$ (or order $k$) and $\beta$, the sequence of modified Dyson index $\beta'(k)$ (or $\beta'(m)$) obtained using the sum of absolute differences between the cumulative distribution functions method (denoted as $D(\beta')$) is unique. Also, for a given $k$, the distribution tends to the corresponding $k$-th order Poisson statistics in the limit $m\rightarrow \infty$. The quantum chaotic kicked top model for various Hilbert space dimensions is studied, and it is found to satisfy our conjecture. This involves the numerical verification of $m=2$ case of COE results. Our result can be used as a tool for the characterization of a system and to determine the symmetry structure of the system without desymmetrization of the spectra. Additionally, the comparative study of the higher-order spacing and ratio distributions in both $m=1$ and $m=2$ cases of COE as well as GOE is performed within and across these ensembles numerically using the $D(\beta')$ method. This study is carried out both by varying the dimension and keeping the number of realizations constant, and vice-versa. The same asymptotic higher-order statistics are observed across COE and GOE in terms of a given spectral fluctuation measure. But, within a given ensemble of COE or GOE, the results of higher-order spacing and ratio distributions agree with each other only up to some lower $k$, and beyond that, they start deviating from each other. Further, the spectral fluctuations of the intermediate map of various dimensions are studied. Various important observations and discussions from the analysis of our extensive numerical computations are presented.
\end{abstract}
 
\maketitle
 
\section{Introduction}
\label{sec:Introduction} 
Random matrix theory (RMT) was introduced in physics to understand the statistical properties of the spectra of heavy atomic nuclei \cite{porter1965AcademicPress}. Later on, it has been used successfully in various branches of physics \cite{mehta2004random,akemann2011oxford,forrester2010log,guhr1998random,haake2010quantum,weidenmuller2009random,mitchell2010random}. Among them are nuclear physics \cite{gomez2011many}, atomic physics \cite{rosenzweig1960repulsion,camarda1983statistical}, and systems having a single particle to many interacting particles studied in quantum chaos \cite{stockmann2000quantum} and condensed matter physics \cite{RevModPhys.69.731,alhassid2000statistical}, respectively. Apart from physics, it has also been successfully used in various fields such as economics and finance, number theory (Riemann Zeta function), analysis of atmospheric and weather data, biology, wireless communication, complex networks \cite{akemann2011oxford,rao2020higher,kwapien2012physical,santhanam2001statistics,laloux1999noise,plerou1999universal,couillet2011random,jalan2007random,tulino2004random,mishra2024exploring} etc. Recently, RMT has found application in Machine Learning (ML) and brain mapping \cite{lawrence2025applications}, there, they have discussed how an algorithm originated from RMT can be used as a tool in ML for detecting correlations between functional areas of the brain.

The information about the physical characteristics (various phases) of complex quantum systems can be
revealed from their spectral fluctuations by using the theoretical setup of RMT.
For example, the integrable or chaotic phase of systems with or without a classical limit can be studied \cite{stockmann2000quantum,reichl2004transition,haake1987classical}.
This includes systems like coupled oscillators \cite{atkins1995quantum}, billiards \cite{PhysRevLett.52.1,hurt2013quantum},
many-body interacting systems \cite{kjall2018many,canovi2012many,vosk2015theory,varma2017energy,sierant2017many,imai2025quantum}, and various other systems \cite{weidenmuller2009random,izrailev1990simple,stockmann1990quantum,delande1986quantum,kota2014embedded,friedrich1989hydrogen}.
RMT is also used to study metallic or
insulating phases in tight-binding models and crystalline lattices \cite{hasegawa2000random,nishigaki1999level,zhong1998level,siringo1998mobility}, many-body localized and
thermal phases of interacting spin chains \cite{geraedts2016many,rao2020higher,rao2021critical,rao2022random,oganesyan2007localization} and for other purposes \cite{modak2014universal,serbyn2016spectral,riddell2024no}. 

Among all the statistical measures of spectral fluctuations, those act as probes to detect quantum chaos in RMT, the most well-known is the nearest neighbor spacing (NNS), defined as $s_{i}=E_{i+1}-E_{i}$, where $E_{i},\,\,\, i=1,2,\ldots$ are the eigenvalues of the corresponding matrix. The conjecture relating random matrices to chaotic quantum systems, is known as Bohigas-Giannoni-Schmit (BGS) conjecture, which states that the spectral fluctuation of a quantum system whose classical limit is chaotic agrees with the random matrix under suitable symmetry consideration \cite{PhysRevLett.52.1}. This has been verified extensively in experiments \cite{delande1986quantum,friedrich1989hydrogen,ellegaard1996symmetry}, simulations \cite{baranger1994mesoscopic,bandyopadhyay2002testing,jacquod2003quantum}, and supported by some theoretical studies \cite{muller2004semiclassical,muller2005periodic,berry1985semiclassical,sieber2001correlations,muller2009periodic}.

But, for the correct characterization of the system, the spectra must be symmetry-deduced. Because if there exists any additional symmetry, the Hamiltonian becomes block diagonal in the eigenbasis of the operator, corresponding to that symmetry, or we can say that the Hilbert space of the system splits into invariant subspaces, i.e.,  $H=H_1 \oplus H_2 \oplus H_3 \oplus \ldots H_m$, $i=1,2, \ldots m$ characterized by good quantum numbers corresponding to the respective symmetries \cite{stockmann2000quantum}. If we ignore symmetries and the eigenvalues from different blocks get superposed, then the true correlation between the eigenvalues is lost due to the near degeneracies, resulting in level clustering, which is misleading \cite{mehta2004random,guhr1998random,tekur2020symmetry,dyson1962statistical3,gunson1962proof,anderson2010introduction,
giraud2022probing,santos2010localization,gubin2012quantum,tkocz2012tensor,smaczynski2013extremal,tkocz2013note}. Because, integrable system shows level clustering as a spectral signature and the spacing distribution of its eigenvalues follows the Poisson distribution, $P(s) = \exp(-s)$, which is known as Berry and Tabor's conjecture \cite{berry1977level}. Thus, the spectra drawn from the same subspace can only provide the correct fluctuation property. Therefore, the presence or absence of symmetries has a great impact on the spectral correlations.

Also, symmetry consideration is a crucial aspect in RMT \cite{mehta2004random,forrester2010log}. One of the best-known examples is the construction of classical Gaussian ensembles by considering time-reversal symmetry. Depending on the symmetry of the system under consideration, it (especially the quantum Hamiltonian) can be modeled by one of the three classes of Gaussian random matrix ensembles. The three classes are Gaussian Orthogonal Ensemble (GOE), Gaussian Unitary Ensemble (GUE), and Gaussian Symplectic Ensemble (GSE) corresponding to Dyson index $\beta=1, 2$, and $4$ and consists of real symmetric, complex hermitian, and quaternion self-dual matrices, respectively \cite{dumitru2002models,dyson1962statistical1,dyson1962statistical2,dyson1962statistical3}. The GOE is suitable for systems having time-reversal symmetry, and besides that, are either consist of integer spin particles or invariant under rotation. The GUE is suitable for systems without having time-reversal symmetry irrespective of rotational symmetry, and the GSE for systems with time-reversal symmetry, but not rotational symmetry, and having half-integer spin particles \cite{brody1981random,pandey2019quantum}. This is known as Dyson's three-fold way.

Dyson introduced a new class of random matrix ensembles known as the circular ensemble, which are measures in the spaces of unitary matrices. They are Circular Orthogonal Ensemble (COE), Circular Unitary Ensemble (CUE), and Circular Symplectic Ensemble (CSE) corresponding to the Dyson index $\beta=1, 2$, and $4$, respectively. The symmetries for defining circular ensembles are the same as those of the respective Gaussian ensembles. They have found applications in condensed matter physics, optical physics \cite{akemann2011oxford}, scattering from a disorder cavity \cite{forrester2010log} and description of Floquet operators \cite{dyson1962statistical1}.

There are two theorems that relate circular ensembles. One that relates COE to CUE was proposed by Dyson \cite{dyson1962statistical3} and later proved by Gunson \citep{gunson1962proof}. According to this theorem, if we take alternate eigenvalues from the superposed spectra of two equal-dimensional COE, then these constitute the spectra of CUE. Another one relates COE with CSE, stating that the alternate eigenvalues of an even-dimensional COE belong to that of CSE \cite{mehta1963statistical}.
As a corollary of the above theorems, one can say that the level statistics of CUE and CSE can be obtained from COE. Analogous to these interrelations of circular ensembles, similar results were established by Forrester and Rains for the Gaussian, Laguerre, and Jacobi ensembles \cite{forrester2001interrelationships, forrester2001interrelationships3}. They classified all weight functions for which these relations hold true for a finite dimensional matrix.

When we have no prior idea about the symmetry structure of the system, or it's very difficult to split the model into symmetry sectors, how can the system be characterized correctly? Several attempts have been made in this direction. Some studies can be used to determine $m$, defined as the number of blocks in the Hamiltonian matrix or the number of independent sets of spectra, from any composite spectra. However, these are based on the complicated two-level cluster function, requiring regression methods, and most importantly, require unfolding at the initial step \cite{brody1981random,guhr1998random,porter1965AcademicPress,mehta2004random,guhr1990correlations,leviandier1986fourier,french1988statistical,molina2007power}.
But there is another work \cite{giraud2022probing}, in which they have derived analytical surmises using the spectral gap ratio for Gaussian ensembles consisting of several independent blocks. This result can be used to detect the number and size of independent symmetry subspaces. The most important aspect of this method is that it does not require unfolding, there is no constraint on the number of blocks or independent symmetry subspaces, and even without the assumption of chaos in the system under consideration. In another study, Ref. \cite{santos2020speck}, they have focused on two indicators of chaos namely the correlation hole and the distribution of off-diagonal elements of local observables. These indicators were restricted to detect {\it only} chaos, which requires no unfolding and no desymmetrization of eigenvalues. 

On the other hand, some methods are straightforward, no unfolding and no desymmetrization of the spectra are required, and only numerical calculation of higher-order spacing ratio (HOSR) is required. One of these is Ref.~\cite{tekur2020symmetry}.
There, the authors have studied the HOSR distribution in the superposed spectra of equal-dimensional GOE matrices. They have shown that the $m$-th order spacing ratio in the superposed spectra of $m$ GOE matrices are the same as the NN-spacing ratio distribution of GOE with modified Dyson index $\beta'=m$. Also, they used that result to find symmetries in spin chains, quantum billiards, and experimentally measured nuclear resonances. Another work is based on HOSR distributions in the superposed spectra of all three classes of circular ensembles \cite{bhosale2021superposition}. They have studied HOSR distributions extensively for larger $k$ and conjectured scaling relations in the case of COE and CSE that relate $\beta'$ with $k$ and $m$. It is also conjectured that for given $m(k)$ and $\beta$, the obtained value of $\beta'$ as a function of $k(m)$ is unique. These results can not only be used as a stringent test to determine symmetry structure of the system but also give true fluctuation characteristics without desymmetrizing the spectra, subject to the condition that the dimensions of independent subspaces are equal and are of equal class.

However, there are no studies on higher-order spacing (HOS) distributions in the superposed spectra of random matrices, be it numerical or analytical. Thus, our main objective is to study HOS distributions in superposed spectra of all the three classes of circular ensembles numerically. The advantage of circular ensembles is that their spectra can be easily unfolded. Here, for our study, we consider the i.i.d. spectra from the same jpdf with equal dimension because it is a simpler case to start with numerically. This study can give information about the long-range spectral correlation that complements the HOSR study for the same case \cite{bhosale2021superposition}. It can also be used as a tool to characterize the system and give symmetry information of the system. Further, earlier studies have demonstrated that the scaling relation Eq.~(\ref{Eq:ScalingRelation1}) (discussed in Sec.~\ref{sec:Preliminaries}), which relates $k$, $\beta$, and $\beta'$, is valid for both HOS and HOSR distributions \cite{rao2020higher,tekur2018higher,bhosale2023universal} in the case of non-superposed Gaussian random matrix ensemble. This led to our curiosity about exploring this aspect in the case of superposed spectra of random matrices. Are the obtained values of $\beta'$ for both HOS and HOSR distributions, the same or different in each of the superposed circular and Gaussian ensembles? Moreover, circular and Gaussian ensembles have the same asymptotic nearest level spacing distribution in the bulk \cite{mehta2004random} and circular ensembles follow the relation Eq.~(\ref{Eq:ScalingRelation1}) at large matrix dimension \cite{tekur2018higher}. In these directions, there is no such work on the superposed matrices and the circular-Gaussian correspondence in terms of HOS in non-superposed matrices. Hence, we would like to explore these aspects numerically. 

The rest of this paper is structured as follows:
In Sec.~\ref{sec:Preliminaries}, various spectral fluctuation measures in RMT and their applications in various systems in the distant and recent past are discussed. In Sec.~\ref{sec:SuperpositionCE}, the results (the values of $\beta'$) for the three classes of circular ensembles using the measure HOS are tabulated, and some of them are plotted in figures. In Sec.~\ref{sec:numerical method}, the statistical method we adopted in obtaining our results is discussed. In Sec.~\ref{sec:CompStudy}, we have studied comparatively HOS and HOSR distributions (with and without superposition), both for COE and GOE, in two ways. One with varying matrix dimensions ($N$), keeping the number of realizations ($n$) constant, and the other is by varying $n$, keeping $N$ constant. In Sec.~\ref{sec:Testingsystems}, we have verified our results by applying them to physical systems. In Sec.~\ref{sec:SomeImpObs}, some important observations and discussions are addressed. Finally, in Sec.~\ref{sec:SummConclsn}, summary and conclusions of our work are presented, mentioning some open questions and future directions as well.    
 
\section{Preliminaries}
\label{sec:Preliminaries}
In this section, we will discuss various measures for quantifying spectral fluctuations with their applications. The distribution of NNS (defined in Sec.~\ref{sec:Introduction}) for random matrices is given by \cite{mehta2004random}:
\begin{equation}
P(s,\beta)=A_{\beta}s^{\beta}\exp(-C_{\beta}s^2),\,\,\,\,\beta=1, 2, 4,
\label{Eq:PSBeta}
\end{equation}
where $A_{\beta}$ and $C_{\beta}$ are normalization constants that depend on $\beta$. However, to study the nearest neighbor spacing distribution (NNSD), spectral unfolding is required, which removes the system-dependent spectral features \cite{mehta2004random,haake2010quantum,porter1965AcademicPress,bruus1997energy,abul2014unfolding,guhr1998random}. This procedure is cumbersome and non-trivial, especially in many-body physics. There, it is ambiguous to write the closed form of the average level density due to its irregular pattern and the finite size of the Hilbert space \cite{bruus1997energy,haake2010quantum,oganesyan2009energy,gomez2002misleading}. Therefore, another measure was introduced and is known as nearest neighbor spacing ratio (NNSR), which is independent of the local DOS and hence doesn't require unfolding \cite{oganesyan2007localization}. The expression for NNSR is given by:
\begin{equation}
\label{Eq:NNratio}
r_{i}=\frac{s_{i+1}}{s_{i}},\,\,\,\,i=1,2,3,\ldots
\end{equation}
The distribution of $r_i$, denoted by $P(r)$, has been obtained for Gaussian ensembles and is given as follows: 
\cite{atas2013distribution,atas2013joint}:
\begin{equation}
P(r,\beta)=\frac{1}{Z_{\beta}} \frac{(r+r^2)^\beta}{(1+r+r^2)^{(1+3\beta/2)}},\,\,\,\,\beta=1,2,4
\label{Eq:PRBeta}
\end{equation}
where $Z_{\beta}$ is the normalization constant that depends on $\beta$. Whereas, for the Poisson case, the distribution is $P(r)=1/(1+r)^2$ \cite{oganesyan2007localization,atas2013distribution}. This quantity has been applied in various areas, such as in the context of many-body localization (MBL) \cite{buijsman2019random,oganesyan2009energy,pal2010many,iyer2013many,geraedts2016many,regnault2016floquet}, quantum chaos in Sachdev-Ye-Kitaev models \cite{sun2020periodic,sun2020classification,nosaka2020quantum}, finding symmetries in variety of complex quantum systems \cite{giraud2022probing,tekur2020symmetry,bhosale2021superposition}, in triangular billiards \cite{lozej2022quantum}, in the Hessian matrices of artificial neural networks \cite{baskerville2022appearance}, and in the study of quantum many-body scars \cite{imai2025quantum}. 

Both the NNS and NNSR quantify short-range level correlations. However, level correlation at a large spectral interval is useful in many cases. For example, probing short-time dynamics in chaotic quantum systems with classical limit \cite{stockmann2000quantum} and especially study concerning the MBL transition phenomena \cite{sierant2019level,agarwal2017rare}. Generally, in a random matrix, long-range correlations are described by the number variance $\Sigma^{2}$ or the Dyson-Mehta $\Delta_{3}$ statistics \cite{brody1981random}. But, both of them are strongly sensitive to the kind of unfolding procedure used, and some of the standard unfolding procedures can give misleading results \cite{gomez2002misleading}. In the same paper, it is shown that long-range correlations are more sensitive to the unfolding procedure employed than short-range correlations. 
However, HOS (provided uniform or/ and known close form of the average spectral density) and HOSR are simpler, and it's numerically easier to compute and analyze their distributions \cite{tekur2018higher}. Many studies are based on these higher-order measures \cite{engel1998higher,abul1999wigner,sakhr2006wigner,tekur2018higher,rao2020higher,sierant2019level,sierant2020model,
atas2013joint,tekur2018exact}. There is a recent work \cite{astaneh2025average}, where the importance of HOS can be seen. There, the authors have investigated the spread complexity to study the influence of energy level statistics, comparing both integrable and chaotic systems. Another recent work \cite{akhshani2025statistical} employs the HOSR in pseudointegrable systems. 

The non-overlapping $k$-th order spacing ratio $r_{i}^{(k)}$ where only one eigenvalue is shared 
between the spacings of the numerator and denominator \cite{tekur2018higher}) and $k$-th order spacing $s_{i}^{(k)}$ \cite{rao2020higher}, are defined as follows:
\begin{equation}
\label{Eq:HOSR}
r_{i}^{(k)}=\frac{s_{i+k}^{(k)}}{s_{i}^{(k)}}=\frac{E_{i+2k}-E_{i+k}}{E_{i+k}-E_{i}},\,\,\,\,  \\ \newline
\end{equation}
\begin{equation}
\mbox{and}\,\,\,\,   s_{i}^{(k)}=E_{i+k}-E_{i},\,\,\,\, i, k=1, 2, 3,\ldots,
\label{Eq:HOS}
\end{equation}
where $E_{i}$'s are the eigenvalues of a given matrix. Now, we give a comparison of numerical and analytical studies of spacing and spacing ratio distributions.
The analytical derivation of the HOS distribution is
known \cite{rao2020higher}, whereas no such derivation exists for the HOSR except for partial results \cite{bhosale2023universal}. 

The numerical study is comparatively easier in the case of NNSR and HOSR distribution. Because no unfolding procedure is required. The HOSR given in Eq.~(\ref{Eq:HOSR}) has found applications in the Gaussian \cite{tekur2018higher}, circular \cite{tekur2018higher}, and Wishart ensembles \cite{bhosale2018scaling}. There it is applied to various physical systems like spectra of spin chains, Floquet systems, atmospheric and weather data, and observed stock market. Also, a scaling relation is proposed as follows:
\begin{equation}
P^{k}(r,\beta,m=1)=P(r,\beta'),\,\,\,\,\beta=1, 2, 4
\label{Eq:GenDistribution1}
\end{equation}
and
\begin{equation}
\beta'=\frac{k(k+1)}{2}\;\beta + (k-1),\,\,\,\,k \geq 1.
\label{Eq:ScalingRelation1}
\end{equation}
The Eq.~(\ref{Eq:GenDistribution1}) implies that for a given ensemble corresponding to $\beta$, the distribution of $k$-th order spacing ratio in $m$ superposed spectra $P^{k}(r,\beta, m=1)$ is the same as that of the NNSR distribution of the ensemble corresponding to $\beta'$, i.e. $P(r,\beta')$. Here, $m=1$ represents spectra without superposition, i.e., spectra from a {\it single} random matrix. The Eq.~(\ref{Eq:GenDistribution1}) can be considered as a generalization of the Wigner surmise. The scaling relation in Eq.~(\ref{Eq:ScalingRelation1}) has been proved analytically, but in the asymptotic limits of $r^{(k)}\rightarrow0$ and $r^{(k)}\rightarrow\infty$
\cite{bhosale2023universal}.
The same scaling relation is proved analytically, but for the HOS distribution \cite{rao2020higher}. 
There, a generalized Wigner-Dyson distribution is given as follows:
\begin{equation}
P^{k}(s,\beta,m=1)=P(s,\beta'),\,\,\,\,\beta=1, 2, 4
\label{Eq:GenDistribution2}
\end{equation}
and
\begin{equation}
\beta'=\frac{k(k+1)}{2}\;\beta + (k-1),\,\,\,\,k \geq 1.
\label{Eq:ScalingRelation2}
\end{equation}
According to Eq.~(\ref{Eq:GenDistribution2}), $k$-th order spacing distribution in the $m$ superposed spectra for a given ensemble $\beta$, i.e., $P^{k}(s,\beta,m=1)$ is the same as that of the NNSD of the ensemble corresponding to $\beta'$, i.e. $P(s,\beta')$. Here, $m=1$ represents non-superposed spectra or spectra of a {\it single} random matrix. Also, the numerical evidence through simulations of random spin systems and nontrivial zeros of the Riemann $\zeta$ function are provided. 

Further, a similar kind of generalized distribution for ratio, i.e., Eq.~(\ref{Eq:GenDistribution3}) defined below, has been used to find symmetry structure in various complex quantum systems with the help of superposed spectra of random matrices \cite{tekur2020symmetry,bhosale2021superposition}.
\begin{equation}
P^{k}(r,\beta,m)=P(r,\beta'),\,\,\,\,\beta=1, 2, 4,
\label{Eq:GenDistribution3}
\end{equation}
where in Ref.~\cite{tekur2020symmetry}, they have shown that $\beta'=m=k$. The Eq.~(\ref{Eq:GenDistribution3}) implies the same as that of Eq.~(\ref{Eq:GenDistribution1}), but for any general $m$, where $m$ is a positive integer. But in Ref.~\cite{bhosale2021superposition}, there is no restriction on $k$, and the author has given the sequences of $\beta'$ for various values of $k$, which are unique for a given $\beta$ and $m$. Our aim here is to fill the gap by studying HOS in the superposed spectra of circular random matrices using the following generalized spacing distribution:
\begin{equation}
P^{k}(s,\beta,m)=P(s,\beta'),\,\,\,\,\beta=1, 2, 4.
\label{Eq:GenDistribution4}
\end{equation}
The Eq.~(\ref{Eq:GenDistribution4}) implies the same as that of Eq.~(\ref{Eq:GenDistribution2}), but for any general $m$, where $m$ is a positive integer.

In the subsequent sections, the results obtained using numerical simulations are presented.

\section{HOS in the superposed spectra of circular random matrices}
\label{sec:SuperpositionCE}
Circular ensemble comes into the picture when a system is not characterized by a Hamiltonian but by a unitary matrix. For example, in quantum mechanics, scattering matrices, and Floquet operators can be modeled by circular ensembles. Details about these are provided in Sec.~\ref{sec:Introduction}. The jpdf of eigenvalues for circular ensembles is given by the following expression:
\begin{equation} 
Q_{N,\beta}[\{ \theta_i\}] = C_{N,\beta} \prod_{k>j}^{N} |\exp(i\theta_j) - \exp(i\theta_k)|^{\beta},
\label{CJPDF}
\end{equation}
where $N$ and $C_{\beta,N}=(2\pi)^{-N}\{\Gamma(1+\beta/2)\}^{N}\{\Gamma(1+N\beta/2)\}^{-1}$ are the dimension and the normalization 
constant, respectively \cite{mehta2004random,forrester2010log}. The eigenvalues $\exp(i\theta_{\mu})$, $\mu=1,2,3...., N$ are distributed uniformly on the unit circle in the complex plane and show level repulsion, according to the Dyson index $\beta$ \cite{mehta2004random}. With increasing $\beta$, the repulsion increases. If we put $\beta=0$ in Eq.~(\ref{CJPDF}), we can find that all the eigenvalues become independent. Such uncorrelated eigenvalues follow Poisson statistics.
\subsection{COE case}
\label{subsec:SuperpositionCOE}
The matrices of the circular orthogonal ensemble ($\beta=1$) are symmetric and unitary in nature. The system that possesses time-reversal  and rotational symmetry, or has time-reversal symmetry and integral spin can be characterized by COE \cite{mehta2004random}. In this subsection, we study the HOS distribution, represented by $P^{k}(s, \beta, m)$ in the superposed spectra of $m=2$ to $7$ COE and for various values of $k$. We compare the distribution $P^{k}(s, \beta, m)$ with $P(s, \beta')$ as given in Eq.(\ref{Eq:PSBeta}) with modified Dyson index $\beta'$. We tabulate the value of $\beta'$ (see Table~\ref{Table:COEtable1}) for which both the distributions agree very well with each other numerically. The best fit is determined based on the value of $D(\beta')$, which is defined in Sec.~\ref{subsec:Estimation of parameter}. Here, the $k$-th order distribution $P^{k}(s, \beta, m)$ is considered equivalent to the nearest neighbor distribution $P(s,\beta')$, where $\beta'$ can be any number. This approach is adopted in the entire numerical calculation of this paper, wherever $\beta'$ is calculated. 


From Table~\ref{Table:COEtable1}, we observe that, except for some values of $\beta'$ (generally for lower $k$), all others are whole numbers for a given $m$. In this whole work, we have tried to give the value of best fit up to two decimal places, especially for those cases where the analytical distribution doesn't fit properly with the histogram or is not visually satisfactory as a proper fit for whole numbers. 
It can be seen that for given $m$, the maximum value of $k$, for which there is a positive non-integer $\beta'$,
increases with $m$.
For each $m$, the value of $\beta'$ increases with $k$, and for a given $k$, the value of $\beta'$ decreases as $m$ increases. Similar behavior was observed while studying HOSR distributions \cite{bhosale2021superposition}.
Further, the $k=1$ case is studied in detail in Sec.~\ref{subsec:Simul.Comp.HOS}. The obtained results (corresponding to both positive integer and non-integer $\beta'$ values) are plotted in Figs. \ref{fig:k_2_4_6_8_10_12_m2_COE} and \ref{fig:k_2_to_7_COE_m5}. More results are illustrated in the supplementary material \cite{supplementry2025}. In figures, the insets show the variation of $D(\beta')$ with $\beta'$, where the minima give the values of $\beta'$ for which both the  distributions fit very well with each other.
\begin{table}[t]
\renewcommand{\arraystretch}{1.5} 
\setlength{\tabcolsep}{6pt}  
\begin{center}
\begin{tabular}{|c|c|c|c|c|c|c|}
\hline 
$k$ & $m=2$ & $m=3$ & $m=4$ & $m=5$ & $m=6$& $m=7$\\
&$\beta'$&$\beta'$&$\beta'$&$\beta'$&$\beta'$&$\beta'$\\ 
\hline
1&$P$&$P$&$P$&$P$&$P$&$P$\\
2&2&1.25&1&0.8&0.69&0.61\\
3&4.28&3&2.40&2&1.79&1.62\\
4&7&5&4 &3.54&3.10&2.80\\
5&11&8&6&5.3&4.66&4.21\\
6&15&11 &9&7.25&6.5&5.80\\
7&19&14 &11&9.5&8.35&7.5\\
8&24&17 &14&12&10.5&9.45\\
9&30&21 &17&15&13&11.5\\
10&36&26&21&17&15&14\\
11&42&30&24&21&18&16\\
12&49&35&28&24&21&19\\
13&56&40&32&27&24&21\\
14&64&46&37&31&27&24\\
15&72&52&41&35&30&27\\
16&81&58&46&39&34&30\\
17&90&64&51&43&38&34\\
18&100&71&57&48&42&37\\
19&110&78&62&52&46&41\\
20&120&86&68&57&50&45\\
\hline
\end{tabular}
\end{center}
\caption{\justifying Tabulation of higher-order indices $\beta'$ for spacing distributions of order $k$ in the superposed spectra of $m$ COEs. Here, the dimension of each matrix without superposition is $N=5000$ and the number of realizations without superposition is $n=600, 900, 1000, 1000, 1002$, and $1001$, respectively, for $m=2,3,4,5,6$, and $7$. In the table, $P$ denotes the Poisson distribution $\exp(-s)$.}
\label{Table:COEtable1}
\end{table}
\begin{figure}[tbp]
\begin{center}
\includegraphics*[scale=0.35]{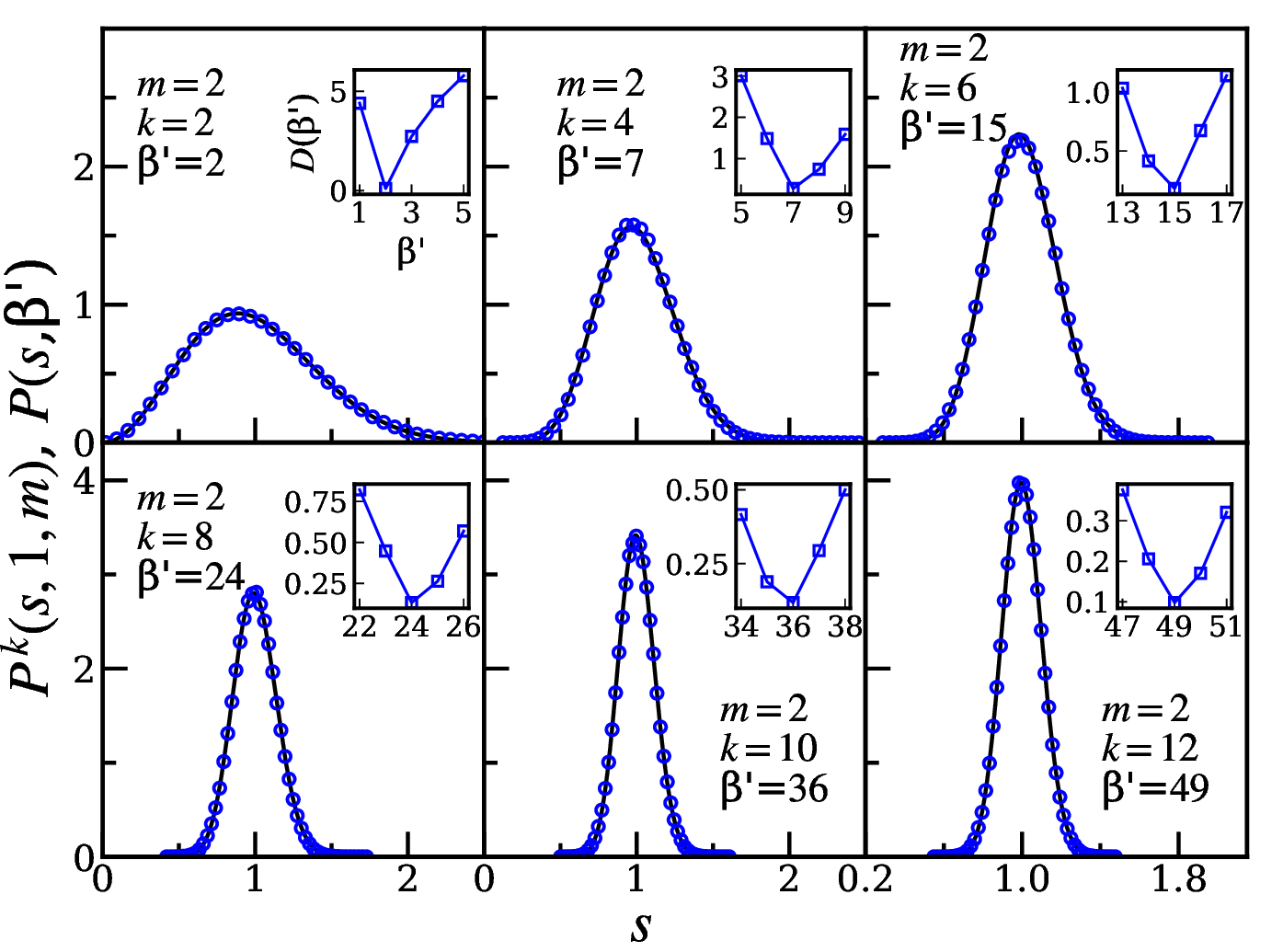}
\caption{\justifying Distributions of the $k$-th order spacing $P^{k}(s,1,m)$ for the superposition of $m=2$ COE spectra (circles). Here, $N=5000$ and $n=600$. The solid curve corresponds to $P(s,\beta')$ as given in Eq.~(\ref{Eq:PSBeta}), in which $\beta$ is replaced by $\beta'$ and $\beta'$ is given in Table~\ref{Table:COEtable1}. The insets show $D(\beta')$ as a function of $\beta'$.}
\label{fig:k_2_4_6_8_10_12_m2_COE}
\end{center}
\end{figure}
\begin{figure}[tbp]
\begin{center}
\includegraphics*[scale=0.35]{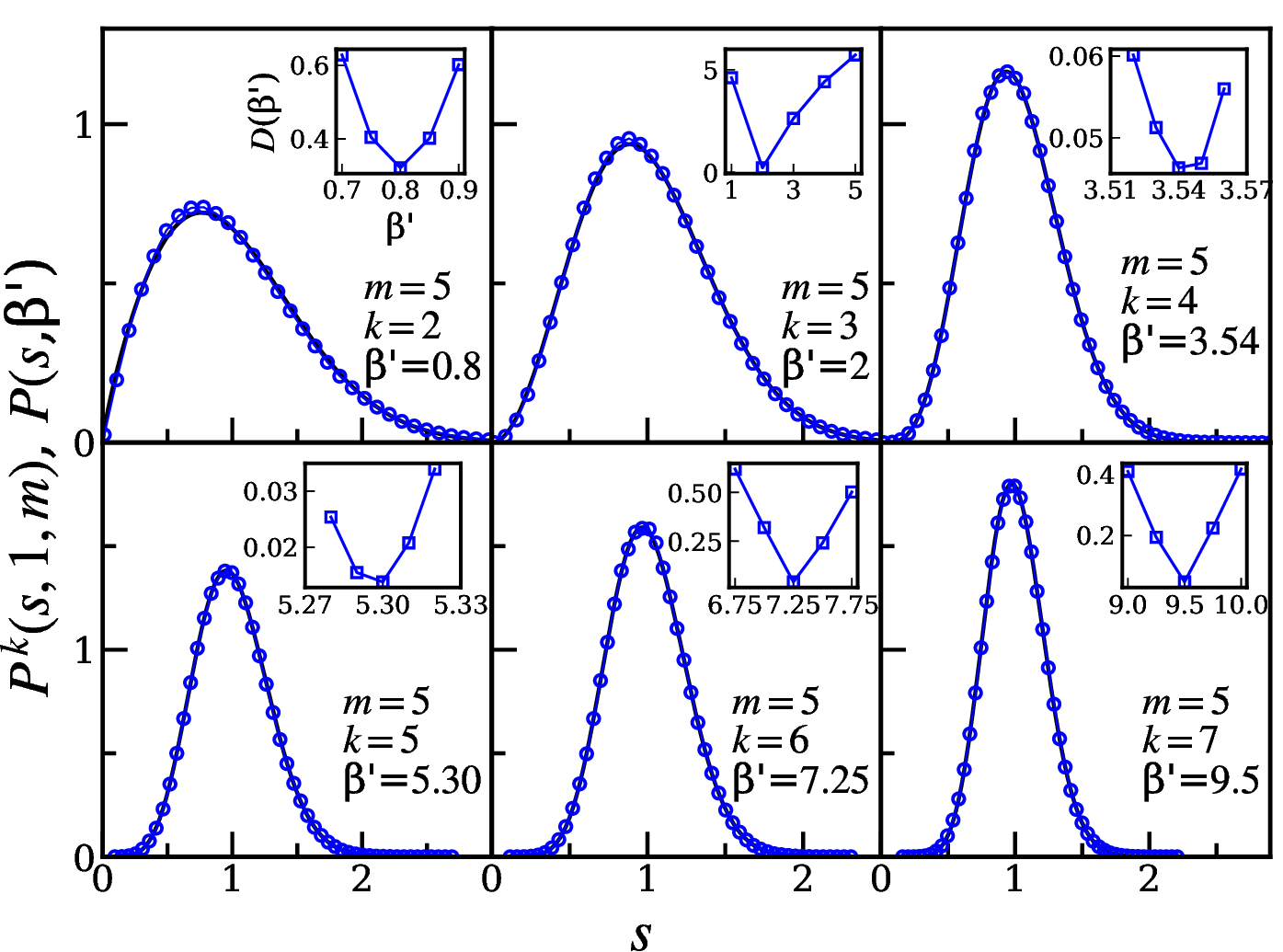}
\caption{\justifying Same as Fig.~\ref{fig:k_2_4_6_8_10_12_m2_COE} but for $m=5$, $n=1000$, and different values of $k$ and $\beta'$.}
\label{fig:k_2_to_7_COE_m5}
\end{center}
\end{figure}
\subsection{CUE case}
\label{subsec:SuperpositionCUE}
\begin{figure}[tbp]
\begin{center}
\includegraphics*[scale=0.35]{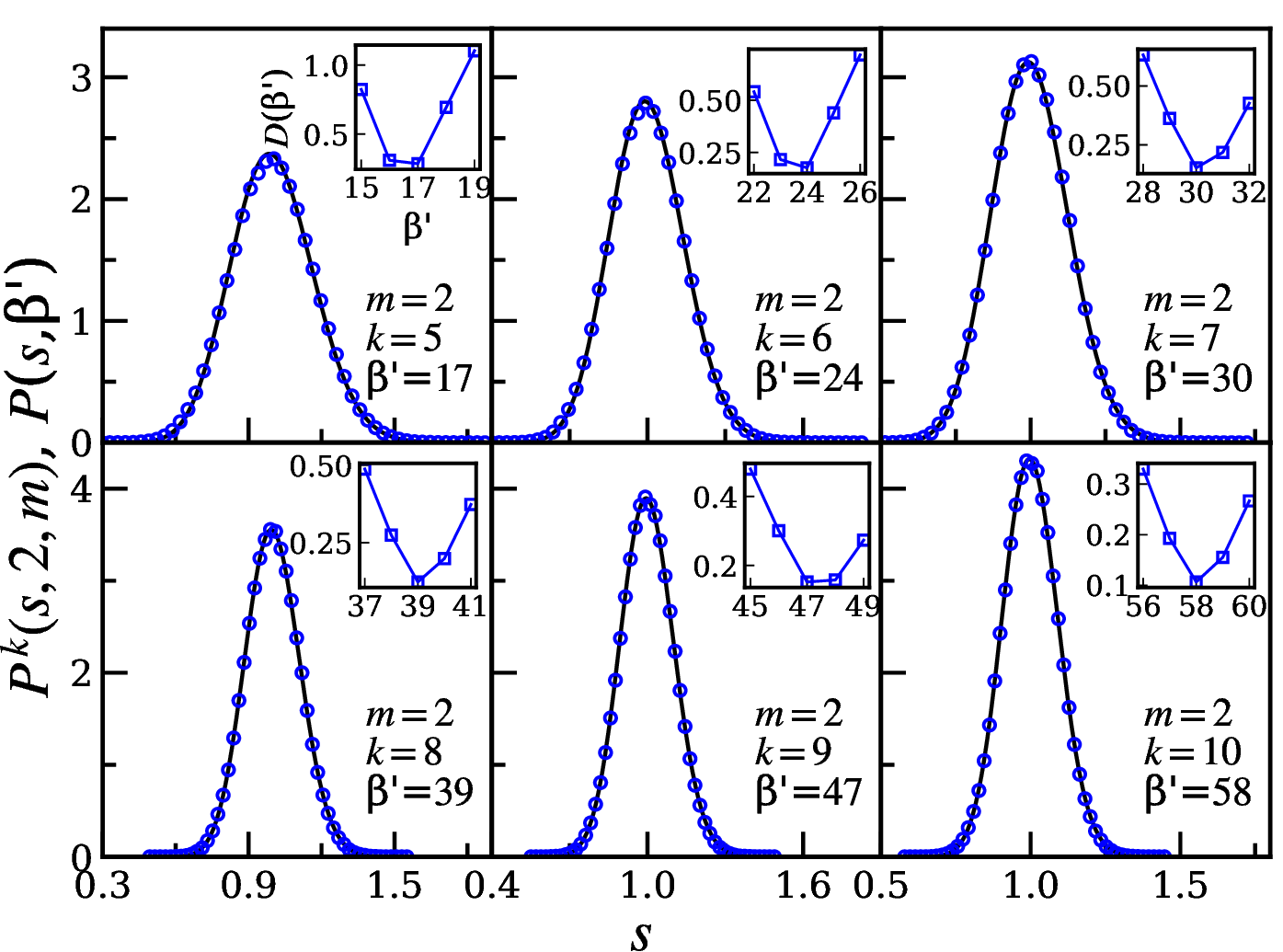}
\caption{\justifying Distributions of HOS $P^{k}(s,2,m)$ corresponding to various order $k$ in the  
$m=2$ CUE spectra (circles). Here, $N=5000$ and $n=600$. The solid curve corresponds to $P(s,\beta')$ as given in Eq.~(\ref{Eq:PSBeta}), where $\beta$ is replaced by $\beta'$ and $\beta'$ is given in Table~\ref{Table:CUEtable1}. The insets show $D(\beta')$ as a function of $\beta'$.} 
\label{fig:k_5_to_10_m2_CUE}
\end{center}
\end{figure}

\begin{figure}[tbp]
\begin{center}
\includegraphics*[scale=0.35]{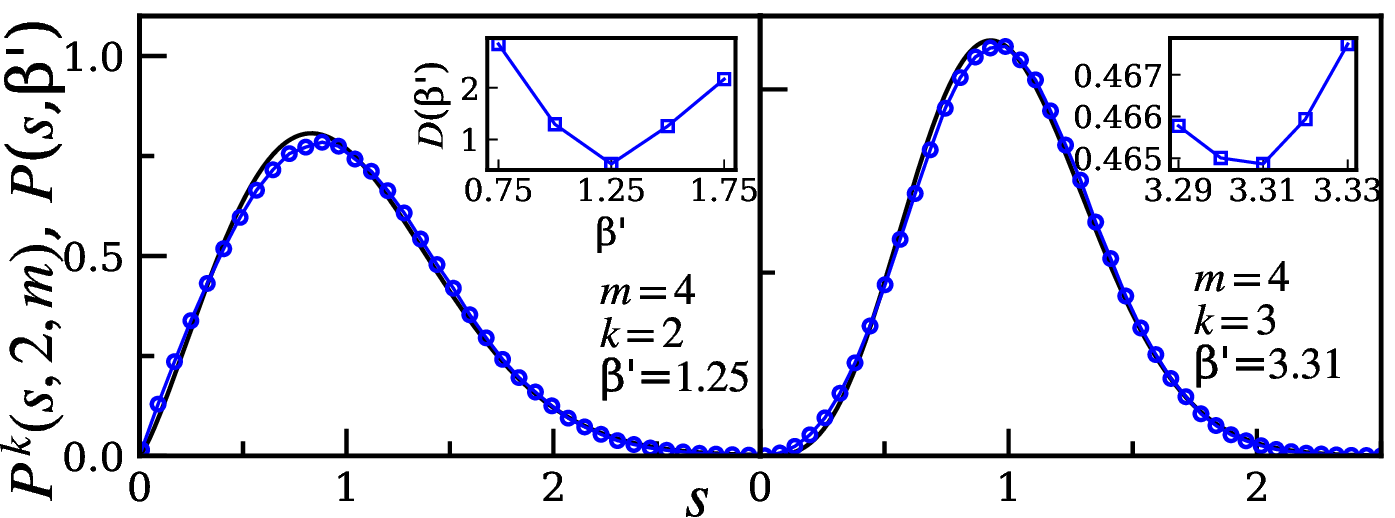}
\caption{\justifying Same as Fig.~\ref{fig:k_5_to_10_m2_CUE} but for different values of $k$, $m=4$, and $n=1000$.}
\label{fig:k_2_to_3_CUE_m4}
\end{center}
\end{figure}

In this subsection, we study HOS distributions in the superposed spectra of the CUE in similar lines to Sec. \ref{subsec:SuperpositionCOE}, where the superposition of COE is studied. The CUE ($\beta=2$) is used to model systems without time-reversal symmetry, irrespective of the rotational symmetry \cite{mehta2004random}. We tabulate our results in Table~\ref{Table:CUEtable1} for $m=2$ to $7$ and various values of $k$. In this case also, all values of $\beta'$ are whole numbers except few. Here, we can see that for lower values of $k$, especially for $k$ less than $m$, we get positive non-integer values. For higher $k$, the analytical distribution fits very well with the numerical data, but for a few lower values of $k$, it doesn't. It can be seen that for each $m$, the value of $\beta'$ increases with $k$, and for a given $k$, the value of $\beta'$ decreases as $m$ increases. Further, the $k=1$ case is studied in detail in Sec. \ref{subsec:Simul.Comp.HOS}. The results are plotted in Figs.~\ref{fig:k_5_to_10_m2_CUE} and \ref{fig:k_2_to_3_CUE_m4}. Here, the non-integer $\beta'$ values are shown in Fig.~\ref{fig:k_2_to_3_CUE_m4}. More results are plotted in the supplementary material \cite{supplementry2025}.
\begin{table}[t]
\renewcommand{\arraystretch}{1.5} 
\setlength{\tabcolsep}{6pt}  
\begin{center}
\begin{tabular}{|c|c|c|c|c|c|c|}
\hline
$k$ & $m=2$ & $m=3$ & $m=4$ & $m=5$&$m=6$&$m=7$\\
&$\beta'$&$\beta'$&$\beta'$&$\beta'$&$\beta'$&$\beta'$\\
\hline
1&$P$&$P$&$P$&$P$&$P$&$P$\\
2&3.25&1.80&1.25&0.95&0.79&0.68\\
3&6.41&4.57&3.31&2.6&2.14&1.85\\
4&11.5&7.5&6&4.75&3.97&3.43\\
5&17&11.25&9&7.38&6.25&5.40\\
6&24&16&12&10&9&7.69\\
7&30&21&16&13&12&10\\
8&39&26&21&17&15&13\\
9&47&33&25&21&18&16\\
10&58&39&31&25&22&19\\
11&68&47&36&30&26&23\\
12&80&55&43&35&30&26\\
13&92&63&49&40&35&31\\
14&106&72&56&46&40&35\\
15&119&82&63&52&45&39\\
16&134&92&71&59&50&44\\
17&149&102&79&65&56&49\\
18&166&113&87&72&62&54\\
19&182&125&96&80&68&60\\
20&201&137&106&87&75&66\\
\hline
\end{tabular}
\caption{\justifying Tabulation of higher-order indices $\beta'$ of spacing distributions for various $k$ in the superposed spectra of $m$ CUEs, each having $N=5000$. Here, $n=600, 900, 1000, 1000, 1002$, and $1001$, respectively, for $m=2, 3, 4, 5, 6$, and $7$. In the table, $P$ denotes the Poisson distribution $\exp(-s)$.}
\label{Table:CUEtable1}
\end{center}
\end{table}
\subsection{CSE case}
\label{subsec:SuperpositionCSE}
\begin{figure}[tbp]
\begin{center}
\includegraphics*[scale=0.35]{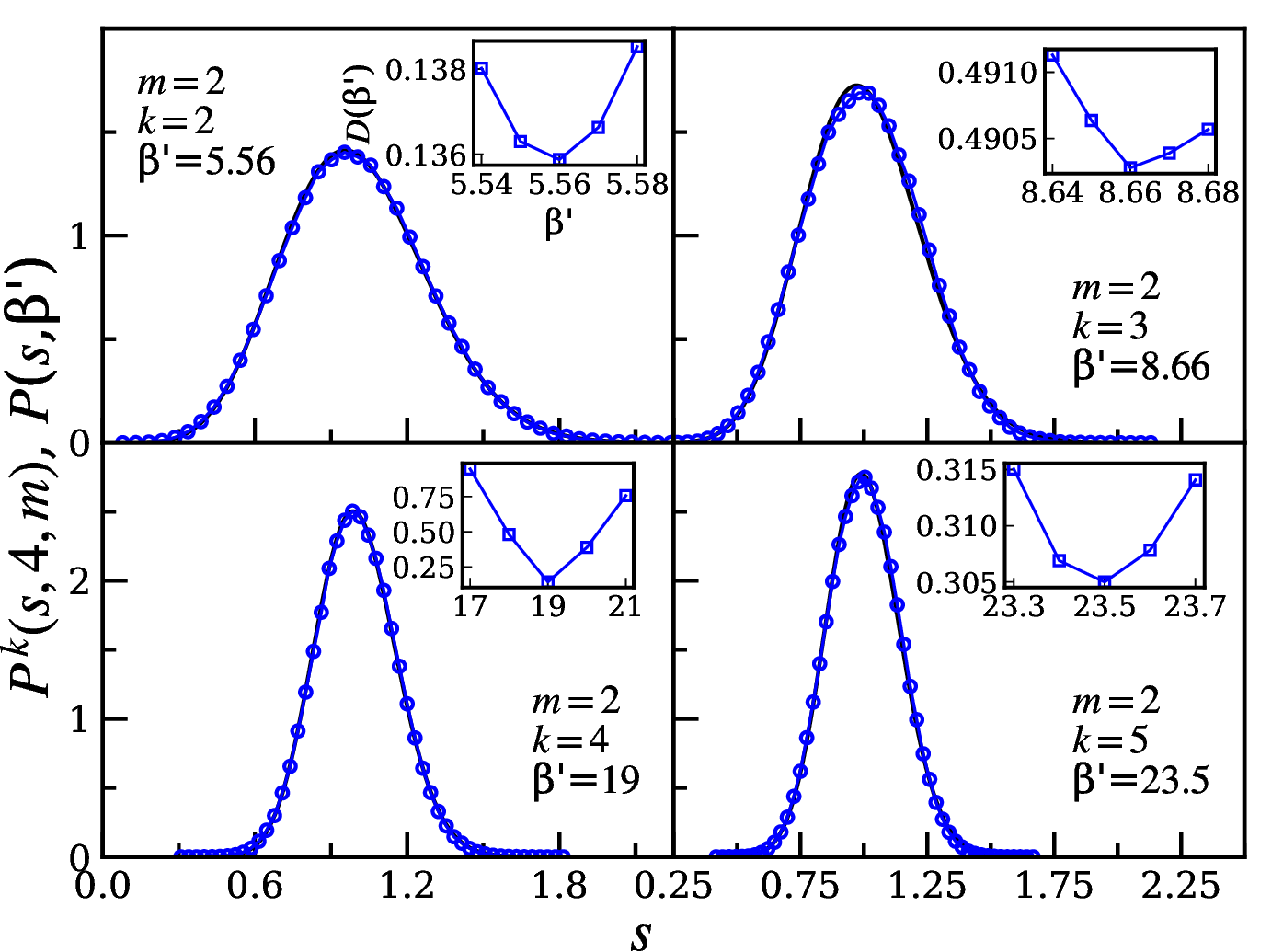}
\caption{\justifying Distributions of the $k$-th order spacings $P^{k}(s,4,m)$ in the $m=2$ CSE spectra (circles). Here, $N=5000$ and $n=600$. The solid curve corresponds to 
$P(s,\beta')$ as given in Eq.~(\ref{Eq:PSBeta}), where $\beta$ is replaced by $\beta'$ and $\beta'$ is given in  Table~\ref{Table:CSEtable1}. The insets show $D(\beta')$ as a function of $\beta'$.}
\label{fig:k_2_to_5_CSE_m2}
\end{center}
\end{figure}

\begin{figure}[tbp]
\begin{center}
\includegraphics*[scale=0.35]{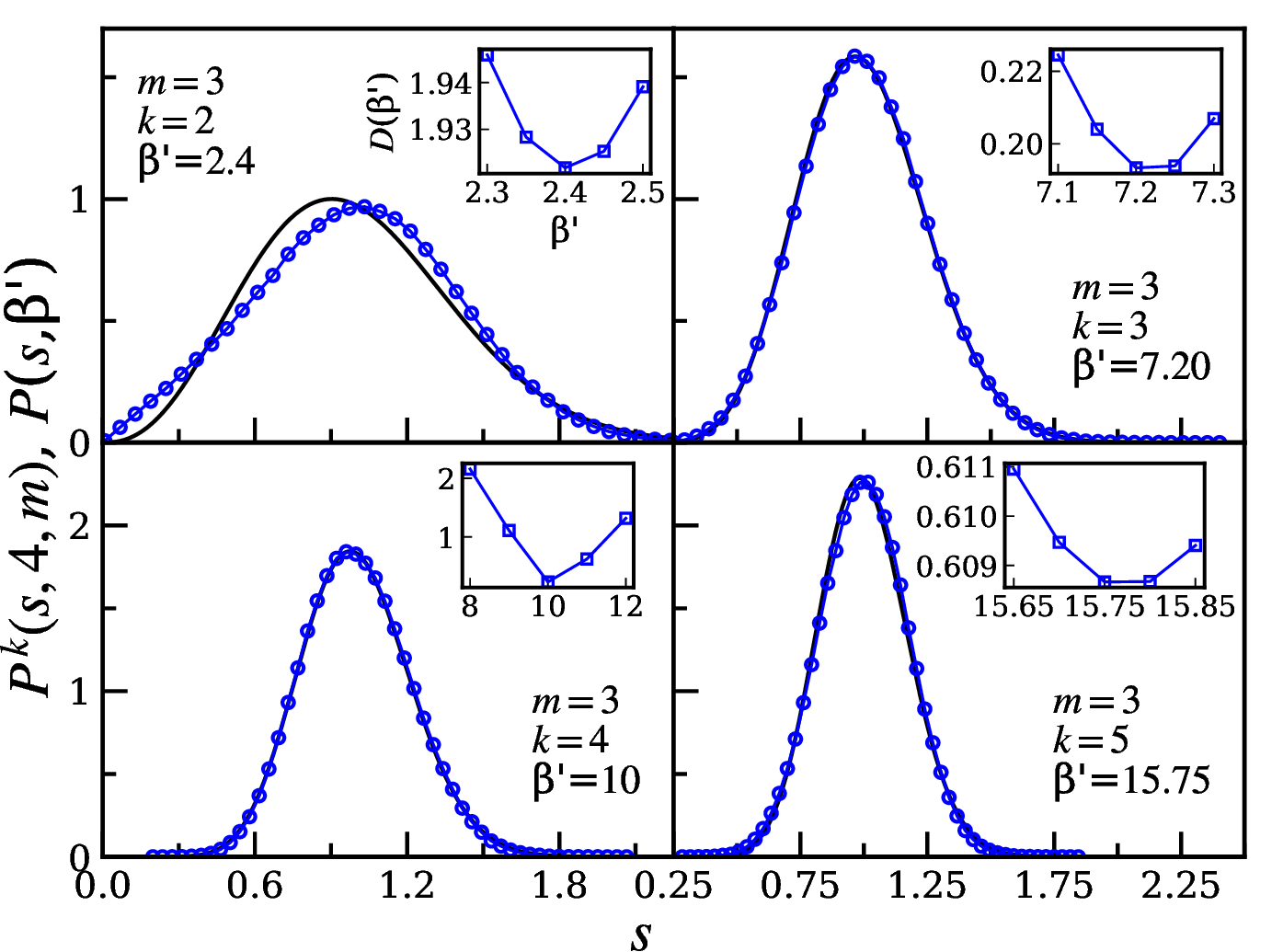}
\caption{\justifying Same as Fig.~\ref{fig:k_2_to_5_CSE_m2} but for different values of $m$.}
\label{fig:k_2_to_5_CSE_m3}
\end{center}
\end{figure}

In this subsection, we study HOS distributions in the superposed spectra of CSE in a similar line to that of  Sec.~\ref{subsec:SuperpositionCOE}. The CSE ($\beta=4$) is used to model systems having time-reversal symmetry, a half-integral spin interaction, and no rotational symmetry \cite{mehta2004random,zyczkowski2005random}. We tabulate the results in Table~\ref{Table:CSEtable1} for $m=2$ to $7$ and various values of $k$. The obtained results (corresponding to both positive integer and non-integer $\beta'$ values) are plotted in Figs.~\ref{fig:k_2_to_5_CSE_m2}-\ref{fig:k_2_to_5_CSE_m3}. In this case also, except few, all values of $\beta'$ are whole numbers. Here, we can see that for higher $k$, the analytical distribution fits very well with the numerical data, but for a few lower $k$, it doesn't. It can be seen that for each $m$, the value of $\beta'$ increases with $k$, and for a given $k$, the value of $\beta'$ decreases as $m$ increases. Further, the $k=1$ case is studied in detail in Sec.~\ref{subsec:Simul.Comp.HOS}. More results are plotted in the supplementary material \cite{supplementry2025}.
\begin{table}[t]                                           
\renewcommand{\arraystretch}{1.5} 
\setlength{\tabcolsep}{6pt}  
\begin{center}
\begin{tabular}{|c|c|c|c|c|c|c|}
\hline 
$k$ & $m=2$ & $m=3$ & $m=4$ & $m=5$ & $m=6$& $m=7$  \\
&$\beta'$&$\beta'$&$\beta'$&$\beta'$&$\beta'$&$\beta'$ \\ 
\hline
1&$P$&$P$&$P$&$P$&$P$&$P$\\
2&5.56&2.4&1.43&1&0.82&0.71\\
3&8.66&7.20&4.5&3&2.40&2\\
4&19&10&9&6.54&4.93&3.97\\
5&23.5&15.75&12&11&8.55&6.85\\
6&39&25 &16&14&13&10.59\\
7&45&30 &23&17.54&16&15\\
8&65&38 &31&22.6&19&18\\
9&71&51 &37&30&23.54&21\\
10&97&58&43&38&29.44&25\\
11&104&69&53&43&37&30\\
12&134&86&65&49&44&36\\
13&142&94&72&56&50&44\\
14&177&108&80&67&55&51\\
15&185&129&93&78&62&57\\
16&225&138&109&86&71&62\\
17&233&154&118&93&81&68\\
18& &179&128&104&92&76\\
19& &190&145&117&100&85\\
20& &209&163&132&107&96\\
\hline
\end{tabular}
\caption{\justifying Tabulation of higher-order indices $\beta'$ of spacing distributions for various $k$ in the superposed spectra of $m$ CSEs, each having $N=5000$. Here, $n=600, 900, 1000, 1000, 1002$, and $1001$, respectively, for $m=2,3,4,5,6$, and $7$. In the table, $P$ denotes the Poisson distribution $\exp(-s)$.}
\label{Table:CSEtable1}
\end{center}
\end{table}

\section{Numerical methods}
\label{sec:numerical method}
\subsection{Estimation of parameter: Finding the best fit}
\label{subsec:Estimation of parameter}
In this work, among the various statistical measures, $D(\beta')$ is chosen to determine the best fits with the numerical data quantitatively. In this paper, we have studied HOS and HOSR distributions in both superposed and non-superposed random matrix spectra. For HOS distribution of the $m$-superposed spectra, $D(\beta')$ is defined as \cite{tekur2018higher,bhosale2018scaling,tekur2020symmetry,bhosale2021superposition}:
\begin{equation}
D(\beta')=\sum_{i=1}^n\left|F^{k}(s_i, \beta, m)-F(s_i, \beta')\right|,
\label{Eq:DBeta_HOS_Superposed}
\end{equation}
And for HOSR distribution of the $m$-superposed spectra, $D(\beta')$ is defined as:
\begin{equation}
D(\beta')=\sum_{i=1}^n\left|F^{k}(r_i, \beta, m)-F(r_i, \beta')\right|,
\label{Eq:DBeta_HOSR_Superposed}
\end{equation}
where $F^{k}(s, \beta, m)$ and $F^{k}(r, \beta, m)$ denote cumulative distribution functions corresponding to the observed histograms $P^{k}(s, \beta, m)$ and $P^{k}(r, \beta, m)$, respectively. Whereas, $F(s, \beta')$ and $F(r, \beta')$ denote the cumulative distribution functions corresponding to the postulated functions $P(s, \beta')$ and $P(r, \beta')$ respectively, which are used as the fitting functions. Here, the running index $i$ corresponds to the bins of the histogram. We have fixed the number of bins to $200$ in all cases throughout this paper. These definitions of $D(\beta')$ have already been used in earlier works \cite{tekur2018higher,bhosale2018scaling,tekur2020symmetry,bhosale2021superposition}. Depending on the range of $i$, $D(\beta')$ can attain any positive value; however, it is minimum only for that value of $\beta'$, for which $P(s, \beta')$ or $P(r, \beta')$ is a best fit for the corresponding observed histogram. 
Such values of $\beta'$ are tabulated in Tables \ref{Table:COEtable1}-\ref{Table:GunsonResulttable} and are illustrated in various figures of this paper. In these figures, the variation of $D(\beta')$ with $\beta'$ is shown in the insets. The $\beta'$ at which the minimum occurs, the corresponding $P(s, \beta')$ ($P(r, \beta')$) 
is the best fit for the histogram of spacing (ratio) and it is shown in the main plot.
\subsection{Assessment of goodness-of-fit: Kolmogorov-Smirnov test}
\label{subsec: Kolmogorov-Smirnov test}
After finding the best fit using $D(\beta')$ for a given spacing data, we then examine its goodness-of-fit.
For this, we chose Kolmogorov-Smirnov test \cite{boes1974introduction}.
For a very large data set, the K-S test becomes very sensitive, and it gives very small $p$-value. 
As our data sets (the spacings of superposed spectra of $m$ circular ensembles with a large number of realizations) are very large,
we have only mentioned the Kolmogorov–Smirnov distance ($D_{KS}$), which is still meaningful as a distance measure. The $D_{KS}$ value gives information about the closeness of the two distributions; the lower the $D_{KS}$ value is, the higher the closeness, and vice versa. The $D_{KS}$ is defined as:
\begin{equation}
D_{KS} = \sup_x \left| F(x) - G(x) \right|  ,
\end{equation}
where $F(x)$ and $G(x)$ are cumulative distribution functions corresponding to the observed histogram of spacings or spacing ratios and the theoretical distribution function corresponding to spacings or spacing ratios, respectively. We have taken the distribution function for which $D(\beta')$ is minimum as the theoretical distribution function. From Tables~\ref{Table:KS_D_COE}-\ref{Table:KS_D_CSE}, we can observe that most of the $D_{KS}$ values are of order $10^{-3}-10^{-4}$, except for very few lower $k$ values, which are of order $10^{-2}$. Such an order of $D_{KS}$ values was also observed in the RMT literature \cite{akemann2022spacing}. The visual agreement observed from the figures and the corresponding $D_{KS}$ values indicates the goodness-of-fit clearly. For a few lower $k$ values, slightly higher $D_{KS}$ values indicate deviation from the Wigner-Dyson distribution and their convergence towards the corresponding $k$-th order Poisson statistics, which are also evident from the Figs.~\ref{fig: coe_s_2_tending_to_poisson}-\ref{fig: cse_r_2_tending_to_poisson}. 

We have also performed the K-S test using {\it only} one realization of the superposed spectra of given $m$ COEs, CUEs, and CSEs each separately, which are shown in the supplementary material \cite{supplementry2025}, taking into account the length of the data set. 
There, we observe that the values of $p$ are of the order of $10^{-1}$ or $10^{-2}$ and $D_{KS}$ are of order $10^{-3}$ or $10^{-4}$, except for small number of cases. Hence, we can conclude that the values of $p$ are not very small (greater than the significance level of $0.05$) and the $D_{KS}$ values are not very large, which indicates that there is no strong discrepancy between the data and the model function. 
\begin{table}[t!]                                           
\renewcommand{\arraystretch}{1.6} 
\setlength{\tabcolsep}{6pt}  
\begin{center}
\begin{tabular}{|c|c|c|c|c|c|c|}
\hline 
$k$&$m=2$&$m=3$&$m=4$&$m=5$&$m=6$&$m=7$  \\
&$D_{KS}$&$D_{KS}$&$D_{KS}$&$D_{KS}$&$D_{KS}$&$D_{KS}$ \\ 
\hline
2&1.761&3.157&3.759&5.916&8.605&10.700\\
3&1.393&2.472&1.310&3.162&4.319&6.103\\
4&4.349&4.448&5.113&0.917&2.163&3.144\\
5&4.160&5.110&4.345&0.326&0.908&1.589\\
6&3.055&4.854&6.902&0.963&1.916&0.838\\
7&2.483&2.484&2.000&0.870&0.469&0.679\\
8&2.361&3.880&1.662&1.414&0.792&0.368\\
9&2.513&2.949&2.074&3.901&2.381&0.725\\
10&2.304&2.420&3.208&3.310&2.630&2.208\\
11&2.108&2.412&1.895&3.271&1.389&1.554\\
12&1.894&2.042&1.784&1.777&1.600&2.182\\
13&2.154&2.249&2.035&2.191&1.617&2.768\\
14&1.699&1.995&2.105&1.651&1.572&2.048\\
15&1.752&1.965&1.806&1.754&2.571&1.958\\
16&1.629&1.801&1.906&1.719&1.351&2.390\\
17&1.705&1.886&1.776&1.843&1.662&1.538\\
18&1.536&1.559&1.985&1.811&1.807&1.792\\
19&1.574&1.691&1.798&1.847&1.660&1.604\\
20&1.398&1.715&1.884&1.579&1.475&1.903\\
\hline
\end{tabular}
\caption{\justifying Tabulation of K-S distance ($D_{KS}$) for spacing distributions of order $k$ in the superposed spectra of $m$ COEs. Here, the spacing data are the same as those used in Table~\ref{Table:COEtable1}. The theoretical distributions are taken corresponding to the $\beta'$ values as per Table~\ref{Table:COEtable1}, for a given $k$ and $m$. 
Here, all $D_{KS}$'s are in the unit of $10^{-3}$.}
\label{Table:KS_D_COE}
\end{center}
\end{table}
\begin{table}[t!]                                           
\renewcommand{\arraystretch}{1.6} 
\setlength{\tabcolsep}{6pt}  
\begin{center}
\begin{tabular}{|c|c|c|c|c|c|c|}
\hline 
$k$ & $m=2$ & $m=3$ & $m=4$ & $m=5$ & $m=6$& $m=7$  \\
&$D_{KS}$&$D_{KS}$&$D_{KS}$&$D_{KS}$&$D_{KS}$&$D_{KS}$ \\ 
\hline
2&2.600&11.847&9.113&4.418&2.158&5.863\\
3&5.459&3.461&8.252&8.337&6.166&3.337\\
4&3.133&3.582&3.687&6.911&7.023&6.129\\
5&4.896&4.900&3.603&3.628&5.411&6.060\\
6&3.457&3.084&4.404&3.544&4.111&4.373\\
7&3.158&3.205&4.307&4.353&5.169&4.528\\
8&2.571&3.918&3.254&3.665&4.467&2.774\\
9&2.662&2.701&3.340&3.248&3.372&2.834\\
10&2.256&2.779&3.308&3.425&3.613&3.056\\
11&2.451&2.918&3.030&2.499&3.199&3.659\\
12&2.059&2.545&2.860&2.591&2.790&3.741\\
13&1.388&2.250&2.372&2.900&2.593&2.856\\
14&1.891&2.297&2.491&2.720&2.905&2.385\\
15&1.831&2.305&2.521&2.348&2.887&2.678\\
16&1.741&2.110&2.266&2.510&2.700&2.296\\
17&1.783&2.253&2.016&2.426&2.663&2.366\\
18&1.641&2.209&2.036&2.271&2.409&2.635 \\
19&1.393&1.888&2.090&2.238&2.123&2.354\\
20&1.250&1.785&2.186&2.001&2.237&2.421 \\
\hline
\end{tabular}
\caption{\justifying Same as Table~\ref{Table:KS_D_COE} but for CUE. Here, the data and the theoretical distributions used are taken from Table~\ref{Table:CUEtable1}.
}
\label{Table:KS_D_CUE}
\end{center}
\end{table}
\begin{table}[t!]                                           
\renewcommand{\arraystretch}{1.6} 
\setlength{\tabcolsep}{6pt}  
\begin{center}
\begin{tabular}{|c|c|c|c|c|c|c|}
\hline 
$k$ & $m=2$ & $m=3$ & $m=4$ & $m=5$ & $m=6$& $m=7$  \\
&$D_{KS}$&$D_{KS}$&$D_{KS}$&$D_{KS}$&$D_{KS}$&$D_{KS}$ \\ 
\hline
2&2.746&31.092&23.948&14.240&7.316&3.749 \\
3&9.494&4.009&22.902&23.630&18.399&12.716\\
4&2.851&3.346&4.458&17.572&20.753&18.742\\
5&5.984&11.950&2.038&4.722&13.660&17.828\\
6&2.741&3.164&6.868&1.230&5.153&11.468\\
7&5.054&2.531&9.846&3.337&0.761&5.511\\
8&2.232&7.474&3.559&7.917&2.839&0.991\\
9&3.522&2.718&2.402&8.121&5.632&1.869\\
10&1.843&1.556&4.778&3.403&7.839&3.793\\
11&2.895&6.014&6.382&0.872&6.462&6.619\\
12&1.871&2.243&3.015&2.699&3.162&7.635\\
13&2.933&1.087&0.803&5.149&1.069&5.698\\
14&1.749&4.289&3.175&5.295&1.544&3.155\\
15&2.491&1.973&4.927&2.781&3.673&1.488\\
16&1.775&1.003&2.355&0.980&5.062&1.033\\
17&2.225&3.549&0.884&1.757&4.615&2.526\\
18&&1.831&2.607&3.904&2.623&4.273\\
19&&0.925&3.772&3.972&1.013&4.797\\
20&&3.199&1.827&2.286&1.236&3.989\\
\hline
\end{tabular}
\caption{\justifying Same as Table~\ref{Table:KS_D_COE} but for CSE. Here, the data and the theoretical distributions used are taken from Table~\ref{Table:CSEtable1}.
}
\label{Table:KS_D_CSE}
\end{center}
\end{table} 

\subsection{Uncertainty quantification: Based on ensemble sampling}
\label{subsec: Uncertainty quantification}
In this subsection, we quantify statistical uncertainty or sampling error that occurs when repeating the numerical experiment a large number of times. For this, we generate various data sets for $m$ COEs, each having the same number of realizations and each realization having the same dimension. 
Then, we obtain the values of $\beta'$ for sets corresponding to a given $k$, $\beta$, and $m$. Further, the uncertainty is calculated in the observed $\beta'$ as $\beta'_{\mbox{avg}}$$\pm$S.E., where S.E.$=$ Standard Error $=$ $\sigma/\sqrt{R}$. Here, $\sigma$ is the standard deviation, and $R$ is the number of $\beta'$, which is the same as the number of data sets.

Here, we have considered the following representative cases only, taking into account the computation time. For the case of $m=2$ COEs, $k=2$ to $20$ is taken, and the results are shown in the Table.~\ref{Table:uncertainty_m2}. Also, for $m = 2, 3, 4, 5, 6,$ and $7$ COEs, $k=5, 10, 15,$ and $20$ are taken, and the results are shown in the Table.~\ref{Table:uncertainty_m2_to_m_7}. From these results, especially from the very small values of S.E., it can be concluded that upon repeated sampling, for a given $m$ and $\beta$ by varying $k$, or for a given $k$ and $\beta$ by varying $m$, the sequence of obtained $\beta'$ using $D(\beta')$ method remains unique.

\begin{table*}[t!]
\renewcommand{\arraystretch}{1.5} 
\setlength{\tabcolsep}{15.5pt}  
\begin{center}
\begin{tabular}{|c|c|c|c|c|c|}
\hline 
order & $m=2$ & order & $m=2$ &order& $m=2$\\
$k$ & $\beta'_{\mbox{avg}}\pm$ S.E. & $k$ & $\beta'_{\mbox{avg}}\pm$ S.E.& $k$ & $\beta'_{\mbox{avg}}\pm$ S.E.\\
\hline
2&2$\pm$0.00&9&30$\pm$0.00&16&81.02$\pm$$\num{8.943e-03
}$\\
3&4.27$\pm$\num{1.093e-16}&10&36$\pm$0.00&17&90.09$\pm$$\num{1.809e-02}$\\
4&7$\pm$0.00&11&42$\pm$0.00&18&99.94$\pm$$\num{1.517e-02}$\\
5&11$\pm$0.00&12&49$\pm$0.00&19&109.91$\pm$$\num{1.897e-02}$\\
6&15$\pm$0.00&13&56.14$\pm$$\num{2.215e-02}$&20&120.10$\pm$$\num{2.031e-02}$\\
7&19$\pm$0.00&14&64.02$\pm$$\num{8.015e-03}$&&\\
8&24$\pm$0.00&15&72.36$\pm$$\num{3.059e-02}$&&\\
\hline
\end{tabular}
\end{center}
\caption{\justifying Tabulation of average values of $\beta'$ for $k=1$ to $20$, along with their corresponding standard error. Here, they are averaged over $250$ sets of $m=2$ COEs, and each set has $n=600$ and $N=5000$. When S.E. is less than $\num{e-16}$, we have taken it as zero.}
\label{Table:uncertainty_m2}
\end{table*}

\begin{table*}[t!]
\renewcommand{\arraystretch}{1.5} 
\setlength{\tabcolsep}{2.5pt}  
\begin{center}
\begin{tabular}{|c|c|c|c|c|c|c|}
\hline 
order & $m=2$ & $m=3$ & $m=4$ & $m=5$ & $m=6$ & $m=7$\\
$k$ & $\beta'_{\mbox{avg}}\pm$ S.E. & $\beta'_{\mbox{avg}}\pm$ S.E. & $\beta'_{\mbox{avg}}\pm$ S.E.& $\beta'_{\mbox{avg}}\pm$ S.E. & $\beta'_{\mbox{avg}}\pm$ S.E. & $\beta'_{\mbox{avg}}\pm$ S.E.\\
\hline
5&11$\pm$0.00&8$\pm$0.00&6$\pm$0.00&5.29$\pm$$\num{5.794e-04}$&4.66$\pm$$\num{5.937e-04}$&4.20$\pm$$\num{5.232e-04}$\\
10&36$\pm$0.00&26$\pm$0.00&20.96$\pm$$\num{1.616e-02}$&17$\pm$0.00&15$\pm$0.00&14$\pm$0.00\\
15&72.36$\pm$$\num{3.059e-02}$&52$\pm$0.00&41$\pm$0.00&35$\pm$0.00&30.19$\pm$$\num{3.230e-02}$&27$\pm$0.00\\
20&120.10$\pm$$\num{2.031e-02}$&85.76$\pm$$\num{3.364e-02}$&68$\pm$0.00&57$\pm$0.00&50$\pm$0.00&44.40$\pm$$\num{4.020e-02}$\\
\hline
\end{tabular}
\end{center}
\caption{\justifying Tabulation of average values of $\beta'$ for $k=5, 10, 15,$ and $20$, along with their corresponding standard error. Here, they are averaged over $250, 165, 150, 120, 150,$ and $150$ sets of $m=2, 3, 4, 5, 6,$ and $7$ COEs, respectively, and each set has $n=600, 900, 1000, 1000, 1002,$ and $1001$ for $m=2, 3, 4, 5, 6,$ and $7$, respectively, and every realization has $N=5000$. When S.E. is less than $\num{e-16}$, we have taken it as zero.}
\label{Table:uncertainty_m2_to_m_7}
\end{table*}
\section{Comparative study of HOS and HOSR distributions: A numerical investigation}
\label{sec:CompStudy}
In this section, we aim to study comparatively both HOS and HOSR distributions of the spectra of COE and GOE for both $m=1$ and $m=2$ cases. The readers who are not interested in this study can directly move to Sec.~\ref{sec:Testingsystems} for the application of our results (shown in Sec.~\ref{sec:SuperpositionCE}) to physical systems. We will be comparing the results of COE with GOE in the following two ways. Firstly, the effect of dimension on the obtained results, keeping the number of realizations constant, and secondly, the effect of the number of realizations, keeping dimension constant. In the case of superposition, here, only the $m=2$ case of COE and GOE is considered. Here, we are considering the $m=1$ case also, although the HOS and HOSR distributions have already been studied for the same case analytically and numerically. As the existing works have shown the numerical results only up to some values of $k$, we want to check them for higher values of $k$. This reproduction by using our numerical approach can check the robustness of the results and will also provide a base to analyze our results of the $m=2$ case.
\subsection{Dimensional analysis: Effect of dimensions}
\label{subsec:DimAnalysis}
In this subsection, we have studied the effect of dimension $N$ on the observed value of $\beta'$ for both HOS and HOSR distributions. The cases of COE and GOE, both with and without superposition are considered. 
The motivation for this dimensional analysis comes from the result, where similar asymptotic behavior is observed for the nearest neighbor statistics of both circular and the Gaussian ensembles. 

\begin{figure}[tbp]
\begin{center}
\includegraphics*[scale=0.35]{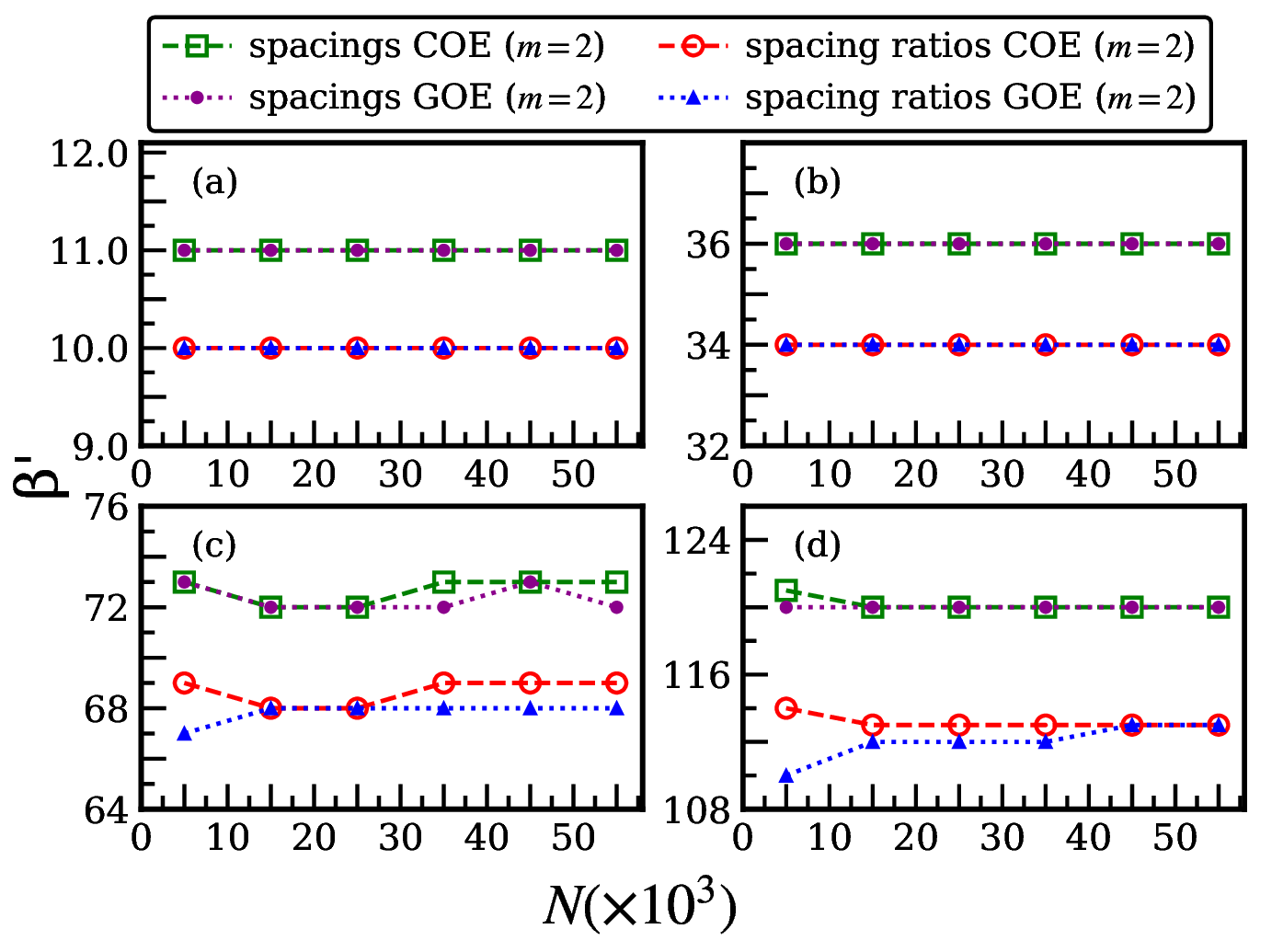}
\caption{\justifying Variation of $\beta'$ as a function of $N$ for the $m=2$ case of both COE and GOE, as given in the dimensional analysis tables for $m=2$ case of both COE and GOE from supplementary material \cite{supplementry2025}, keeping $n=300$ for all $N$. Here, the subplots (a), (b), (c), and (d) corresponds to $k=5, 10, 15$, and $20$, respectively.}
\label{fig: DimAnalysis_COE_GOE_m2}
\end{center}
\end{figure}

In Ref.~\cite{tekur2018higher}, the authors have studied finite-size effects for the HOSR distribution in the case of the Gaussian ensemble and the GOE spin chain. It is observed that as $N$ increases, the obtained $\beta'$ converges to the predicted value. The convergence is faster for smaller $k$.
They also claim that the predicted value agrees very well for both circular and the Gaussian ensembles.
In Ref.~\cite{rao2020higher}, the HOS distribution is studied both numerically and analytically for the Gaussian ensembles. In these works, numerical exploration was restricted to some $k$.
In our present work, we numerically explore in depth for large values of $k$, specifically for $k=1$ to $20$ using the $D(\beta')$ method.
By analyzing our results from Tables~\ref{Table: COE_m2_DimAnalysistable}-\ref{Table: GOE_m2_detailTable} and the Figs.~\ref{fig: DimAnalysis_COE_GOE_m2}-\ref{fig: DimAnalysis_COE_GOE} (For convenience, refer the dimensional analysis tables in the supplementary material \cite{supplementry2025} for the values of $\beta'$ ), we observe the following: \\

1. The case of COE and $m=2$:\\
In the $m=2$ case of COE, the observed values of the
$\beta'$ (refer to Fig.~\ref{fig: DimAnalysis_COE_GOE_m2} or the dimensional analysis table for the $m=2$ case of COE in the supplementary material \cite{supplementry2025}) remain almost the same except in some cases, where they differ by $\pm 1$, both for the distributions of spacing and spacing ratio as we increase $N$. Here, we increase $N$ from $1000$ to $55000$, keeping $n=300$ for each $N$ and a given $k$. Further, from Table~\ref{Table: COE_m2_DimAnalysistable} (or the dimensional analysis table for the $m=2$ case of COE in the supplementary material \cite{supplementry2025}), comparing both spacing and spacing ratio, we find that the values of $\beta'$ are the same for both up to $k=4$, and after $k\geq 5$, they started deviating from each other for a given $N$. Also, as $k$ increases, this deviation increases for a given $N$.\\

2. The case of GOE and $m=2$:\\ 
In the $m=2$ case of GOE, the observed values of the $\beta'$ for spacing distributions remain almost the same, except in some cases, where they differ by $\pm 1$, as we increase $N$. But for spacing ratio distributions, the values of $\beta'$ increases with $N$. Here, we increase $N$ from $1000$ to $55000$, keeping $n=300$ for each $N$ and given $k$.
For a given $k$, except for a few cases, the value of $\beta'$ becomes saturated beyond a certain $N$ as far as our results are concerned, but if we further increase $N$, the saturated value may change.
From Table~\ref{Table: GOE_m2_detailTable} and Fig.~\ref{fig: DimAnalysis_COE_GOE_m2} (or the dimensional analysis table for the $m=2$ case of GOE in the supplementary material \cite{supplementry2025}), comparing both the distributions of spacing and spacing ratio, we find that the values of $\beta'$ are the same up to $k=4$, and after $k\geq 5$, they started deviating from each other. Further, this deviation increases with $k$ for a given $N$.\\

3. The case of COE and $m=1$:\\
In the case of COE without superposition, it can be observed from Fig.~\ref{fig: DimAnalysis_COE_GOE} (or the dimensional analysis table for $m=1$ case of COE in the supplementary material \cite{supplementry2025}) that the observed values of $\beta'$ for both the distributions of spacing and spacing ratio remain almost the same, except for some cases, where they differ by $\pm 1$, or in some rare cases $\pm 2$, as we increase $N$. Here, we increase $N$ from $1000$ to $55000$, keeping $n=300$ for each $N$ and given $k$. From Table~\ref{Table: COEAndGOEDimAnalysistable} and Fig.~\ref{fig: DimAnalysis_COE_GOE} (or the dimensional analysis table for $m=1$ case of COE in the supplementary material \cite{supplementry2025}), comparing both spacing and spacing ratio results, we find that the values of $\beta'$ are same up to $k=3$ and after $k\geq 4$, they started deviating from each other for a given $N$. As $k$ increases, the deviation increases for both. The observed values of $\beta'$ match the predicted values according to Eq.~(\ref{Eq:ScalingRelation1}) upto $k=3$ and $k=8$ for spacing and spacing ratio, respectively.\\

4. The case of GOE and $m=1$:\\
In the case of GOE without superposition, it can be observed from Table~\ref{Table: COEAndGOEDimAnalysistable} and Fig.~\ref{fig: DimAnalysis_COE_GOE} (or the dimensional analysis table for $m=1$ case of GOE from the supplementary material \cite{supplementry2025}) that the observed values of $\beta'$ for spacing distributions remain almost the same, except for some cases where they differ by $\pm 1$ or in some rare cases $\pm 2$
as we increase $N$ from $1000$ to $55000$ for a given $k$. Here, $n=300$ for each $N$ and given $k$. 
For the corresponding case of the HOSR distribution, $\beta'$ increases with $N$ for a given $k$. 
But, it appears to be tending towards that of the corresponding value of the HOSR of COE. Comparing both spacing and spacing ratio distributions, we find that the values of $\beta'$ are the same up to $k=3$, and after $k\geq 4$, they start deviating from each other. As $k$ increases, the deviation also increases for given $N$. The observed values of $\beta'$ match the predicted values according to Eq.~(\ref{Eq:ScalingRelation1}) up to $k=3$ and $k=7$ for the distributions of spacing and spacing ratio, respectively. And for ratio, this agreement increases with $N$ beyond $k\geq 7$.\\

Hence, we can conclude from the above that for the $m=2$ case of COE and GOE, for smaller $k$ (up to $k=4$), the results of spacing and ratio distributions are the same within the ensemble and across these ensembles. For higher $k$ (i.e., $k \geq 4$), the results of spacing and ratio distributions start deviating from each other, and the deviation increases with $k$ within each ensemble. Also, we find that $N$ does not have a very significant effect on the results of HOS distributions of COE and GOE, and HOSR distributions of COE after a certain $N$, which is small. The results of HOS distribution of both COE and GOE are found to be nearly the same, and the results of HOSR distributions of COE aren't the same as that of the corresponding HOSR distributions of GOE for a particular $N$. But, as $N$ increases, the HOSR distribution results of GOE seem to tend towards the HOSR of COE. We also find that for a particular $N$, the HOSR distribution results of both the unfolded eigenvalues of GOE and without unfolding the eigenvalues of COE are same, with $\pm 1$ difference for some cases at higher $k$, which can be neglected, because we think they are due to statistical fluctuations. Similar behaviors are also observed for the $m=1$ case of both COE and GOE. But, here both COE and GOE follow the scaling relation up to slightly higher $k$ for HOSR distribution than HOS. As $k$ increases further, the ratio distribution results are found to be getting more away from the scaling relation than spacing. The possible reasons for deviation from the scaling relation for higher values of $k$ in the case of both spacings and ratios will be discussed in Sec.~\ref{Gunson's result} and Sec.~\ref{subsec:Dist.Func.Anal}.
\begin{table*}      
\renewcommand{\arraystretch}{1.5} 
\setlength{\tabcolsep}{8.5pt}  
\begin{center}
\begin{tabular}{|c|cc|cc|cc|cc|cc|cc|}
\hline
\rule{0pt}{12pt}  
Order&\multicolumn{2}{c|}{$N=5000$}&\multicolumn{2}{c|}{$N=15000$}&\multicolumn{2}{c|}{$N=25000$}&\multicolumn{2}{c|}{$N=35000$}&\multicolumn{2}{c|}{$N=45000$}&\multicolumn{2}{c|}{$N=95000$}\\ [1.5ex]  
\cline{2-13}  
\rule{0pt}{10pt}  
&HOS&\multicolumn{1}{c|}{HOSR}&HOS&\multicolumn{1}{c|}{HOSR}&HOS&\multicolumn{1}{c|}{HOSR}&HOS&\multicolumn{1}{c|}{HOSR}&HOS&\multicolumn{1}{c|}{HOSR}&HOS&\multicolumn{1}{c|}{HOSR}\\
$k$&$\beta^\prime$&$\beta^\prime$&$\beta^\prime$&$\beta^\prime$&$\beta^\prime$&$\beta^\prime$&$\beta^\prime$&$\beta^\prime$&$\beta^\prime$&$\beta^\prime$&$\beta^\prime$&$\beta^\prime$\\
\hline
1&$P$&\multicolumn{1}{c|}{$P$}&$P$&\multicolumn{1}{c|}{$P$}&$P$&\multicolumn{1}{c|}{$P$}&$P$&\multicolumn{1}{c|}{$P$}&$P$&\multicolumn{1}{c|}{$P$}&$P$&\multicolumn{1}{c|}{$P$}\\
2&2&\multicolumn{1}{c|}{2}&2&\multicolumn{1}{c|}{2}&2&\multicolumn{1}{c|}{2}&2&\multicolumn{1}{c|}{2}&2&\multicolumn{1}{c|}{2}&2&\multicolumn{1}{c|}{2}\\
3&4.28&\multicolumn{1}{c|}{4}&4.28&\multicolumn{1}{c|}{4}&4.28&\multicolumn{1}{c|}{4}&4.29&\multicolumn{1}{c|}{4}&4.28&\multicolumn{1}{c|}{4}&4.29&\multicolumn{1}{c|}{4}\\
4&7&\multicolumn{1}{c|}{7}&7&\multicolumn{1}{c|}{7}&7&\multicolumn{1}{c|}{7}&7&\multicolumn{1}{c|}{7}&7&\multicolumn{1}{c|}{7}&7&\multicolumn{1}{c|}{7}\\
5&11&\multicolumn{1}{c|}{10}&11&\multicolumn{1}{c|}{10}&11&\multicolumn{1}{c|}{10}&11&\multicolumn{1}{c|}{10}&11&\multicolumn{1}{c|}{10}&11&\multicolumn{1}{c|}{10}\\
6&15&\multicolumn{1}{c|}{14}&15&\multicolumn{1}{c|}{14}&15&\multicolumn{1}{c|}{14}&15&\multicolumn{1}{c|}{14}&15&\multicolumn{1}{c|}{14}&15&\multicolumn{1}{c|}{14}\\
7&19&\multicolumn{1}{c|}{18}&19&\multicolumn{1}{c|}{18}&19&\multicolumn{1}{c|}{18}&19&\multicolumn{1}{c|}{18}&19&\multicolumn{1}{c|}{18}&19&\multicolumn{1}{c|}{18}\\
8&24&\multicolumn{1}{c|}{23}&24&\multicolumn{1}{c|}{23}&24&\multicolumn{1}{c|}{23}&24&\multicolumn{1}{c|}{23}&24&\multicolumn{1}{c|}{23}&24&\multicolumn{1}{c|}{23}\\
9&30&\multicolumn{1}{c|}{28}&30&\multicolumn{1}{c|}{28}&30&\multicolumn{1}{c|}{28}&30&\multicolumn{1}{c|}{28}&30&\multicolumn{1}{c|}{28}&30&\multicolumn{1}{c|}{28}\\
10&36&\multicolumn{1}{c|}{34}&36&\multicolumn{1}{c|}{34}&36&\multicolumn{1}{c|}{34}&36&\multicolumn{1}{c|}{34}&36&\multicolumn{1}{c|}{34}&36&\multicolumn{1}{c|}{34}\\
11&42&\multicolumn{1}{c|}{40}&42&\multicolumn{1}{c|}{40}&42&\multicolumn{1}{c|}{40}&42&\multicolumn{1}{c|}{40}&42&\multicolumn{1}{c|}{40}&42&\multicolumn{1}{c|}{40}\\
12&49&\multicolumn{1}{c|}{46}&49&\multicolumn{1}{c|}{46}&49&\multicolumn{1}{c|}{47}&49&\multicolumn{1}{c|}{47}&49&\multicolumn{1}{c|}{47}&49&\multicolumn{1}{c|}{47}\\
13&56&\multicolumn{1}{c|}{53}&56&\multicolumn{1}{c|}{53}&56&\multicolumn{1}{c|}{53}&56&\multicolumn{1}{c|}{53}&56&\multicolumn{1}{c|}{53}&56&\multicolumn{1}{c|}{53}\\
14&64&\multicolumn{1}{c|}{61}&64&\multicolumn{1}{c|}{61}&64&\multicolumn{1}{c|}{61}&64&\multicolumn{1}{c|}{61}&64&\multicolumn{1}{c|}{61}&64&\multicolumn{1}{c|}{61}\\
15&72&\multicolumn{1}{c|}{68}&72&\multicolumn{1}{c|}{68}&72&\multicolumn{1}{c|}{68}&73&\multicolumn{1}{c|}{69}&73&\multicolumn{1}{c|}{69}&72&\multicolumn{1}{c|}{69}\\
16&81&\multicolumn{1}{c|}{76}&81&\multicolumn{1}{c|}{77}&81&\multicolumn{1}{c|}{77}&81&\multicolumn{1}{c|}{77}&81&\multicolumn{1}{c|}{77}&81&\multicolumn{1}{c|}{77}\\
17&90&\multicolumn{1}{c|}{85}&90&\multicolumn{1}{c|}{85}&90&\multicolumn{1}{c|}{85}&90&\multicolumn{1}{c|}{85}&90&\multicolumn{1}{c|}{85}&90&\multicolumn{1}{c|}{85}\\
18&100&\multicolumn{1}{c|}{94}&100&\multicolumn{1}{c|}{94}&100&\multicolumn{1}{c|}{94}&100&\multicolumn{1}{c|}{94}&100&\multicolumn{1}{c|}{94}&100&\multicolumn{1}{c|}{94}\\
19&110&\multicolumn{1}{c|}{103}&110&\multicolumn{1}{c|}{104}&110&\multicolumn{1}{c|}{103}&110&\multicolumn{1}{c|}{103}&110&\multicolumn{1}{c|}{103}&110&\multicolumn{1}{c|}{103}\\
20&120&\multicolumn{1}{c|}{113}&120&\multicolumn{1}{c|}{113}&120&\multicolumn{1}{c|}{113}&120&\multicolumn{1}{c|}{113}&120&\multicolumn{1}{c|}{113}&120&\multicolumn{1}{c|}{113}\\
21&131&\multicolumn{1}{c|}{124}&131&\multicolumn{1}{c|}{123}&131&\multicolumn{1}{c|}{123}&131&\multicolumn{1}{c|}{123}&131&\multicolumn{1}{c|}{123}&131&\multicolumn{1}{c|}{123}\\
22&142&\multicolumn{1}{c|}{134}&142&\multicolumn{1}{c|}{134}&142&\multicolumn{1}{c|}{134}&142&\multicolumn{1}{c|}{134}&142&\multicolumn{1}{c|}{134}&142&\multicolumn{1}{c|}{134}\\
23&154&\multicolumn{1}{c|}{145}&154&\multicolumn{1}{c|}{145}&154&\multicolumn{1}{c|}{144}&154&\multicolumn{1}{c|}{145}&154&\multicolumn{1}{c|}{145}&154&\multicolumn{1}{c|}{145}\\
24&166&\multicolumn{1}{c|}{156}&166&\multicolumn{1}{c|}{156}&166&\multicolumn{1}{c|}{156}&166&\multicolumn{1}{c|}{156}&166&\multicolumn{1}{c|}{156}&166&\multicolumn{1}{c|}{156}\\
25&179&\multicolumn{1}{c|}{168}&179&\multicolumn{1}{c|}{168}&178&\multicolumn{1}{c|}{167}&178&\multicolumn{1}{c|}{168}&179&\multicolumn{1}{c|}{168}&178&\multicolumn{1}{c|}{167}\\
\hline
\end{tabular}
\caption{\justifying Tabulation of higher-order indices $\beta'$ corresponding to both spacings and spacing ratios for various $k$ of the COE ($m=2$ case). Here, $N=5000, 15000, 25000, 35000, 45000$, and $95000$, having $n=600, 600, 300, 300, 300,$ and $50$, respectively. In the table, $P$ denotes the Poisson distribution. For spacings, it is $\exp(-s)$ and ratios, it is $1/(1+r)^2$.}
\label{Table: COE_m2_DimAnalysistable}
\end{center}
\end{table*}
\begin{table*}
\renewcommand{\arraystretch}{2} 
\setlength{\tabcolsep}{4.2pt}  
\centering
    \begin{tabular}{|c|c|*{2}{cc}|*{2}{cc}|*{2}{cc}|*{2}{cc}|} 
    
        \hline
       
        Order &  \makecell{According to \\ the scaling \\relation} & \multicolumn{4}{c|}{$N=5000$} & \multicolumn{4}{c|}{$N=15000$} & \multicolumn{4}{c|}{$N=45000$}& \multicolumn{4}{c|}{$N=95000$} \\ 
       \cline{3-18}
        &Eq.~(\ref{Eq:ScalingRelation1}) & \multicolumn{2}{c|}{HOS} & \multicolumn{2}{c|}{HOSR} & \multicolumn{2}{c|}{HOS} & \multicolumn{2}{c|}{HOSR} & \multicolumn{2}{c|}{HOS} & \multicolumn{2}{c|}{HOSR}& \multicolumn{2}{c|}{HOS} & \multicolumn{2}{c|}{HOSR} \\ 
        \cline{3-18}
        & & COE & \multicolumn{1}{c|}{GOE} & COE & \multicolumn{1}{c|}{GOE} & COE & \multicolumn{1}{c|}{GOE} & COE & \multicolumn{1}{c|}{GOE}& COE & \multicolumn{1}{c|}{GOE} & COE & \multicolumn{1}{c|}{GOE} & COE & \multicolumn{1}{c|}{GOE}& COE & \multicolumn{1}{c|}{GOE}\\ 
      
        $k$& $\beta'$ & $\beta'$ & \multicolumn{1}{c|}{$\beta'$} & $\beta'$ & \multicolumn{1}{c|}{$\beta'$} & $\beta'$ & \multicolumn{1}{c|}{$\beta'$} & $\beta'$ & \multicolumn{1}{c|}{$\beta'$} & $\beta'$ & \multicolumn{1}{c|}{$\beta'$} & $\beta'$ & \multicolumn{1}{c|}{$\beta'$}& $\beta'$ & \multicolumn{1}{c|}{$\beta'$} & $\beta'$ & \multicolumn{1}{c|}{$\beta'$} \\ 
        \hline
        1&1&1&\multicolumn{1}{c|}{1}&1&\multicolumn{1}{c|}{1}&1&\multicolumn{1}{c|}{1}&1&\multicolumn{1}{c|}{1}&1&\multicolumn{1}{c|}{1}&1&\multicolumn{1}{c|}{1}&1&\multicolumn{1}{c|}{1}&1&\multicolumn{1}{c|}{1}\\
        2&4&4&\multicolumn{1}{c|}{4}&4&\multicolumn{1}{c|}{4}&4&\multicolumn{1}{c|}{4}&4&\multicolumn{1}{c|}{4}&4&\multicolumn{1}{c|}{4}&4&\multicolumn{1}{c|}{4}&4&\multicolumn{1}{c|}{4}&4&\multicolumn{1}{c|}{4}\\
        3&8&8&\multicolumn{1}{c|}{8}&8&\multicolumn{1}{c|}{8}&8&\multicolumn{1}{c|}{8}&8&\multicolumn{1}{c|}{8}&8&\multicolumn{1}{c|}{8}&8&\multicolumn{1}{c|}{8}&8&\multicolumn{1}{c|}{8}&8&\multicolumn{1}{c|}{8}\\
        4&13&14&\multicolumn{1}{c|}{14}&13&\multicolumn{1}{c|}{13}&14&\multicolumn{1}{c|}{14}&13&\multicolumn{1}{c|}{13}&14&\multicolumn{1}{c|}{14}&13&\multicolumn{1}{c|}{13}&14&\multicolumn{1}{c|}{14}&13&\multicolumn{1}{c|}{13}\\
        5&19&20&\multicolumn{1}{c|}{20}&19&\multicolumn{1}{c|}{19}&20&\multicolumn{1}{c|}{20}&19&\multicolumn{1}{c|}{19}&20&\multicolumn{1}{c|}{20}&19&\multicolumn{1}{c|}{19}&20&\multicolumn{1}{c|}{20}&19&\multicolumn{1}{c|}{19}\\
        6&26&27&\multicolumn{1}{c|}{27}&26&\multicolumn{1}{c|}{26}&27&\multicolumn{1}{c|}{27}&26&\multicolumn{1}{c|}{26}&27&\multicolumn{1}{c|}{27}&26&\multicolumn{1}{c|}{26}&27&\multicolumn{1}{c|}{27}&26&\multicolumn{1}{c|}{26}\\
        7&34&36&\multicolumn{1}{c|}{36}&34&\multicolumn{1}{c|}{34}&36&\multicolumn{1}{c|}{36}&34&\multicolumn{1}{c|}{34}&36&\multicolumn{1}{c|}{36}&34&\multicolumn{1}{c|}{34}&36&\multicolumn{1}{c|}{36}&34&\multicolumn{1}{c|}{34}\\
        8&43&45&\multicolumn{1}{c|}{45}&43&\multicolumn{1}{c|}{42}&45&\multicolumn{1}{c|}{45}&43&\multicolumn{1}{c|}{43}&46&\multicolumn{1}{c|}{45}&43&\multicolumn{1}{c|}{43}&45&\multicolumn{1}{c|}{45}&43&\multicolumn{1}{c|}{43}\\
        9&53&55&\multicolumn{1}{c|}{55}&53&\multicolumn{1}{c|}{52}&55&\multicolumn{1}{c|}{55}&53&\multicolumn{1}{c|}{52}&55&\multicolumn{1}{c|}{55}&52&\multicolumn{1}{c|}{52}&55&\multicolumn{1}{c|}{55}&53&\multicolumn{1}{c|}{52}\\
        10&64&67&\multicolumn{1}{c|}{67}&63&\multicolumn{1}{c|}{62}&67&\multicolumn{1}{c|}{67}&63&\multicolumn{1}{c|}{63}&67&\multicolumn{1}{c|}{67}&63&\multicolumn{1}{c|}{63}&67&\multicolumn{1}{c|}{67}&63&\multicolumn{1}{c|}{63}\\
        11&76&79&\multicolumn{1}{c|}{79}&74&\multicolumn{1}{c|}{73}&79&\multicolumn{1}{c|}{79}&74&\multicolumn{1}{c|}{74}&79&\multicolumn{1}{c|}{79}&74&\multicolumn{1}{c|}{74}&79&\multicolumn{1}{c|}{79}&74&\multicolumn{1}{c|}{74}\\
        12&89&92&\multicolumn{1}{c|}{92}&87&\multicolumn{1}{c|}{84}&92&\multicolumn{1}{c|}{92}&87&\multicolumn{1}{c|}{85}&92&\multicolumn{1}{c|}{92}&86&\multicolumn{1}{c|}{86}&92&\multicolumn{1}{c|}{92}&87&\multicolumn{1}{c|}{86}\\
        13&103&105&\multicolumn{1}{c|}{106}&100&\multicolumn{1}{c|}{96}&105&\multicolumn{1}{c|}{105}&99&\multicolumn{1}{c|}{98}&105&\multicolumn{1}{c|}{105}&99&\multicolumn{1}{c|}{99}&105&\multicolumn{1}{c|}{105}&99&\multicolumn{1}{c|}{99}\\
        14&118&120&\multicolumn{1}{c|}{120}&113&\multicolumn{1}{c|}{109}&120&\multicolumn{1}{c|}{120}&113&\multicolumn{1}{c|}{111}&120&\multicolumn{1}{c|}{120}&113&\multicolumn{1}{c|}{112}&120&\multicolumn{1}{c|}{120}&113&\multicolumn{1}{c|}{113}\\
        15&134&136&\multicolumn{1}{c|}{136}&128&\multicolumn{1}{c|}{122}&136&\multicolumn{1}{c|}{136}&128&\multicolumn{1}{c|}{125}&136&\multicolumn{1}{c|}{136}&128&\multicolumn{1}{c|}{127}&136&\multicolumn{1}{c|}{136}&128&\multicolumn{1}{c|}{127}\\
        16&151&152&\multicolumn{1}{c|}{152}&143&\multicolumn{1}{c|}{136}&152&\multicolumn{1}{c|}{152}&143&\multicolumn{1}{c|}{140}&152&\multicolumn{1}{c|}{152}&143&\multicolumn{1}{c|}{142}&152&\multicolumn{1}{c|}{152}&143&\multicolumn{1}{c|}{142}\\
        17&169&169&\multicolumn{1}{c|}{169}&159&\multicolumn{1}{c|}{150}&169&\multicolumn{1}{c|}{169}&159&\multicolumn{1}{c|}{156}&169&\multicolumn{1}{c|}{169}&159&\multicolumn{1}{c|}{157}&169&\multicolumn{1}{c|}{169}&159&\multicolumn{1}{c|}{158}\\
        18&188&187&\multicolumn{1}{c|}{187}&176&\multicolumn{1}{c|}{165}&187&\multicolumn{1}{c|}{187}&176&\multicolumn{1}{c|}{172}&187&\multicolumn{1}{c|}{187}&176&\multicolumn{1}{c|}{174}&187&\multicolumn{1}{c|}{187}&176&\multicolumn{1}{c|}{175}\\
        19&208&206&\multicolumn{1}{c|}{206}&193&\multicolumn{1}{c|}{180}&206&\multicolumn{1}{c|}{206}&193&\multicolumn{1}{c|}{188}&206&\multicolumn{1}{c|}{206}&194&\multicolumn{1}{c|}{191}&206&\multicolumn{1}{c|}{206}&194&\multicolumn{1}{c|}{192}\\
        20&229&226&\multicolumn{1}{c|}{226}&212&\multicolumn{1}{c|}{196}&226&\multicolumn{1}{c|}{226}&212&\multicolumn{1}{c|}{206}&226&\multicolumn{1}{c|}{226}&212&\multicolumn{1}{c|}{209}&226&\multicolumn{1}{c|}{225}&212&\multicolumn{1}{c|}{210}\\    
        \hline
    \end{tabular}
    \caption{\justifying Tabulation of higher-order indices $\beta'$ corresponding to both spacings and spacing ratios for various $k$ of the COE and GOE ($m=1$ case). They have $N=5000, 15000, 45000$, and $95000$, and $n=1000, 700, 300$, and ($50$ and $57$ for COE and GOE) respectively.}
\label{Table: COEAndGOEDimAnalysistable}    
\end{table*}
\begin{table}[t]     
\renewcommand{\arraystretch}{1.5} 
\setlength{\tabcolsep}{6.3pt}  
\begin{center}
\begin{tabular}{|c|cc|cc|cc|}
\hline
\rule{0pt}{12pt}  
Order&\multicolumn{2}{c|}{$N=5000$}&\multicolumn{2}{c|}{$N=45000$}&\multicolumn{2}{c|}{$N=95000$}\\ [1.5ex]  
\cline{2-7}  
\rule{0pt}{10pt}  
&HOS&\multicolumn{1}{c|}{HOSR}&HOS&\multicolumn{1}{c|}{HOSR}&HOS&\multicolumn{1}{c|}{HOSR}\\
$k$&$\beta^\prime$&\multicolumn{1}{c|}{$\beta^\prime$}&$\beta^\prime$&\multicolumn{1}{c|}{$\beta^\prime$}&$\beta^\prime$&\multicolumn{1}{c|}{$\beta^\prime$}\\
\hline
1&$P$&\multicolumn{1}{c|}{$P$}&$P$&\multicolumn{1}{c|}{$P$}&$P$&\multicolumn{1}{c|}{$P$}\\
2&2&\multicolumn{1}{c|}{2}&2&\multicolumn{1}{c|}{2}&2&\multicolumn{1}{c|}{2}\\
3&4&\multicolumn{1}{c|}{4}&4&\multicolumn{1}{c|}{4}&4&\multicolumn{1}{c|}{4}\\
4&7&\multicolumn{1}{c|}{7}&7&\multicolumn{1}{c|}{7}&7&\multicolumn{1}{c|}{7}\\
5&11&\multicolumn{1}{c|}{10}&11&\multicolumn{1}{c|}{10}&11&\multicolumn{1}{c|}{10}\\
6&15&\multicolumn{1}{c|}{14}&15&\multicolumn{1}{c|}{14}&15&\multicolumn{1}{c|}{14}\\
7&19&\multicolumn{1}{c|}{18}&19&\multicolumn{1}{c|}{18}&19&\multicolumn{1}{c|}{18}\\
8&24&\multicolumn{1}{c|}{23}&24&\multicolumn{1}{c|}{23}&24&\multicolumn{1}{c|}{23}\\
9&30&\multicolumn{1}{c|}{28}&30&\multicolumn{1}{c|}{28}&30&\multicolumn{1}{c|}{28}\\
10&36&\multicolumn{1}{c|}{34}&36&\multicolumn{1}{c|}{34}&36&\multicolumn{1}{c|}{34}\\
11&42&\multicolumn{1}{c|}{40}&42&\multicolumn{1}{c|}{40}&42&\multicolumn{1}{c|}{40}\\
12&49&\multicolumn{1}{c|}{46}&49&\multicolumn{1}{c|}{46}&49&\multicolumn{1}{c|}{46}\\
13&57&\multicolumn{1}{c|}{53}&56&\multicolumn{1}{c|}{53}&56&\multicolumn{1}{c|}{53}\\
14&64&\multicolumn{1}{c|}{60}&64&\multicolumn{1}{c|}{61}&64&\multicolumn{1}{c|}{61}\\
15&73&\multicolumn{1}{c|}{67}&73&\multicolumn{1}{c|}{68}&73&\multicolumn{1}{c|}{68}\\
16&81&\multicolumn{1}{c|}{75}&81&\multicolumn{1}{c|}{76}&81&\multicolumn{1}{c|}{77}\\
17&91&\multicolumn{1}{c|}{83}&90&\multicolumn{1}{c|}{85}&90&\multicolumn{1}{c|}{85}\\
18&100&\multicolumn{1}{c|}{92}&100&\multicolumn{1}{c|}{94}&100&\multicolumn{1}{c|}{94}\\
19&110&\multicolumn{1}{c|}{101}&110&\multicolumn{1}{c|}{103}&110&\multicolumn{1}{c|}{103}\\
20&120&\multicolumn{1}{c|}{110}&120&\multicolumn{1}{c|}{113}&120&\multicolumn{1}{c|}{113}\\
\hline
\end{tabular}
\caption{\justifying Tabulation of higher-order indices $\beta'$ for both the distributions of spacing and spacing ratio of order $k$ for $m=2$ case of GOE. Here, $n=600, 300,$ and $50$ for $N=5000, 45000,$ and $95000$, respectively. In the table, $P$ denotes the Poisson distribution. For spacings, it is $\exp{(-s)}$, and for ratios, it is $1/(1+r)^2$.}
\label{Table: GOE_m2_detailTable}
\end{center}
\end{table}
\begin{figure}[tbp]
\begin{center}
\includegraphics*[scale=0.35]{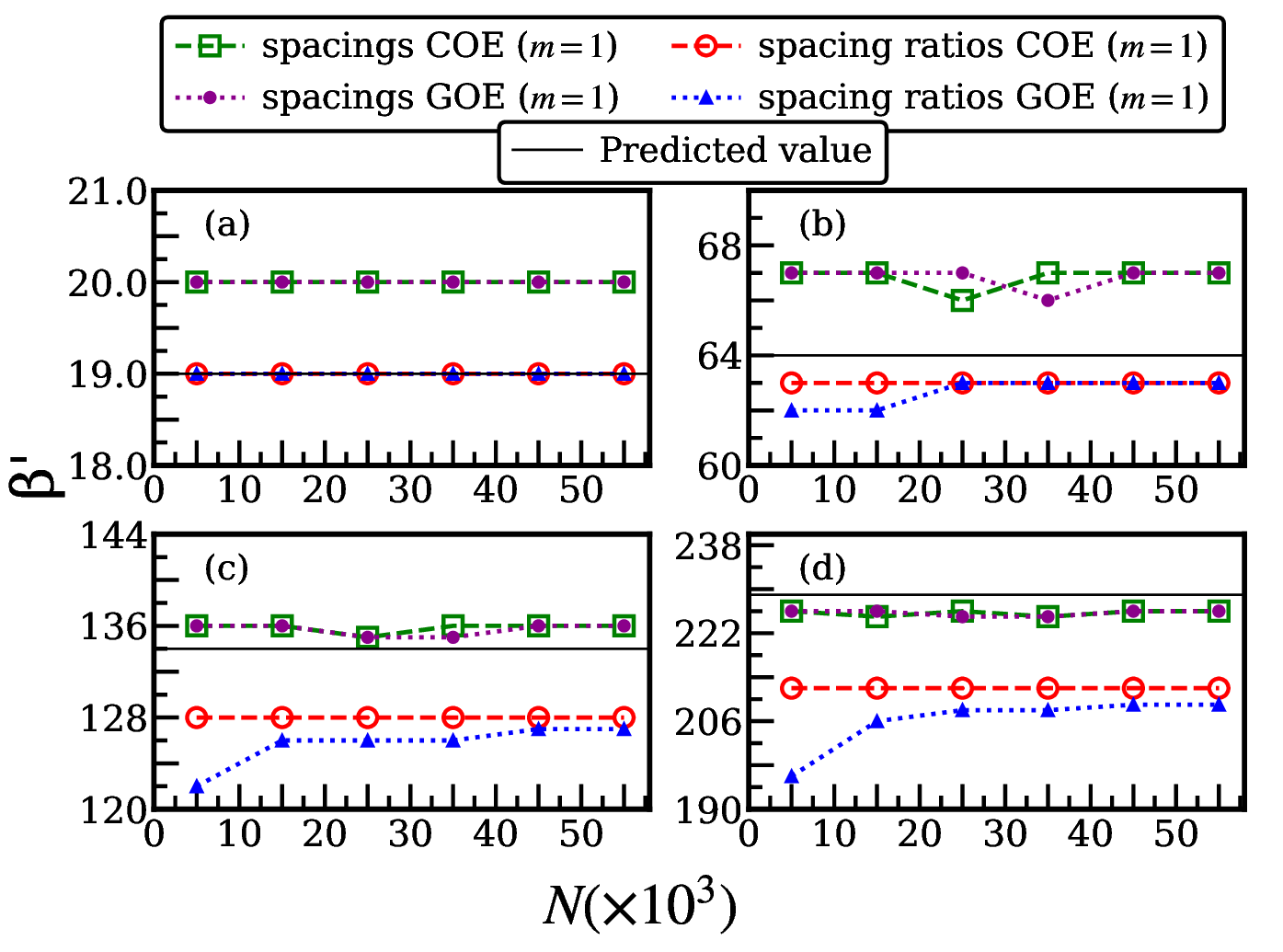}
\caption{\justifying Same as Fig.~\ref{fig: DimAnalysis_COE_GOE_m2} but for $m=1$ and as per the dimensional analysis tables for $m=1$ case of both COE and GOE from supplementary material \cite{supplementry2025}. Here, the solid line corresponds to the value of $\beta'$ according to the scaling relation (\ref{Eq:ScalingRelation1}).}
\label{fig: DimAnalysis_COE_GOE}
\end{center}
\end{figure}
\subsection{Effect of the number of realizations}
\label{subsec:no of realizations}
\begin{figure}[tbp]
\begin{center}
\includegraphics*[scale=0.35]{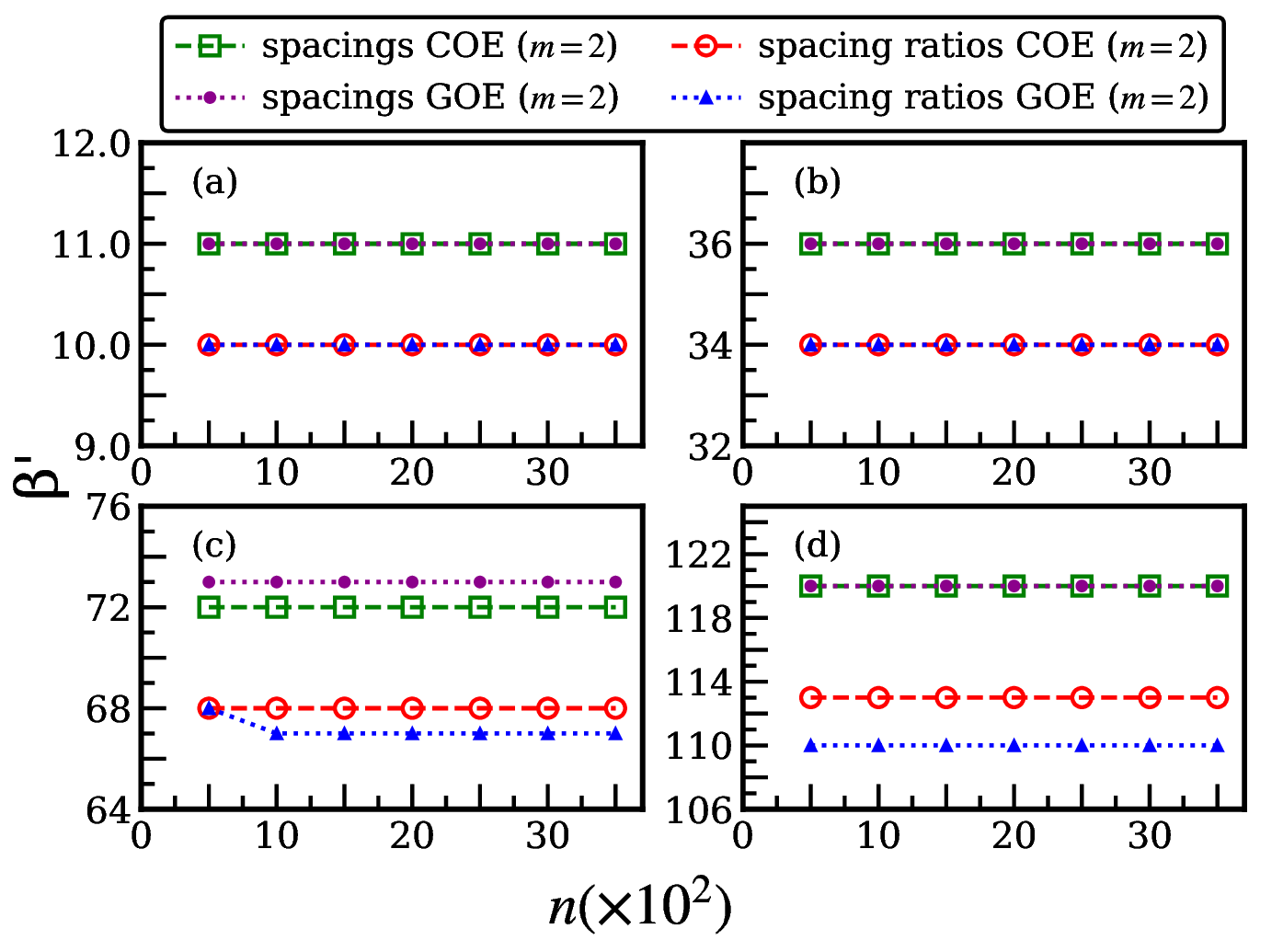}
\caption{\justifying Variation of $\beta'$ as a function of $n$ as given in the tables of effect of the number of realizations section of the supplementary material \cite{supplementry2025} for the $m=2$ case for both COE and GOE, keeping $N$ for both constants to $5000$. Here, the subplots (a), (b), (c), and (d) correspond to $k=5, 10, 15$, and $20$, respectively.}
\label{fig: NumberEffect_COE_GOE_m2}
\end{center}
\end{figure}

\begin{figure}[tbp]
\begin{center}
\includegraphics*[scale=0.35]{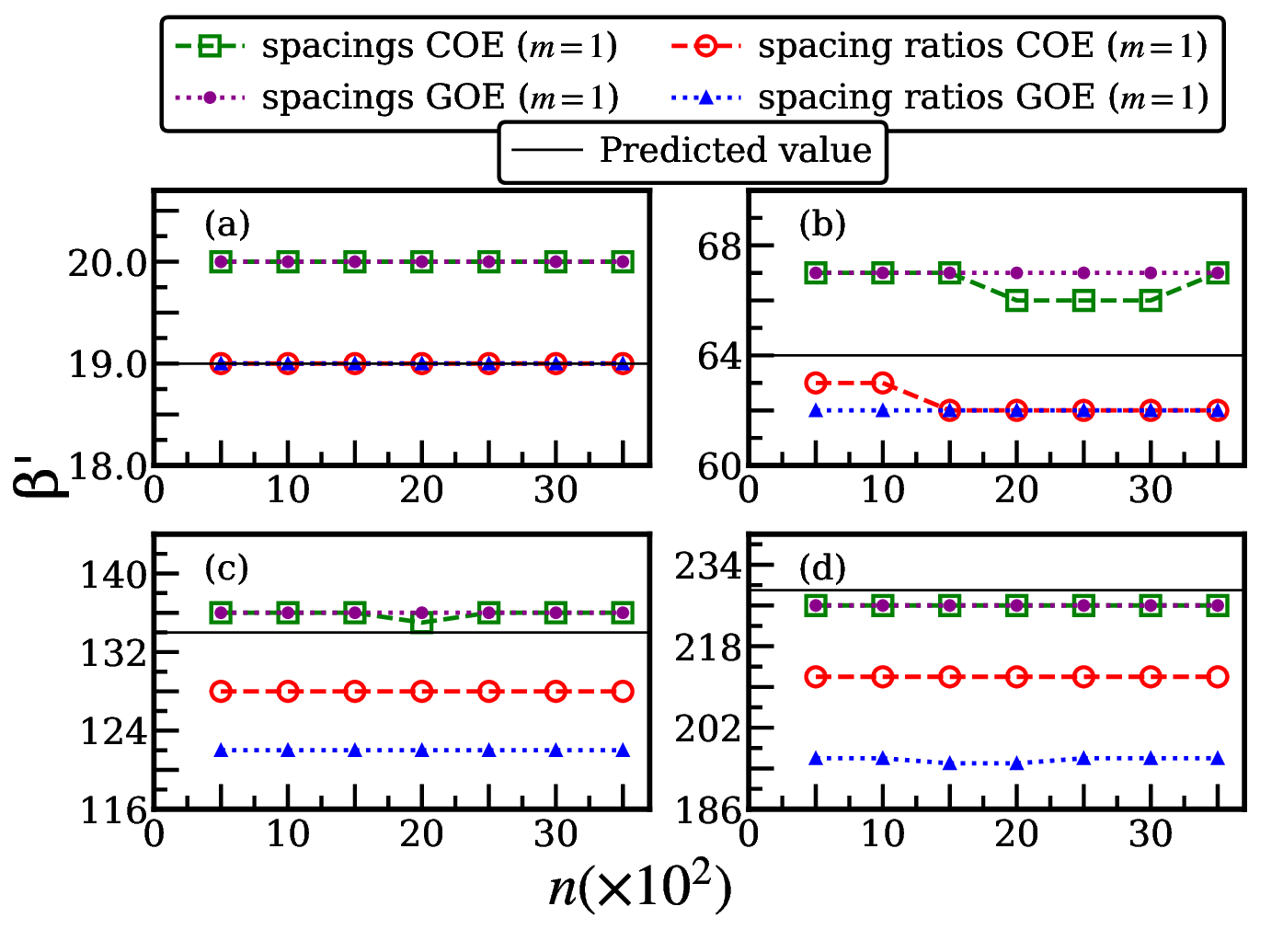}
\caption{\justifying Same as Fig.~\ref{fig: NumberEffect_COE_GOE_m2} but for $m=1$ cases of both COE and GOE. Here, the solid line corresponds to the value of $\beta'$ according to the scaling relation (\ref{Eq:ScalingRelation1}).}
\label{fig: NumberEffect_COE_GOE}
\end{center}
\end{figure}

In this subsection, our motivation is to study the effect of the number of realizations ($n$) on the obtained values of $\beta'$ for both spacing and spacing ratio distributions of the COE and GOE, for a given $N$. Here, we consider
the representative cases of $k=5,10,15$, and $20$. We have studied this by varying $n$ from $500$ to $3500$ for each $k$ in $m=1$ and $m=2$ cases of both COE and GOE and have shown the results in Figs. \ref{fig: NumberEffect_COE_GOE_m2} and \ref{fig: NumberEffect_COE_GOE}. Here, $N=5000$ for each $n$ and $k$. By analyzing these results, we can conclude that in all the cases, after a particular value of $n$ for a given $N$, the $\beta'$ saturates. (To obtain the values of $\beta'$ conveniently, one can refer to  the tables of effect of the number of realizations section in the supplementary material \cite{supplementry2025}.)

Comparing Figs.~\ref{fig: DimAnalysis_COE_GOE_m2}, \ref{fig: DimAnalysis_COE_GOE}, \ref{fig: NumberEffect_COE_GOE_m2}, and \ref{fig: NumberEffect_COE_GOE}, for larger values of $k$, in the case of
the spacing ratio distributions of GOE, we find that even if the number of eigenvalues are nearly the same in the case of both studying dimensional analysis and the effect of the number of realizations, the results are close but not exactly the same in both cases. For example, consider the case of GOE ($m=1$) and $k=20$. For this, let's take two cases: one in which $N=55000$, $n=300$ (Fig.~\ref{fig: DimAnalysis_COE_GOE}), and the other in which $N=5000$, $n=3500$ (Fig.~\ref{fig: NumberEffect_COE_GOE}). Even though in the later case, eigenvalues are more, the $\beta'=209$ of the first case is closer to the predicted value $229$, as per Eq.~(\ref{Eq:ScalingRelation1}), than the later case where $\beta'=196$. Thus, large $N$ and small $n$ is the preferable case over large $n$ and small $N$ for a particular number of eigenvalues. In our work, we have considered both $n$ and $N$ sufficiently large.

\section{Application to physical systems}
\label{sec:Testingsystems}
In this section, we have studied higher-order spectral statistics of the spectra for two physical systems
and verify our results from the previous sections. One of them is the intermediate map, and other one is the quantum kicked top (QKT). We will now present our study on these systems in the subsequent subsections.
\subsection{Intermediate map}
\label{subsec:Intmap}
In this subsection, we have studied HOS and HOSR distributions in the arguments of the eigenvalues of the unitary operator corresponding to the intermediate map \cite{dubertrand2015multifractality}.
The matrix form of the unitary operator corresponding to the quantum version of this map can be written as  follows:
\begin{equation}
U_{ab}=\dfrac{\exp{(-i\phi_{a})}}{N}\dfrac{1-\exp{[i2\pi\gamma N]}}{1-\exp{[i2\pi(a-b+\gamma N)/N]}}\;\ ,
\label{Eq:IM}
\end{equation}
where $N$ is the dimension of the Hilbert space, $\phi_{a}$ is the uniformly distributed random variable between $[0,2\pi]$, and $\gamma$ is any irrational number. The spectral statistics of this map are found to be of the CUE type. This map has been used to study HOSR Ref.~\cite{tekur2018higher}. There, the authors have presented the results up to $k=4$. These results are reproduced here for completeness. Here, in our work, we have studied extensively for various $N$, each for the same $n$, and larger $k$. The objective here is to study how the statistics of this map get affected by the dimension. The eigenvalues of this unitary matrix for $N=6000$, $12000$, $18000$, $24000$, $30000$, and $36000$ are generated by taking $\gamma=\sqrt{3}$. Here, for each $N$, $n=80$. We have studied HOS and HOSR distributions up to $k=17$ for each $N$. The obtained values of $\beta'$ are tabulated in Table ~\ref{Table:IMtable}, and the representative figures for spacing and spacing ratio distributions are illustrated in Figs. \ref{fig: k_1_to_6_spacings_IM} and \ref{fig: k_1_to_6_sratios_IM}, respectively. See supplementary material \cite{supplementry2025} for more such figures. From the analysis of the obtained results, we observe that: for all $N$, the HOS distributions follow the scaling relation as per Eq.~(\ref{Eq:ScalingRelation2}) up to $k=2$ and for higher $k$, the difference between the values of $\beta'$ increases with $k$. For HOSR distribution, the scaling relation is followed up to slightly higher $k$ compared to spacings.
The maximum value of $k$ for which there is such agreement is different for different $N$.
Here, for a given $k$ and increasing $N$, we don't observe any monotonic pattern in the obtained value of $\beta'$. Such as increasing or decreasing from the results of $N=6000$, or tending towards the predicted value according to the scaling relation. Rather, they seem to be fluctuating. This observation is made based on the parameters taken by us.
\begin{table*}     
\renewcommand{\arraystretch}{1.5} 
\setlength{\tabcolsep}{6.5pt}  
\begin{center}
\begin{tabular}{|c|c|cc|cc|cc|cc|cc|cc|}
\hline
\rule{0pt}{12pt}  
Order&CUE&\multicolumn{2}{c|}{$N=6000$}&\multicolumn{2}{c|}{$N=12000$}&\multicolumn{2}{c|}{$N=18000$}&\multicolumn{2}{c|}{$N=24000$}&\multicolumn{2}{c|}{$N=30000$}&\multicolumn{2}{c|}{$N=36000$}\\ [1.5ex]  
\cline{3-14}  
\rule{0pt}{10pt}  
$k$&Eq.~(\ref{Eq:ScalingRelation2})&HOS&\multicolumn{1}{c|}{HOSR}&HOS&\multicolumn{1}{c|}{HOSR}&HOS&\multicolumn{1}{c|}{HOSR}&HOS&\multicolumn{1}{c|}{HOSR}&HOS&\multicolumn{1}{c|}{HOSR}&HOS&\multicolumn{1}{c|}{HOSR}\\
&$\beta^\prime$&$\beta^\prime$&\multicolumn{1}{c|}{$\beta^\prime$}&$\beta^\prime$&\multicolumn{1}{c|}{$\beta^\prime$}&$\beta^\prime$&\multicolumn{1}{c|}{$\beta^\prime$}&$\beta^\prime$&\multicolumn{1}{c|}{$\beta^\prime$}&$\beta^\prime$&\multicolumn{1}{c|}{$\beta^\prime$}&$\beta^\prime$&\multicolumn{1}{c|}{$\beta^\prime$}\\
\hline
1&2&2&\multicolumn{1}{c|}{2}&2&\multicolumn{1}{c|}{2}&2&\multicolumn{1}{c|}{2}&2&\multicolumn{1}{c|}{2}&2&\multicolumn{1}{c|}{2}&2&\multicolumn{1}{c|}{2}\\
2&7&7&\multicolumn{1}{c|}{7}&7&\multicolumn{1}{c|}{7}&7&\multicolumn{1}{c|}{7}&7&\multicolumn{1}{c|}{7}&7&\multicolumn{1}{c|}{7}&7&\multicolumn{1}{c|}{7}\\
3&14&15&\multicolumn{1}{c|}{14}&15&\multicolumn{1}{c|}{14}&15&\multicolumn{1}{c|}{14}&15&\multicolumn{1}{c|}{14}&15&\multicolumn{1}{c|}{14}&15&\multicolumn{1}{c|}{14}\\
4&23&24&\multicolumn{1}{c|}{23}&24&\multicolumn{1}{c|}{23}&26&\multicolumn{1}{c|}{24}&24&\multicolumn{1}{c|}{23}&24&\multicolumn{1}{c|}{23}&24&\multicolumn{1}{c|}{23}\\
5&34&36&\multicolumn{1}{c|}{34}&36&\multicolumn{1}{c|}{34}&39&\multicolumn{1}{c|}{37}&36&\multicolumn{1}{c|}{34}&36&\multicolumn{1}{c|}{34}&36&\multicolumn{1}{c|}{34}\\
6&47&50&\multicolumn{1}{c|}{47}&50&\multicolumn{1}{c|}{47}&55&\multicolumn{1}{c|}{52}&49&\multicolumn{1}{c|}{46}&49&\multicolumn{1}{c|}{47}&50&\multicolumn{1}{c|}{47}\\
7&62&66&\multicolumn{1}{c|}{62}&65&\multicolumn{1}{c|}{61}&73&\multicolumn{1}{c|}{70}&64&\multicolumn{1}{c|}{60}&65&\multicolumn{1}{c|}{61}&65&\multicolumn{1}{c|}{61}\\
8&79&84&\multicolumn{1}{c|}{78}&82&\multicolumn{1}{c|}{77}&94&\multicolumn{1}{c|}{91}&81&\multicolumn{1}{c|}{76}&82&\multicolumn{1}{c|}{77}&83&\multicolumn{1}{c|}{78}\\
9&98&105&\multicolumn{1}{c|}{97}&101&\multicolumn{1}{c|}{94}&115&\multicolumn{1}{c|}{111}&100&\multicolumn{1}{c|}{93}&101&\multicolumn{1}{c|}{95}&102&\multicolumn{1}{c|}{96}\\
10&119&127&\multicolumn{1}{c|}{117}&122&\multicolumn{1}{c|}{114}&137&\multicolumn{1}{c|}{133}&120&\multicolumn{1}{c|}{112}&121&\multicolumn{1}{c|}{114}&124&\multicolumn{1}{c|}{116}\\
11&142&151&\multicolumn{1}{c|}{139}&145&\multicolumn{1}{c|}{134}&162&\multicolumn{1}{c|}{157}&143&\multicolumn{1}{c|}{133}&143&\multicolumn{1}{c|}{134}&147&\multicolumn{1}{c|}{137}\\
12&167&179&\multicolumn{1}{c|}{162}&169&\multicolumn{1}{c|}{156}&186&\multicolumn{1}{c|}{182}&167&\multicolumn{1}{c|}{155}&167&\multicolumn{1}{c|}{156}&171&\multicolumn{1}{c|}{161}\\
13&194&206&\multicolumn{1}{c|}{185}&196&\multicolumn{1}{c|}{180}&213&\multicolumn{1}{c|}{208}&193&\multicolumn{1}{c|}{178}&192&\multicolumn{1}{c|}{179}&198&\multicolumn{1}{c|}{186}\\
14&223&236&\multicolumn{1}{c|}{211}&224&\multicolumn{1}{c|}{205}&240&\multicolumn{1}{c|}{233}&220&\multicolumn{1}{c|}{203}&220&\multicolumn{1}{c|}{205}&227&\multicolumn{1}{c|}{213}\\
15&254&269&\multicolumn{1}{c|}{236}&255&\multicolumn{1}{c|}{232}&269&\multicolumn{1}{c|}{261}&250&\multicolumn{1}{c|}{230}&249&\multicolumn{1}{c|}{232}&256&\multicolumn{1}{c|}{242}\\
16&287&302&\multicolumn{1}{c|}{264}&286&\multicolumn{1}{c|}{260}&297&\multicolumn{1}{c|}{288}&281&\multicolumn{1}{c|}{257}&280&\multicolumn{1}{c|}{260}&289&\multicolumn{1}{c|}{273}\\
17&322&338&\multicolumn{1}{c|}{294}&320&\multicolumn{1}{c|}{290}&328&\multicolumn{1}{c|}{317}&313&\multicolumn{1}{c|}{285}&312&\multicolumn{1}{c|}{290}&322&\multicolumn{1}{c|}{306}\\
\hline
\end{tabular}
\caption{\justifying Tabulation of higher-order indices $\beta'$ as per the scaling relation Eq.~(\ref{Eq:ScalingRelation2}) and for the intermediate map for various values of $k$ and $N$. Here, $n=80$ for all the values of $N$.}
\label{Table:IMtable}
\end{center}
\end{table*}
\begin{figure}[tbp]
\begin{center}
\includegraphics*[scale=0.35]{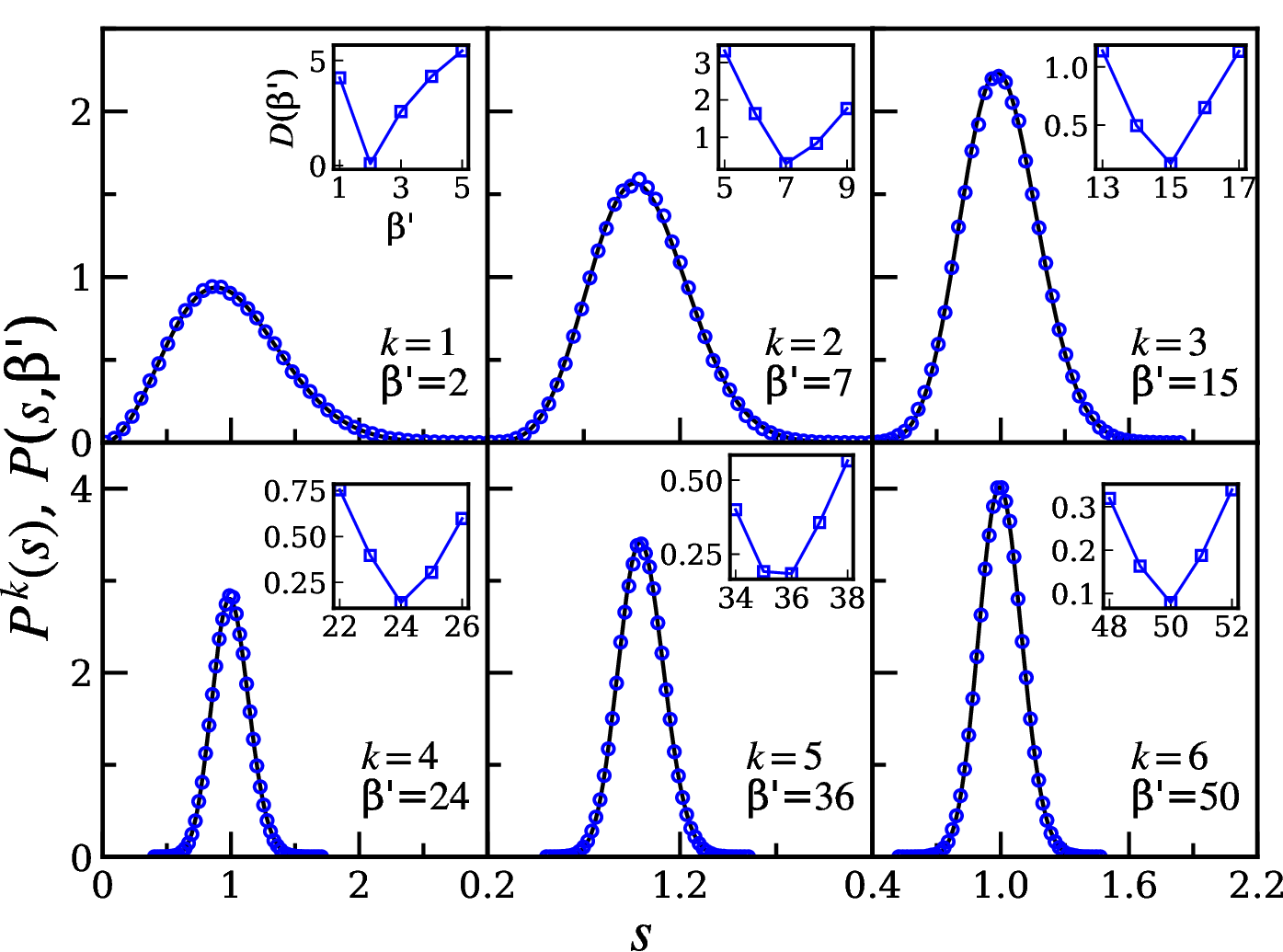}
\caption{\justifying HOS distribution $P^{k}(s)$ of eigenangles of the intermediate map (circles). Solid line corresponds to $P(s, \beta')$. The solid curve corresponds to $P(s,\beta')$ as given in Eq.~(\ref{Eq:PSBeta}), where $\beta$ is replaced by $\beta'$ and $\beta'$ is given in Table~\ref{Table:IMtable}. The insets show $D(\beta')$ as a function of $\beta'$. Here, we have taken $N=12000$ and $n=80$.}
\label{fig: k_1_to_6_spacings_IM}
\end{center}
\end{figure}
\begin{figure}[tbp]
\begin{center}
\includegraphics*[scale=0.35]{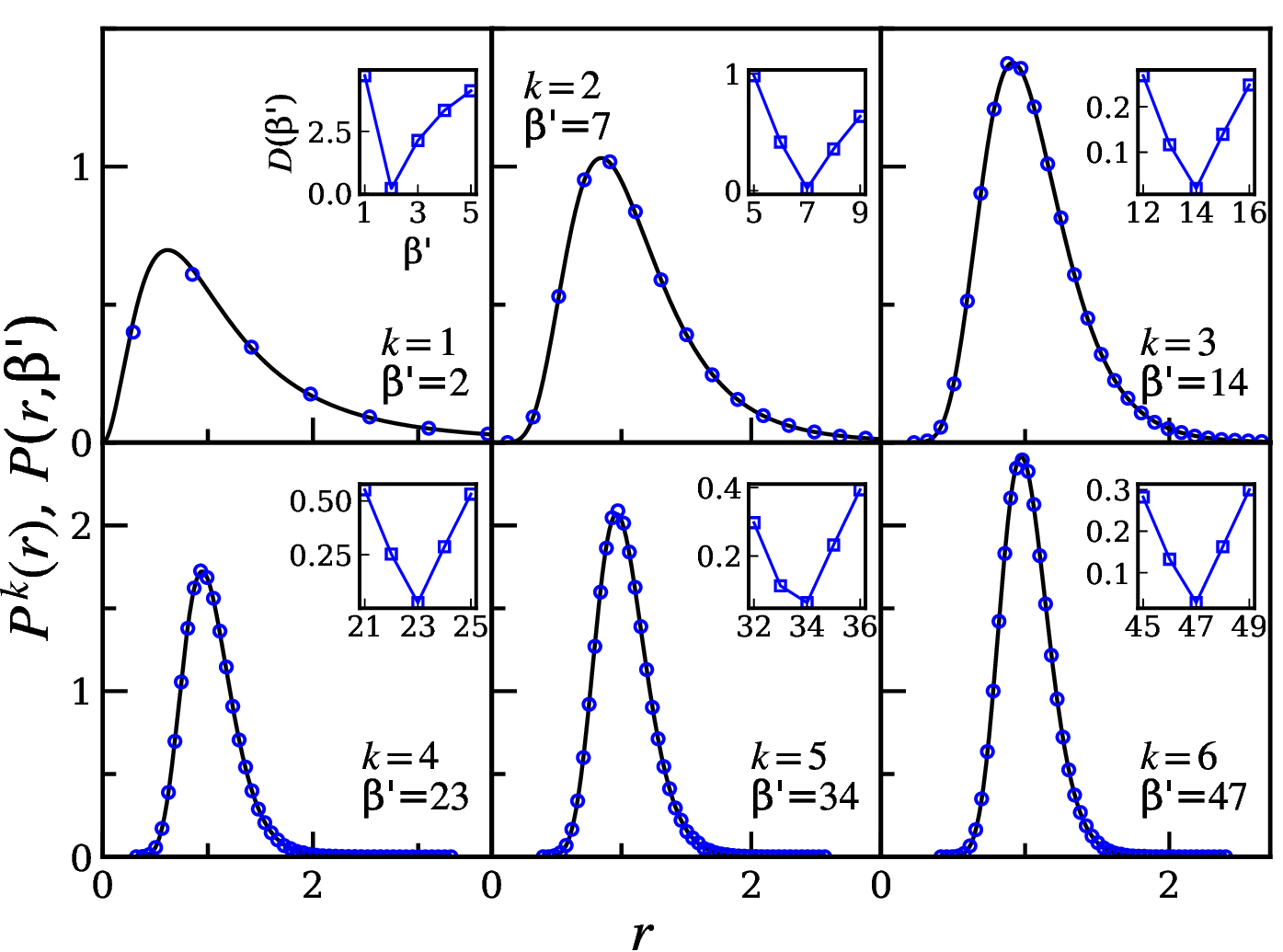}
\caption{\justifying Same as Fig.~\ref{fig: k_1_to_6_spacings_IM} but for spacing ratio.}
\label{fig: k_1_to_6_sratios_IM}
\end{center}
\end{figure}
\subsection{Quantum Kicked Top}
\label{subsec:qkt}
In this subsection, we have verified our $m=2$ case of COE results on the QKT model. For a chaotic Hamiltonian system, this is a basic and significant time-dependent model \cite{haake1987classical}. It has been studied extensively both theoretically and experimentally \cite{madhok2018quantum,dogra2019quantum,haake1987classical,meier2019exploring,munoz2020simulating,kumar2020wishart,xu2020does,bhosale2021superposition,tekur2018higher,bhosale2017signatures,bhosale2018periodicity,krithika2019nmr,ruebeck2017entanglement,neill2016ergodic,zakrzewski1991distributions,alicki1996quantum,weinstein2002border,lombardi2011entanglement,puchala2016distinguishability,chaudhury2009quantum,demkowicz2004global}. 
It has been implemented in various experimental setups, such as in a two-qubit NMR system \cite{krithika2019nmr}, three coupled superconducting qubits \cite{neill2016ergodic}, and hyperfine states of cold atoms \cite{chaudhury2009quantum}. It has been studied from the angle of RMT and quantum information. Some studies show the effect of the underlying phase space on various measures of quantum correlations \cite{miller1999signatures,neill2016ergodic,bhosale2017signatures,ruebeck2017entanglement}. 
The NN-spectral statistics of desymmetrized spectra of QKT are the same as those of the COE ensemble, provided its classical limit is fully chaotic \cite{haake1987classical}.

The QKT is described by an angular momentum vector ${\bf{J}} = (J_x, J_y, J_z)$ and its components obey the standard algebra of angular momentum. The unitary time evolution operator for QKT is represented as follows \cite{haake1987classical}:
\begin{equation}
 \widehat{U}=\exp\left(-ipJ_y \right) \exp\left(-i\dfrac{\tilde{k}}{2j}J_z^2 \right).
 \label{Eq: QKT}
\end{equation}
The first term represents free precession of the top around $y$-axis with angular frequency $p$, and the second term represents periodic $\delta$ kicks applied to the top. Here, $\tilde{k}$ is the kick strength or chaos parameter. For $\tilde{k}=0$, the top is integrable, and for $\tilde{k}>0$, as it increases, the top becomes increasingly chaotic.

For a given $j$, the dimension of the Hilbert space is equal to $2j+1$. For $p \neq \pi/2$, which is the relevant case for us, there exist a symmetry in the QKT  Ref.~\cite{haake1987classical} such that $\widehat{U}$ commutes with $\hat{R_{y}}$, which has two eigenvalues. As a result, the matrix representation of $\widehat{U}$ in the basis of $\hat{R_{y}}$ is block diagonal, having two blocks of dimensions $j$ and $j+1$. For the fully chaotic case, the spectral fluctuations of $\widehat{U}$ in each block satisfies COE statistics \cite{haake1987classical}. Hence, if we take the eigenvalues together for studying fluctuation statistics, it will be an ideal case for validating our $m=2$ case of COE results, provided the value of $j$ should be large, so that $j$ and $j+1$ become very close to each other or equivalent.

This model has been used in Ref.~\cite{bhosale2021superposition} for the verification of the obtained results for the $m=2$ case of COE in the study of HOSR distributions. Here, we have taken $j=1000, 1500$, and $2500$ and calculated the eigenvalues of $n=50$ such realizations corresponding to $\tilde{k}=10$ to $59$, for each $j$. We study both HOS and HOSR distributions for all three cases of $j$. The results are tabulated in Table ~\ref{Table: QKTtable}, and some of them are shown in Fig.~\ref{fig: QKT_s_1} for spacing and in Fig.~\ref{fig: QKT_r_1} for spacing ratio. See supplementary material \cite{supplementry2025} for more such figures. Analyzing the results, we find that up to $k=8$ and for all $N$, the $\beta'$s of QKT match with the corresponding $m=2$ results of COE consistently (refer to Table~\ref{Table: COE_m2_DimAnalysistable}). 
But as $k$ increases, the results at times agree with the $m=2$ COE case, and at other times differ by $\pm1$, or $\pm2$, irrespective of $N$, as far as our numerical results are concerned.   

 \begin{table}[t]
\renewcommand{\arraystretch}{1.5} 
\setlength{\tabcolsep}{5pt}  
\begin{tabular}{|c|cc|cc|cc|}
\hline
\rule{0pt}{12pt}  
Order&\multicolumn{2}{c|}{$N=2001$}&\multicolumn{2}{c|}{$N=3001$}&\multicolumn{2}{c|}{$N=5001$}\\ [1.5ex]   
\cline{2-7}  
\rule{0pt}{10pt}  
&HOS&\multicolumn{1}{c|}{HOSR}&HOS&\multicolumn{1}{c|}{HOSR}&HOS&\multicolumn{1}{c|}{HOSR}\\[1ex]
$k$&$\beta'$&\multicolumn{1}{c|}{$\beta'$}&$\beta'$&\multicolumn{1}{c|}{$\beta'$}&$\beta'$&\multicolumn{1}{c|}{$\beta'$} \\ 
\hline
1&$P$&\multicolumn{1}{c|}{$P$}&$P$&\multicolumn{1}{c|}{$P$}&$P$&\multicolumn{1}{c|}{$P$}\\
2&2&\multicolumn{1}{c|}{2}&2&\multicolumn{1}{c|}{2}&2&\multicolumn{1}{c|}{2}\\
3&4.28&\multicolumn{1}{c|}{4}&4.26&\multicolumn{1}{c|}{4}&4.28&\multicolumn{1}{c|}{4} \\
4&7&\multicolumn{1}{c|}{7}&7&\multicolumn{1}{c|}{7}&7&\multicolumn{1}{c|}{7} \\
5&11&\multicolumn{1}{c|}{10}&11&\multicolumn{1}{c|}{10}&11&\multicolumn{1}{c|}{10}\\
6&15&\multicolumn{1}{c|}{14}&15&\multicolumn{1}{c|}{14}&15&\multicolumn{1}{c|}{14} \\
7&19&\multicolumn{1}{c|}{18}&19&\multicolumn{1}{c|}{18}&19&\multicolumn{1}{c|}{18}\\
8&24&\multicolumn{1}{c|}{23}&24&\multicolumn{1}{c|}{23}&24&\multicolumn{1}{c|}{23}\\
9&30&\multicolumn{1}{c|}{28}&29&\multicolumn{1}{c|}{28}&30&\multicolumn{1}{c|}{28}\\
10&36&\multicolumn{1}{c|}{34}&35&\multicolumn{1}{c|}{33}&35&\multicolumn{1}{c|}{34}\\
11&42&\multicolumn{1}{c|}{40}&41&\multicolumn{1}{c|}{39}&42&\multicolumn{1}{c|}{40}\\
12&49&\multicolumn{1}{c|}{47}&48&\multicolumn{1}{c|}{46}&49&\multicolumn{1}{c|}{46}\\
13&57&\multicolumn{1}{c|}{54}&55&\multicolumn{1}{c|}{52}&56&\multicolumn{1}{c|}{53}\\
14&65&\multicolumn{1}{c|}{61}&63&\multicolumn{1}{c|}{59}&63&\multicolumn{1}{c|}{60}\\
15&73&\multicolumn{1}{c|}{69}&71&\multicolumn{1}{c|}{67}&71&\multicolumn{1}{c|}{68}\\
16&82&\multicolumn{1}{c|}{78}&80&\multicolumn{1}{c|}{76}&80&\multicolumn{1}{c|}{76}\\
17&91&\multicolumn{1}{c|}{86}&89&\multicolumn{1}{c|}{84}&89&\multicolumn{1}{c|}{84}\\
18&101&\multicolumn{1}{c|}{96}&99&\multicolumn{1}{c|}{93}&98&\multicolumn{1}{c|}{93}\\
19&111&\multicolumn{1}{c|}{105}&108&\multicolumn{1}{c|}{103}&108&\multicolumn{1}{c|}{102}\\
20&121&\multicolumn{1}{c|}{114}&118&\multicolumn{1}{c|}{112}&118&\multicolumn{1}{c|}{111}\\
\hline
\end{tabular}
\caption{\justifying Tabulation of higher-order indices $\beta'$ for the distributions of spacing and spacing ratio of QKT for various values of $k$. Here, $n=50$ such that $\tilde{k}=10$ to $59$ and dimension of unitary operators $\widehat{U}$ are $N=2001$, $3001$, and $5001$. In the table, $P$ denotes the Poisson distribution. For spacings, it is $\exp(-s)$ and ratios, it is $1/(1+r)^2$.}
\label{Table: QKTtable}
\end{table}
\begin{figure}[tbp]
\begin{center}
\includegraphics*[scale=0.35]{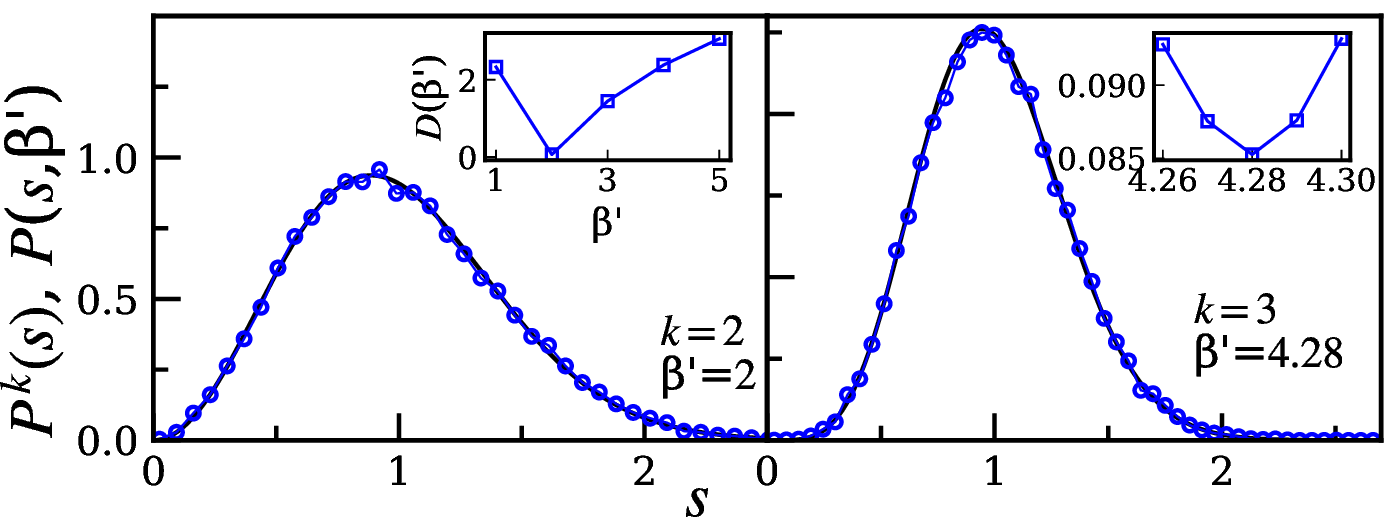}
\caption{\justifying HOS distribution $P^{k}(s)$ of eigenangles of QKT (circles) for $j=1000$, i.e., $N=2001$ and $n=50$, such that $\tilde{k}=10$ to $59$. Here, the solid line corresponds to $P(s, \beta')$ as given in Eq.~(\ref{Eq:PSBeta}), where $\beta$ is replaced by $\beta'$ and $\beta'$ is given in Table~\ref{Table: QKTtable}. The insets show $D(\beta')$ as a function of $\beta'$.}.
\label{fig: QKT_s_1}
\end{center}
\end{figure}
\begin{figure}[tbp]
\begin{center}
\includegraphics*[scale=0.35]{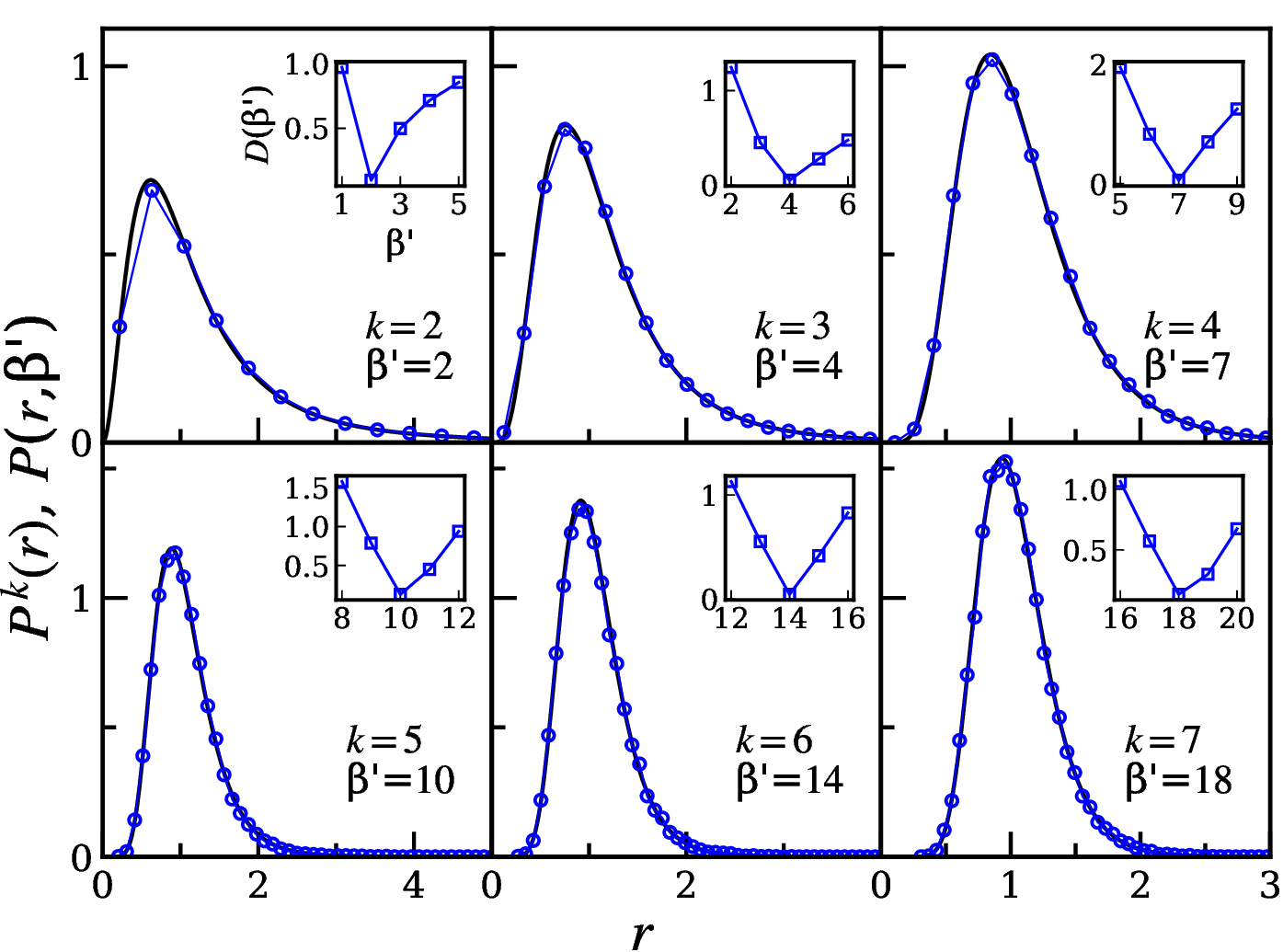}
\caption{\justifying Plot of HOSR distribution $P^{k}(r)$ of QKT (circles). Solid line corresponds to $P(r, \beta')$ as given in Eq.~(\ref{Eq:PRBeta}), where $\beta$ is replaced by $\beta'$ and $\beta'$ is given in Table~\ref{Table: QKTtable}. Here, $N=2001$ ($j=1000$) and $n=50$ such that $\tilde{k}=10$ to $59$.}
\label{fig: QKT_r_1}
\end{center}
\end{figure}

\section{Some important observations and discussions}
\label{sec:SomeImpObs}
In this section, we have mentioned some important observations based on the analysis of our obtained results and discussions in support of our results.
\subsection{Simultaneous comparison of HOS distributions and their convergence to the Poisson distribution as $m \rightarrow \infty $ for the cases of COE, CUE, and CSE}
\label{subsec:Simul.Comp.HOS}
In this subsection, we aim to simultaneously compare the HOS distributions of the three classes of Dyson's circular ensemble. It is found that for a particular $k$ and $m$, the value of $\beta'$ is largest for CSE and smallest for COE. For a particular $m$, the value of $\beta'$ increases with an increase in $k$, and for lower values of $k$, most of the values of $\beta'$s are positive non-integers. Similar behavior is also observed in the study of HOSR distributions of the superposed spectra of circular ensembles \cite{bhosale2021superposition}. In all the three classes, we observe that for a particular $k$, as we increase $m$, the value of $\beta'$ decreases gradually.

After a certain $m$, the distributions are not getting fitted by the generalized Wigner-Dyson distribution and start deviating from it with an increase in $m$, except for some in between $m$'s where they don't deviate, which can be easily seen from the figures. For example, in the case of CSE, such cases are $k=2$ of $m=3$ (see Fig.~\ref{fig:k_2_to_5_CSE_m3}); $k=2$ and $3$ of $m=4$; $k=3$ of $m=5$; $k=3,$ and $4$ of $m=6$; and $k=3$ and $4$ of $m=7$ (refer to supplementary material \cite{supplementry2025} for figures of later cases). In the case of CUE, such cases are $k=2$ of $m=3$; $k=2,$ and $3$ of $m=4$; $k=3$ of $m=5$; and $k=2$ of $m=7$. One of which is shown in Fig.~\ref{fig:k_2_to_3_CUE_m4}, and other cases are shown in the supplementary material \cite{supplementry2025}. 
In the case of COE, such cases are $k=2$ and $3$ for $m=5$, $6$, and $7$. One of which is shown in Fig. \ref{fig:k_2_to_7_COE_m5} and other cases are shown in the supplementary material \cite{supplementry2025}.

\begin{figure}[tbp]
\begin{center}
\includegraphics*[scale=0.34]{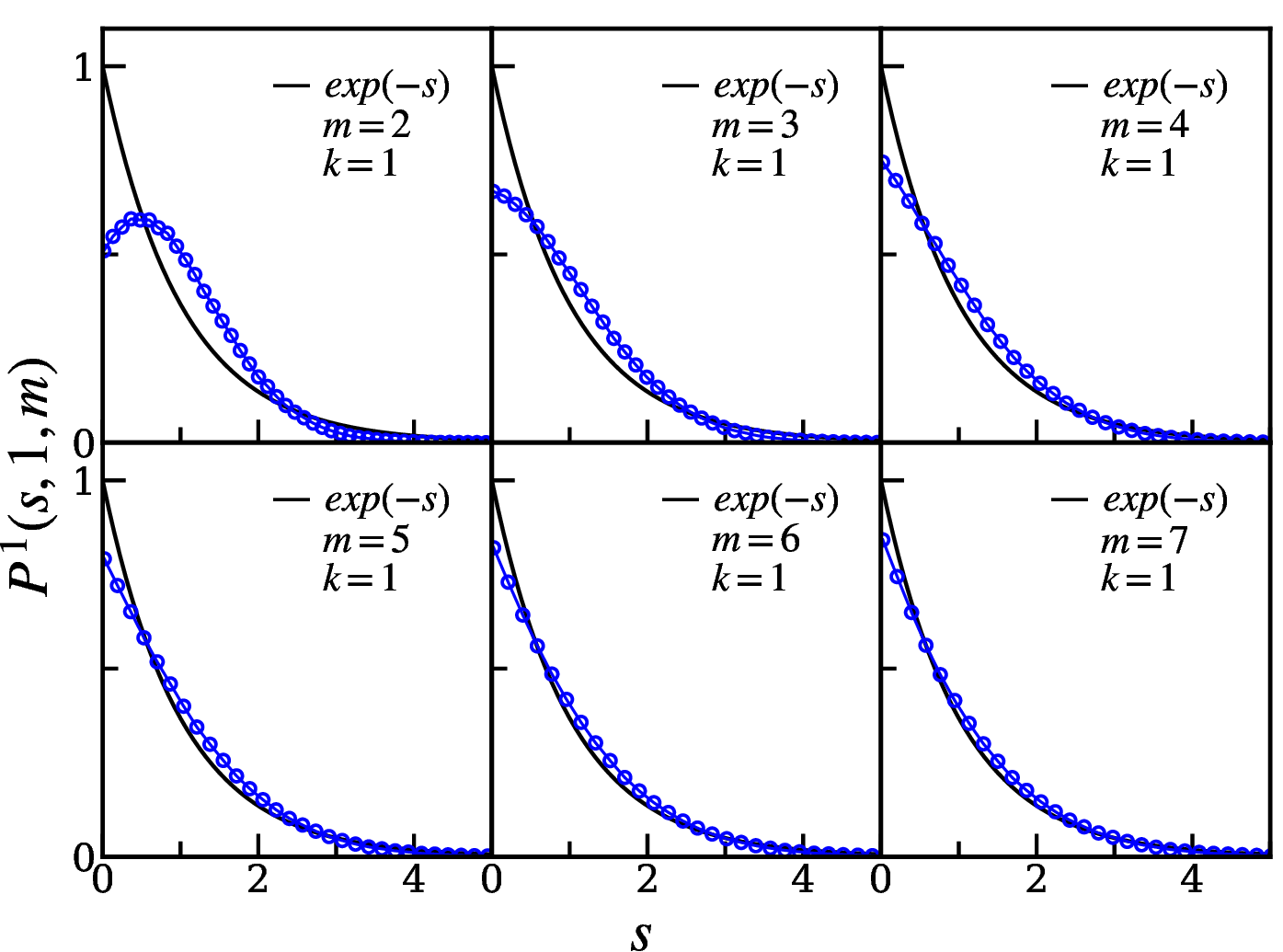}
\caption{\justifying Plot of the distribution of $s^{(1)}$ for $m=2$ to $7$ COE spectra $P^{1}(s, 1, m)$, denoted as circles. The black solid curve corresponds to the nearest neighbor Poisson spacing distribution. Here, in each case $N= 5000$ and $n= 600, 900, 1000, 1000, 1002$, and $1001$, respectively, for $m= 2, 3, 4, 5, 6$, and $7$.}
\label{fig: coe_s_1_tending_to_poisson}
\end{center}
\end{figure}

\begin{figure}[tbp]
\begin{center}
\includegraphics*[scale=0.34]{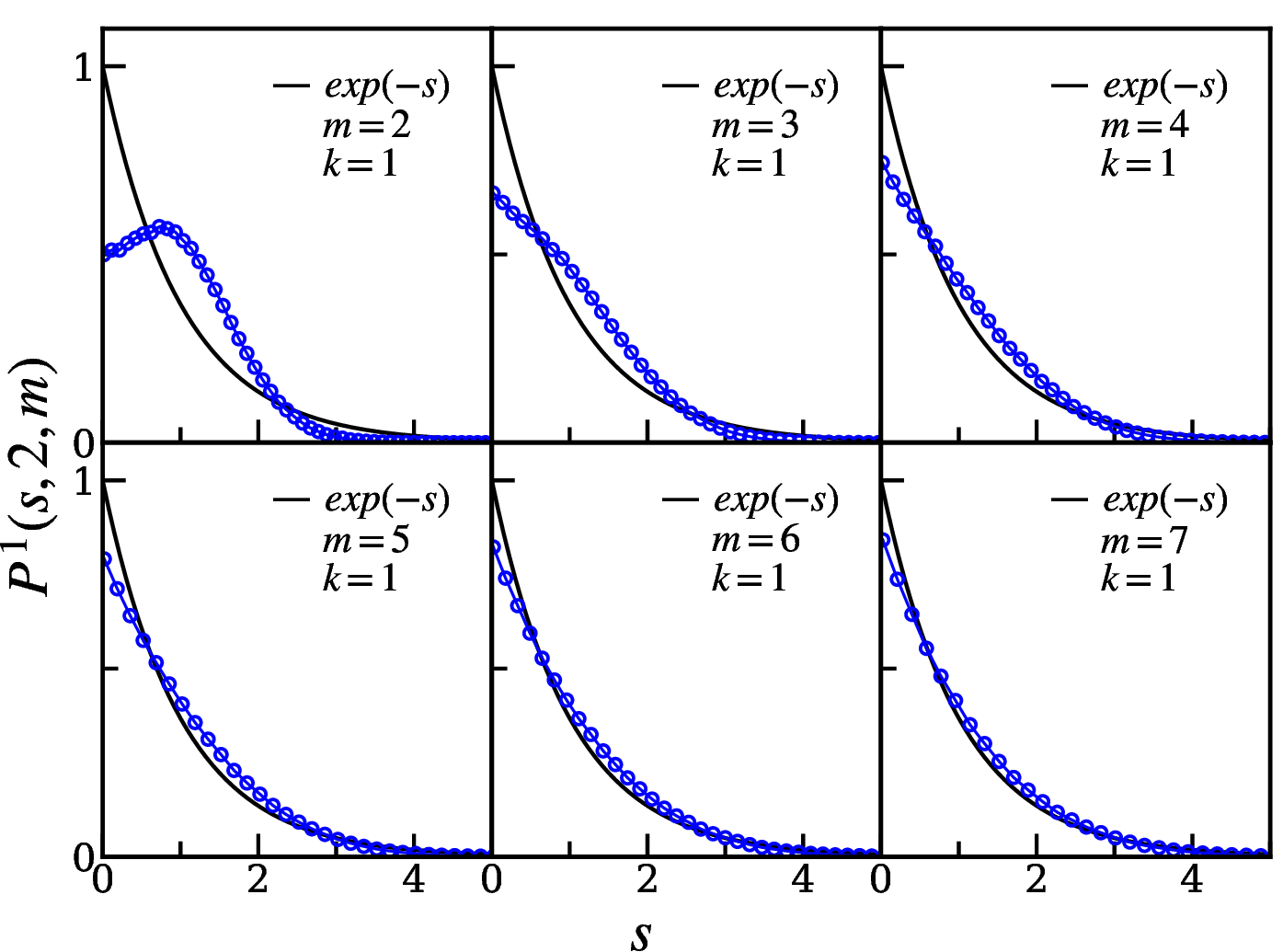}
\caption{\justifying Same as Fig.~\ref{fig: coe_s_1_tending_to_poisson} but for CUE. }
\label{fig: cue_s_1_tending_to_poisson}
\end{center}
\end{figure}

\begin{figure}[tbp]
\begin{center}
\includegraphics*[scale=0.34]{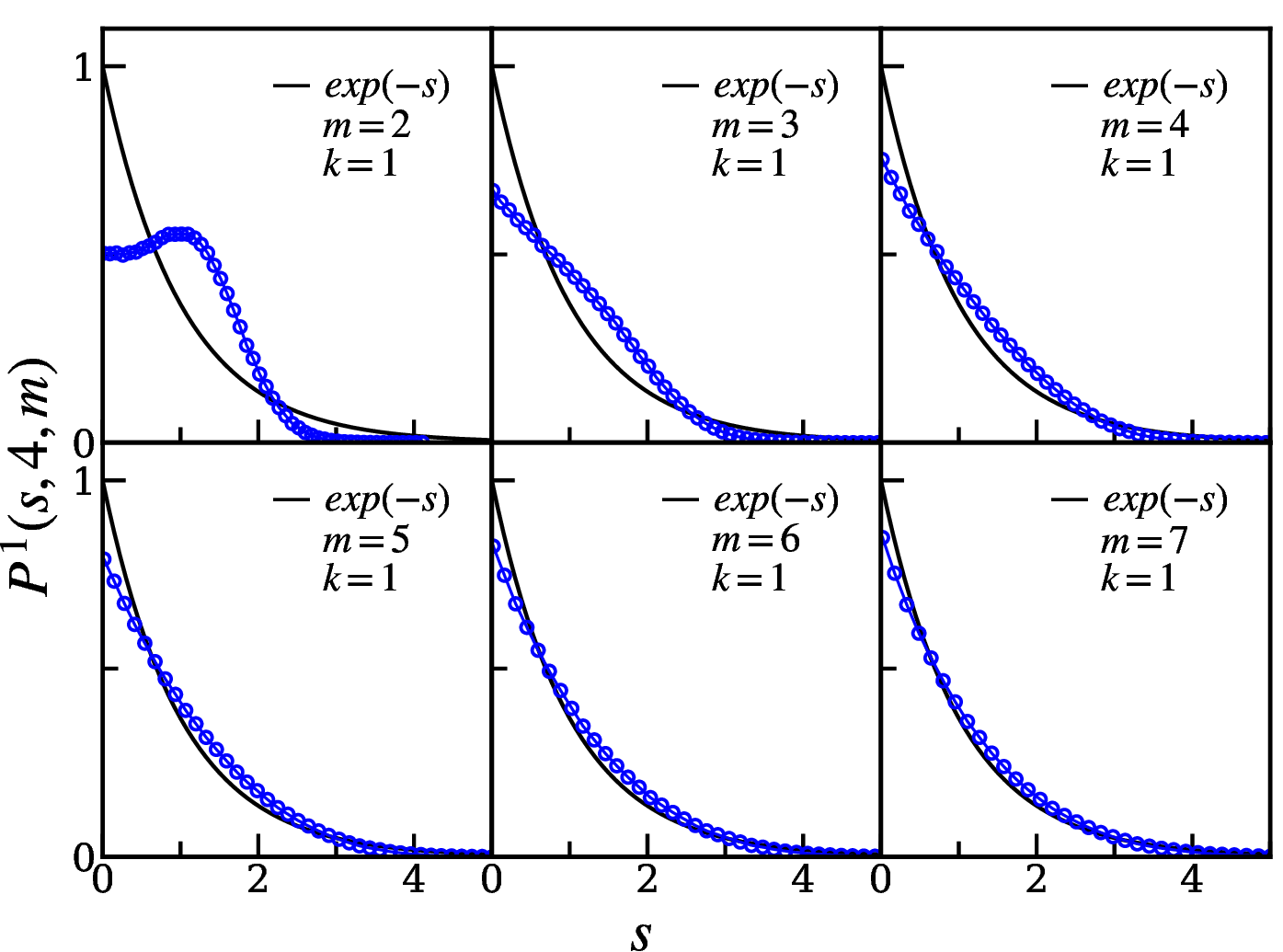}
\caption{\justifying Same as Fig.~\ref{fig: coe_s_1_tending_to_poisson} but for CSE.}
\label{fig: cse_s_1_tending_to_poisson}
\end{center}
\end{figure}

\begin{figure}[tbp]
\begin{center}
\includegraphics*[scale=0.34]{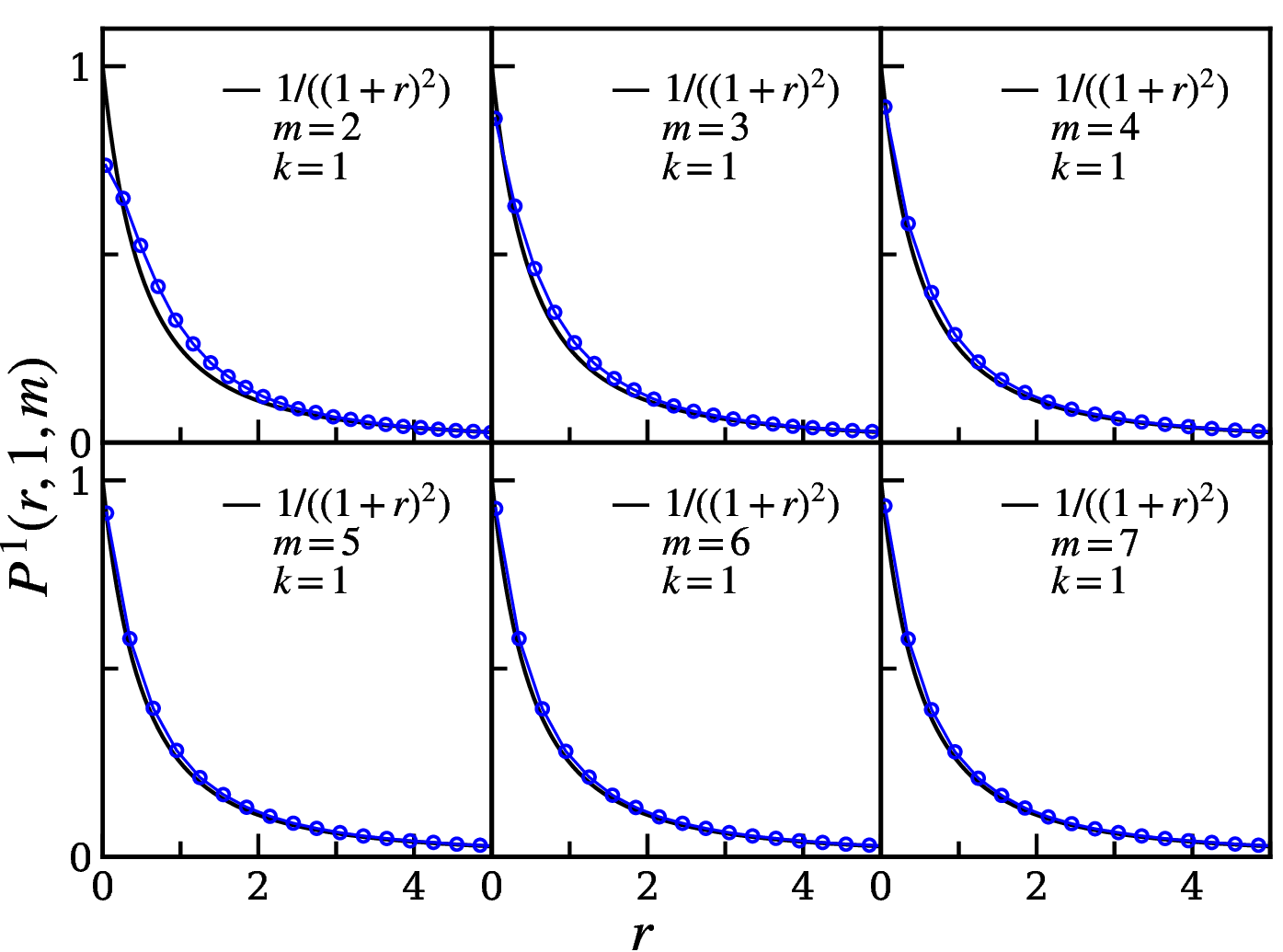}
\caption{\justifying Same as Fig.~\ref{fig: coe_s_1_tending_to_poisson} but for spacing ratio $r^{(1)}$.}
\label{fig: coe_r_1_tending_to_poisson}
\end{center}
\end{figure}

\begin{figure}[tbp]
\begin{center}
\includegraphics*[scale=0.34]{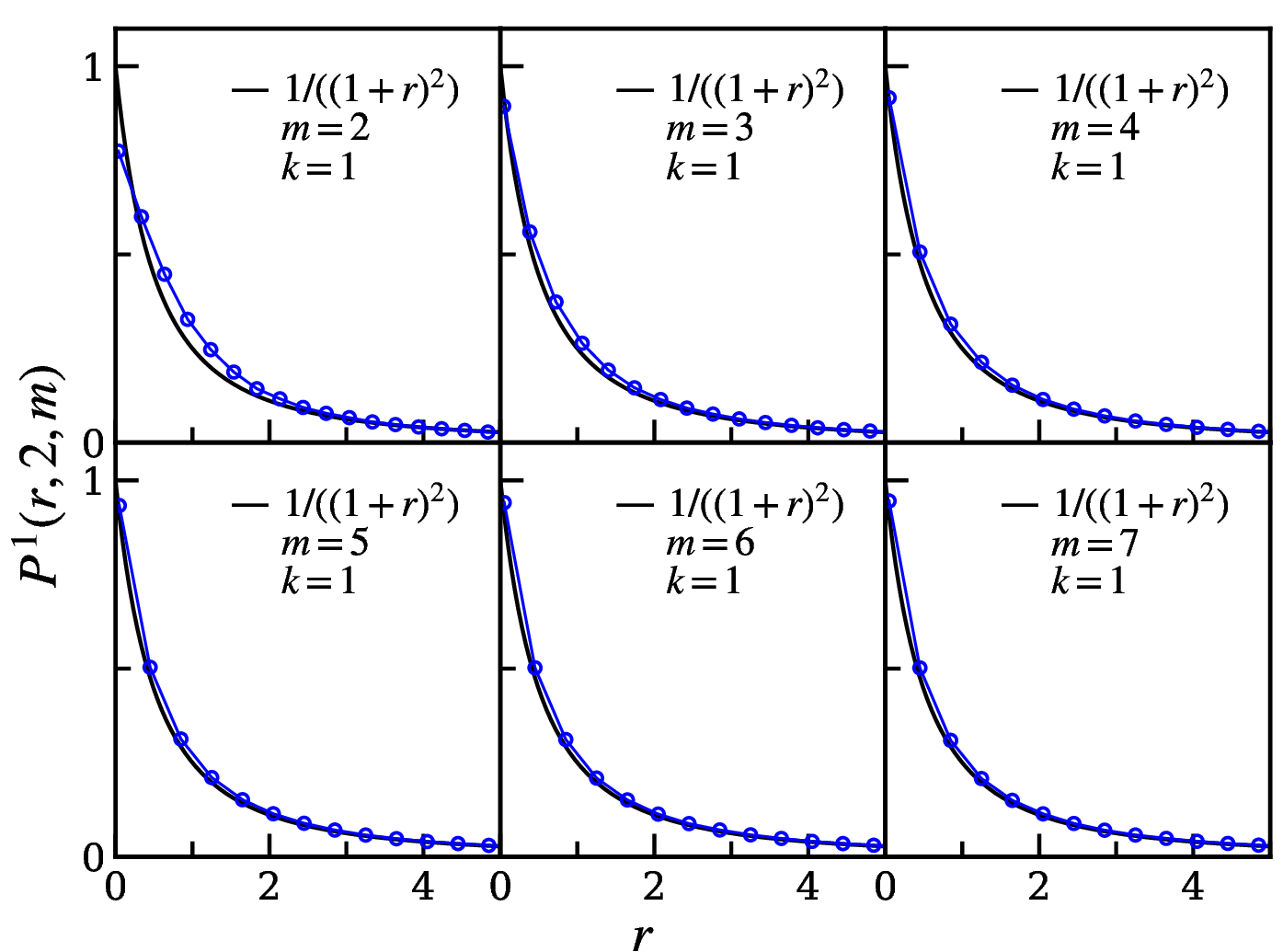}
\caption{\justifying Same as Fig.~\ref{fig: coe_r_1_tending_to_poisson} but for CUE.}
\label{fig: cue_r_1_tending_to_poisson}
\end{center}
\end{figure}

\begin{figure}[tbp]
\begin{center}
\includegraphics*[scale=0.34]{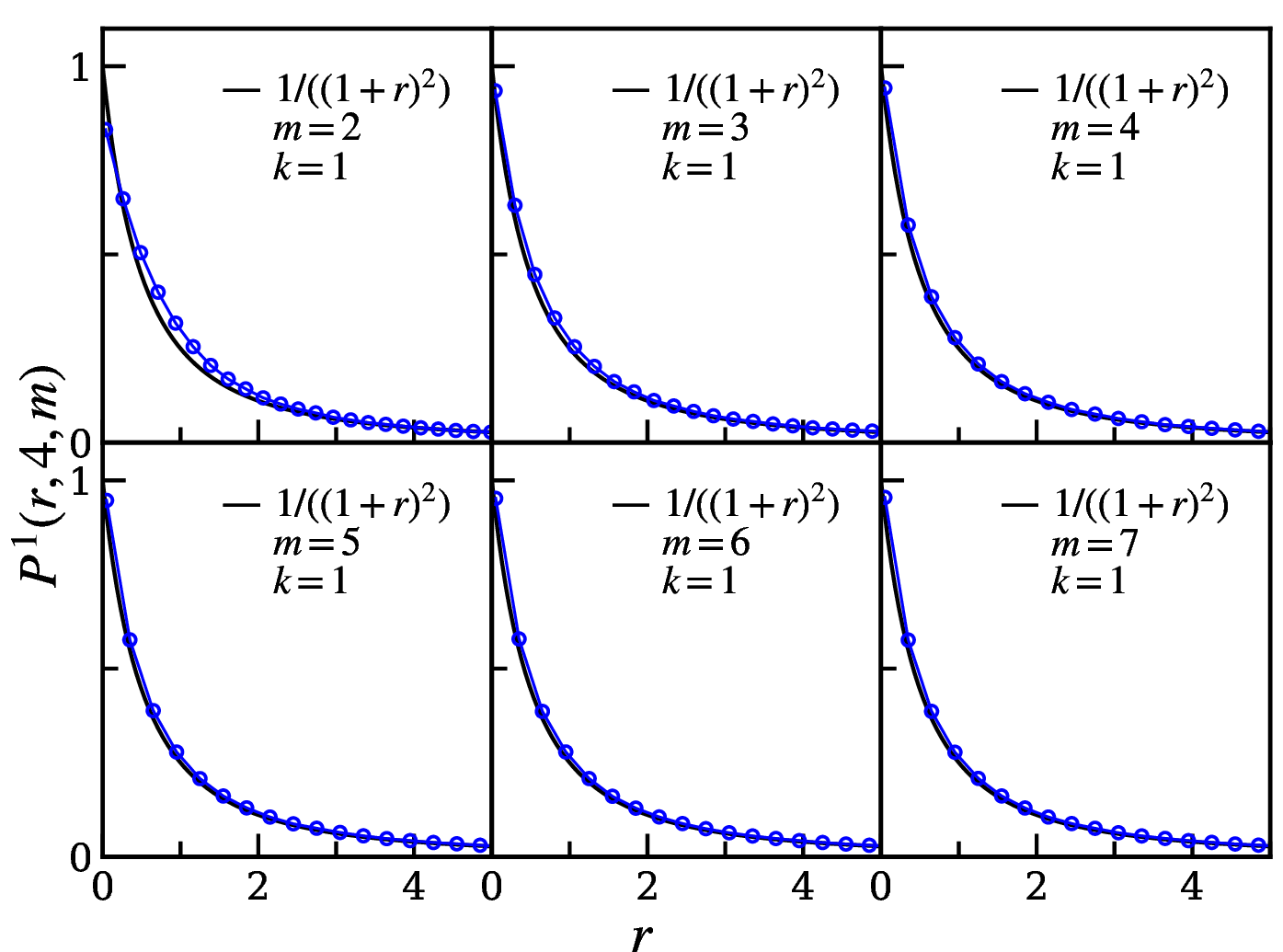}
\caption{\justifying Same as Fig.~\ref{fig: coe_r_1_tending_to_poisson} but for CSE.}
\label{fig: cse_r_1_tending_to_poisson}
\end{center}
\end{figure}

Further, we have found that for a given $k$, as $m$ increases the spacing or ratio distribution in the superposed spectra of the circular ensemble tends to the corresponding $k$-th order spacing or ratio distribution of the Poisson ensemble respectively. The expression for higher order spacing distribution in the Poisson ensemble is given by \cite{rao2020higher}:
\begin{equation}
P_{P}^{k}(s)=\frac{k^k}{(k-1)!}s^{k-1}e^{-ks}
\label{Eq: HOS_Poisson}
\end{equation}
Also, the expression for higher order spacing ratio distribution in the Poisson ensemble is given by \cite{tekur2020symmetry}:
\begin{equation}
P_{P}^{k}(r)=\frac{(2k-1)!}{[(k-1)!]^2}\frac{r^{k-1}}{(1+r)^{2k}}
\label{Eq: HOSR_Poisson}
\end{equation}
Thus, we conjecture that for a given $k$ in all the three classes of circular ensemble with Dyson index $\beta$, the distribution tends to the corresponding $k$-th order Poisson distribution as $m$ tends to infinity. 
Hence, we can conclude that at large $m$, the spectral statistics of COE, CUE, and CSE are the same. The special case of this conjecture with $\beta=2$ and $k=1$ is addressed analytically in Ref.~\cite{tkocz2012tensor}. Similarly, it is shown analytically that for $k=1$ case of any Gaussian ensemble, and $m \rightarrow \infty$ the spacing ratio distribution tends to be Poisson \cite{giraud2022probing}. We have found that this convergence is faster in the case of ratio than that of the spacing in all three classes of circular ensemble (For example, see Figs. \ref{fig: coe_s_1_tending_to_poisson}, \ref{fig: cue_s_1_tending_to_poisson}, \ref{fig: cse_s_1_tending_to_poisson}, and compare them with Figs. \ref{fig: coe_r_1_tending_to_poisson}, \ref{fig: cue_r_1_tending_to_poisson}, \ref{fig: cse_r_1_tending_to_poisson} respectively for $k=1$ case). Our conjecture, based on numerics, generalizes the result of Ref.~\cite{tkocz2012tensor}.

\begin{figure}[tbp]
\begin{center}
\includegraphics*[scale=0.34]{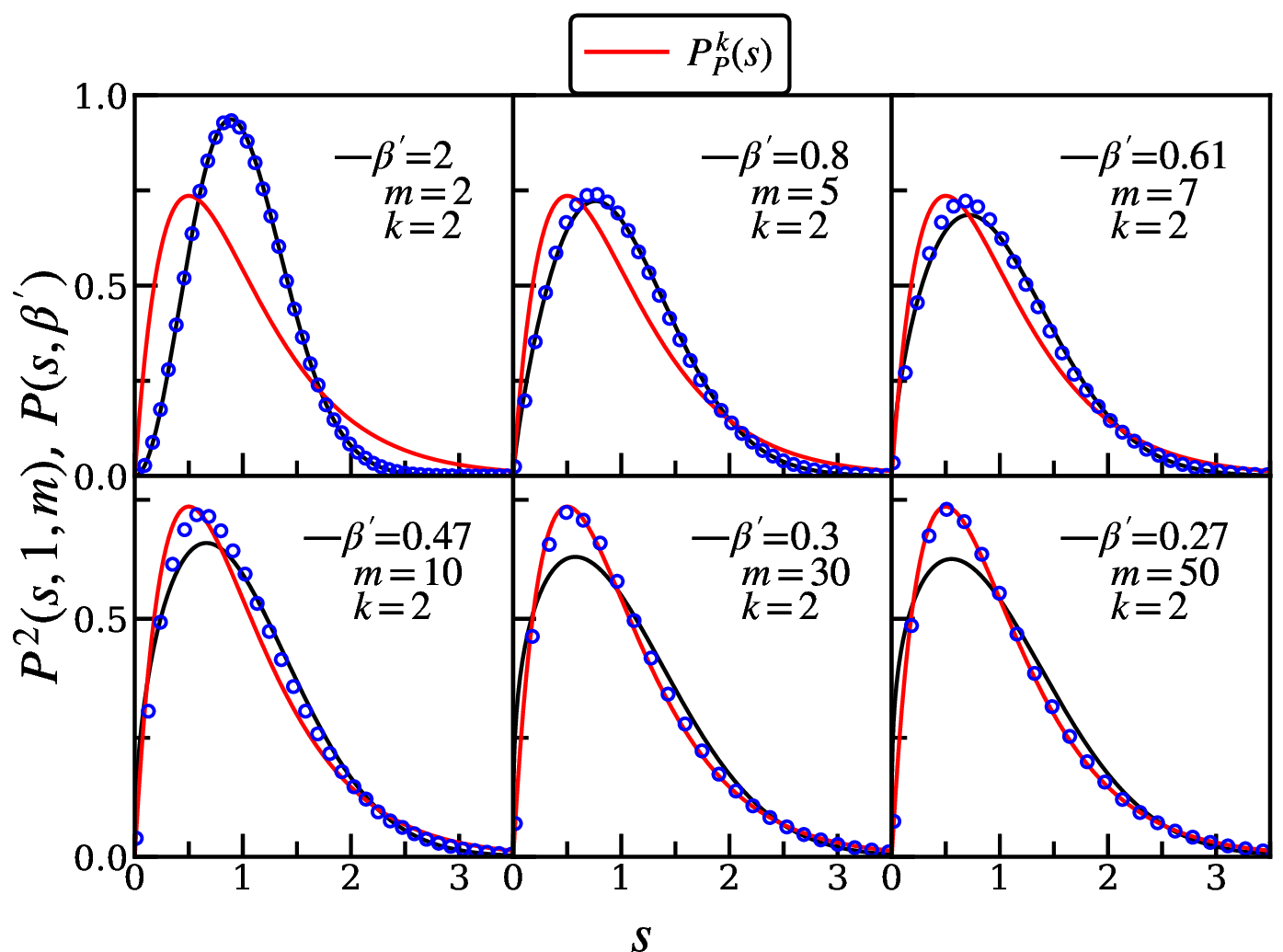}
\caption{\justifying Second order spacing distribution $P^{2}(s, 1, m)$ in the $m$ superposed spectra of COEs, denoted by circles. Here, the black solid line corresponds to the generalized Wigner-Dyson distribution corresponding to $\beta'$ for spacings $P(s, \beta')$. Red solid line corresponds to the second order ($k= 2$) spacing distribution of the Poisson ensemble $P^{k}_{P}(s)$ as per Eq.~(\ref{Eq: HOS_Poisson}). Here, in each case $N= 5000$ and $n= 600, 1000, 1001, 900, 900$, and $900$ respectively for $m= 2, 5, 7, 10, 30$, and $50$.}
\label{fig: coe_s_2_tending_to_poisson}
\end{center}
\end{figure}

\begin{figure}[tbp]
\begin{center}
\includegraphics*[scale=0.34]{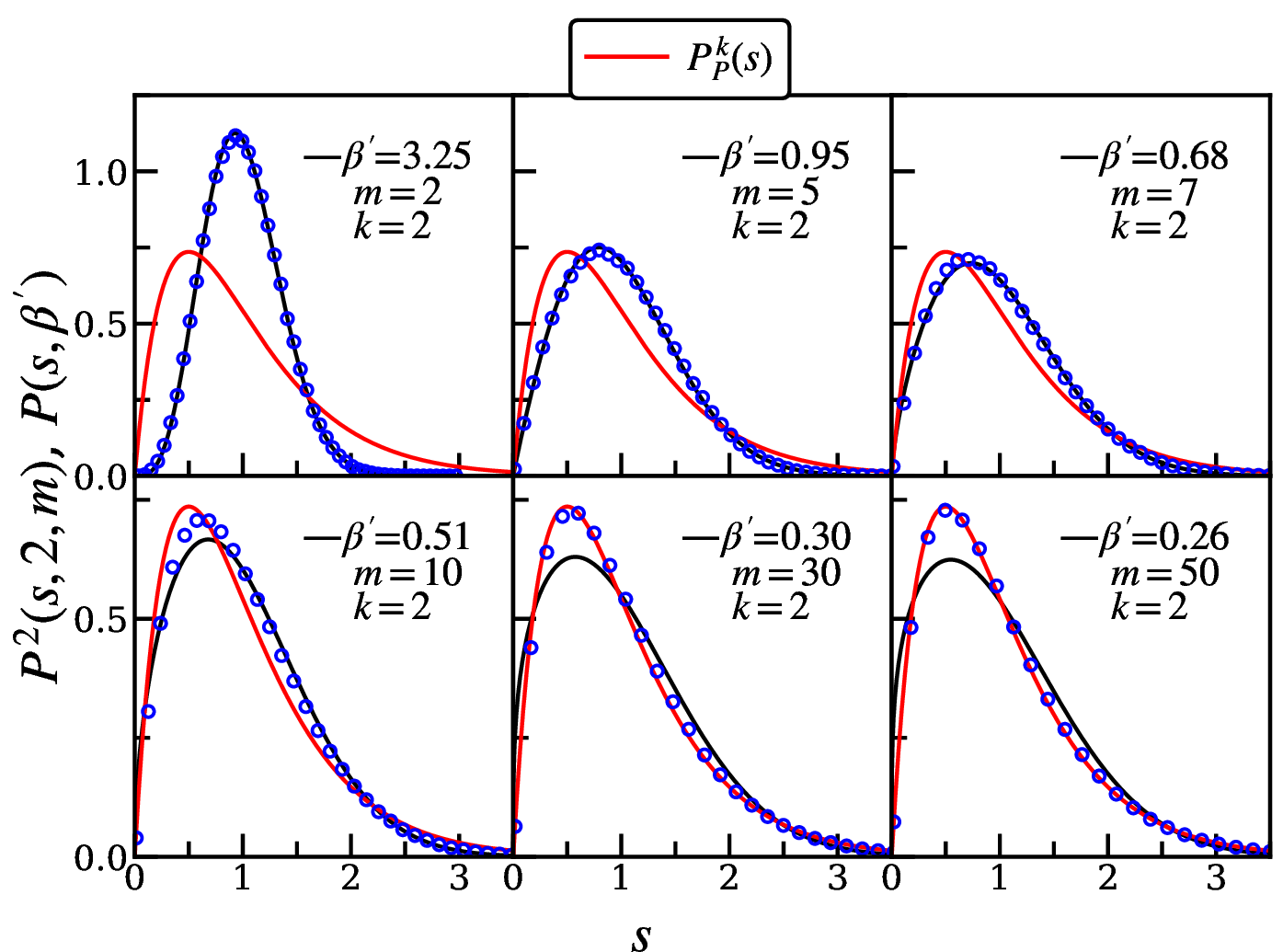}
\caption{\justifying Same as Fig.~\ref{fig: coe_s_2_tending_to_poisson} but for CUE. Here, in each case $N= 5000$ and $n= 600, 1000, 1001, 1000, 990$, and $1000$ respectively for $m= 2, 5, 7, 10, 30$, and $50$.}
\label{fig: cue_s_2_tending_to_poisson}
\end{center}
\end{figure}

\begin{figure}[tbp]
\begin{center}
\includegraphics*[scale=0.34]{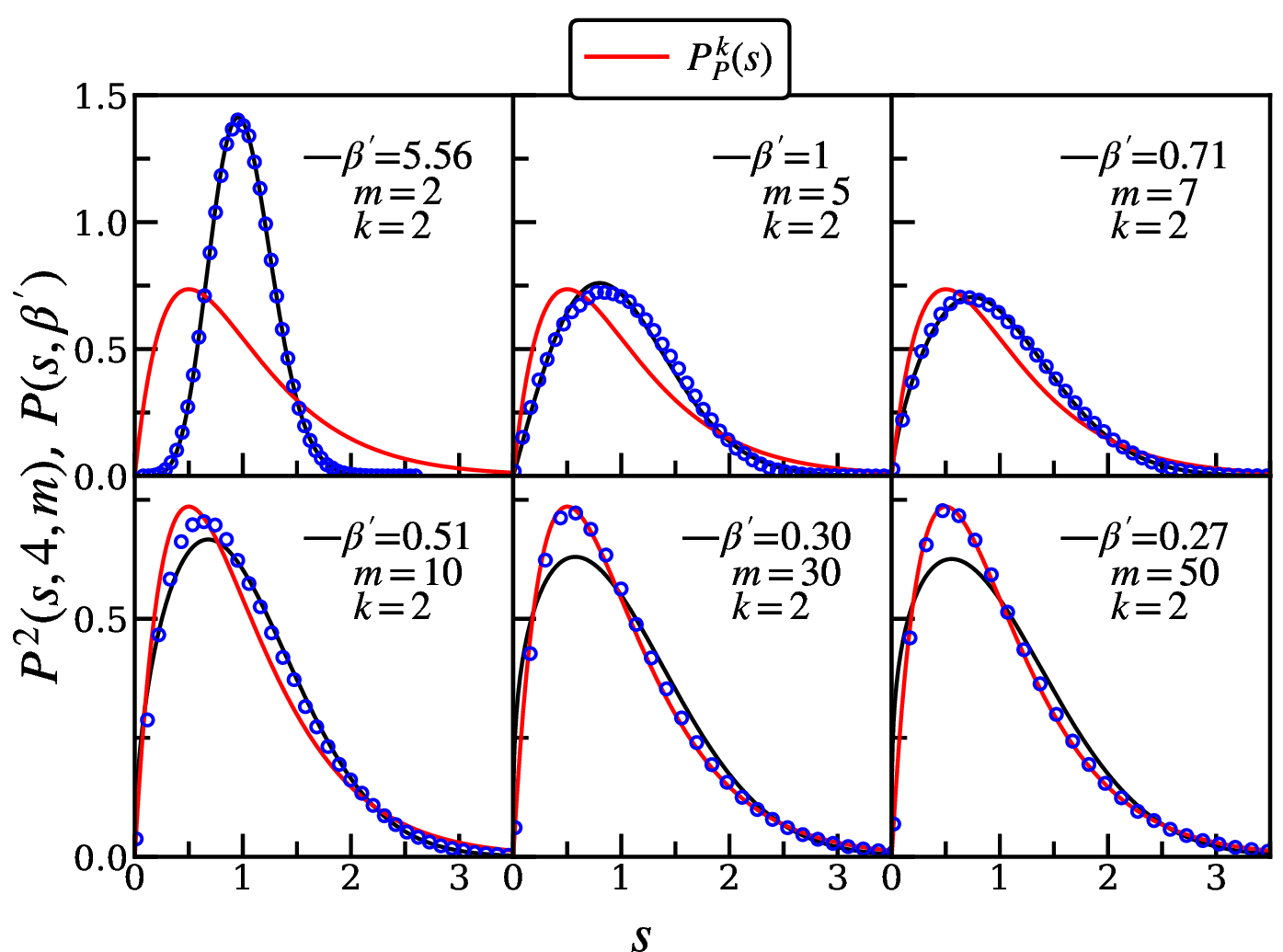}
\caption{\justifying Same as Fig.~\ref{fig: coe_s_2_tending_to_poisson} but for CSE.}
\label{fig: cse_s_2_tending_to_poisson}
\end{center}
\end{figure}

\begin{figure}[tbp]
\begin{center}
\includegraphics*[scale=0.34]{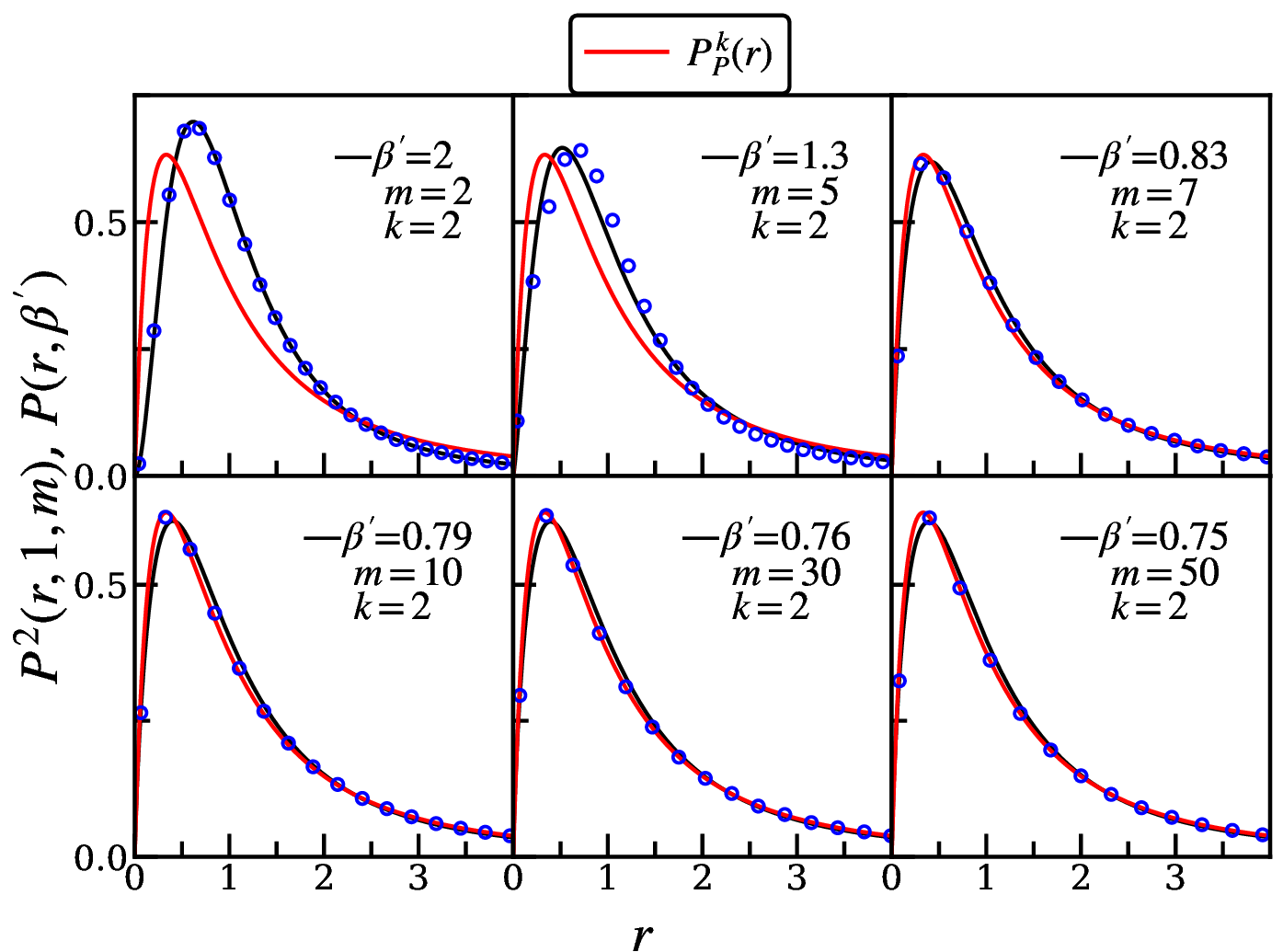}
\caption{\justifying Second order spacing ratio distribution $P^{2}(r, 1, m)$ in the $m$ superposed spectra of COEs, denoted by circles. Here, the black solid line corresponds to the generalized Wigner-Dyson distribution corresponding to $\beta'$ for spacing ratios $P(r, \beta')$. The red solid line corresponds to the second order ($k= 2$) spacing ratio distribution of the Poisson ensemble $P^{k}_{P}(r)$ as per Eq.~(\ref{Eq: HOSR_Poisson}). Here, in each case $N= 5000$ and $n= 500, 995, 994, 900, 900$, and $900$, respectively, for $m= 2, 5, 7, 10, 30$, and $50$.}
\label{fig: coe_r_2_tending_to_poisson}
\end{center}
\end{figure}

\begin{figure}[tbp]
\begin{center}
\includegraphics*[scale=0.34]{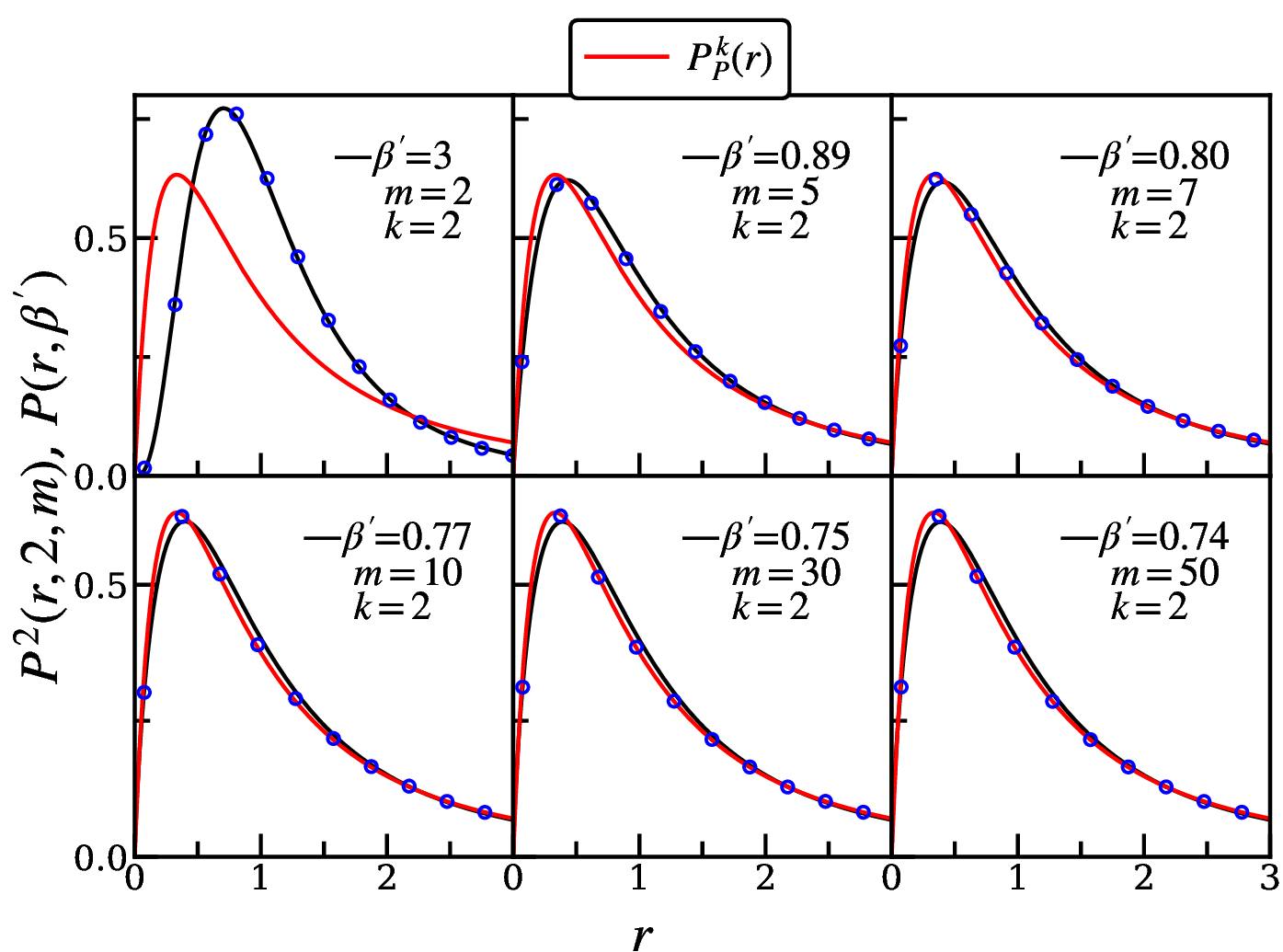}
\caption{\justifying Same as Fig.~\ref{fig: coe_r_2_tending_to_poisson} but for CUE. Here, in each case $N= 5000$ and $n= 500, 1000, 1001, 1000, 990$, and $1000$ respectively for $m= 2, 5, 7, 10, 30$, and $50$.}
\label{fig: cue_r_2_tending_to_poisson}
\end{center}
\end{figure}

\begin{figure}[tbp]
\begin{center}
\includegraphics*[scale=0.34]{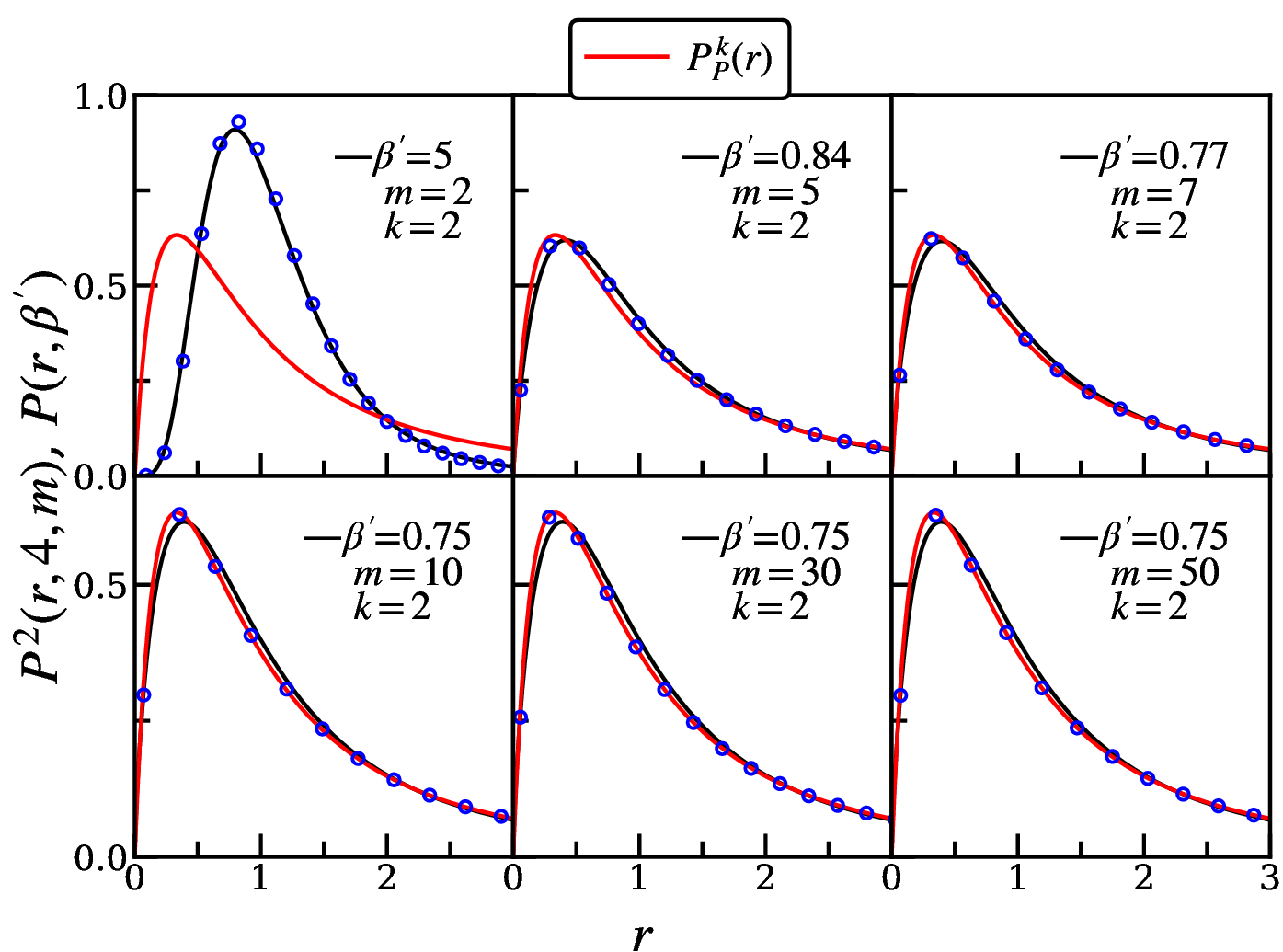}
\caption{\justifying Same as Fig.~\ref{fig: cue_r_2_tending_to_poisson} but for CSE.}
\label{fig: cse_r_2_tending_to_poisson}
\end{center}
\end{figure}

The $k=2$ case of the three circular ensembles is shown in Figs. \ref{fig: coe_s_2_tending_to_poisson}-\ref{fig: cse_s_2_tending_to_poisson} for spacings and in Figs. \ref{fig: coe_r_2_tending_to_poisson}-\ref{fig: cse_r_2_tending_to_poisson} for spacing ratios. Some more figures corresponding to other $k$ values are shown in the supplementary material \cite{supplementry2025}. From the $k$-th order ratio distributions (see both in the main text and supplementary material), it can be seen that as $m$ increases, it seems they are getting fitted by both the generalized Wigner-Dyson and $k$-th order Poisson ratio distribution. But, as it is evident clearly from the $k$-th order spacing distributions that they tend towards the $k$-th order Poisson spacing distribution, hence, from the circumstantial evidence and from the figures, we can conclude that the ratio distribution also tends towards the Poisson ratio distribution.

\subsection{COE and CUE correspondence: Gunson's result}
\label{Gunson's result}
\begin{figure}[tbp]
\begin{center}
\includegraphics*[scale=0.35]{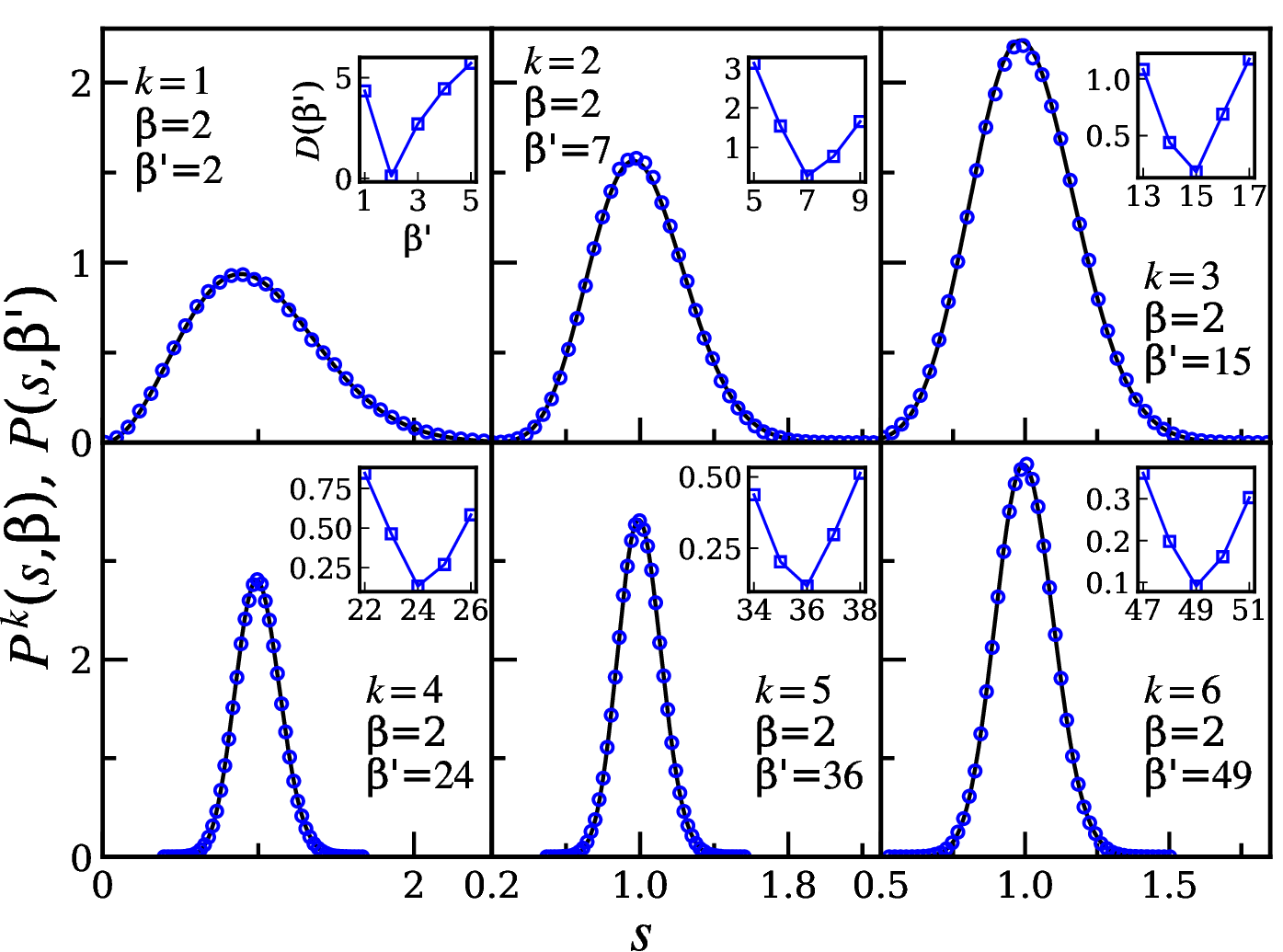}
\caption{\justifying HOS distribution $P^{k}(s,\beta)$ of CUE spectra (circles) for $m=1$, $N=5000$, and $n= 300$. Here, solid line corresponds to $P(s, \beta')$ as given in Eq.~(\ref{Eq:PSBeta}), where $\beta$ is replaced by $\beta'$ and $\beta'$ are given in Table~\ref{Table:GunsonResulttable}. The insets show $D(\beta')$ as a function of $\beta'$.}
\label{fig: cue_s_1}
\end{center}
\end{figure}

\begin{figure}[tbp]
\begin{center}
\includegraphics*[scale=0.35]{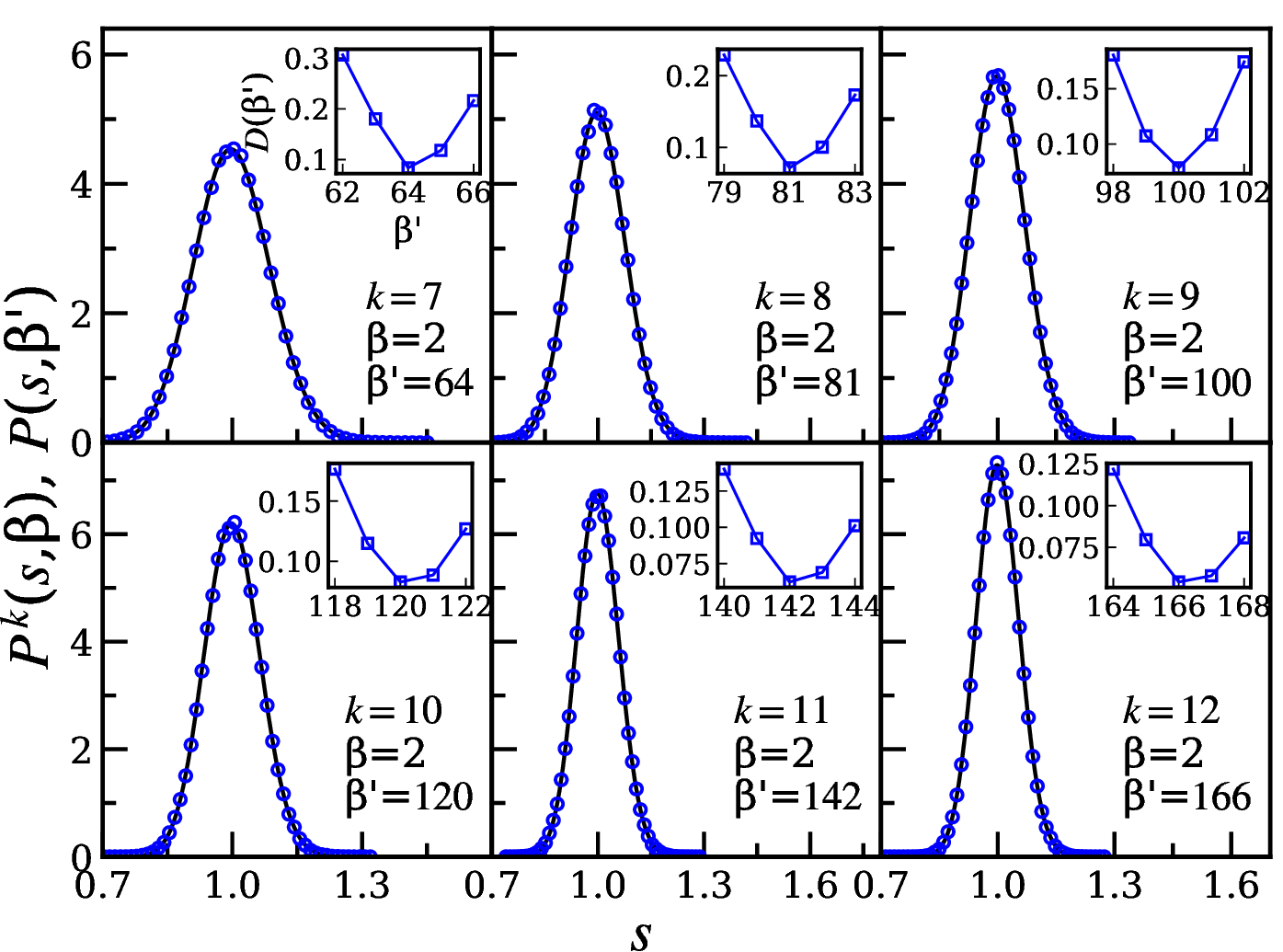}
\caption{\justifying Same as Fig.~\ref{fig: cue_s_1} but for $k=7$ to $12$.}
\label{fig: cue_s_2}
\end{center}
\end{figure}

In this subsection, we want to reproduce the theorem relating COE and CUE as mentioned in the Sec. \ref{sec:Introduction} using HOS as the fluctuation measure and present it here. This study will act as an extra validation to confirm the correctness of our computations and can help in examining the behavior, especially at higher $k$. We have tabulated the results in Table~\ref{Table:GunsonResulttable}, and some of them are plotted in Figs.~\ref{fig: cue_s_1} and \ref{fig: cue_s_2}. By analyzing these results, we find that there is a good agreement between the results (obtained $\beta'$) of CUE ($m= 1$) and COE ($m= 2$). But, for some cases of higher $k$, they differ from each other by $\pm 1$ or $\pm 2$ as far as our results are concerned, despite the analytical result proved at the level of the jpdf of the eigenvalues. 

But, here we find that in the case of CUE, up to $k= 2$ the distributions of spacing obey the scaling relation Eq.~(\ref{Eq:ScalingRelation1}), and the ratio follow the same relation for a bit larger $k$ than the spacings. But as $k$ increases, we can see that the $\beta'$ for ratio are found to be highly deviated from the scaling relation than the corresponding spacings. And the deviation among the results corresponding to spacing and ratio increases with $k$. Here, we can also see that $N$ has little effect on the obtained results ($\beta'$ for HOS and HOSR distributions) of CUE (refer Table~\ref{Table:GunsonResulttable}). Similar behaviors are observed in the $m= 1$ case of COE (refer to Sec.~\ref{subsec:DimAnalysis}). We have also studied the same for the $m= 1$ case of GUE (not shown here), and the behaviors are of a similar kind as those of the $m = 1$ case of GOE.  

The HOS distribution for the $m= 1$ case obeys the scaling relation Eq.~(\ref{Eq:ScalingRelation1}) with a slight deviation of one to three at higher $k$. This is observed despite the fact that there is an analytical result on HOS \citep{rao2020higher}. Hence, we claim that these deviations might be due to the statistical fluctuations or computational precision error or the limitation of the numerical method $D(\beta')$ used (also refer to Sec.~\ref{subsec:Dist.Func.Anal} for further insights). Also, for the HOSR distribution in the $m= 1$ case of the Gaussian ensembles, the analytical result exists {\it only} in the asymptotic limits of $r^{(k)}\rightarrow0$ and $r^{(k)}\rightarrow\infty$ \cite{bhosale2023universal}, and the deviations in the numerical fit from the predicted value for larger $k$ needs further numerical and analytical analysis.

\begin{table*}
\renewcommand{\arraystretch}{1.5} 
\setlength{\tabcolsep}{3.5pt}  
\centering
\begin{tabular}{|cc|c|cccc|cccc|}
\hline 
\multicolumn{2}{|c|}{Order}& \makecell{According to \\ the scaling \\relation}&\multicolumn{4}{c|}{$N= 5000$}&\multicolumn{4}{c|}{$N= 45000$}\\
\cline{4-11}  
\multicolumn{2}{|c|}{}&Eq.~(\ref{Eq:ScalingRelation1})&\multicolumn{2}{c|}{HOS}&\multicolumn{2}{c|}{HOSR}&\multicolumn{2}{c|}{HOS}&\multicolumn{2}{c|}{HOSR}\\
\cline{1-11}
CUE&COE ($m= 2$)&GUE&CUE&\multicolumn{1}{c|}{COE ($m= 2$)}&CUE&\multicolumn{1}{c|}{COE ($m= 2$)}&CUE&\multicolumn{1}{c|}{COE ($m= 2$)}&CUE&\multicolumn{1}{c|}{COE ($m= 2$)}\\
$k$&$k$&$\beta'$&$\beta'$&\multicolumn{1}{c|}{$\beta'$}&$\beta'$&$\beta'$&$\beta'$&\multicolumn{1}{c|}{$\beta'$}&$\beta'$&$\beta'$\\
\hline
1&2&2&2&\multicolumn{1}{c|}{2}&2&2&2&\multicolumn{1}{c|}{2}&2&2\\
2&4&7&7&\multicolumn{1}{c|}{7}&7&7&7&\multicolumn{1}{c|}{7}&7&7\\
3&6&14&15&\multicolumn{1}{c|}{15}&14&14&15&\multicolumn{1}{c|}{15}&14&14\\
4&8&23&24&\multicolumn{1}{c|}{24}&23&23&24&\multicolumn{1}{c|}{24}&23&23\\
5&10&34&36&\multicolumn{1}{c|}{36}&34&34&36&\multicolumn{1}{c|}{36}&34&34\\
6&12&47&49&\multicolumn{1}{c|}{49}&46&46&49&\multicolumn{1}{c|}{49}&46&47\\
7&14&62&64&\multicolumn{1}{c|}{64}&61&61&64&\multicolumn{1}{c|}{64}&61&61\\
8&16&79&81&\multicolumn{1}{c|}{81}&76&76&81&\multicolumn{1}{c|}{81}&77&77\\
9&18&98&100&\multicolumn{1}{c|}{100}&94&94&99&\multicolumn{1}{c|}{100}&94&94\\
10&20&119&120&\multicolumn{1}{c|}{120}&113&113&119&\multicolumn{1}{c|}{120}&113&113\\
11&22&142&142&\multicolumn{1}{c|}{142}&134&134&141&\multicolumn{1}{c|}{142}&133&134\\
12&24&167&166&\multicolumn{1}{c|}{166}&156&156&164&\multicolumn{1}{c|}{166}&155&156\\
\hline
\end{tabular}
\caption{\justifying Tabulation of higher-order indices $\beta'$ for various values of $k$ for both the distributions of spacing and spacing ratio in the case of COE ($m= 2$) and CUE ($m= 1$). Here, for $N= 5000$, $n = 600$ and $300$ for COE and CUE respectively. And for $N= 45000$, $n= 300$ and $150$ for COE and CUE respectively.}
\label{Table:GunsonResulttable}    
\end{table*}
\subsection{Analysis of spacing and spacing ratio distribution functions}
\label{subsec:Dist.Func.Anal}
\begin{figure}[tbp]
\begin{center}
\includegraphics*[scale=0.35]{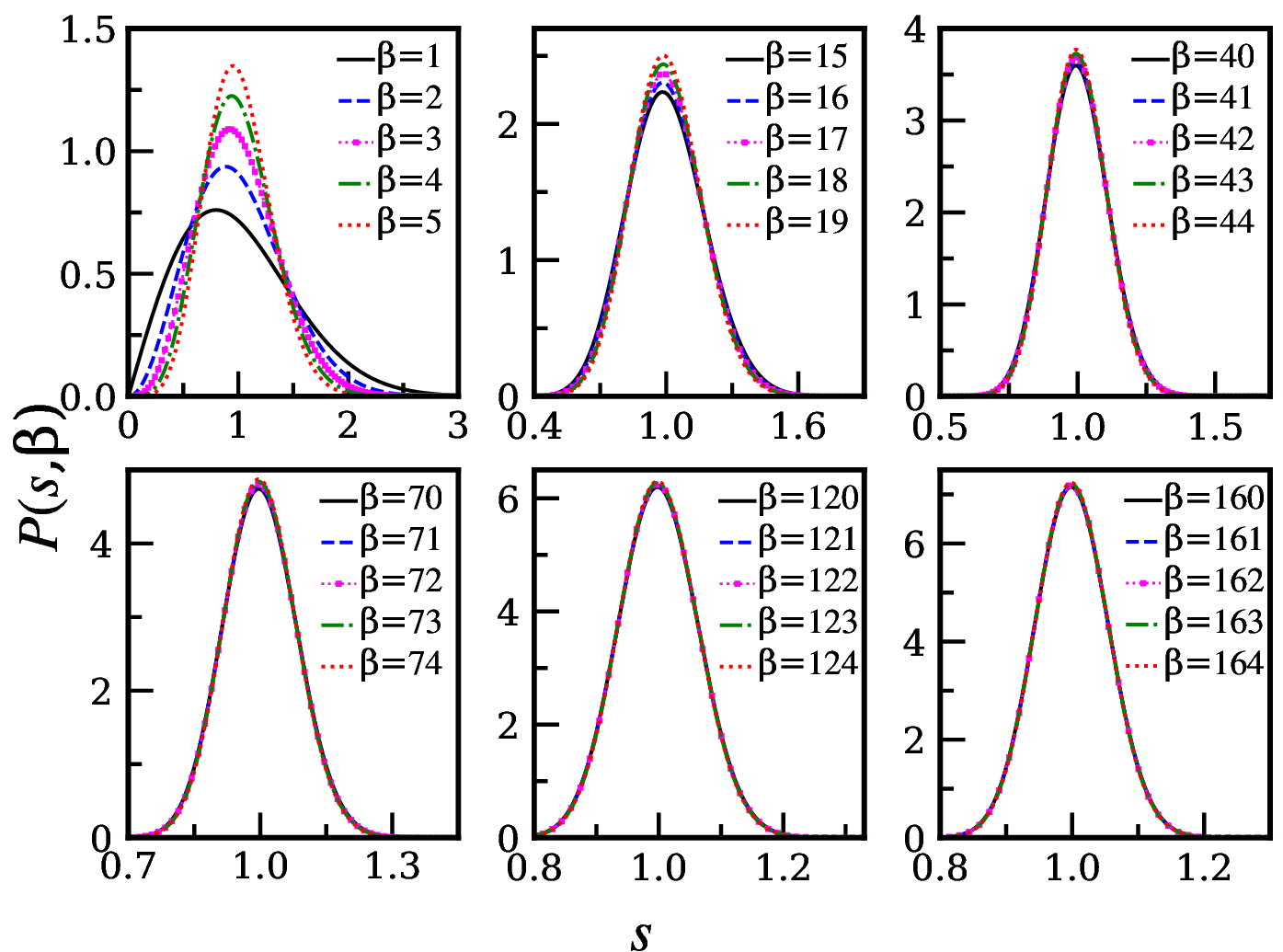}
\caption{\justifying Plot of $P(s,\beta)$ as per Eq.~(\ref{Eq:PSBeta}) for various $\beta$.}
\label{fig:SpacingFuncn}
\end{center}
\end{figure}

\begin{figure}[tbp]
\begin{center}
\includegraphics*[scale=0.35]{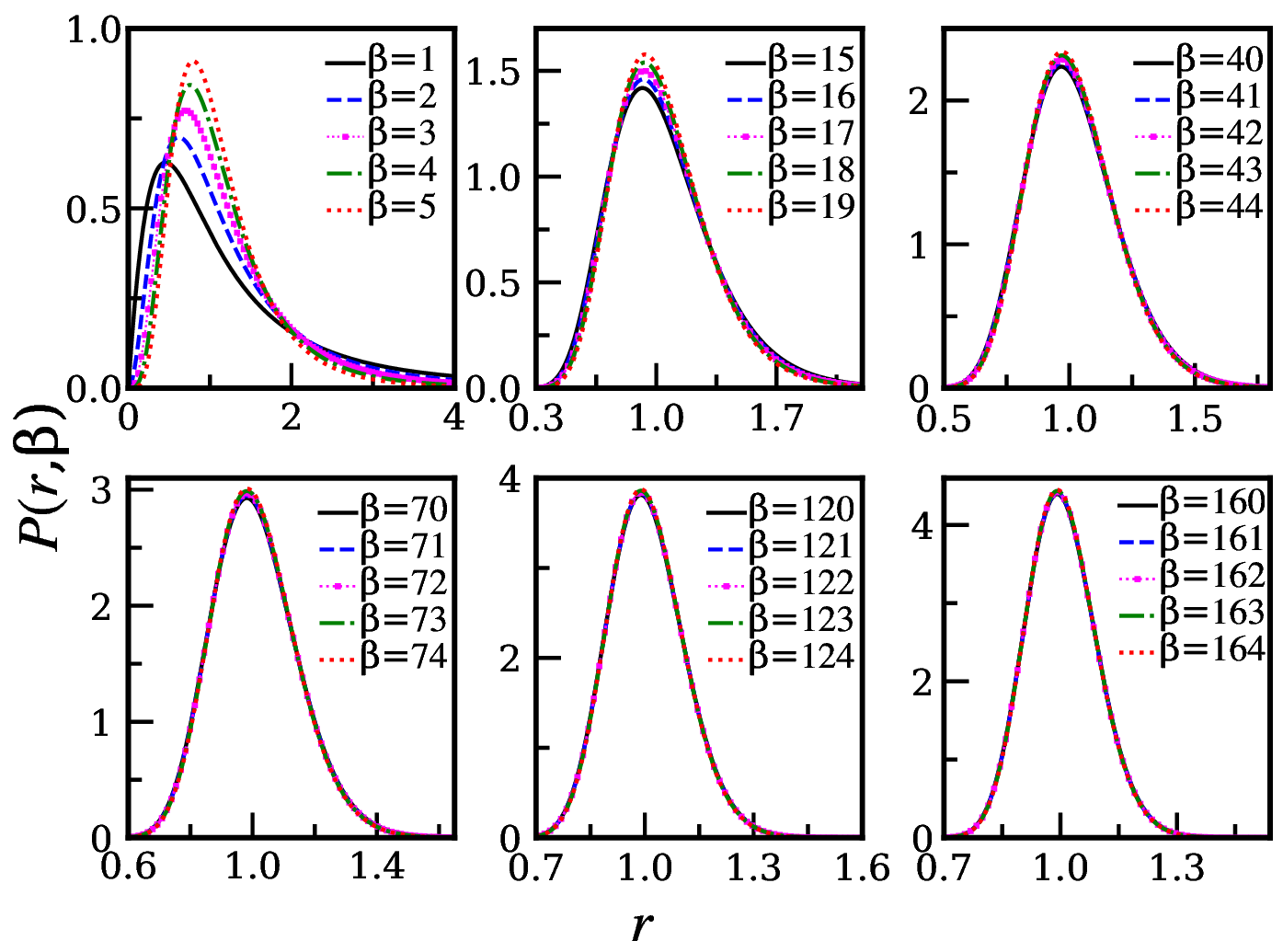}
\caption{\justifying Plot of $P(r,\beta)$ as per Eq.~(\ref{Eq:PRBeta}) for various $\beta$.}
\label{fig:SpacingRatioFuncn}
\end{center}
\end{figure}

In this subsection, we have plotted the analytical distribution functions of both spacing and spacing ratio given by Eqs.~(\ref{Eq:PSBeta}) and (\ref{Eq:PRBeta}) respectively, for various values of $\beta$. The motivation for this study comes from our results for higher values of $k$. Because the results there show deviations from the predicted values in all the cases that we have studied (for example, refer Table~\ref{Table: COEAndGOEDimAnalysistable} and Table~\ref{Table:GunsonResulttable}). In support of explanation to this, we have plotted and analysed 
Eqs.~(\ref{Eq:PSBeta}) and (\ref{Eq:PRBeta}). Our objective here is to easily visualize these distributions and understand their variation with $\beta$. These distributions are shown in Figs.~\ref{fig:SpacingFuncn} and \ref{fig:SpacingRatioFuncn}. Here, we can see that as we increase $\beta$, the widths of the plots are getting narrower, they become sharper, and peak around one.

The plots seem to be very close to each other for higher values of $\beta$. 
It becomes more and more difficult to differentiate plots in the neighborhood of a given $\beta$, as $\beta$
increases (see Figs.~\ref{fig:SpacingFuncn} and \ref{fig:SpacingRatioFuncn}).
We believe that it is difficult for any numerical approach to find the accurate best fit corresponding to such a larger $\beta$. 

\section{Summary and Conclusions}
\label{sec:SummConclsn}
In this paper, HOS distributions of the superposed spectra of circular random matrices is studied. Currently, there are no such studies available on this topic. Here, we have studied HOS distributions in $m=2$ to $m=7$ circular random matrices of the same class and same dimension. The values of $\beta'$ for various $k$ are tabulated, for which the corresponding distribution $P(s, \beta')$ is the best fit according to the numerical method, the sum of difference between the cumulative distributions corresponding to the data and the theoretical distribution function, denoted as $D(\beta')$. We conjecture that for a given $m$ (or $k$) and $\beta$, the obtained sequence of $\beta'$ (using the method $D(\beta')$) as a function of $k$ (or $m$) is unique. Our results can act as an additional litmus test not only to predict the true fluctuation characteristics but also to determine the number of symmetry blocks present in the Hamiltonian matrix, when the dimensions of the blocks are equal or nearly equal and they belong to the same RMT symmetry class. As a consequence, we can determine whether the system is time-reversal invariant or not (with or without rotational symmetry and spin configuration of the system) and get the idea of symmetry structure of the system. 

It is observed that for a particular $k$ and $m$, the value of $\beta'$ is the largest for CSE and the smallest for COE. For a particular $\beta$ and $k$, as we increase $m$, the value of $\beta'$ decreases. After certain $m$, the distribution starts deviating from the generalized Wigner-Dyson distribution and tend towards the corresponding $k$-th order Poisson statistics. Thus, we conjecture that for a given $k$ and circular ensemble with Dyson index $\beta$, the distributions of spacing and ratio tend to the corresponding $k$-th order Poisson statistics as $m$ tends to infinity. Hence, the spectral statistics of three ensembles are the same as $m$ tends to infinity. 
For a particular $m$ and $\beta$, the value of $\beta'$ increases with an increase in $k$, and for lower values of $k$, most of the values of $\beta'$ are positive non-integers. In the case of CSE, the amount of deviation from the analytical distribution function is the largest among the three ensembles, and this occurs for some specific values of $m$ and $k$ that can be easily observed from the figures. We also observe from the plots that as we increase $m$, the distributions of the ratio for a given $k$ converge faster to the Poisson distribution compared to that of the spacings in all the three circular ensembles.

We have verified our $m=2$\; COE results on a physical system known as QKT. We consider the QKT of various dimensions. There, a good agreement between QKT and the random matrix results for both the distributions of spacing and spacing ratio up to a certain $k$ is observed, and beyond that, deviations of $\pm 1$ or $\pm 2$ are observed in the values of $\beta'$ irrespective of $N$.
We can say that, since QKT is described by a circular ensemble, the results are weakly dependent on $N$. 
We have considered another instance of a physical system known as the intermediate map, which comes under the $m=1$ CUE class. There, we have seen that the agreement between the scaling relation and the results of HOSR distributions hold up to slightly higher $k$ than that of HOS, and this value of $k$ is different for different dimensions. Above that certain $k$, the behavior of the results is found to be non-monotonic with respect to increasing $N$. This observation is made based on the parameters $N$, $\gamma$ and $n$ taken by us. Hence, these studies give insights of the the quantitative idea of the dimensions and behavior of spectral statistics of these systems at such dimensions.

From the earlier studies, the nearest neighbor statistics of both circular and the Gaussian ensembles ($m=1$) are found to be the same in the asymptotic limit. Also, both HOS and HOSR distribution for the $m=1$ case of COE and GOE follow the same scaling relation as per the Eq.~(\ref{Eq:ScalingRelation1}). Here, we study these two aspects for higher-order statistics (both HOS and HOSR distributions) in the $m=1$ and $m=2$ case of both COE and GOE in two ways. One is by varying the dimensions, keeping the number of realizations constant, and the other is by varying the number of realizations, keeping dimensions constant. This is to understand whether these two approaches affect the results in the same way. We find numerically that both HOS and HOSR distributions of COE and HOS distributions of  GOE are weakly dependent on $N$ beyond a certain $N$. But, HOSR distributions of GOE are dependent on $N$, and as $N$ increases significantly, they tend towards the HOSR distributions of COE, as far as our results are concerned. It is also found that the results of HOSR distributions of unfolded eigenvalues of GOE are the same as that of HOSR distributions of eigenvalues (without unfolded) of COE. At times, for higher values of $k$, there are a difference of $\pm 1$ in the values of $\beta'$, which can be neglected. 

But, we find from our numerical results that both spacing and ratio distributions in the $m=1$ case of both COE and GOE follow the same scaling relation as per the Eq.~(\ref{Eq:ScalingRelation1}) up to some $k$, and beyond that, they start deviating from each other within each ensemble. The ratio follow the same relation for a bit larger $k$ than the spacings. But as $k$ increases, we can see that the $\beta'$ for ratio are found to be highly deviated from the scaling relation than the corresponding spacings. Also, for both $m=1$ and $m=2$ case of COE and GOE, the deviation in results ($\beta'$) among the spacing and ratio increase with $k$ within each ensemble. Hence, we can say that the corresponding higher-order spectral statistics (HOS and HOSR distributions) of both COE and GOE are the same in the asymptotic limit across these ensembles, but within each ensemble, the corresponding HOS and HOSR distribution results are not the same beyond a certain $k$.

Also, for a particular $N$, the results remain unaffected beyond a certain $n$. Another interesting finding is that even if the number of eigenvalues are same, either by large dimension and small number of realizations or by small dimension and large number of realizations, the results (the obtained values of $\beta'$) are not the same in both cases. Rather, a larger $N$ and a fixed $n$ is preferred.

We observe small deviations in all of the studies concerning HOS from the scaling relation at larger $k$, the small deviations for the cases where we verify our results on physical systems, and in the verification of the Gunson's result on the COE-CUE correspondence. To understand this, we have plotted analytical distributions of both spacing and ratio and studied them as a function of $\beta$. There, it can be seen that as $\beta$ increases, it is becoming difficult to distinguish the functions in the neighborhood of a given $\beta$. 
As a result, we claim that these deviations in results might be due to the statistical fluctuations or computational precision error, or the limitation of the numerical method $D(\beta')$ used. But, for any numerical approach, this will become challenging for getting the accurate $\beta'$ (the best fit as per the scaling relation) for such a large $k$. This led to a further question: Is it possible for any other numerical method, which would be able to give the expected $\beta'$ (as per the scaling relation) for higher $k$ by taking our results into consideration? Also, the significant deviations in the HOSR ($m=1$ case) results of COE and GOE (for larger $k$) from the scaling relation needs further analysis.

Hence, while applying the obtained RMT results to physical systems,
the results agree with each other up to some lower values of $k$ (generally $k \leq 4$, but that may vary from system to system). Above that $k$, the results start to differ since the fluctuation measures are now probing the 
global spectral features. These values of $k$ will also be different for different fluctuation measures. Also, for some systems, they depend on $N$ and for others, they do not. It is possible from our results, even for these lower values of $k$, to characterize the system correctly by adopting our numerical method $D(\beta')$. Because, for all three classes and various $m$, the results, i.e., the sequences of obtained $\beta'$, are unique.

Our numerical studies have opened up new future directions, which are discussed as follows. The full analytical derivation for the HOSR distribution in the bulk for $m \geq 1$ cases needs to be derived. The analytical derivation of the HOS distribution of superposed random matrices is too warranted. Here, we have studied the superposition of matrices of the same dimensions and of the same symmetry classes. It will be interesting to study the superposition of matrices of various dimensions of the same or different symmetry classes \cite{yan2025spacing} along with other directions as mentioned in \cite{bhosale2021superposition}. Our conjecture, based on numerics, on the higher order spectral fluctuations tending to the Poisson distribution as the number of superposed blocks tends to infinity can be addressed analytically \cite{tkocz2012tensor}.

\section{Acknowledgments}
\label{acknowledgement}
We acknowledge National Super computing Mission (NSM) for providing computing resources of ‘PARAM SMRITI’ at NABI, Mohali, which is implemented by C-DAC and supported by the Ministry of Electronics and Information Technology (MeitY) and Department of Science and Technology (DST), Government of India. We thank M. S. Santhanam for his cooperation in getting access to the same. We are grateful to the Department of Science and Technology (DST) for their generous financial support through sanctioned Project No. SR/FST/PSI/2017/5(C) to the Department of Physics of VNIT, Nagpur. We also thank R. Prakash for the useful discussions. S. R. would like to thank A. Purohit, H. Sharma, P. Solanki and N. Patra for their help with system level issues, as well as familiarizing with various software and computational tools.

%% file: SUPPLEMENT.tex
\vspace*{1cm}
\onecolumngrid

\begin{center}
\textbf{\large Supplementary Material for\\ 
``\textit{Higher-order spacings in the superposed spectra of random matrices with comparison to spacing ratios and application to 
complex systems}''}
\end{center}

\tableofcontents

\newpage

\suppsection{Introduction}
\label{intro}
This Supplementary Material presents a few more figures corresponding to our higher-order spacing (HOS) distributions in the superposed spectra of matrices of all three classes of circular ensemble. The HOS and higher-order spacing ratio (HOSR) distributions of eigenangles of the intermediate map and the quantum kicked top model are plotted for some values of $k$. These plots are additional support for our results presented in the main text of this paper. Further, we present the results in tabular form corresponding to dimensional analysis, along with the effect of the number of realizations. We have added two more sections, one for figures corresponding to the convergence to the Poisson distribution, and the other for tables corresponding to the Kolmogorov-Smirnov test for only one realization of the superposed spectra.

\begin{figure}[H]
\begin{center}
\includegraphics*[scale=0.33]{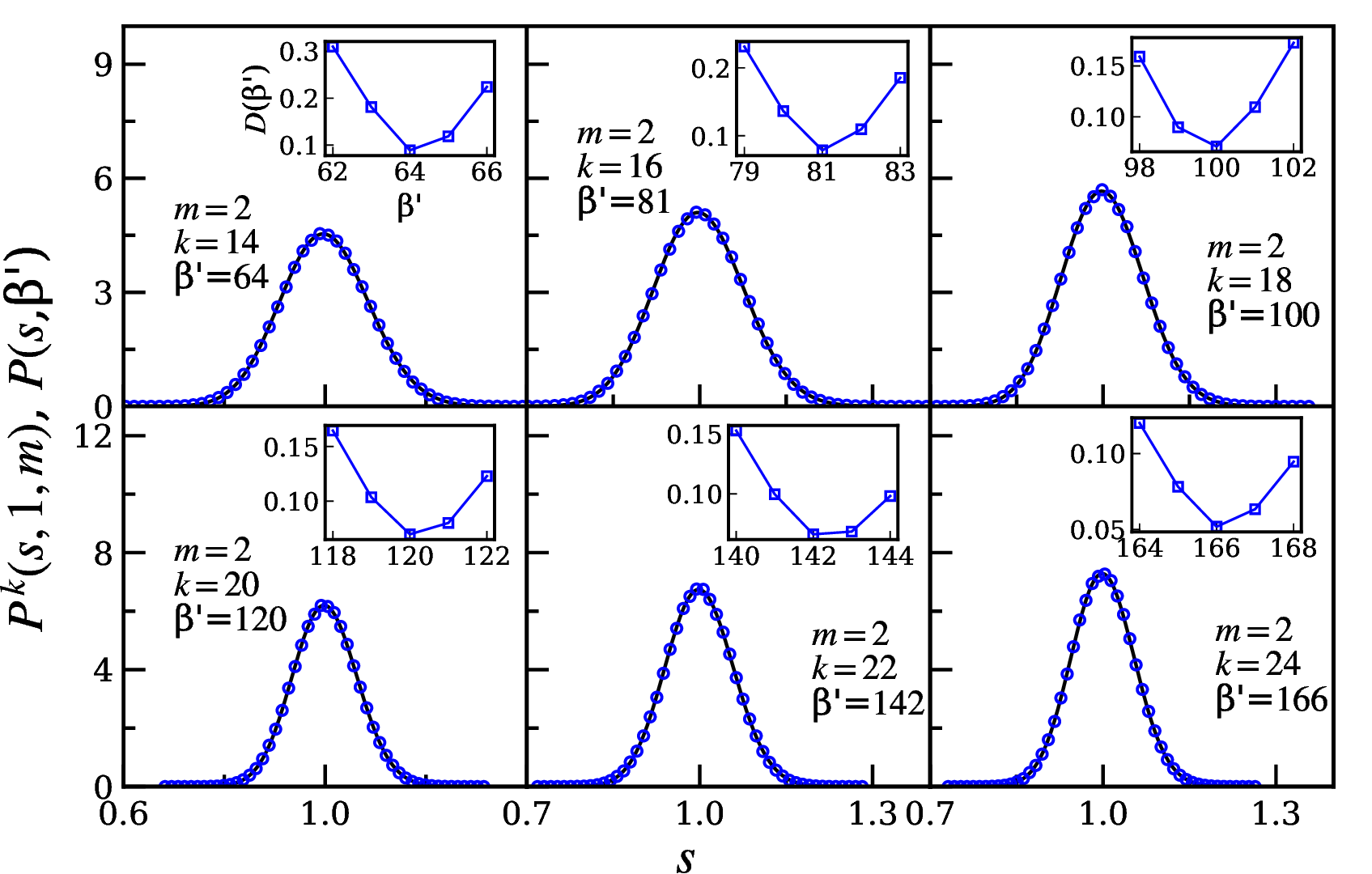}
\caption{\justifying Distribution of $k$-th order spacings $P^{k}(s, 1, m)$ in the $m= 2$ COE spectra (circles). Here, the dimension of the matrix without superposition is $N= 5000$, and the number of realizations without superposition is $n= 600$. The solid curve corresponds to $P(s,\beta')$ as per the equation $P(s,\beta)=A_{\beta}s^{\beta}\exp(-C_{\beta}s^2)$, in which $\beta$ is replaced by $\beta'$ and $\beta'$ is given in the main text. The insets show $D(\beta')$ as a function of $\beta'$.}
\label{fig:k_14_16_18_20_22_24_m2_COE}
\end{center}
\end{figure}

\begin{figure}[H]
\begin{center}
\includegraphics*[scale=0.33]{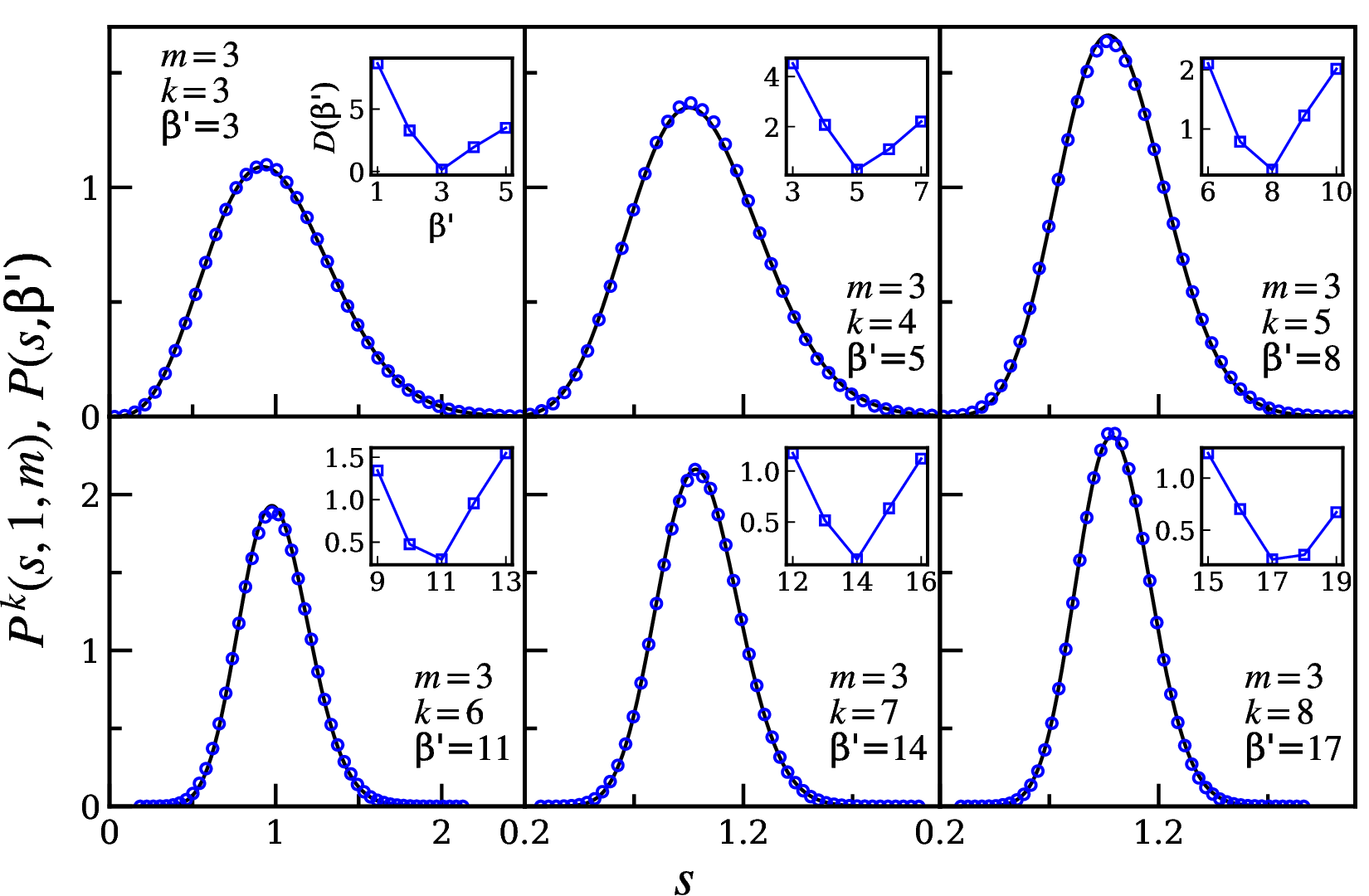}
\caption{\justifying Same as Fig.~\ref{fig:k_14_16_18_20_22_24_m2_COE} but for $m= 3$, $n= 900$, and different values of $k$ and $\beta'$.}
\label{fig:k_3_to_8_m3_COE}
\end{center}
\end{figure}

\begin{figure}[H]
\begin{center}
\includegraphics*[scale=0.335]{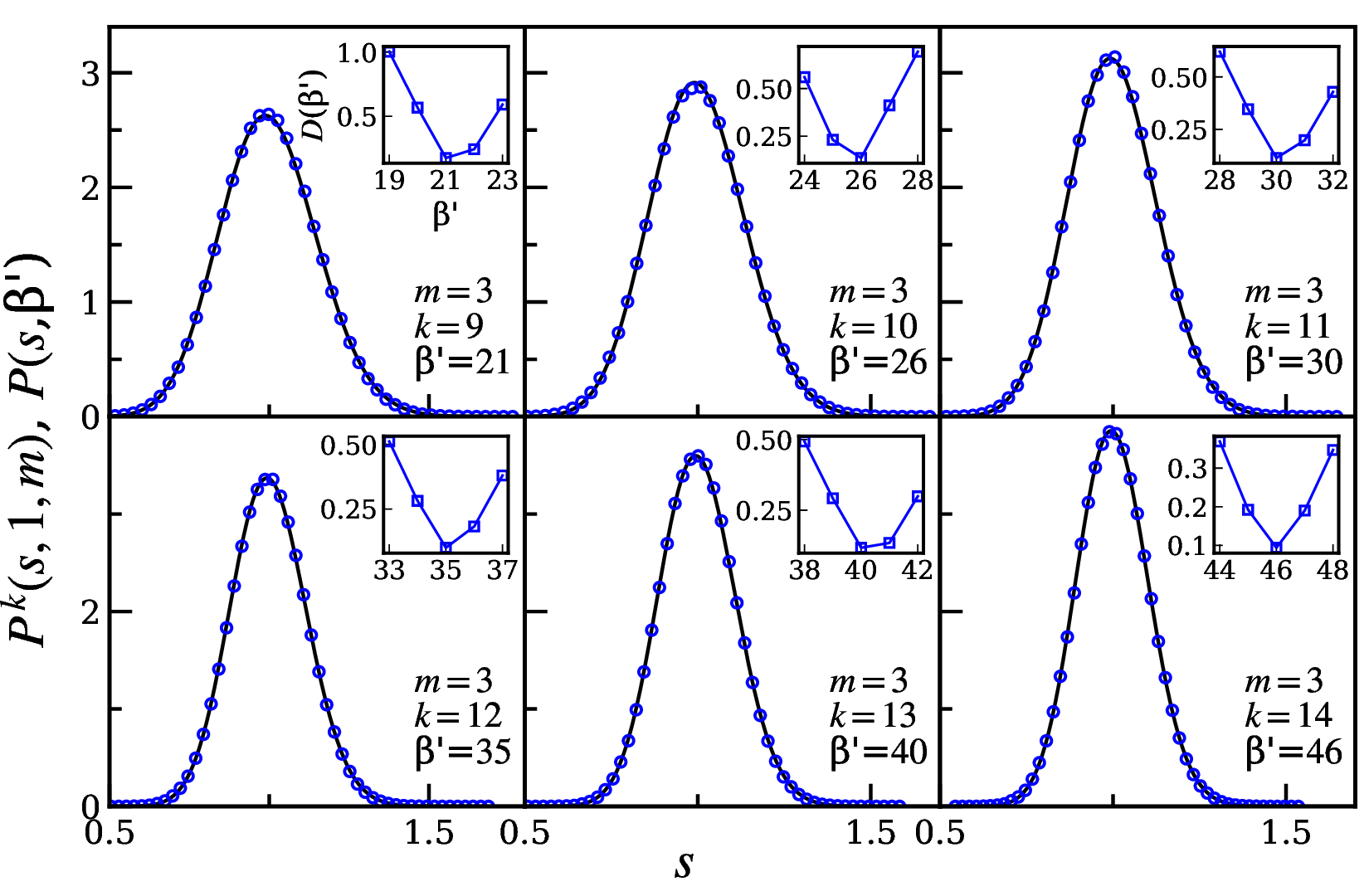}
\caption{\justifying Same as Fig.~\ref{fig:k_3_to_8_m3_COE} but for different values of $k$ and $\beta'$.}
\label{fig:k_9_to_k_14_m3_COE}
\end{center}
\end{figure}

\begin{figure}[H]
\begin{center}
\includegraphics*[scale=0.335]{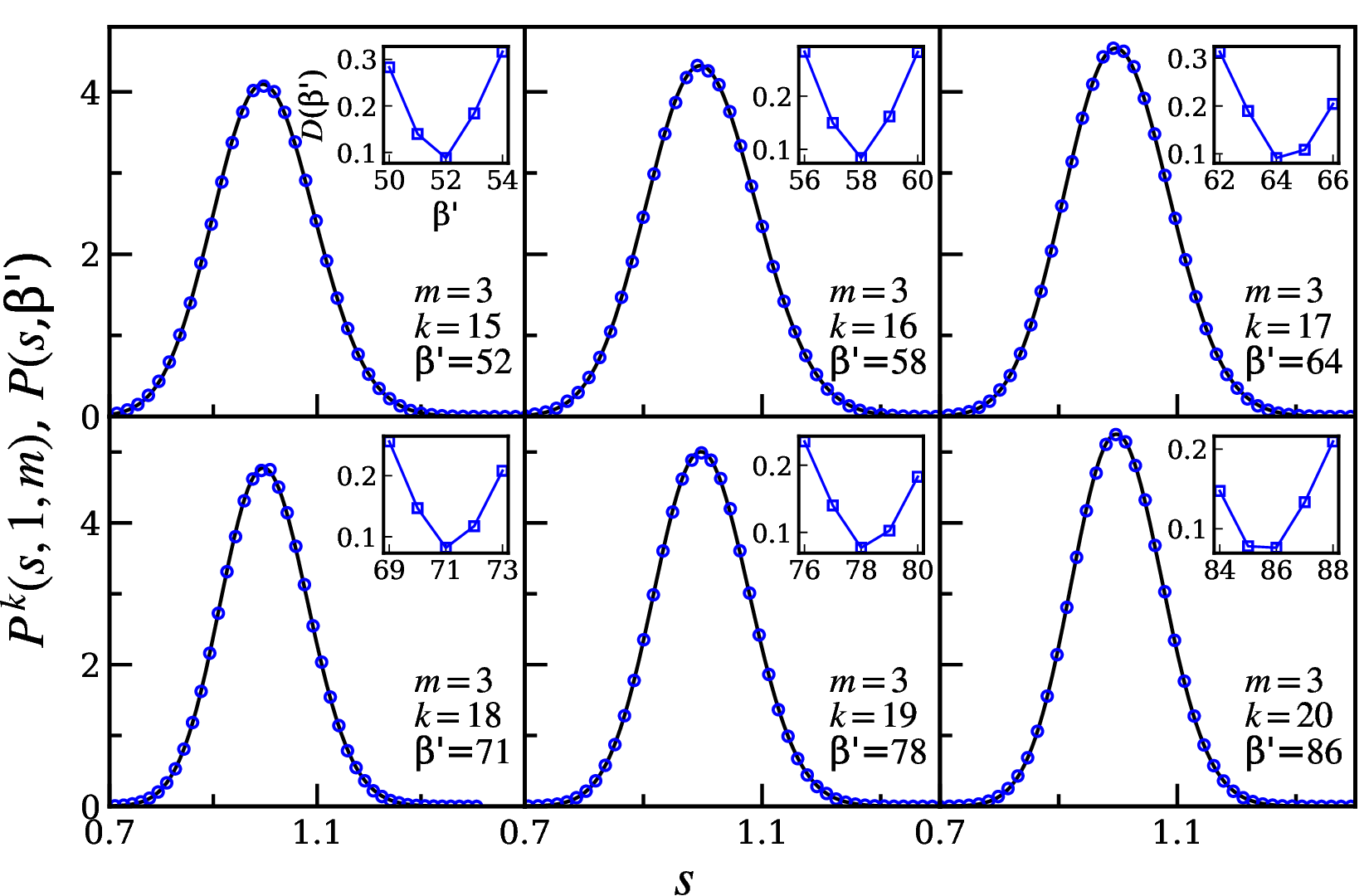}
\caption{\justifying Same as Fig.~\ref{fig:k_3_to_8_m3_COE} but for different values of $k$ and $\beta'$.}
\label{fig:k_15_to_k_20_m3_COE}
\end{center}
\end{figure}

\begin{figure}[H]
\begin{center}
\includegraphics*[scale=0.335]{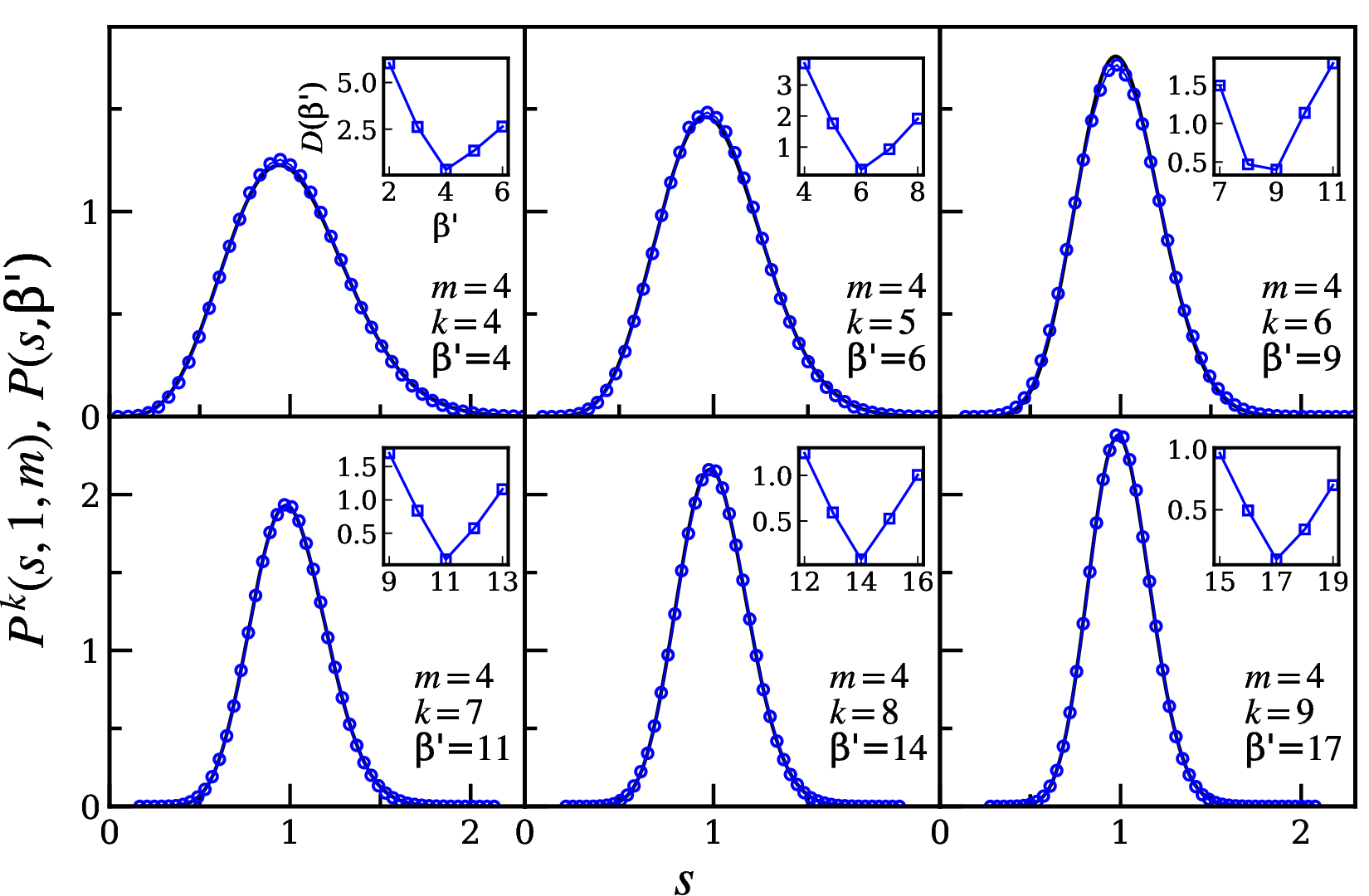}
\caption{\justifying Same as Fig.~\ref{fig:k_14_16_18_20_22_24_m2_COE} but for $m= 4$, $n= 1000$, and different values of $k$ and $\beta'$.}
\label{fig:k_4_to_9_m4_COE}
\end{center}
\end{figure}

\begin{figure}[H]
\begin{center}
\includegraphics*[scale=0.335]{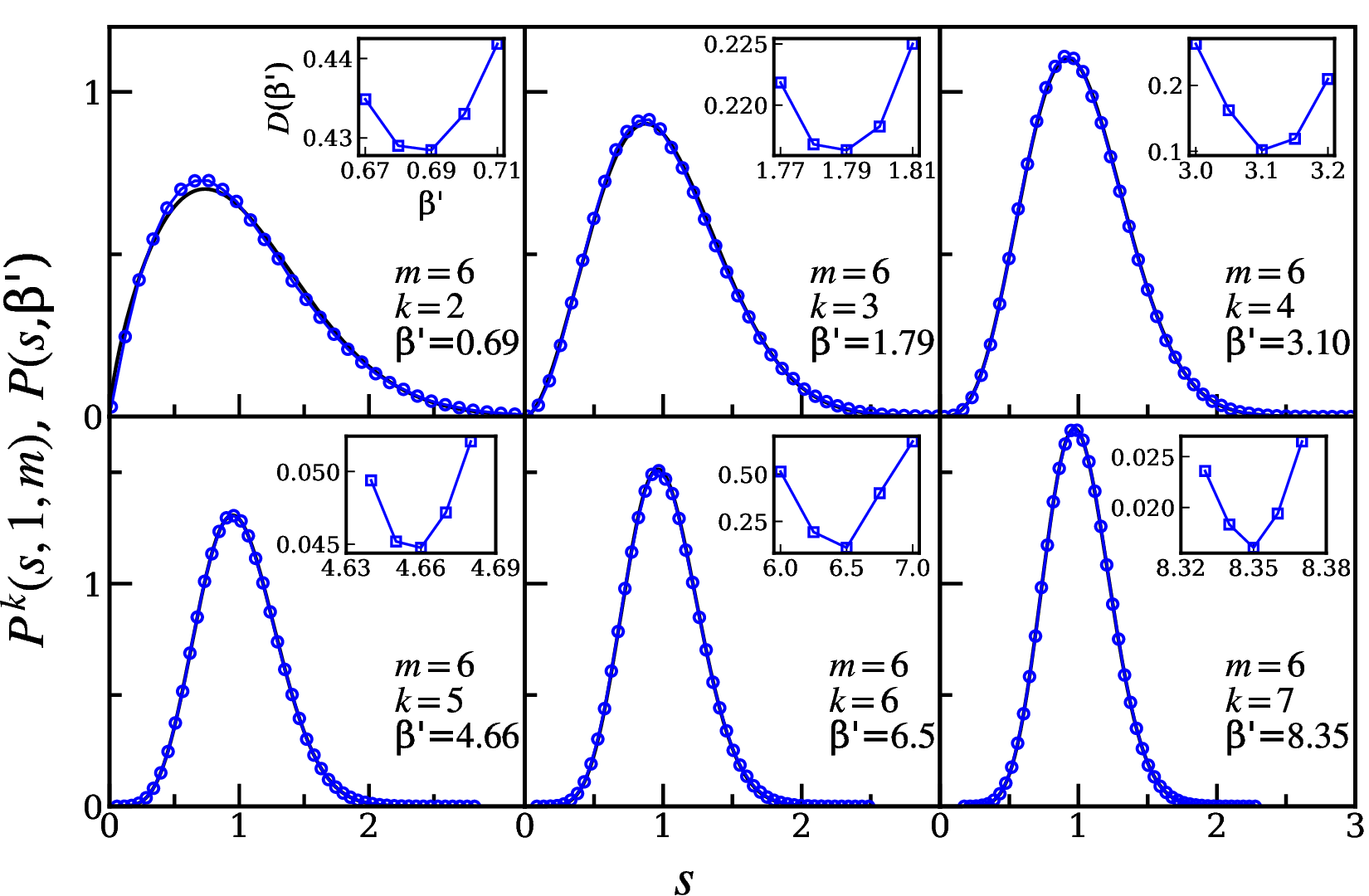}
\caption{\justifying Same as Fig.~\ref{fig:k_14_16_18_20_22_24_m2_COE} but for $m= 6$, $n= 1002$, and different values of $k$ and $\beta'$.}
\label{fig:k_2_to_7_COE_m6}
\end{center}
\end{figure}

\begin{figure}[H]
\begin{center}
\includegraphics*[scale=0.335]{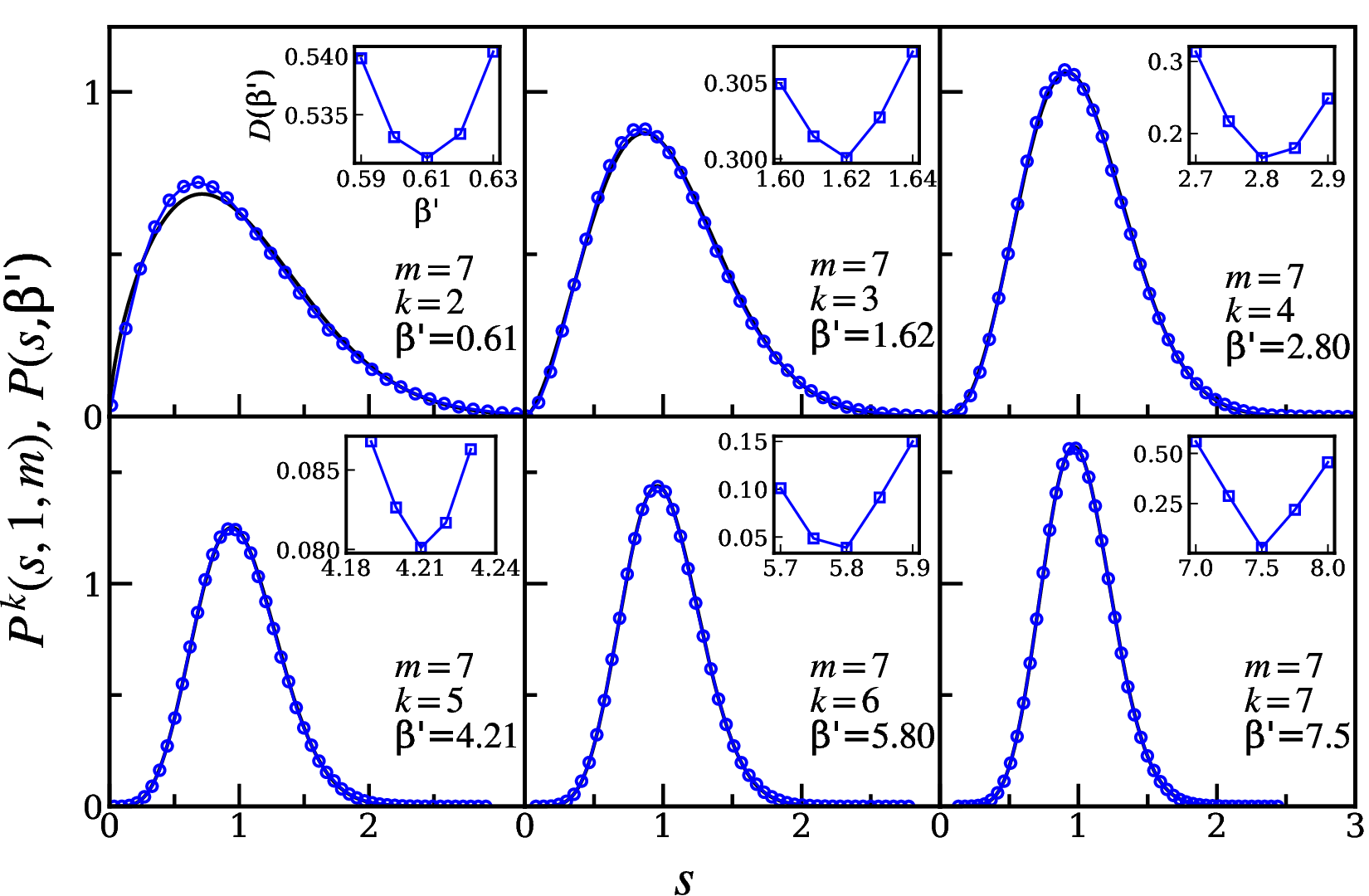}
\caption{\justifying Same as Fig. \ref{fig:k_14_16_18_20_22_24_m2_COE} but for $m= 7$, $n= 1001$, and different values of $k$ and $\beta'$.}
\label{fig:k_2_to_7_COE_m7}
\end{center}
\end{figure}

\begin{figure}[H]
\begin{center}
\includegraphics*[scale=0.335]{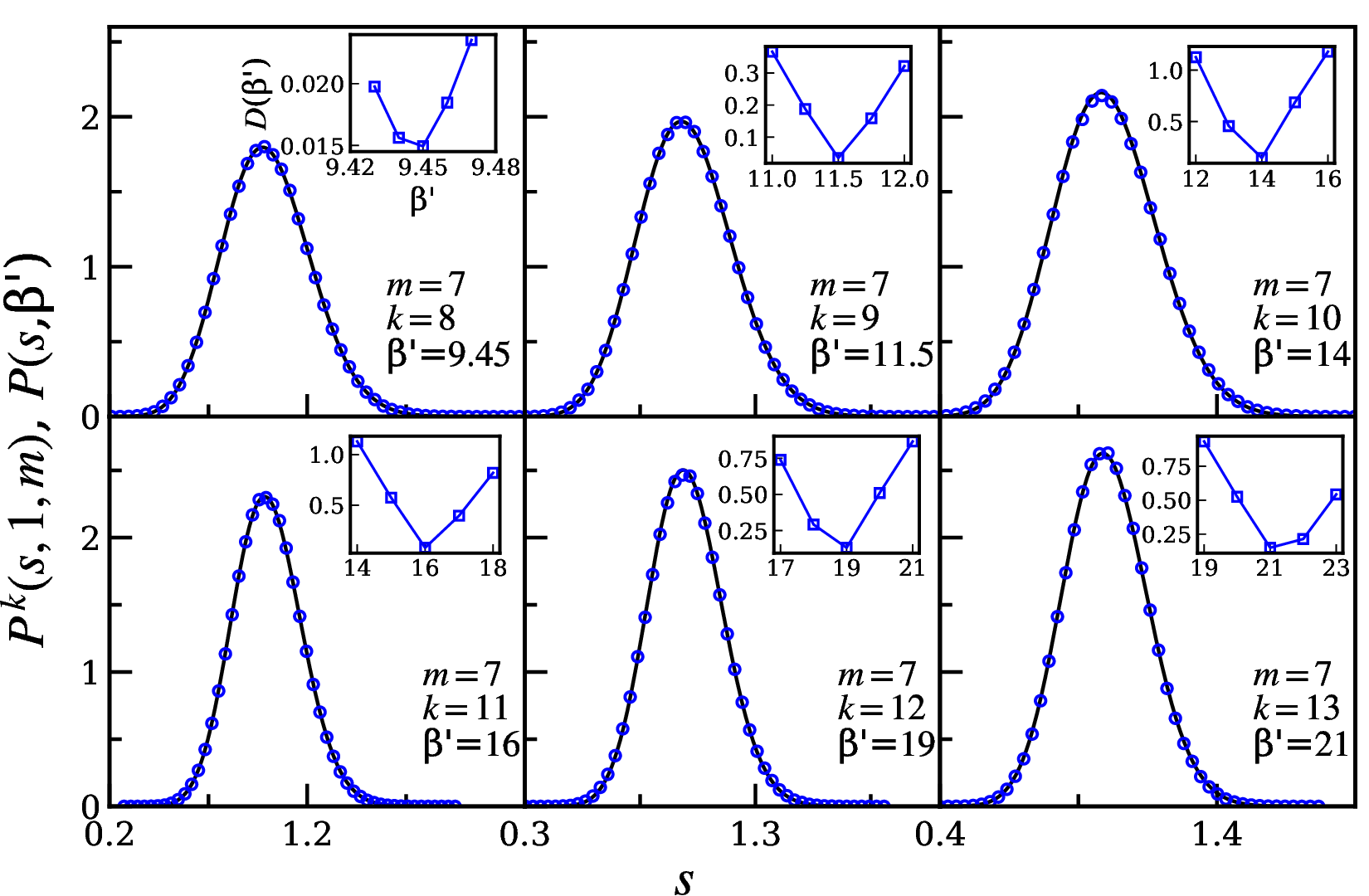}
\caption{\justifying Same as Fig. \ref{fig:k_2_to_7_COE_m7} but for different values of $k$ and $\beta'$.}
\label{fig:k_8_to_13_COE_m7}
\end{center}
\end{figure}

\begin{figure}[H]
\begin{center}
\includegraphics*[scale=0.365]{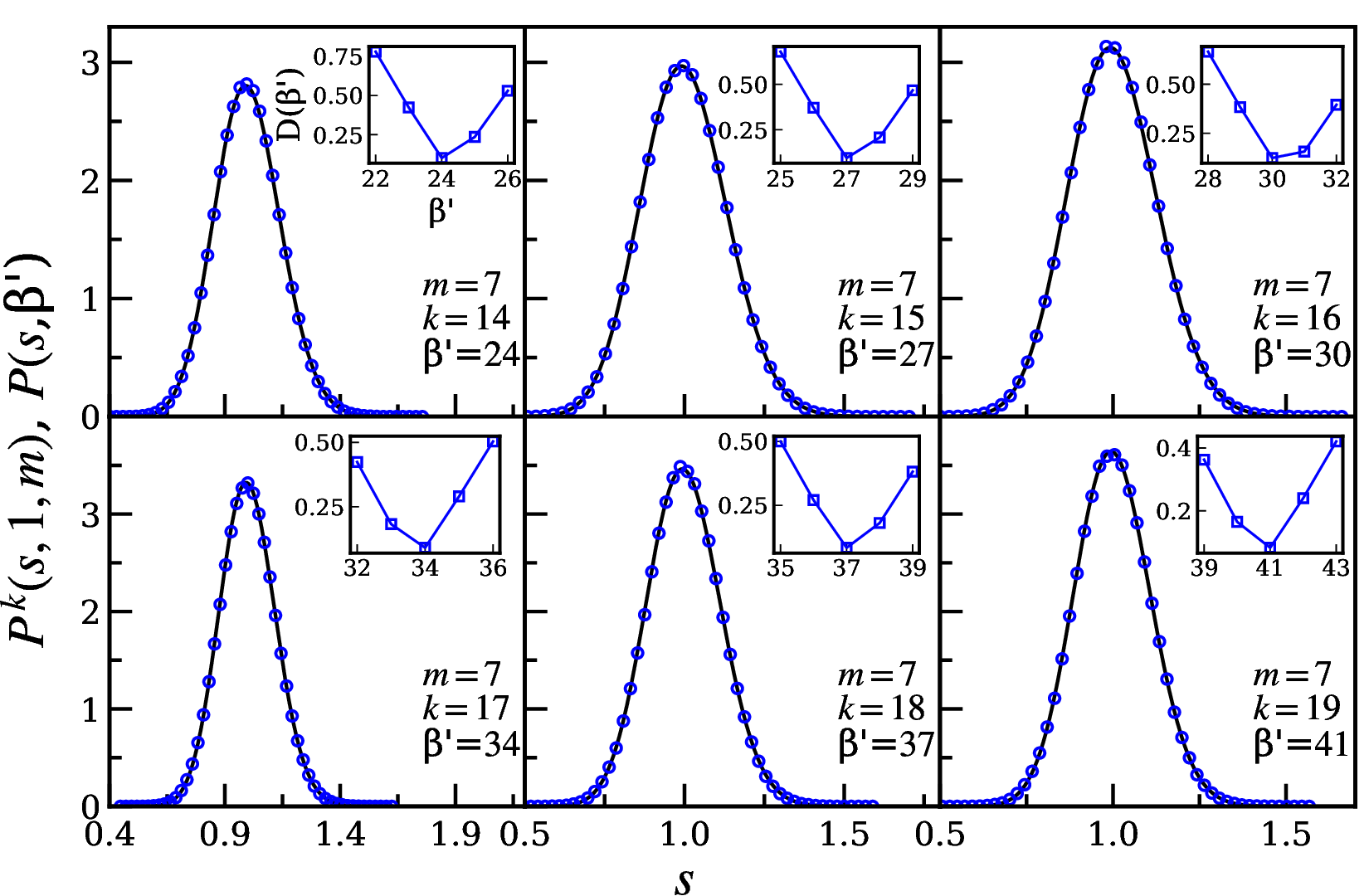}
\caption{\justifying Same as Fig.~\ref{fig:k_2_to_7_COE_m7} but for different values of $k$ and $\beta'$.}
\label{fig:k_14_to_19_COE_m7}
\end{center}
\end{figure}

\begin{figure}[H]
\begin{center}
\includegraphics*[scale=0.365]{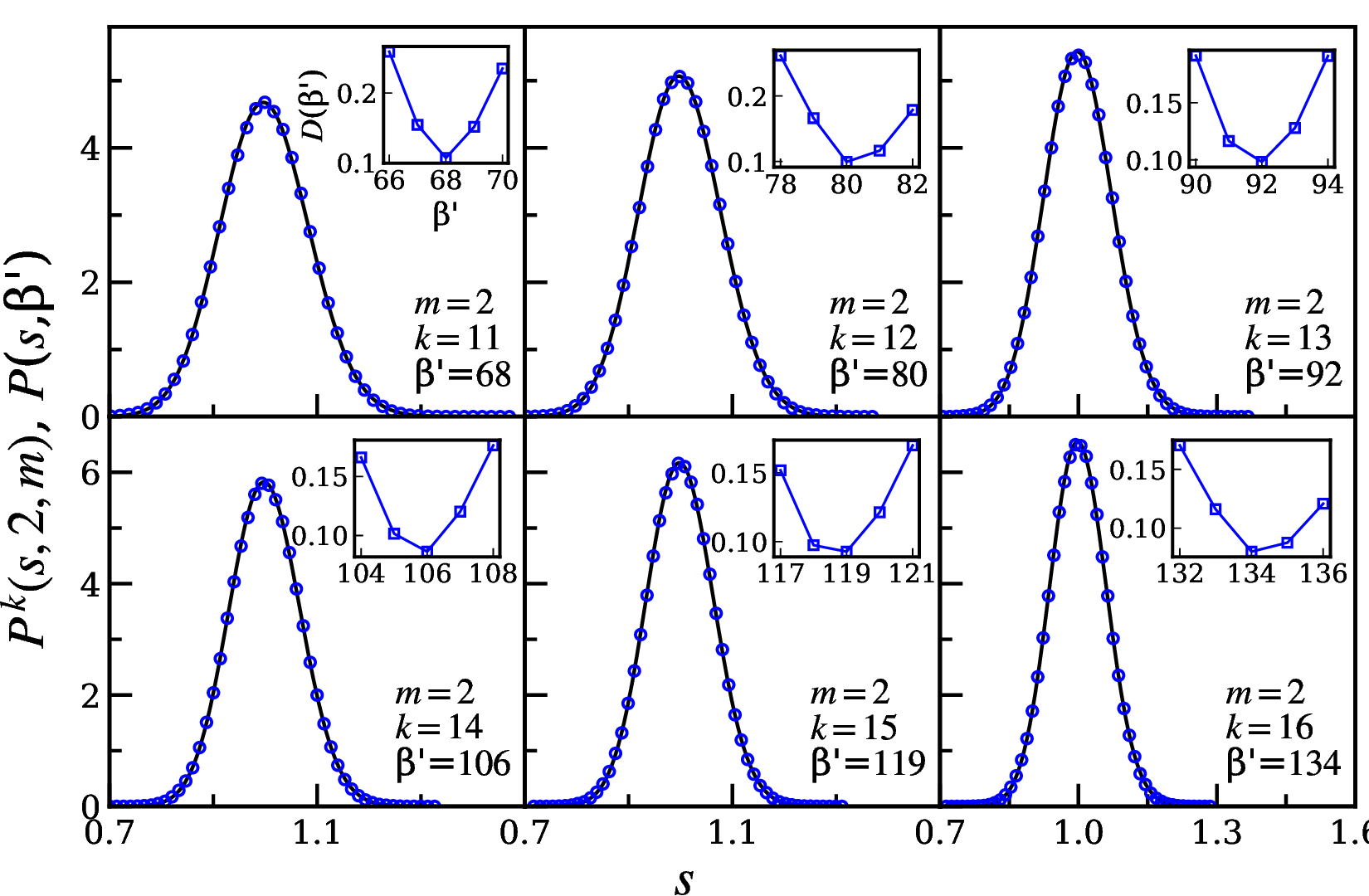}
\caption{\justifying Distribution of HOS $P^{k}(s, 2, m)$ for various $k$ in the $m= 2$ CUE spectra (circles). Here, $N= 5000$ and $n= 600$. The solid curve corresponds to $P(s,\beta')$ as per the equation $P(s,\beta)= A_{\beta}s^{\beta}\exp(-C_{\beta}s^2)$, where $\beta$ is replaced by $\beta'$ and the values of $\beta'$ are given in the main text. The insets show $D(\beta')$ as a function of $\beta'$.}
\label{fig:k_11_to_16_m2_CUE}
\end{center}
\end{figure}

\begin{figure}[H]
\begin{center}
\includegraphics*[scale=0.335]{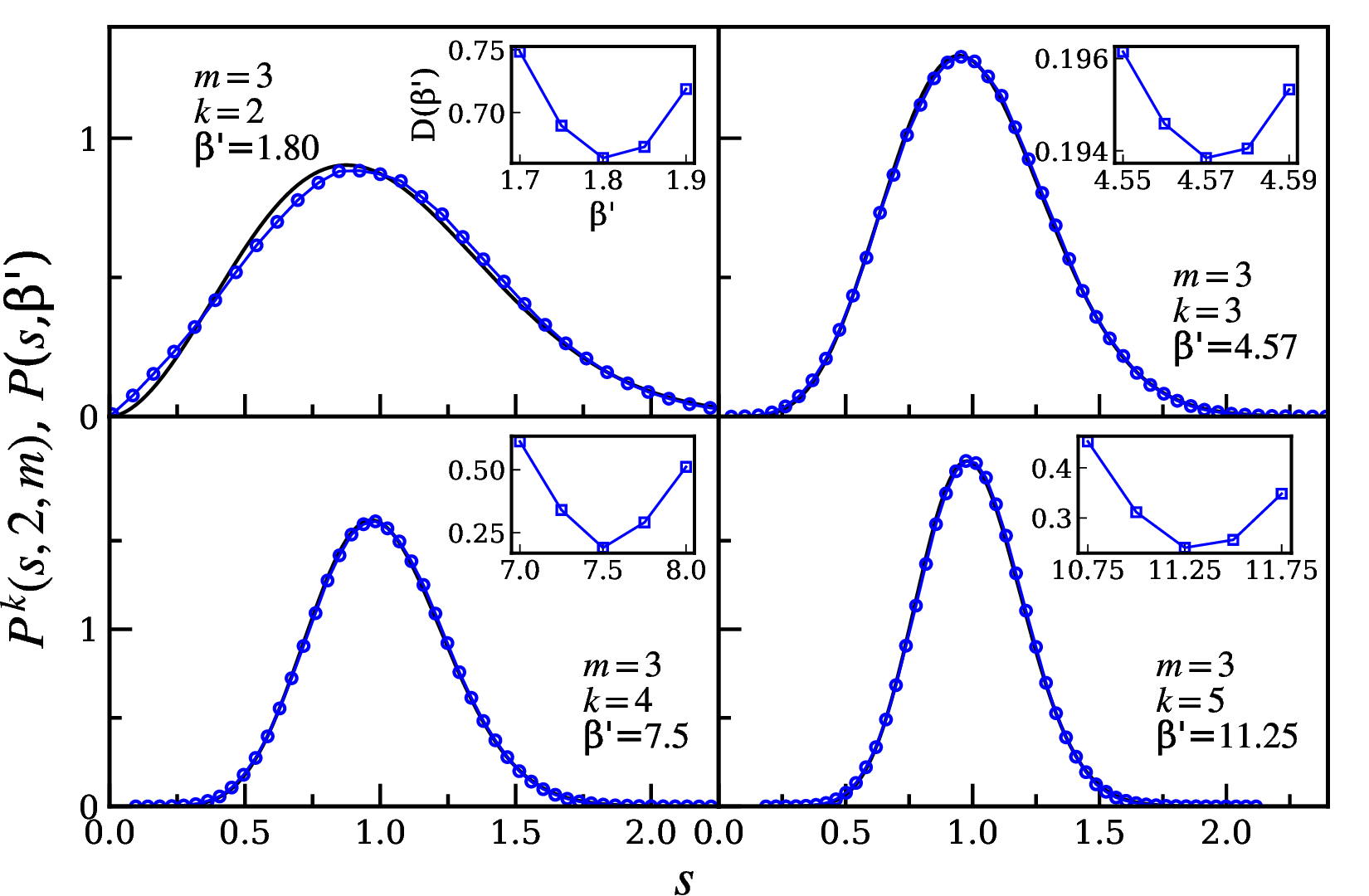}
\caption{\justifying  Same as Fig.~\ref{fig:k_11_to_16_m2_CUE} but for $m= 3$, $n= 900$, and different values of $k$ and $\beta'$ .}
\label{fig:k_2_to_5_CUE_m3}
\end{center}
\end{figure}

\begin{figure}[H]
\begin{center}
\includegraphics*[scale=0.335]{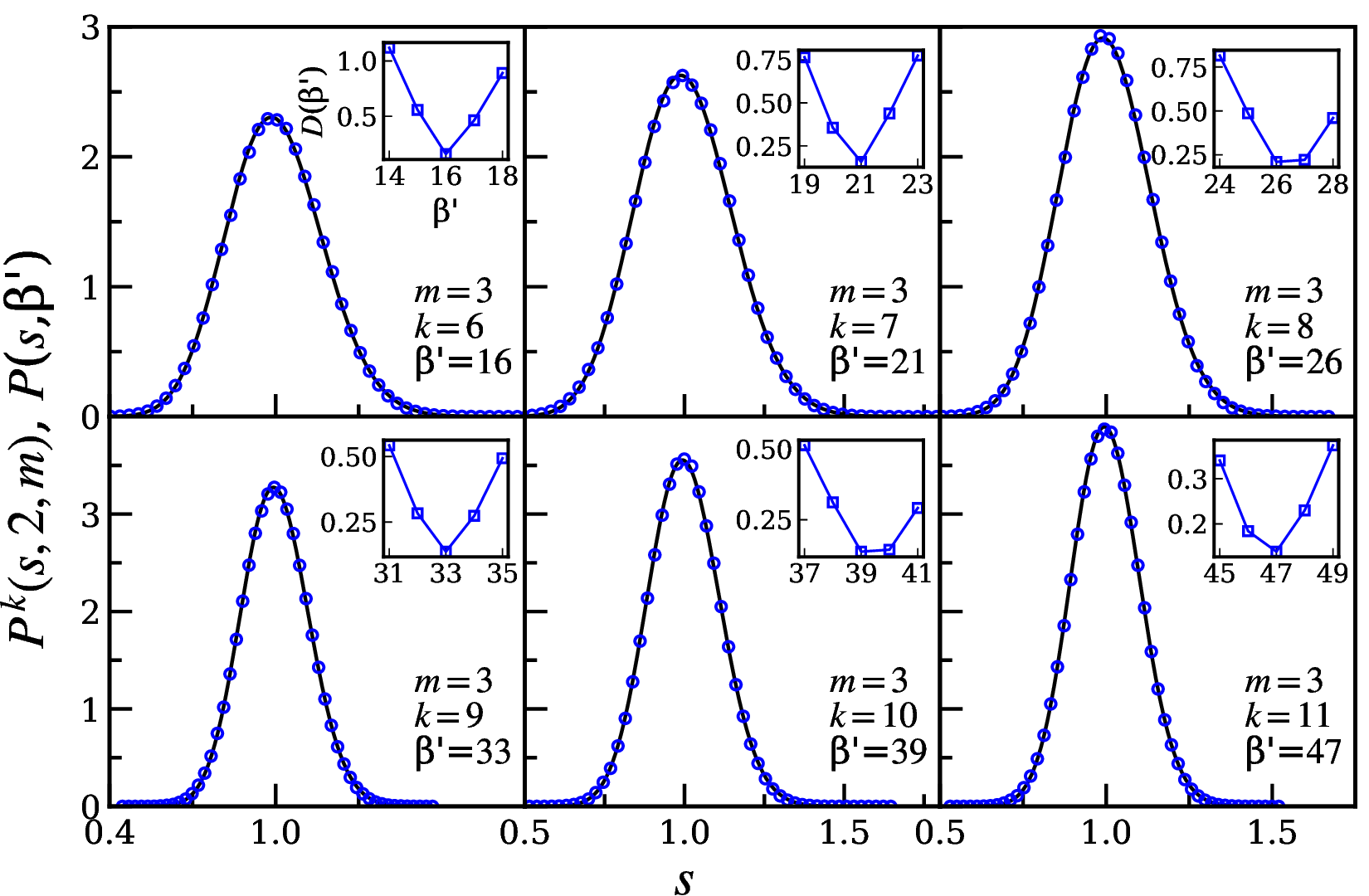}
\caption{\justifying Same as Fig.~\ref{fig:k_2_to_5_CUE_m3} but for different $k$ and $\beta'$.}
\label{fig:k_6_to_11_m3_CUE}
\end{center}
\end{figure}

\begin{figure}[H]
\begin{center}
\includegraphics*[scale=0.335]{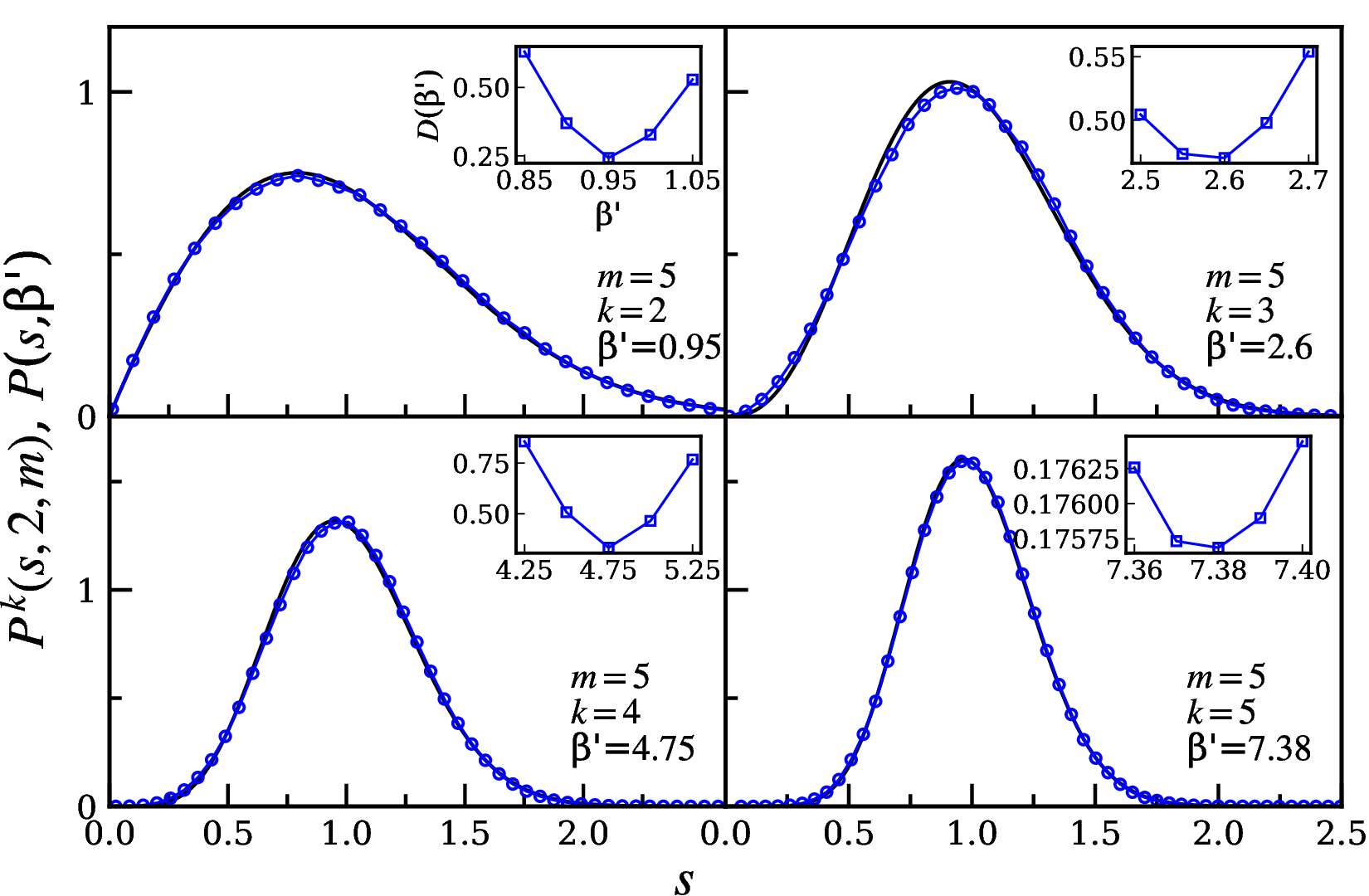}
\caption{\justifying Same as Fig.~\ref{fig:k_11_to_16_m2_CUE} but for $m= 5$, $n= 1000$, and different values of $k$ and $\beta'$.}
\label{fig:k_2_to_5_CUE_m5}
\end{center}
\end{figure}

\begin{figure}[H]
\begin{center}
\includegraphics*[scale=0.335]{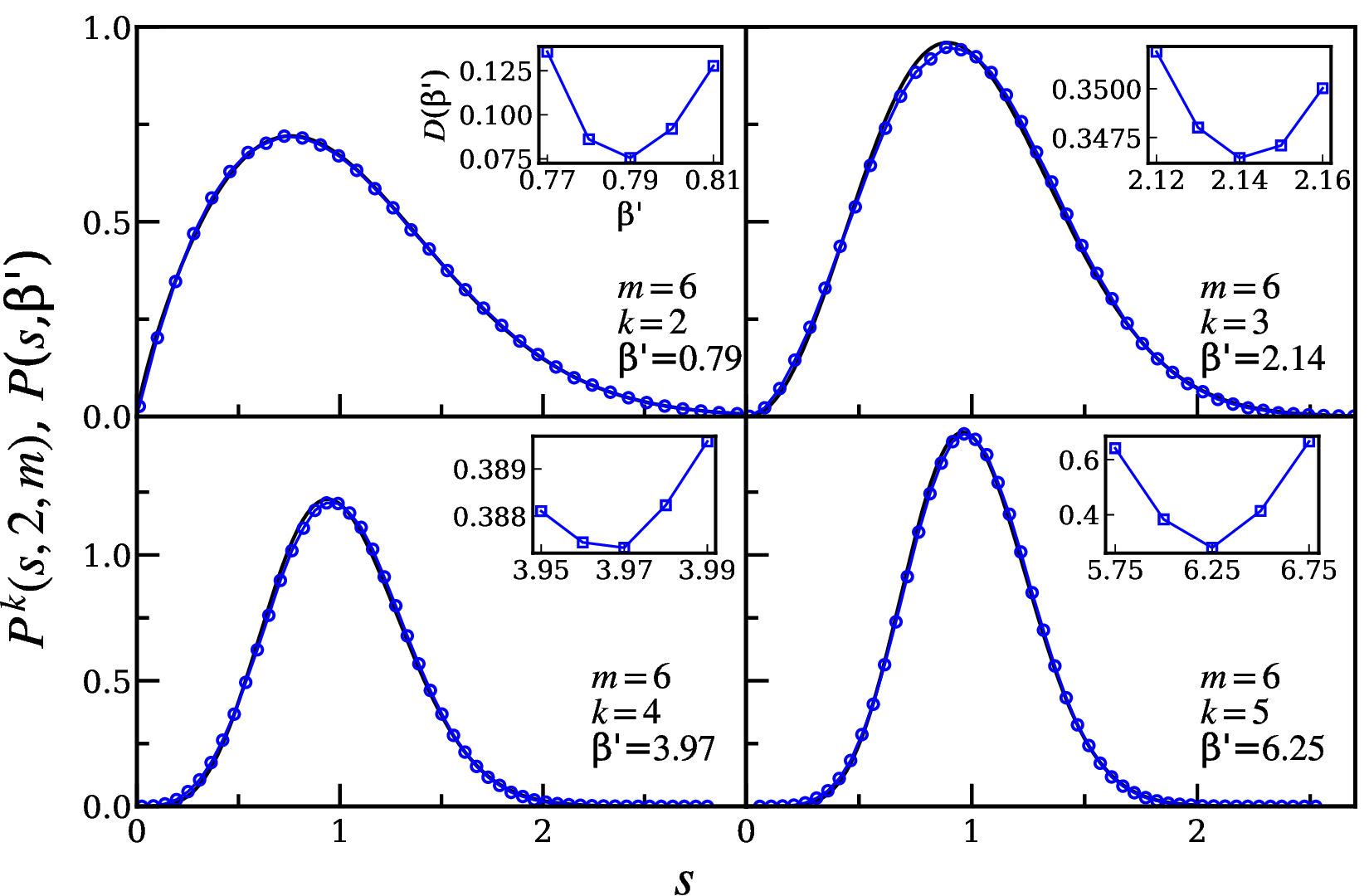}
\caption{\justifying Same as Fig.~\ref{fig:k_11_to_16_m2_CUE} but for $m= 6$, $n= 1002$, and different values of $k$ and $\beta'$.}
\label{fig:k_2_to_5_CUE_m6}
\end{center}
\end{figure}

\begin{figure}[H]
\begin{center}
\includegraphics*[scale=0.335]{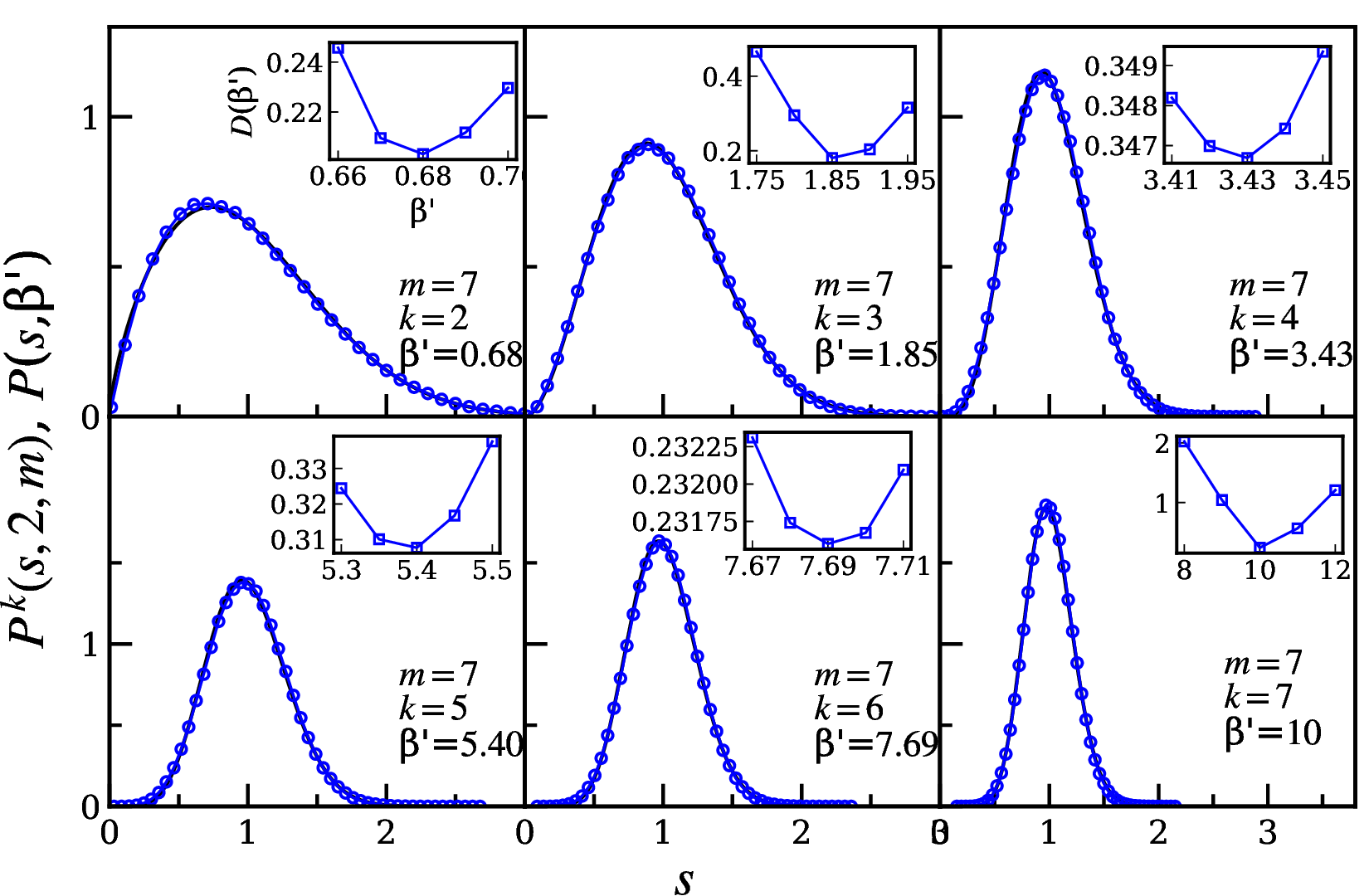}
\caption{\justifying Same as Fig.~\ref{fig:k_11_to_16_m2_CUE} but for $m= 7$, $n= 1001$, and different values of $k$ and $\beta'$.}
\label{fig:k_2_to_7_m7_CUE}
\end{center}
\end{figure}

\begin{figure}[H]
\begin{center}
\includegraphics*[scale=0.335]{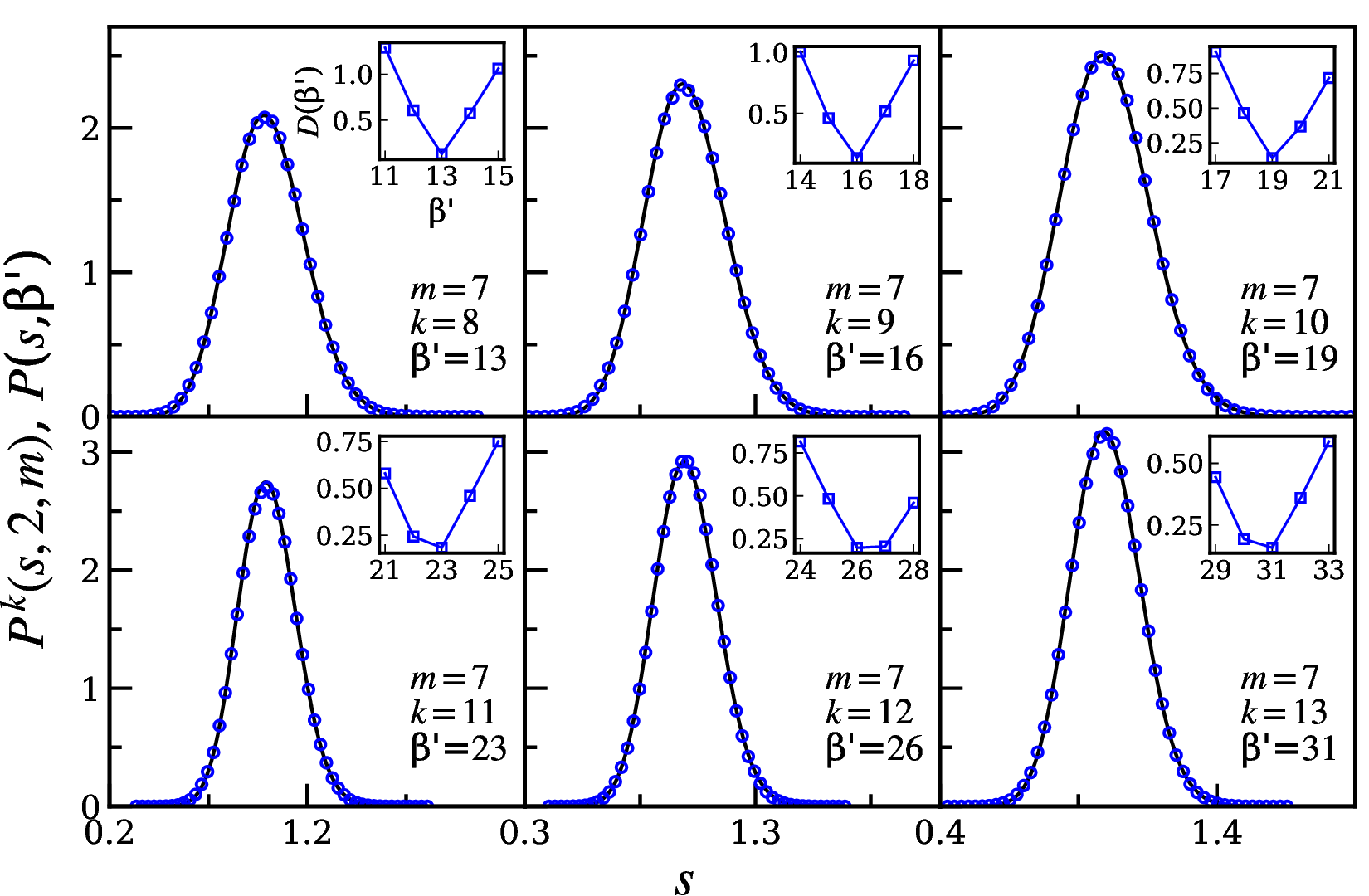}
\caption{\justifying Same as Fig.~\ref{fig:k_2_to_7_m7_CUE} but for different values of $k$ and $\beta'$.}
\label{fig:k_8_to_13_m7_CUE}
\end{center}
\end{figure}

\begin{figure}[H]
\begin{center}
\includegraphics*[scale=0.365]{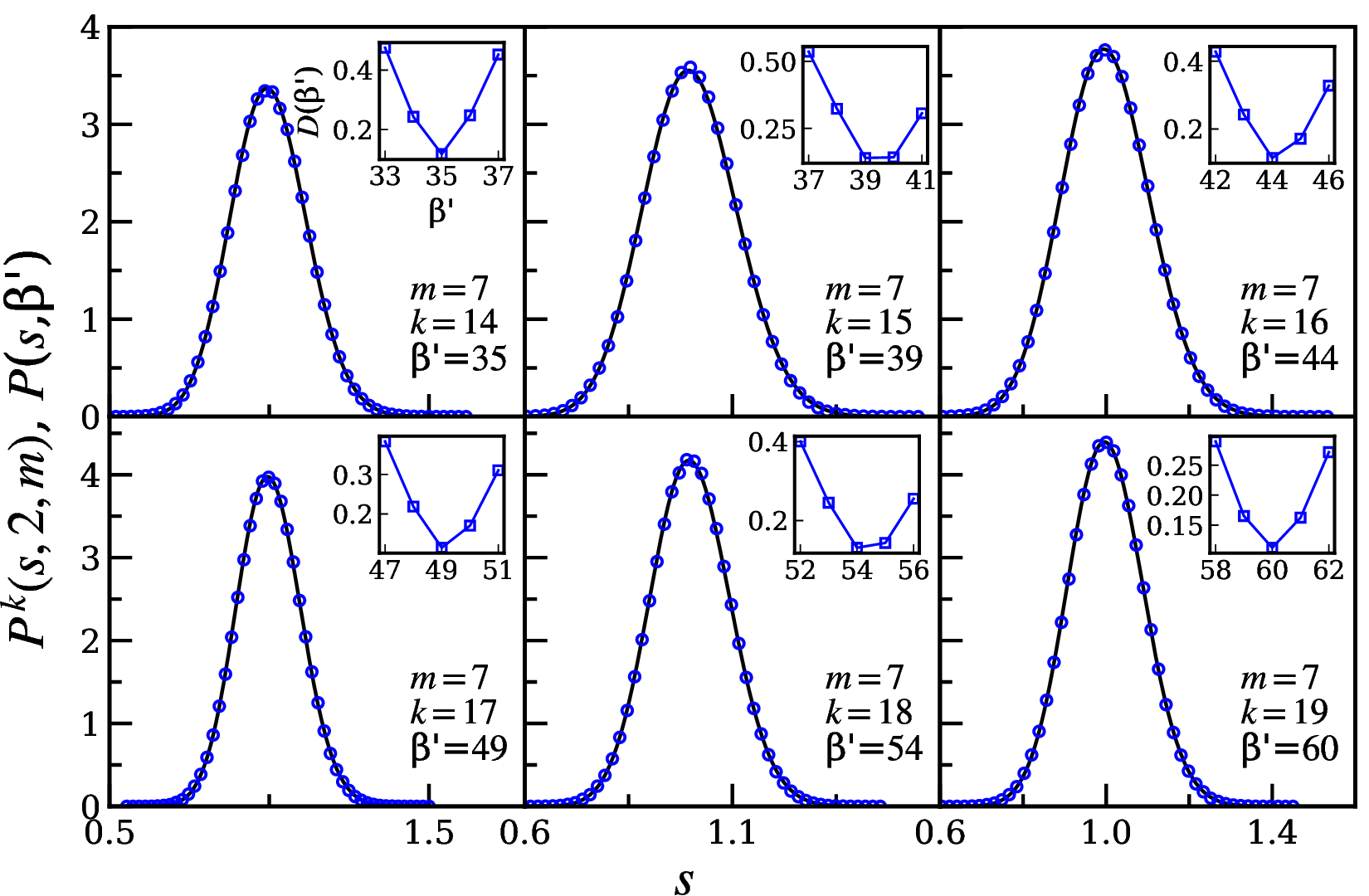}
\caption{\justifying  Same as Fig.~\ref{fig:k_2_to_7_m7_CUE} but for different values of $k$ and $\beta'$.}
\label{fig:k_14_to_19_m7_CUE}
\end{center}
\end{figure}

\suppsection{Illustration of our results through some more plots}
\label{sec:more_plots_CE}
\suppsubsection{The case of COE}
\label{subsec:more_plots_COE}
In this subsection, we have given few more plots of the HOS in the superposed spectra of COE. These are Fig.~\ref{fig:k_14_16_18_20_22_24_m2_COE} to Fig.~\ref{fig:k_14_to_19_COE_m7}.
\suppsubsection{The case of CUE}
\label{subsec:more_plots_CUE}
In this subsection, we have given few more plots of the HOS for the superposed spectra of CUE. These are  Fig.~\ref{fig:k_11_to_16_m2_CUE} to Fig.~\ref{fig:k_14_to_19_m7_CUE}.
\vspace{15.4cm}

\begin{figure}[H]
\begin{center}
\includegraphics*[scale=0.365]{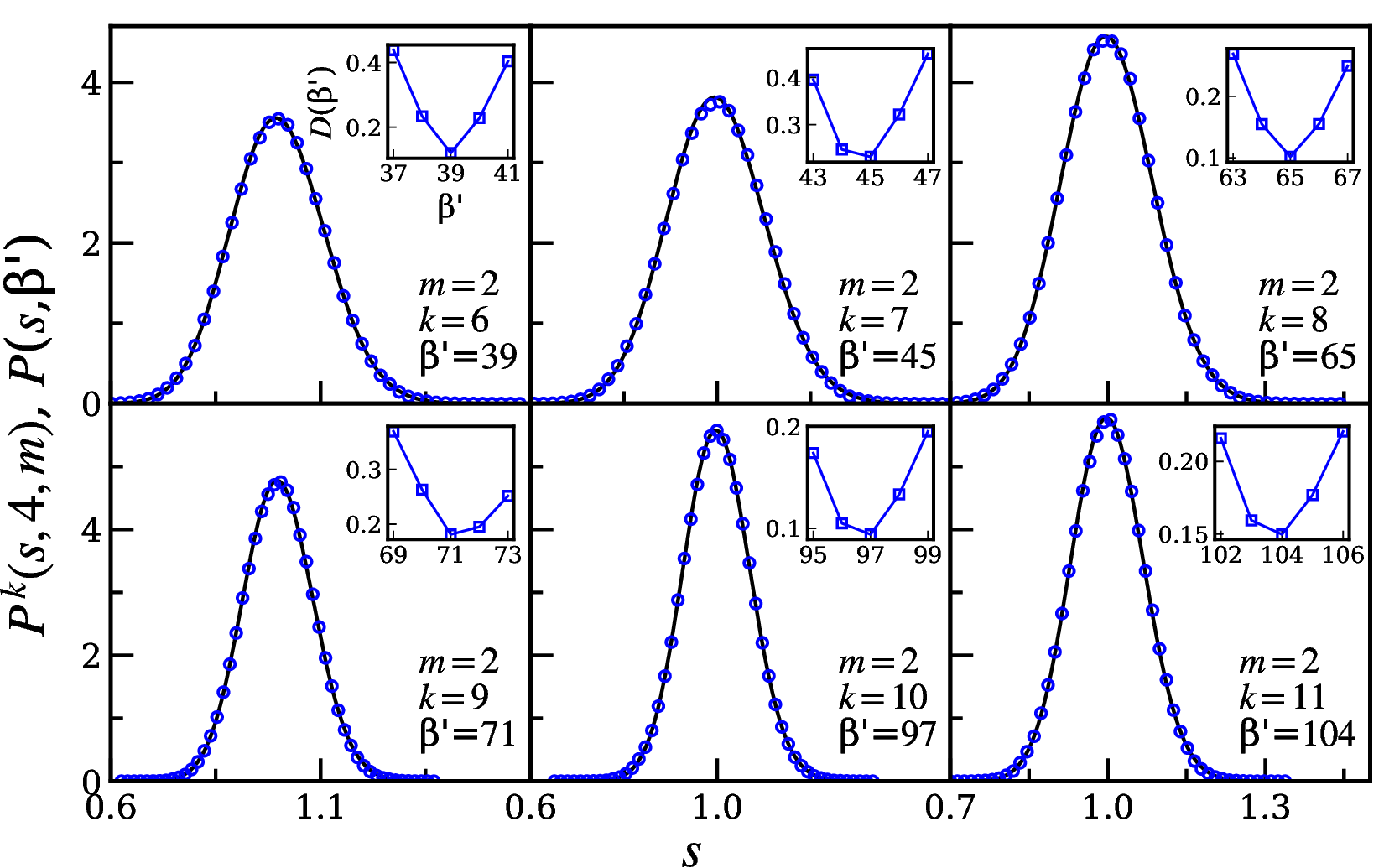}
\caption{\justifying Distributions of the $k$-th order spacing $P^{k}(s, 4, m)$ for various $k$ in $m= 2$ CSE spectra (circles). Here, $N= 5000$ and $n= 600$. The solid curve corresponds to $P(s,\beta')$ as per the equation $P(s,\beta)= A_{\beta}s^{\beta}\exp(-C_{\beta}s^2)$, where $\beta$ is replaced by $\beta'$ and the values of $\beta'$ are given in the main text. The insets show $D(\beta')$ as a function of $\beta'$.}
\label{fig:k_6_to_11_m2_CSE}
\end{center}
\end{figure}

\begin{figure}[H]
\begin{center}
\includegraphics*[scale=0.365]{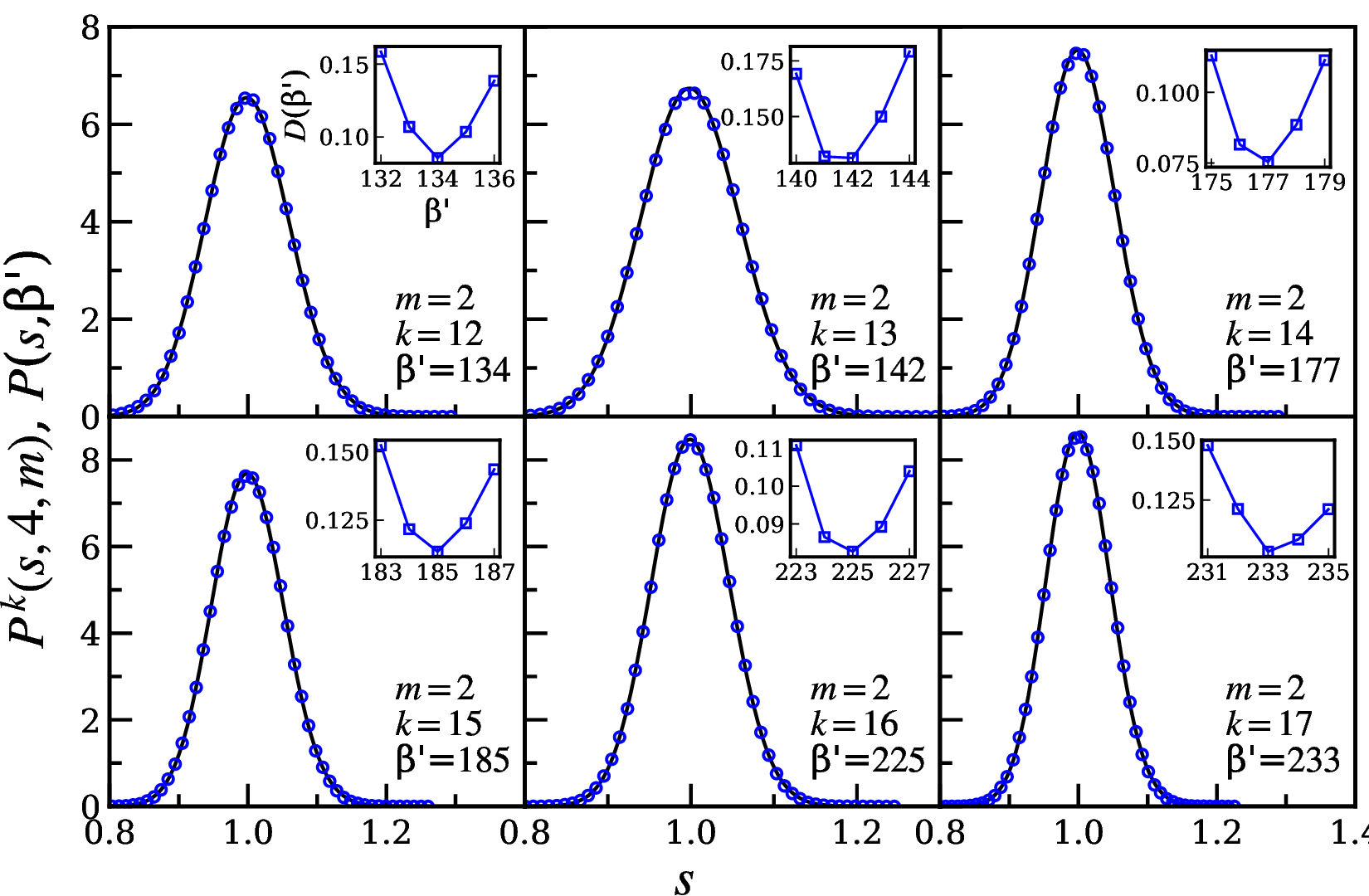}
\caption{\justifying Same as Fig.~\ref{fig:k_6_to_11_m2_CSE} but for different values of $k$ and $\beta'$.}
\label{fig:k_12_to_17_m2_CSE}
\end{center}
\end{figure}

\begin{figure}[H]
\begin{center}
\includegraphics*[scale=0.365]{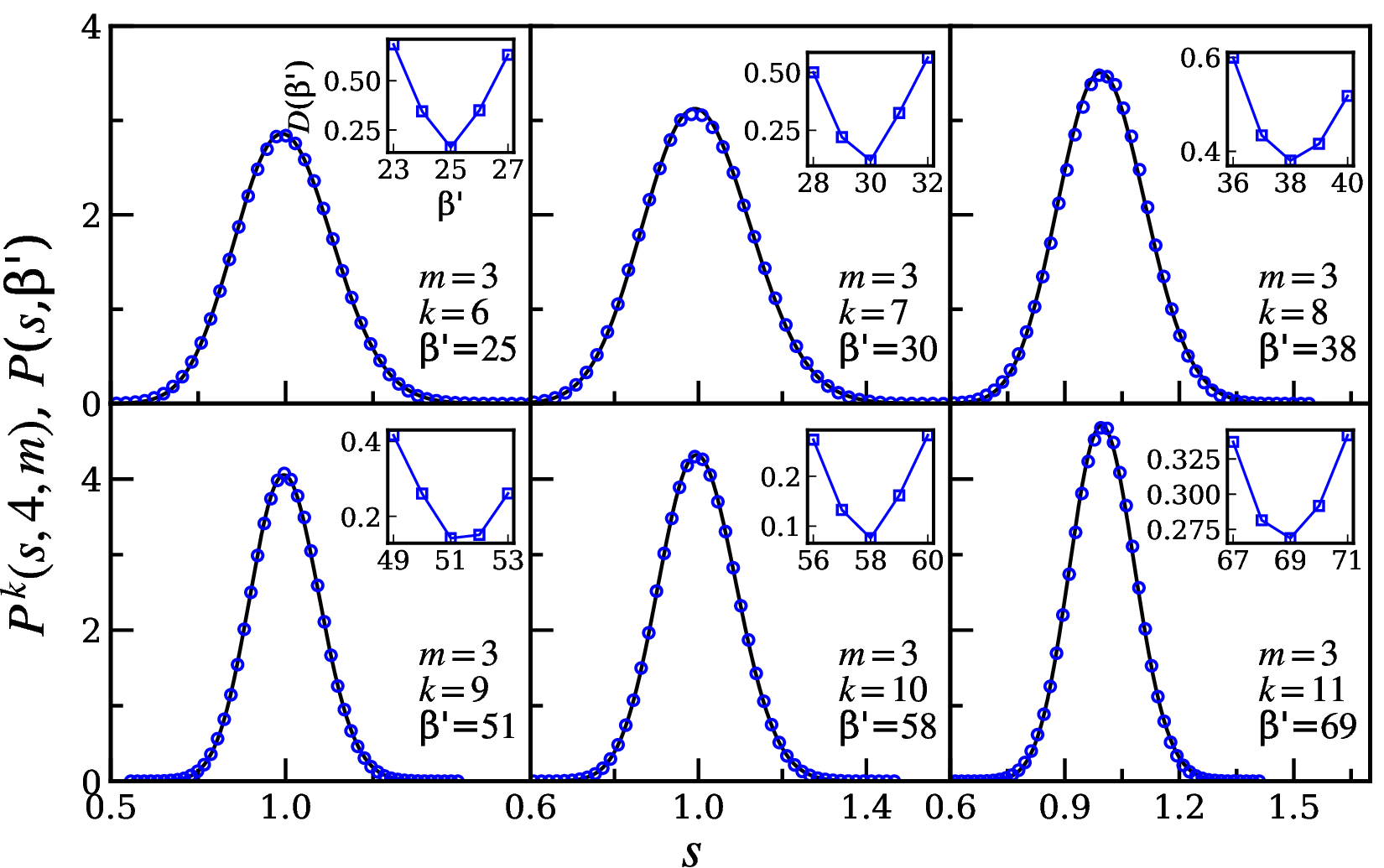}
\caption{\justifying Same as Fig.~\ref{fig:k_6_to_11_m2_CSE} but for $m= 3$, $n= 900$ and different values of $\beta'$.}
\label{fig:k_6_to_11_m3_CSE}
\end{center}
\end{figure}

\begin{figure}[H]
\begin{center}
\includegraphics*[scale=0.365]{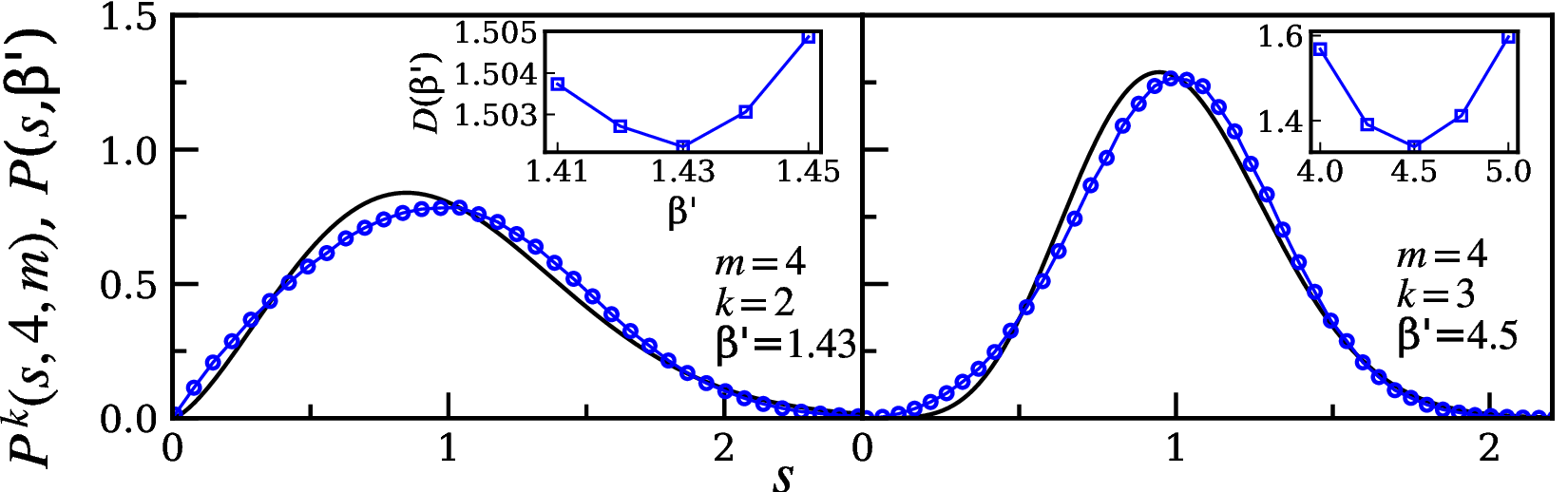}
\caption{\justifying Same as Fig.~\ref{fig:k_6_to_11_m2_CSE} but for $m= 4$, $n= 1000$, and different values of $k$ and $\beta'$.}
\label{fig:k_2_to_3_CSE_m4}
\end{center}
\end{figure}

\begin{figure}[H]
\begin{center}
\includegraphics*[scale=0.38]{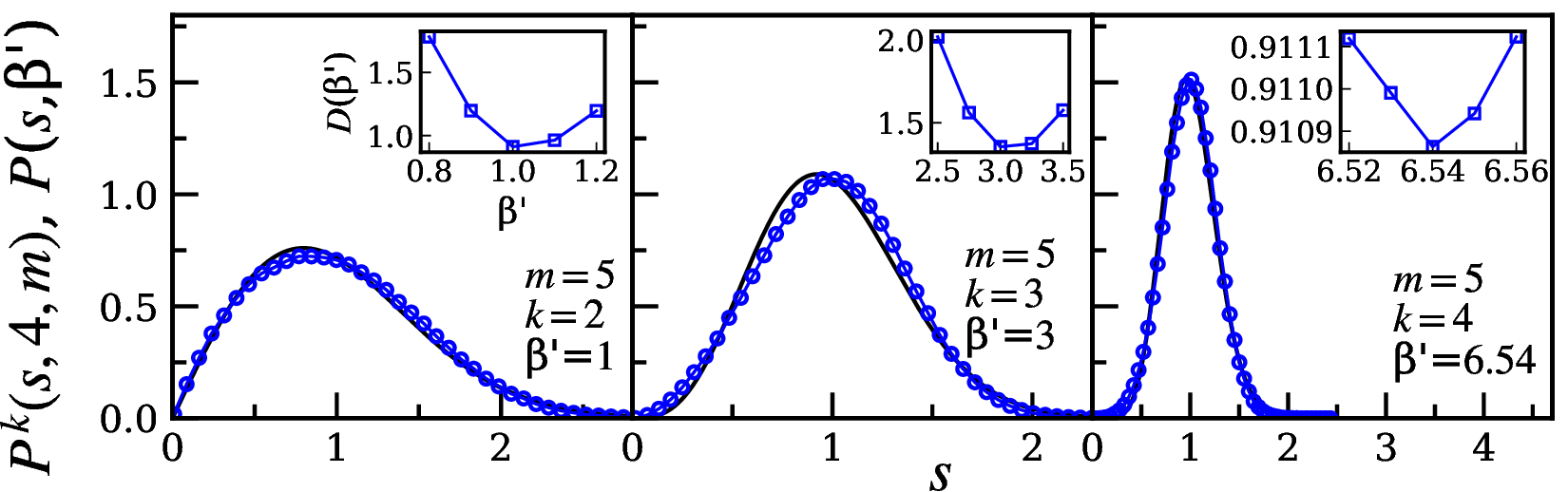}
\caption{\justifying  Same as Fig.~\ref{fig:k_6_to_11_m2_CSE} but for $m= 5$, $n= 1000$, and different values of $k$ and $\beta'$.}
\label{fig:k_2_to_4_CSE_m5}
\end{center}
\end{figure}

\begin{figure}[H]
\begin{center}
\includegraphics*[scale=0.38]{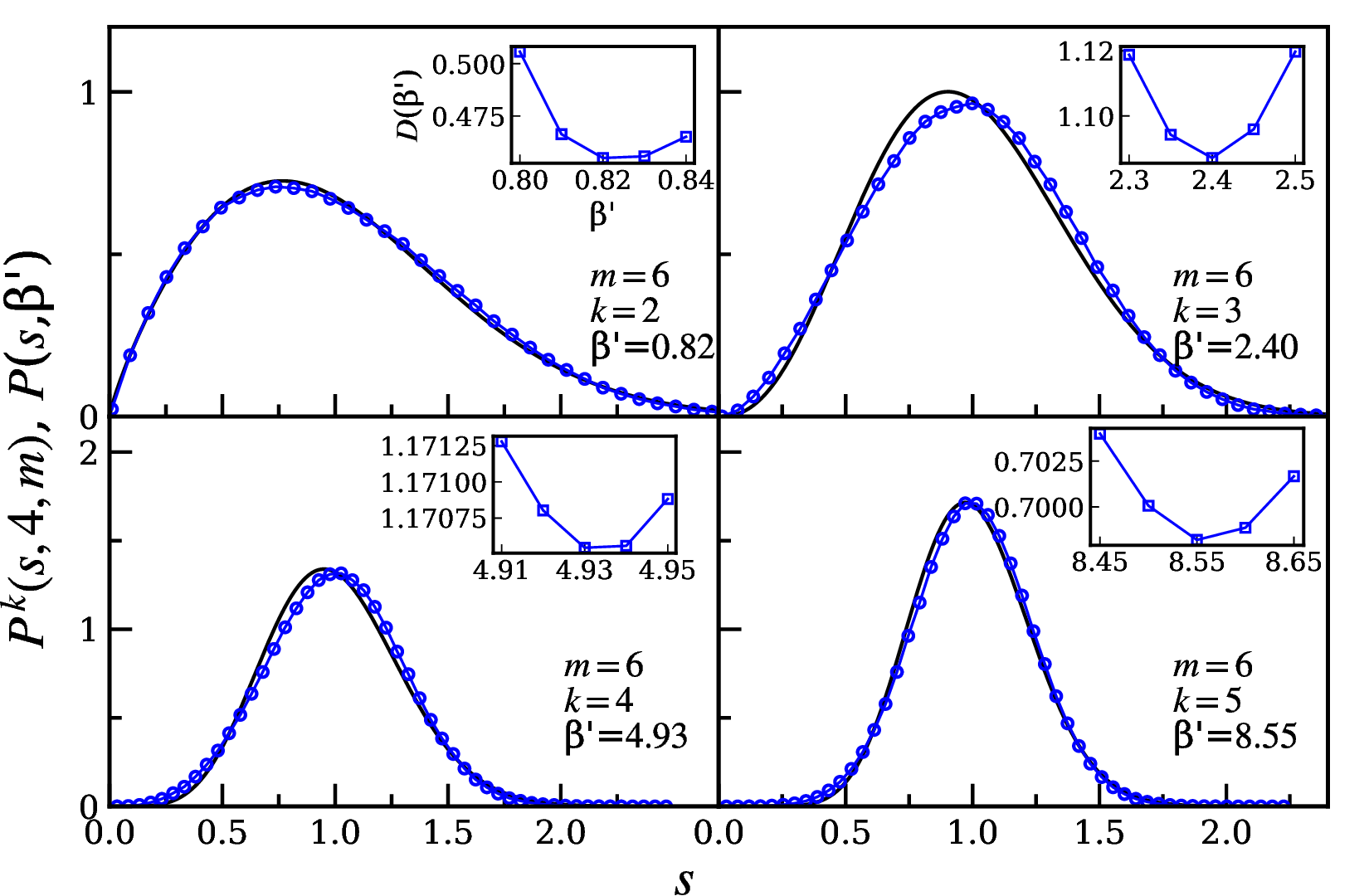}
\caption{\justifying Same as Fig.~\ref{fig:k_6_to_11_m2_CSE} but for $m= 6$, $n= 1002$, and different values of $k$ and $\beta'$.}
\label{fig:k_2_to_5_CSE_m6}
\end{center}
\end{figure}

\begin{figure}[H]
\begin{center}
\includegraphics*[scale=0.38]{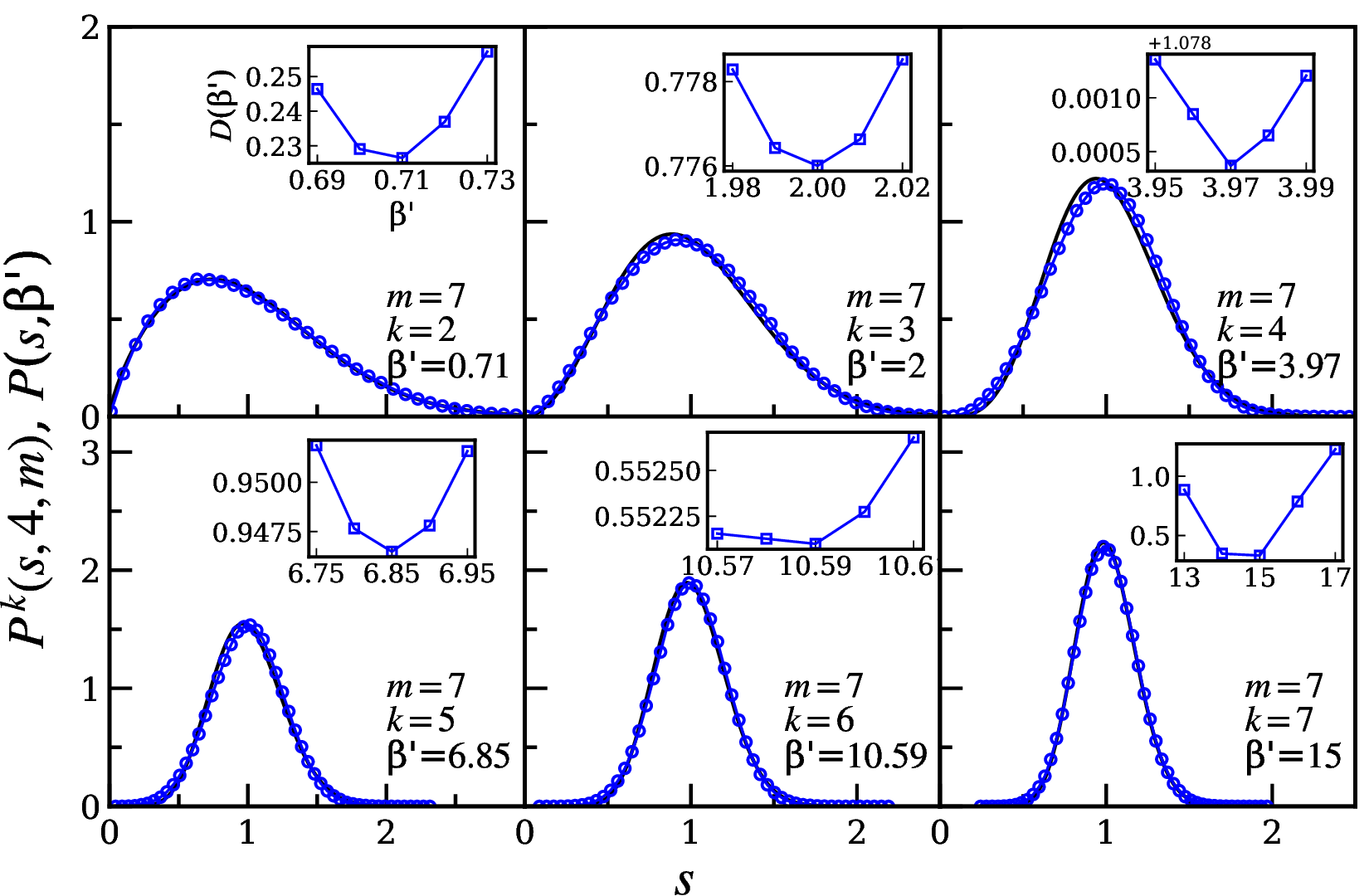}
\caption{\justifying Same as Fig.~\ref{fig:k_6_to_11_m2_CSE} but for $m= 7$, $n= 1001$, and different values of $k$ and $\beta'$.}
\label{fig:k_2_to_7_m7_CSE}
\end{center}
\end{figure}

\begin{figure}[H]
\begin{center}
\includegraphics*[scale=0.33]{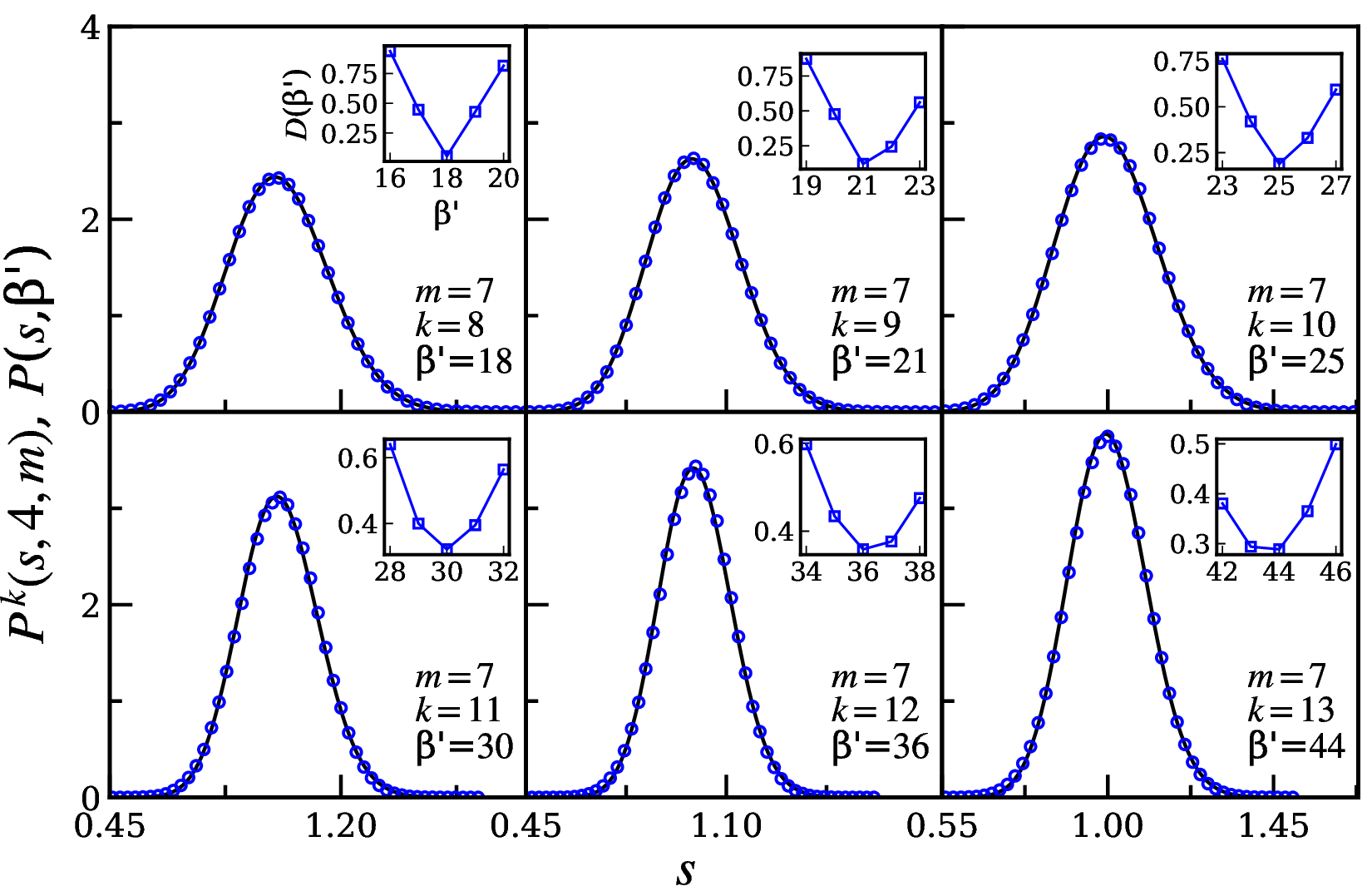}
\caption{\justifying Same as Fig.~\ref{fig:k_2_to_7_m7_CSE} but for different values of $k$ and $\beta'$.}
\label{fig:k_8_to_13_m7_CSE}
\end{center}
\end{figure}

\begin{figure}[H]
\begin{center}
\includegraphics*[scale=0.33]{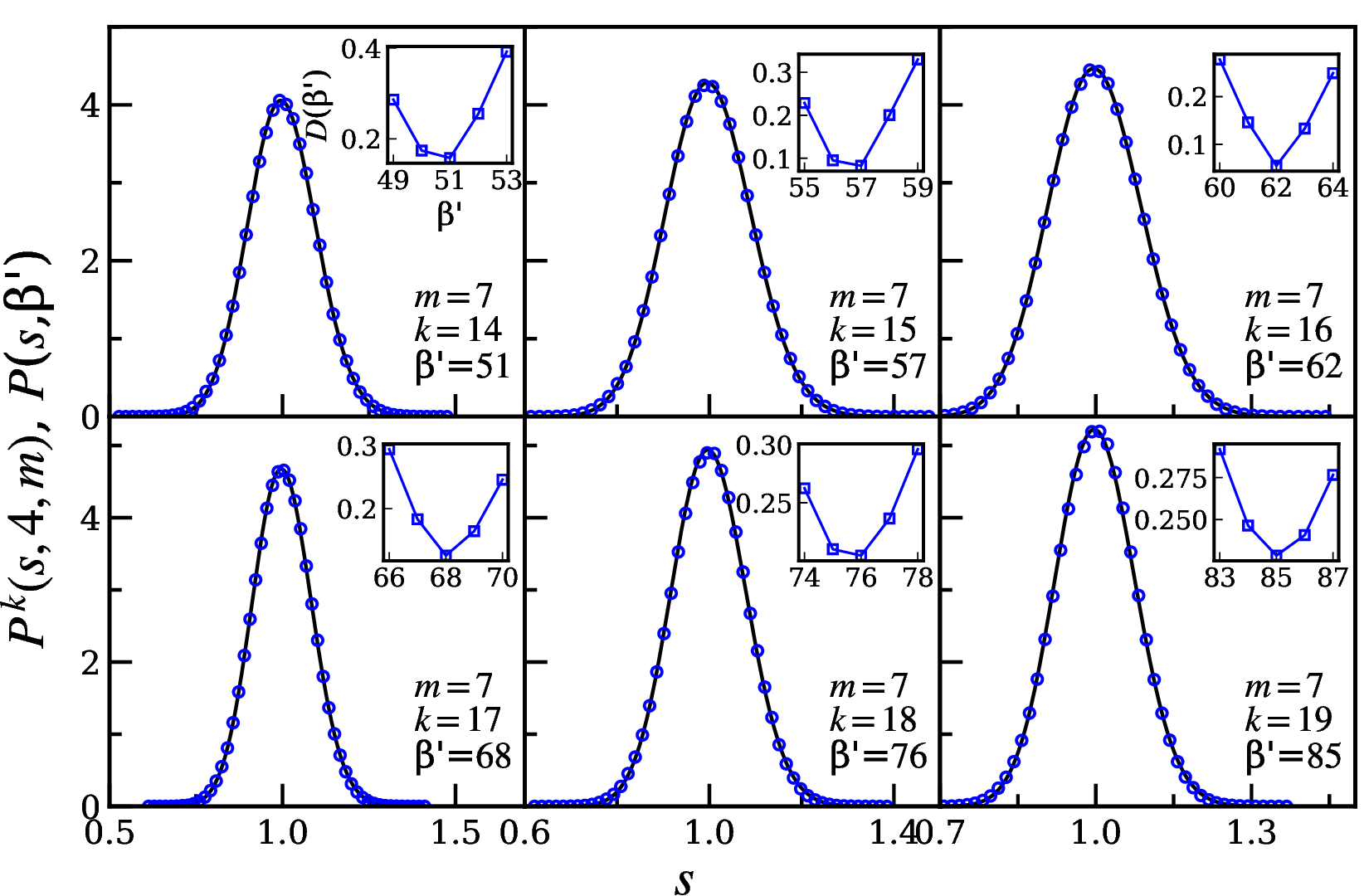}
\caption{\justifying Same as Fig.~\ref{fig:k_2_to_7_m7_CSE} but for different values of $k$ and $\beta'$.}
\label{fig:k_14_to_19_m7_CSE}
\end{center}
\end{figure}

\begin{figure}[H]
\begin{center}
\includegraphics*[scale=0.33]{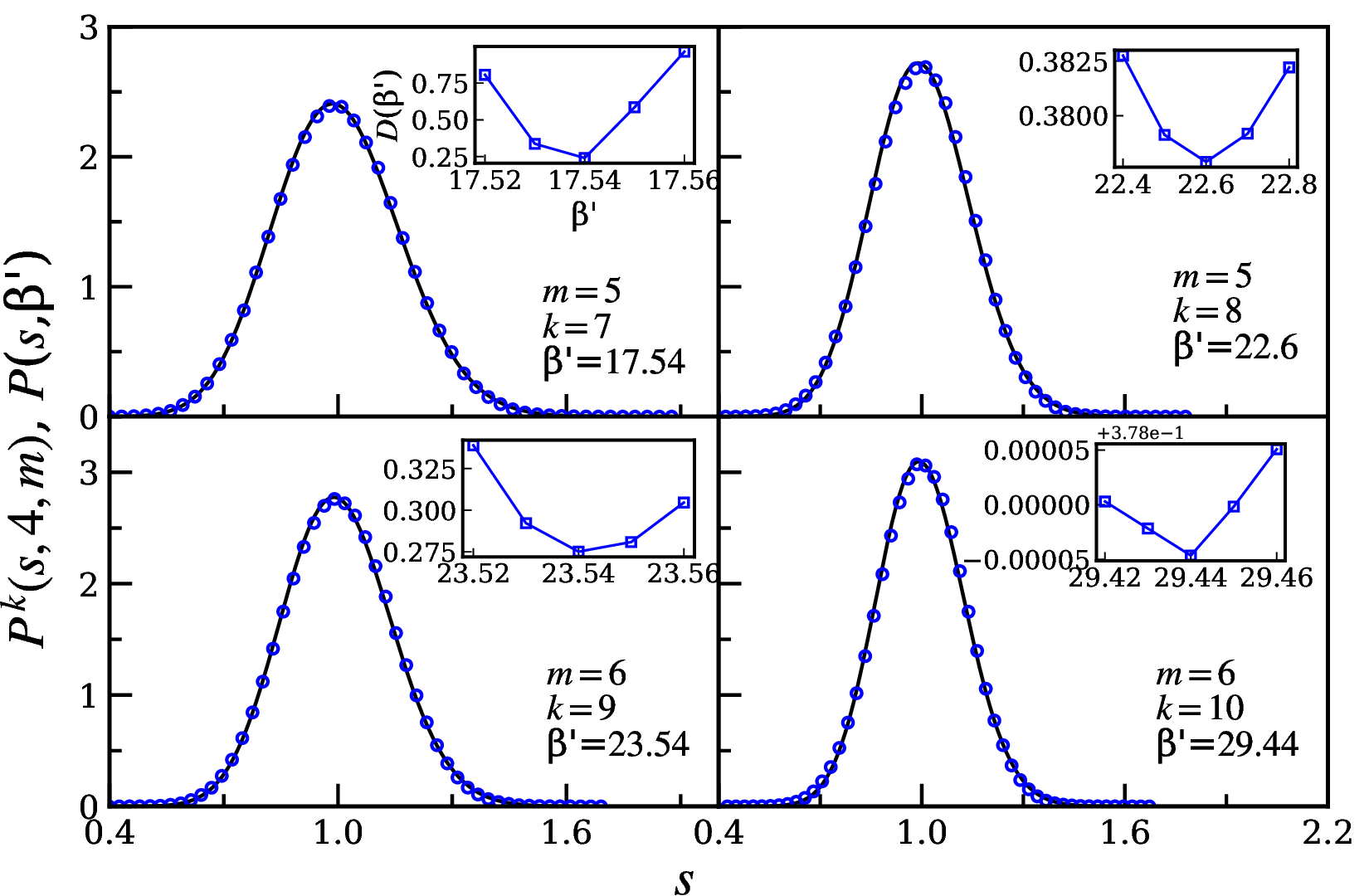}
\caption{\justifying Same as Fig.~\ref{fig:k_6_to_11_m2_CSE} but for $m= 5$, $m= 6$, and different values of $k$ and $\beta'$. Here, $n= 1000$ and $1002$ for $m= 5$ and $6$ respectively.}
\label{fig:k_7_8_m5_and_k_9_10_m6_CSE}
\end{center}
\end{figure}

\suppsubsection{The case of CSE}
\label{subsec:more_plots_CSE}
In this subsection , we have given few more plots of the HOS in the superposed spectra of CSE. These are Figs.~\ref{fig:k_6_to_11_m2_CSE}-\ref{fig:k_7_8_m5_and_k_9_10_m6_CSE}.
Here, Figs.~\ref{fig:k_2_to_3_CSE_m4}, \ref{fig:k_2_to_5_CSE_m6}, and \ref{fig:k_2_to_7_m7_CSE} corresponds to the non-integer values of $\beta'$ .

\suppsection{Dimensional Analysis}
\label{sec:DimAnal}
In this section we have studied the effect of dimension on the obtained results (values of $\beta'$). Here, we have varied dimension from $N= 1000$ to $N= 55000$, keeping number of realizations $n$ constant to $300$ for each $k$. The values of $\beta'$ corresponding to the best fits (analytical distribution functions) to the higher order spacing and ratio distributions for the $m= 1$ and $m= 2$ cases of both COE and GOE are tabulated. Here, only $k= 5, 10, 15$, and $20$ cases are considered. These results are shown in Tables \ref{Table: COE_m2_DimEffectTable}-\ref{Table: GOE_DimEffectTable}. The conclusions are presented in the main text.

\begin{table*}     
\renewcommand{\arraystretch}{1.5} 
\setlength{\tabcolsep}{5.3pt}  
\begin{center}
\begin{tabular}{|c|cc|cc|cc|cc|cc|cc|cc|}
\hline
\rule{0pt}{12pt}  
Order&\multicolumn{2}{c|}{$N=1000$}&\multicolumn{2}{c|}{$N=5000$}&\multicolumn{2}{c|}{$N=15000$}&\multicolumn{2}{c|}{$N=25000$}&\multicolumn{2}{c|}{$N=35000$}&\multicolumn{2}{c|}{$N=45000$}&\multicolumn{2}{c|}{$N=55000$}\\ [1.5ex]  
\cline{2-15}  
\rule{0pt}{10pt}  
&HOS&\multicolumn{1}{c|}{HOSR}&HOS&\multicolumn{1}{c|}{HOSR}&HOS&\multicolumn{1}{c|}{HOSR}&HOS&\multicolumn{1}{c|}{HOSR}&HOS&\multicolumn{1}{c|}{HOSR}&HOS&\multicolumn{1}{c|}{HOSR}&HOS&\multicolumn{1}{c|}{HOSR}\\
$k$&$\beta^\prime$&\multicolumn{1}{c|}{$\beta^\prime$}&$\beta^\prime$&\multicolumn{1}{c|}{$\beta^\prime$}&$\beta^\prime$&\multicolumn{1}{c|}{$\beta^\prime$}&$\beta^\prime$&\multicolumn{1}{c|}{$\beta^\prime$}&$\beta^\prime$&\multicolumn{1}{c|}{$\beta^\prime$}&$\beta^\prime$&\multicolumn{1}{c|}{$\beta^\prime$}&$\beta^\prime$&\multicolumn{1}{c|}{$\beta^\prime$}\\
\hline
5&11&\multicolumn{1}{c|}{10}&11&\multicolumn{1}{c|}{10}&11&\multicolumn{1}{c|}{10}&11&\multicolumn{1}{c|}{10}&11&\multicolumn{1}{c|}{10}&11&\multicolumn{1}{c|}{10}&11&\multicolumn{1}{c|}{10}\\
10&36&\multicolumn{1}{c|}{34}&36&\multicolumn{1}{c|}{34}&36&\multicolumn{1}{c|}{34}&36&\multicolumn{1}{c|}{34}&36&\multicolumn{1}{c|}{34}&36&\multicolumn{1}{c|}{34}&36&\multicolumn{1}{c|}{34}\\
15&72&\multicolumn{1}{c|}{68}&73&\multicolumn{1}{c|}{69}&72&\multicolumn{1}{c|}{68}&72&\multicolumn{1}{c|}{68}&73&\multicolumn{1}{c|}{69}&73&\multicolumn{1}{c|}{69}&73&\multicolumn{1}{c|}{69}\\
20&120&\multicolumn{1}{c|}{113}&121&\multicolumn{1}{c|}{114}&120&\multicolumn{1}{c|}{113}&120&\multicolumn{1}{c|}{113}&120&\multicolumn{1}{c|}{113}&120&\multicolumn{1}{c|}{113}&120&\multicolumn{1}{c|}{113}\\
\hline
\end{tabular}
\caption{\justifying Tabulation of higher-order indices $\beta'$ for both the distributions of spacing and spacing ratio for various values of $k$ in the $m= 2$ case of COE. Here, we have taken $n= 300$ for all $N$.}
\label{Table: COE_m2_DimEffectTable}
\end{center}
\end{table*}
\begin{table*}     
\renewcommand{\arraystretch}{1.5} 
\setlength{\tabcolsep}{5.3pt}  
\begin{center}
\begin{tabular}{|c|cc|cc|cc|cc|cc|cc|cc|}
\hline
\rule{0pt}{12pt}  
Order&\multicolumn{2}{c|}{$N=1000$}&\multicolumn{2}{c|}{$N=5000$}&\multicolumn{2}{c|}{$N=15000$}&\multicolumn{2}{c|}{$N=25000$}&\multicolumn{2}{c|}{$N=35000$}&\multicolumn{2}{c|}{$N=45000$}&\multicolumn{2}{c|}{$N=55000$}\\ [1.5ex]  
\cline{2-15}  
\rule{0pt}{10pt}  
&HOS&\multicolumn{1}{c|}{HOSR}&HOS&\multicolumn{1}{c|}{HOSR}&HOS&\multicolumn{1}{c|}{HOSR}&HOS&\multicolumn{1}{c|}{HOSR}&HOS&\multicolumn{1}{c|}{HOSR}&HOS&\multicolumn{1}{c|}{HOSR}&HOS&\multicolumn{1}{c|}{HOSR}\\
$k$&$\beta^\prime$&\multicolumn{1}{c|}{$\beta^\prime$}&$\beta^\prime$&\multicolumn{1}{c|}{$\beta^\prime$}&$\beta^\prime$&\multicolumn{1}{c|}{$\beta^\prime$}&$\beta^\prime$&\multicolumn{1}{c|}{$\beta^\prime$}&$\beta^\prime$&\multicolumn{1}{c|}{$\beta^\prime$}&$\beta^\prime$&\multicolumn{1}{c|}{$\beta^\prime$}&$\beta^\prime$&\multicolumn{1}{c|}{$\beta^\prime$}\\
\hline
5&11&\multicolumn{1}{c|}{10}&11&\multicolumn{1}{c|}{10}&11&\multicolumn{1}{c|}{10}&11&\multicolumn{1}{c|}{10}&11&\multicolumn{1}{c|}{10}&11&\multicolumn{1}{c|}{10}&11&\multicolumn{1}{c|}{10}\\
10&36&\multicolumn{1}{c|}{33}&36&\multicolumn{1}{c|}{34}&36&\multicolumn{1}{c|}{34}&36&\multicolumn{1}{c|}{34}&36&\multicolumn{1}{c|}{34}&36&\multicolumn{1}{c|}{34}&36&\multicolumn{1}{c|}{34}\\
15&73&\multicolumn{1}{c|}{64}&73&\multicolumn{1}{c|}{67}&72&\multicolumn{1}{c|}{68}&72&\multicolumn{1}{c|}{68}&72&\multicolumn{1}{c|}{68}&73&\multicolumn{1}{c|}{68}&72&\multicolumn{1}{c|}{68}\\
20&120&\multicolumn{1}{c|}{100}&120&\multicolumn{1}{c|}{110}&120&\multicolumn{1}{c|}{112}&120&\multicolumn{1}{c|}{112}&120&\multicolumn{1}{c|}{112}&120&\multicolumn{1}{c|}{113}&120&\multicolumn{1}{c|}{113}\\
\hline
\end{tabular}
\caption{\justifying Same as Table~\ref{Table: COE_m2_DimEffectTable} but for GOE.}
\label{Table: GOE_m2_DimEffectTable}
\end{center}
\end{table*}
\begin{table*}     
\renewcommand{\arraystretch}{1.5} 
\setlength{\tabcolsep}{5.3pt}  
\begin{center}
\begin{tabular}{|c|cc|cc|cc|cc|cc|cc|cc|}
\hline
\rule{0pt}{12pt}  
Order&\multicolumn{2}{c|}{$N=1000$}&\multicolumn{2}{c|}{$N=5000$}&\multicolumn{2}{c|}{$N=15000$}&\multicolumn{2}{c|}{$N=25000$}&\multicolumn{2}{c|}{$N=35000$}&\multicolumn{2}{c|}{$N=45000$}&\multicolumn{2}{c|}{$N=55000$}\\ [1.5ex]  
\cline{2-15}  
\rule{0pt}{10pt}  
&HOS&\multicolumn{1}{c|}{HOSR}&HOS&\multicolumn{1}{c|}{HOSR}&HOS&\multicolumn{1}{c|}{HOSR}&HOS&\multicolumn{1}{c|}{HOSR}&HOS&\multicolumn{1}{c|}{HOSR}&HOS&\multicolumn{1}{c|}{HOSR}&HOS&\multicolumn{1}{c|}{HOSR}\\
$k$&$\beta^\prime$&\multicolumn{1}{c|}{$\beta^\prime$}&$\beta^\prime$&\multicolumn{1}{c|}{$\beta^\prime$}&$\beta^\prime$&\multicolumn{1}{c|}{$\beta^\prime$}&$\beta^\prime$&\multicolumn{1}{c|}{$\beta^\prime$}&$\beta^\prime$&\multicolumn{1}{c|}{$\beta^\prime$}&$\beta^\prime$&\multicolumn{1}{c|}{$\beta^\prime$}&$\beta^\prime$&\multicolumn{1}{c|}{$\beta^\prime$}\\
\hline
5&20&\multicolumn{1}{c|}{19}&20&\multicolumn{1}{c|}{19}&20&\multicolumn{1}{c|}{19}&20&\multicolumn{1}{c|}{19}&20&\multicolumn{1}{c|}{19}&20&\multicolumn{1}{c|}{19}&20&\multicolumn{1}{c|}{19}\\
10&67&\multicolumn{1}{c|}{63}&67&\multicolumn{1}{c|}{63}&67&\multicolumn{1}{c|}{63}&66&\multicolumn{1}{c|}{63}&67&\multicolumn{1}{c|}{63}&67&\multicolumn{1}{c|}{63}&67&\multicolumn{1}{c|}{63}\\
15&137&\multicolumn{1}{c|}{128}&136&\multicolumn{1}{c|}{128}&136&\multicolumn{1}{c|}{128}&135&\multicolumn{1}{c|}{128}&136&\multicolumn{1}{c|}{128}&136&\multicolumn{1}{c|}{128}&136&\multicolumn{1}{c|}{128}\\
20&227&\multicolumn{1}{c|}{213}&226&\multicolumn{1}{c|}{212}&225&\multicolumn{1}{c|}{212}&226&\multicolumn{1}{c|}{212}&225&\multicolumn{1}{c|}{212}&226&\multicolumn{1}{c|}{212}&226&\multicolumn{1}{c|}{212}\\
\hline
\end{tabular}
\caption{\justifying Same as Table~\ref{Table: COE_m2_DimEffectTable} but for $m= 1$.}
\label{Table: COE_DimEffectTable}
\end{center}
\end{table*}
\begin{table*}     
\renewcommand{\arraystretch}{1.5} 
\setlength{\tabcolsep}{5.3pt}  
\begin{center}
\begin{tabular}{|c|cc|cc|cc|cc|cc|cc|cc|}
\hline
\rule{0pt}{12pt}  
Order&\multicolumn{2}{c|}{$N=1000$}&\multicolumn{2}{c|}{$N=5000$}&\multicolumn{2}{c|}{$N=15000$}&\multicolumn{2}{c|}{$N=25000$}&\multicolumn{2}{c|}{$N=35000$}&\multicolumn{2}{c|}{$N=45000$}&\multicolumn{2}{c|}{$N=55000$}\\ [1.5ex]  
\cline{2-15}  
\rule{0pt}{10pt}  
&HOS&\multicolumn{1}{c|}{HOSR}&HOS&\multicolumn{1}{c|}{HOSR}&HOS&\multicolumn{1}{c|}{HOSR}&HOS&\multicolumn{1}{c|}{HOSR}&HOS&\multicolumn{1}{c|}{HOSR}&HOS&\multicolumn{1}{c|}{HOSR}&HOS&\multicolumn{1}{c|}{HOSR}\\
$k$&$\beta^\prime$&\multicolumn{1}{c|}{$\beta^\prime$}&$\beta^\prime$&\multicolumn{1}{c|}{$\beta^\prime$}&$\beta^\prime$&\multicolumn{1}{c|}{$\beta^\prime$}&$\beta^\prime$&\multicolumn{1}{c|}{$\beta^\prime$}&$\beta^\prime$&\multicolumn{1}{c|}{$\beta^\prime$}&$\beta^\prime$&\multicolumn{1}{c|}{$\beta^\prime$}&$\beta^\prime$&\multicolumn{1}{c|}{$\beta^\prime$}\\
\hline
5&20&\multicolumn{1}{c|}{19}&20&\multicolumn{1}{c|}{19}&20&\multicolumn{1}{c|}{19}&20&\multicolumn{1}{c|}{19}&20&\multicolumn{1}{c|}{19}&20&\multicolumn{1}{c|}{19}&20&\multicolumn{1}{c|}{19}\\
10&67&\multicolumn{1}{c|}{58}&67&\multicolumn{1}{c|}{62}&67&\multicolumn{1}{c|}{62}&67&\multicolumn{1}{c|}{63}&66&\multicolumn{1}{c|}{63}&67&\multicolumn{1}{c|}{63}&67&\multicolumn{1}{c|}{63}\\
15&136&\multicolumn{1}{c|}{105}&136&\multicolumn{1}{c|}{122}&136&\multicolumn{1}{c|}{126}&135&\multicolumn{1}{c|}{126}&135&\multicolumn{1}{c|}{126}&136&\multicolumn{1}{c|}{127}&136&\multicolumn{1}{c|}{127}\\
20&227&\multicolumn{1}{c|}{152}&226&\multicolumn{1}{c|}{196}&226&\multicolumn{1}{c|}{206}&225&\multicolumn{1}{c|}{208}&225&\multicolumn{1}{c|}{208}&226&\multicolumn{1}{c|}{209}&226&\multicolumn{1}{c|}{209}\\
\hline
\end{tabular}
\caption{\justifying Same as Table~\ref{Table: COE_m2_DimEffectTable} but for GOE and $m= 1$.}
\label{Table: GOE_DimEffectTable}
\end{center}
\end{table*}

\suppsection{Effect of the number of realizations}
\label{sec:no_effect}
Here, we have tabulated the values of $\beta'$ while varying $n$ from $500$ to $3500$ for each $k$ for both $m=1$ and $m=2$ cases of COE and GOE. Here, $N=5000$ for each $n$ and $k$. These results are shown in Tables \ref{Table: COE_m2_NumberEffectTable}-\ref{Table: GOE_NumberEffectTable}. Here, only $k=5, 10, 15$, and $20$ cases are considered. The conclusions are presented in the main text.

\begin{table*}      
\renewcommand{\arraystretch}{1.5} 
\setlength{\tabcolsep}{5.3pt}  
\begin{center}
\begin{tabular}{|c|cc|cc|cc|cc|cc|cc|cc|}
\hline
\rule{0pt}{12pt}  
Order&\multicolumn{2}{c|}{$n=500$}&\multicolumn{2}{c|}{$n=1000$}&\multicolumn{2}{c|}{$n=1500$}&\multicolumn{2}{c|}{$n=2000$}&\multicolumn{2}{c|}{$n=2500$}&\multicolumn{2}{c|}{$n=3000$}&\multicolumn{2}{c|}{$n=3500$}\\ [1.5ex]  
\cline{2-15}  
\rule{0pt}{10pt}  
&HOS&\multicolumn{1}{c|}{HOSR}&HOS&\multicolumn{1}{c|}{HOSR}&HOS&\multicolumn{1}{c|}{HOSR}&HOS&\multicolumn{1}{c|}{HOSR}&HOS&\multicolumn{1}{c|}{HOSR}&HOS&\multicolumn{1}{c|}{HOSR}&HOS&\multicolumn{1}{c|}{HOSR}\\
$k$&$\beta^\prime$&\multicolumn{1}{c|}{$\beta^\prime$}&$\beta^\prime$&\multicolumn{1}{c|}{$\beta^\prime$}&$\beta^\prime$&\multicolumn{1}{c|}{$\beta^\prime$}&$\beta^\prime$&\multicolumn{1}{c|}{$\beta^\prime$}&$\beta^\prime$&\multicolumn{1}{c|}{$\beta^\prime$}&$\beta^\prime$&\multicolumn{1}{c|}{$\beta^\prime$}&$\beta^\prime$&\multicolumn{1}{c|}{$\beta^\prime$}\\
\hline
5&11&\multicolumn{1}{c|}{10}&11&\multicolumn{1}{c|}{10}&11&\multicolumn{1}{c|}{10}&11&\multicolumn{1}{c|}{10}&11&\multicolumn{1}{c|}{10}&11&\multicolumn{1}{c|}{10}&11&\multicolumn{1}{c|}{10}\\
10&36&\multicolumn{1}{c|}{34}&36&\multicolumn{1}{c|}{34}&36&\multicolumn{1}{c|}{34}&36&\multicolumn{1}{c|}{34}&36&\multicolumn{1}{c|}{34}&36&\multicolumn{1}{c|}{34}&36&\multicolumn{1}{c|}{34}\\
15&72&\multicolumn{1}{c|}{68}&72&\multicolumn{1}{c|}{68}&72&\multicolumn{1}{c|}{68}&72&\multicolumn{1}{c|}{68}&72&\multicolumn{1}{c|}{68}&72&\multicolumn{1}{c|}{68}&72&\multicolumn{1}{c|}{68}\\
20&120&\multicolumn{1}{c|}{113}&120&\multicolumn{1}{c|}{113}&120&\multicolumn{1}{c|}{113}&120&\multicolumn{1}{c|}{113}&120&\multicolumn{1}{c|}{113}&120&\multicolumn{1}{c|}{113}&120&\multicolumn{1}{c|}{113}\\
\hline
\end{tabular}
\caption{\justifying Tabulation of higher-order indices $\beta'$ for various values of $k$ and $n$ for both the distributions of spacing and spacing ratio in the case of COE ($m= 2$). Here, $N= 5000$ for each $n$ and $k$.}
\label{Table: COE_m2_NumberEffectTable}
\end{center}
\end{table*}
\begin{table*}      
\renewcommand{\arraystretch}{1.5} 
\setlength{\tabcolsep}{5.3pt}  
\begin{center}
\begin{tabular}{|c|cc|cc|cc|cc|cc|cc|cc|}
\hline
\rule{0pt}{12pt}  
Order&\multicolumn{2}{c|}{$n=500$}&\multicolumn{2}{c|}{$n=1000$}&\multicolumn{2}{c|}{$n=1500$}&\multicolumn{2}{c|}{$n=2000$}&\multicolumn{2}{c|}{$n=2500$}&\multicolumn{2}{c|}{$n=3000$}&\multicolumn{2}{c|}{$n=3500$}\\ [1.5ex]  
\cline{2-15}  
\rule{0pt}{10pt}  
&HOS&\multicolumn{1}{c|}{HOSR}&HOS&\multicolumn{1}{c|}{HOSR}&HOS&\multicolumn{1}{c|}{HOSR}&HOS&\multicolumn{1}{c|}{HOSR}&HOS&\multicolumn{1}{c|}{HOSR}&HOS&\multicolumn{1}{c|}{HOSR}&HOS&\multicolumn{1}{c|}{HOSR}\\
$k$&$\beta^\prime$&\multicolumn{1}{c|}{$\beta^\prime$}&$\beta^\prime$&\multicolumn{1}{c|}{$\beta^\prime$}&$\beta^\prime$&\multicolumn{1}{c|}{$\beta^\prime$}&$\beta^\prime$&\multicolumn{1}{c|}{$\beta^\prime$}&$\beta^\prime$&\multicolumn{1}{c|}{$\beta^\prime$}&$\beta^\prime$&\multicolumn{1}{c|}{$\beta^\prime$}&$\beta^\prime$&\multicolumn{1}{c|}{$\beta^\prime$}\\
\hline
5&11&10&11&10&11&10&11&10&11&10&11&10&11&10\\
10&36&34&36&34&36&34&36&34&36&34&36&34&36&34\\
15&73&68&73&67&73&67&73&67&73&67&73&67&73&67\\
20&120&110&120&110&120&110&120&110&120&110&120&110&120&110\\
\hline
\end{tabular}
\caption{\justifying Same as Table~\ref{Table: COE_m2_NumberEffectTable} but for GOE.}
\label{Table: GOE_m2_NumberEffectTable}
\end{center}
\end{table*}
\begin{table*}      
\renewcommand{\arraystretch}{1.5} 
\setlength{\tabcolsep}{5.3pt}  
\begin{center}
\begin{tabular}{|c|cc|cc|cc|cc|cc|cc|cc|}
\hline
\rule{0pt}{12pt}  
Order&\multicolumn{2}{c|}{$n=500$}&\multicolumn{2}{c|}{$n=1000$}&\multicolumn{2}{c|}{$n=1500$}&\multicolumn{2}{c|}{$n=2000$}&\multicolumn{2}{c|}{$n=2500$}&\multicolumn{2}{c|}{$n=3000$}&\multicolumn{2}{c|}{$n=3500$}\\ [1.5ex]  
\cline{2-15}  
\rule{0pt}{10pt}  
&HOS&\multicolumn{1}{c|}{HOSR}&HOS&\multicolumn{1}{c|}{HOSR}&HOS&\multicolumn{1}{c|}{HOSR}&HOS&\multicolumn{1}{c|}{HOSR}&HOS&\multicolumn{1}{c|}{HOSR}&HOS&\multicolumn{1}{c|}{HOSR}&HOS&\multicolumn{1}{c|}{HOSR}\\
$k$&$\beta^\prime$&\multicolumn{1}{c|}{$\beta^\prime$}&$\beta^\prime$&\multicolumn{1}{c|}{$\beta^\prime$}&$\beta^\prime$&\multicolumn{1}{c|}{$\beta^\prime$}&$\beta^\prime$&\multicolumn{1}{c|}{$\beta^\prime$}&$\beta^\prime$&\multicolumn{1}{c|}{$\beta^\prime$}&$\beta^\prime$&\multicolumn{1}{c|}{$\beta^\prime$}&$\beta^\prime$&\multicolumn{1}{c|}{$\beta^\prime$}\\
\hline
5&20&19&20&19&20&19&20&19&20&19&20&19&20&19\\
10&67&63&67&63&67&63&66&63&66&63&66&63&67&63\\
15&136&128&136&128&136&128&135&128&136&128&136&128&136&128\\
20&226&212&226&212&226&212&226&212&226&212&226&212&226&212\\
\hline
\end{tabular}
\caption{\justifying Same as Table~\ref{Table: COE_m2_NumberEffectTable} but for $m= 1$.}
\label{Table: COE_NumberEffectTable}
\end{center}
\end{table*}
\begin{table*}      
\renewcommand{\arraystretch}{1.5} 
\setlength{\tabcolsep}{5.3pt}  
\begin{center}
\begin{tabular}{|c|cc|cc|cc|cc|cc|cc|cc|}
\hline
\rule{0pt}{12pt}  
Order&\multicolumn{2}{c|}{$n=500$}&\multicolumn{2}{c|}{$n=1000$}&\multicolumn{2}{c|}{$n=1500$}&\multicolumn{2}{c|}{$n=2000$}&\multicolumn{2}{c|}{$n=2500$}&\multicolumn{2}{c|}{$n=3000$}&\multicolumn{2}{c|}{$n=3500$}\\ [1.5ex]  
\cline{2-15}  
\rule{0pt}{10pt}  
&HOS&\multicolumn{1}{c|}{HOSR}&HOS&\multicolumn{1}{c|}{HOSR}&HOS&\multicolumn{1}{c|}{HOSR}&HOS&\multicolumn{1}{c|}{HOSR}&HOS&\multicolumn{1}{c|}{HOSR}&HOS&\multicolumn{1}{c|}{HOSR}&HOS&\multicolumn{1}{c|}{HOSR}\\
$k$&$\beta^\prime$&\multicolumn{1}{c|}{$\beta^\prime$}&$\beta^\prime$&\multicolumn{1}{c|}{$\beta^\prime$}&$\beta^\prime$&\multicolumn{1}{c|}{$\beta^\prime$}&$\beta^\prime$&\multicolumn{1}{c|}{$\beta^\prime$}&$\beta^\prime$&\multicolumn{1}{c|}{$\beta^\prime$}&$\beta^\prime$&\multicolumn{1}{c|}{$\beta^\prime$}&$\beta^\prime$&\multicolumn{1}{c|}{$\beta^\prime$}\\
\hline
5&20&\multicolumn{1}{c|}{19}&20&\multicolumn{1}{c|}{19}&20&\multicolumn{1}{c|}{19}&20&\multicolumn{1}{c|}{19}&20&\multicolumn{1}{c|}{19}&20&\multicolumn{1}{c|}{19}&20&\multicolumn{1}{c|}{19}\\
10&67&\multicolumn{1}{c|}{62}&67&\multicolumn{1}{c|}{62}&67&\multicolumn{1}{c|}{62}&67&\multicolumn{1}{c|}{62}&67&\multicolumn{1}{c|}{62}&67&\multicolumn{1}{c|}{62}&67&\multicolumn{1}{c|}{62}\\
15&136&\multicolumn{1}{c|}{122}&136&\multicolumn{1}{c|}{122}&136&\multicolumn{1}{c|}{122}&136&\multicolumn{1}{c|}{122}&136&\multicolumn{1}{c|}{122}&136&\multicolumn{1}{c|}{122}&136&\multicolumn{1}{c|}{122}\\
20&226&\multicolumn{1}{c|}{196}&226&\multicolumn{1}{c|}{196}&226&\multicolumn{1}{c|}{195}&226&\multicolumn{1}{c|}{195}&226&\multicolumn{1}{c|}{196}&226&\multicolumn{1}{c|}{196}&226&\multicolumn{1}{c|}{196}\\
\hline
\end{tabular}
\caption{\justifying Same as Table~\ref{Table: COE_m2_NumberEffectTable} but for GOE and $m= 1$.}
\label{Table: GOE_NumberEffectTable}
\end{center}
\end{table*}

\begin{figure}[H]
\begin{center}
\includegraphics*[scale=0.42]{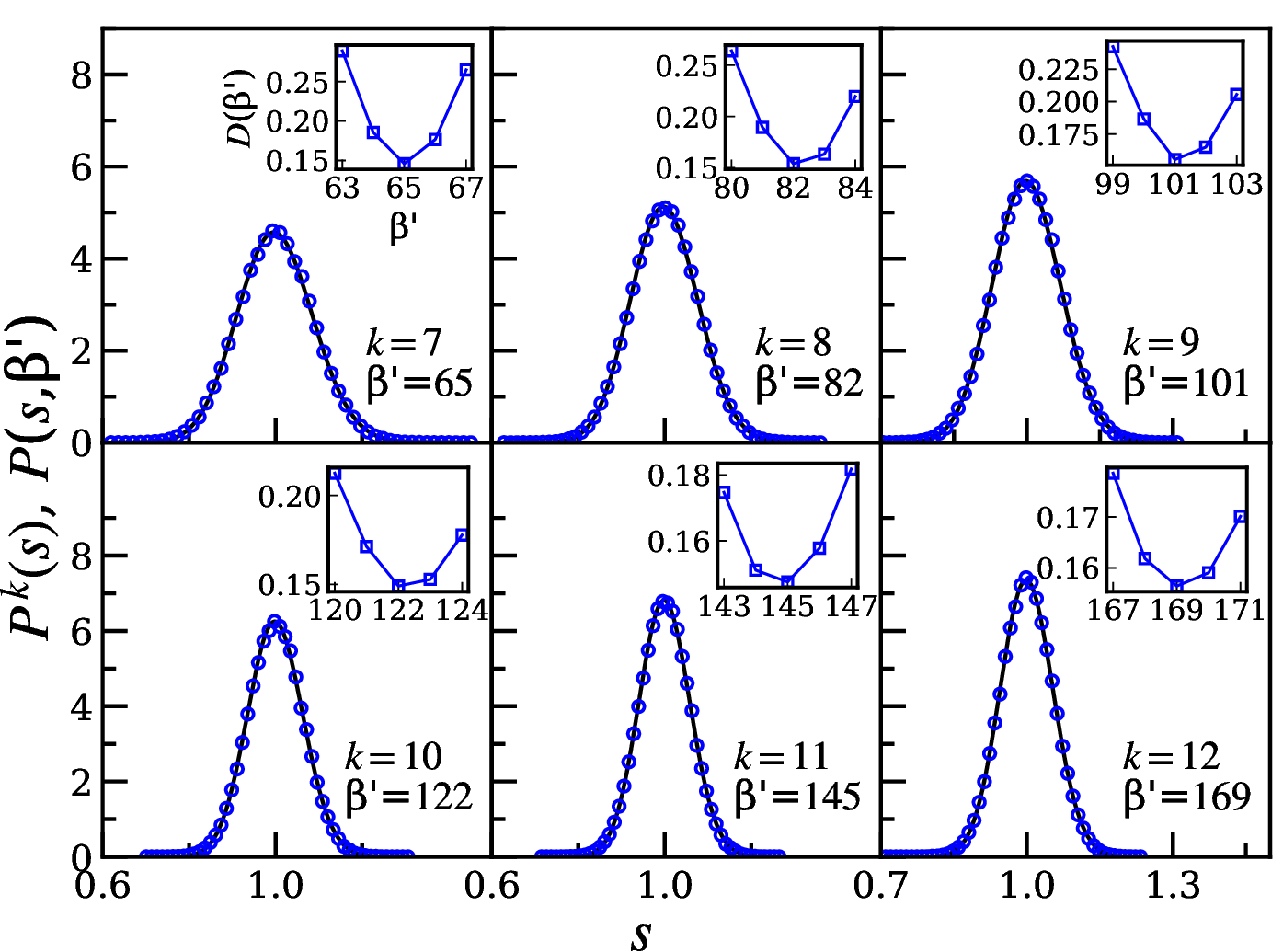}
\caption{\justifying HOS distribution $P^{k}(s)$ of eigenangles of the intermediate map for various $k$ (circles). Here, we have taken $N= 12000$ and $n= 80$. Solid line corresponds to $P(s, \beta')$.}
\label{fig:k_7_to_12_spacings_IM}
\end{center}
\end{figure}

\begin{figure}[H]
\begin{center}
\includegraphics*[scale=0.34]{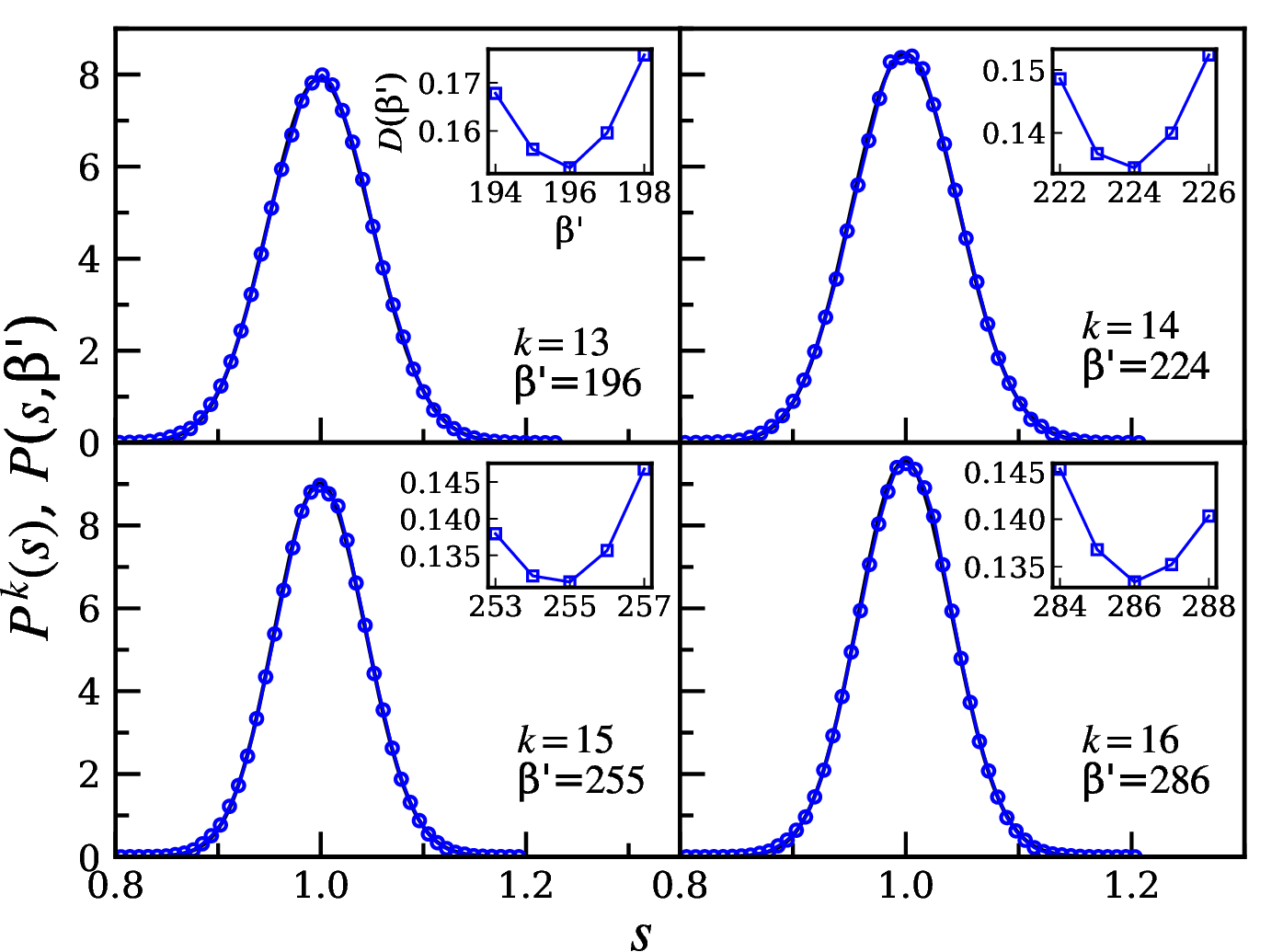}
\caption{\justifying Same as Fig.~\ref{fig:k_7_to_12_spacings_IM} but for different values of $k$ and $\beta'$.}
\label{fig:k_13_to_16_spacings_IM}
\end{center}
\end{figure}

\begin{figure}[H]
\begin{center}
\includegraphics*[scale=0.34]{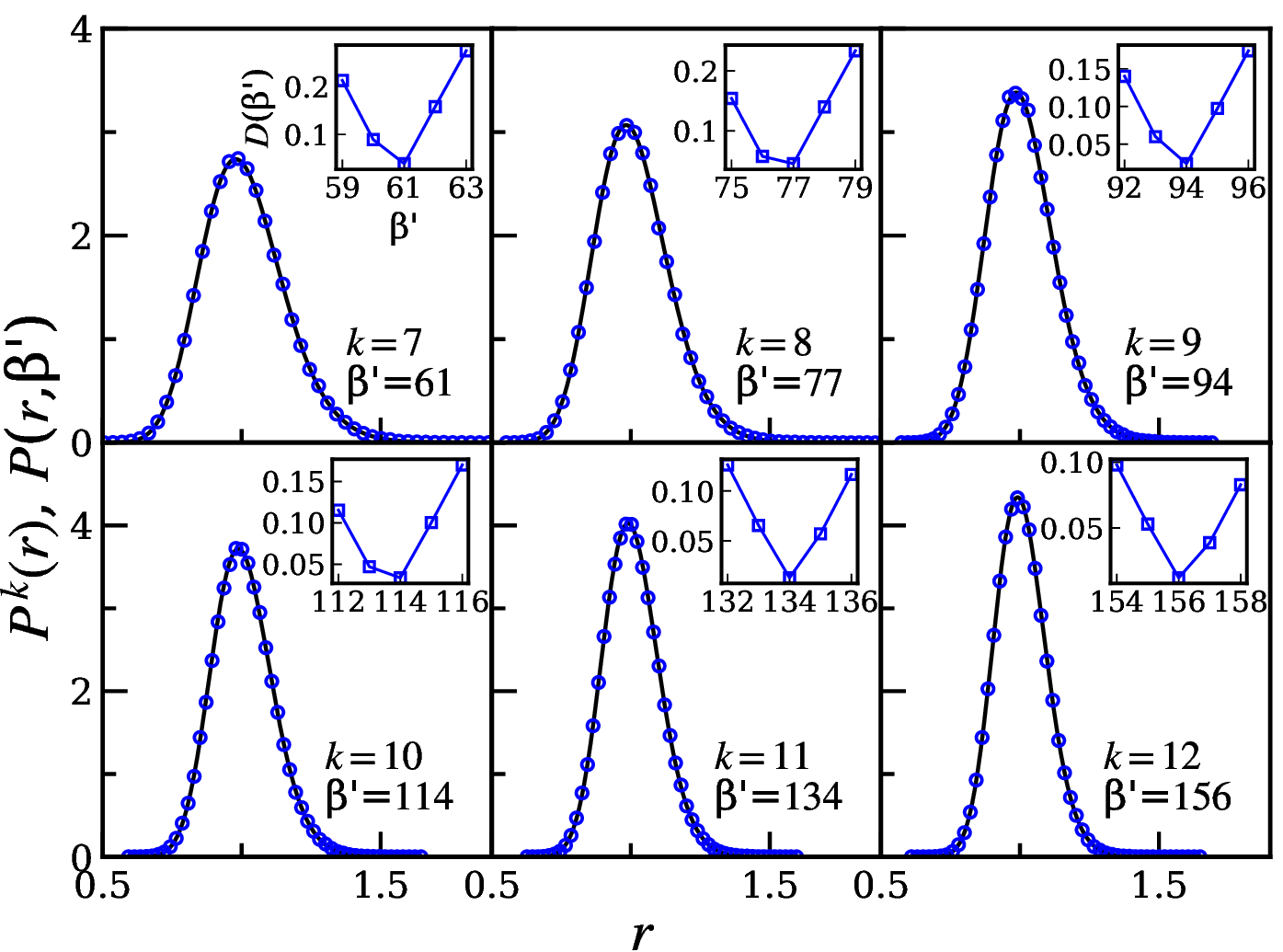}
\caption{\justifying HOSR distribution $P^{k}(r)$ of eigenangles of the intermediate map for various $k$ (circles). Here, we have taken $N= 12000$ and $n= 80$. Solid line corresponds to $P(r, \beta')$.}
\label{fig:k_7_to_12_sratios_IM}
\end{center}
\end{figure}

\begin{figure}[H]
\begin{center}
\includegraphics*[scale=0.34]{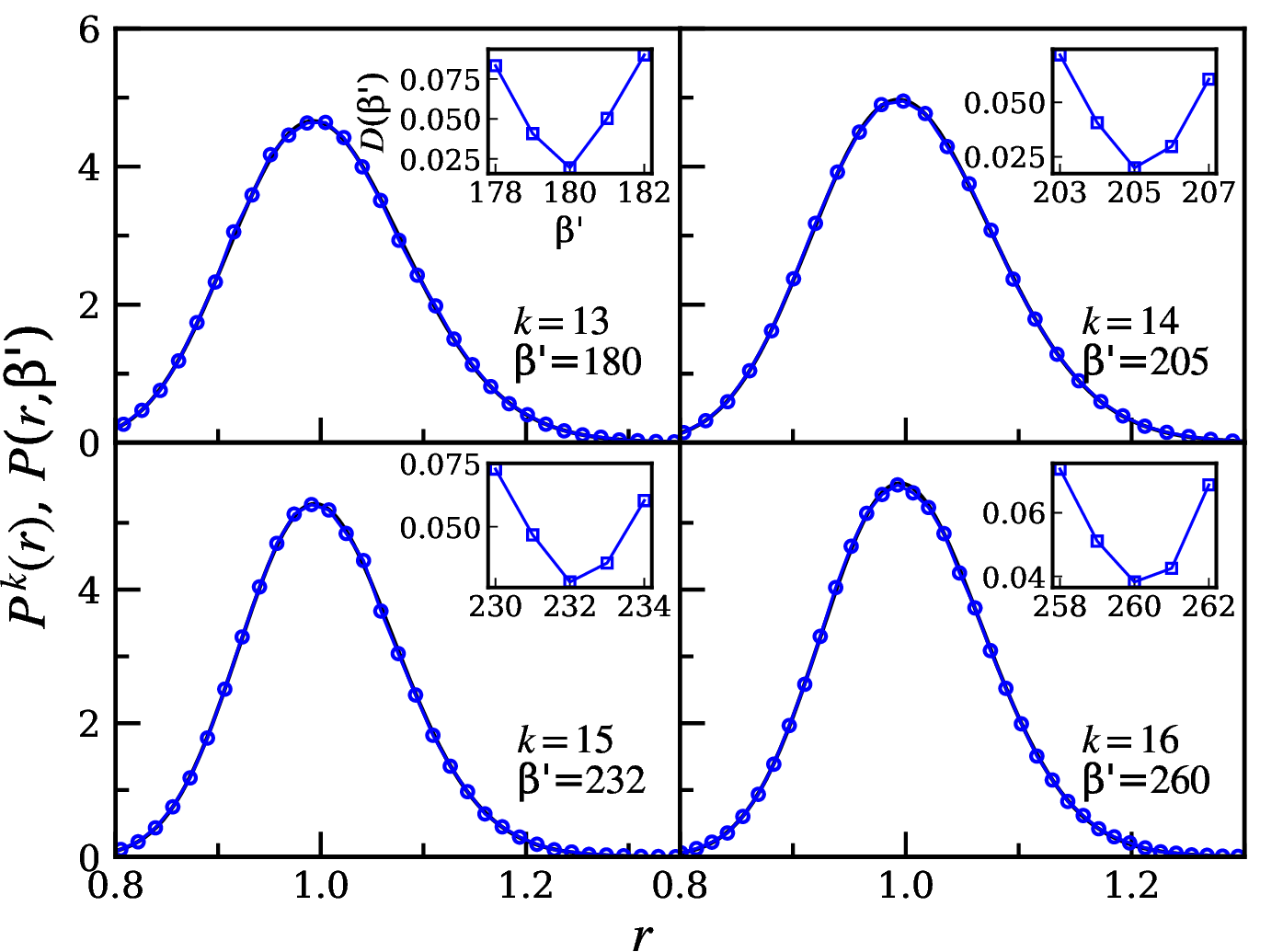}
\caption{\justifying Same as Fig.~\ref{fig:k_7_to_12_sratios_IM} but for different values of $k$ and $\beta'$.}
\label{fig:k_13_to_16_sratios_IM}
\end{center}
\end{figure}

\suppsection{Intermediate map}
\label{sec:more_plots_IM}
In this section, we have presented few more plots corresponding to spacings in Figs.~\ref{fig:k_7_to_12_spacings_IM}-\ref{fig:k_13_to_16_spacings_IM} and ratios in Figs.~\ref{fig:k_7_to_12_sratios_IM}-\ref{fig:k_13_to_16_sratios_IM}
for the intermediate map.

\begin{figure}[H]
\begin{center}
\includegraphics*[scale=0.45]{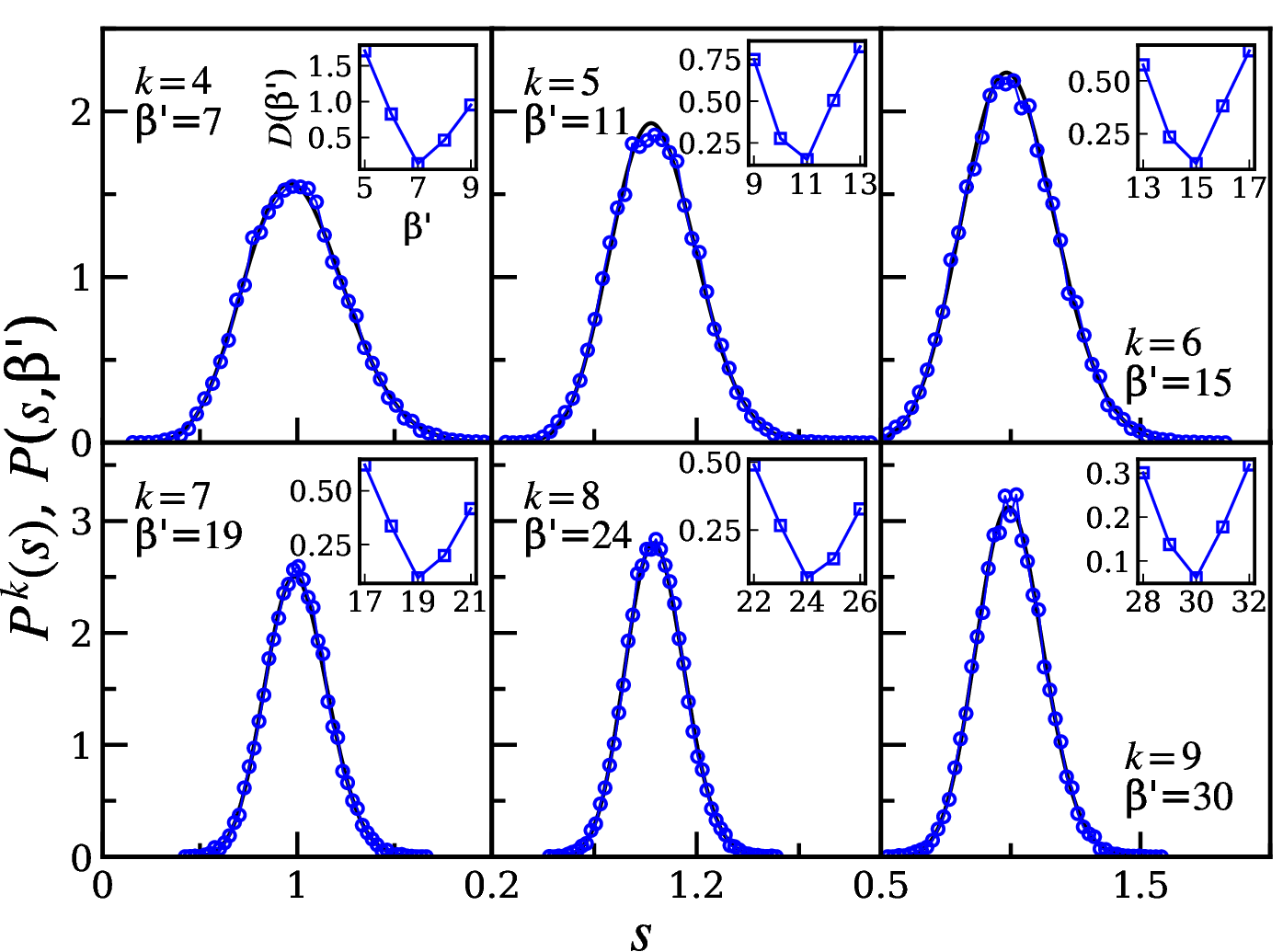}
\caption{\justifying HOS distribution $P^{k}(s)$ of eigenangles of QKT for various $k$ (circles). Here, $j= 1000$ i.e. $N= 2001$ and $n= 50$ such that $\tilde{k}= 10$ to $59$. Solid line corresponds to $P(s, \beta')$.}
\label{fig:k_4_to_9_spacings_QKT}
\end{center}
\end{figure}

\begin{figure}[H]
\begin{center}
\includegraphics*[scale=0.45]{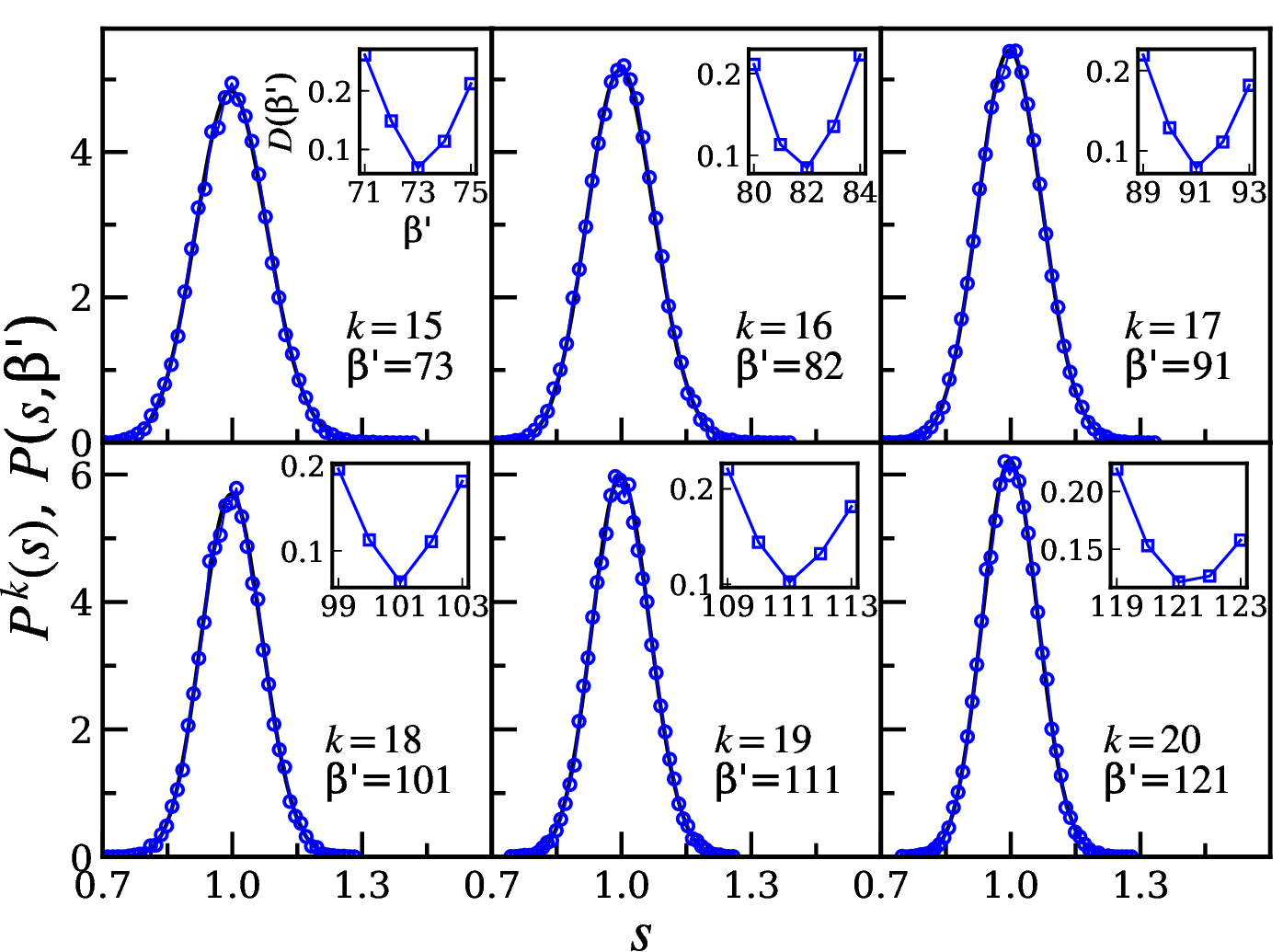}
\caption{\justifying Same as Fig.~\ref{fig:k_4_to_9_spacings_QKT} but for different values of $k$.}
\label{fig:k_15_to_20_spacings_QKT}
\end{center}
\end{figure}

\begin{figure}[H]
\begin{center}
\includegraphics*[scale=0.45]{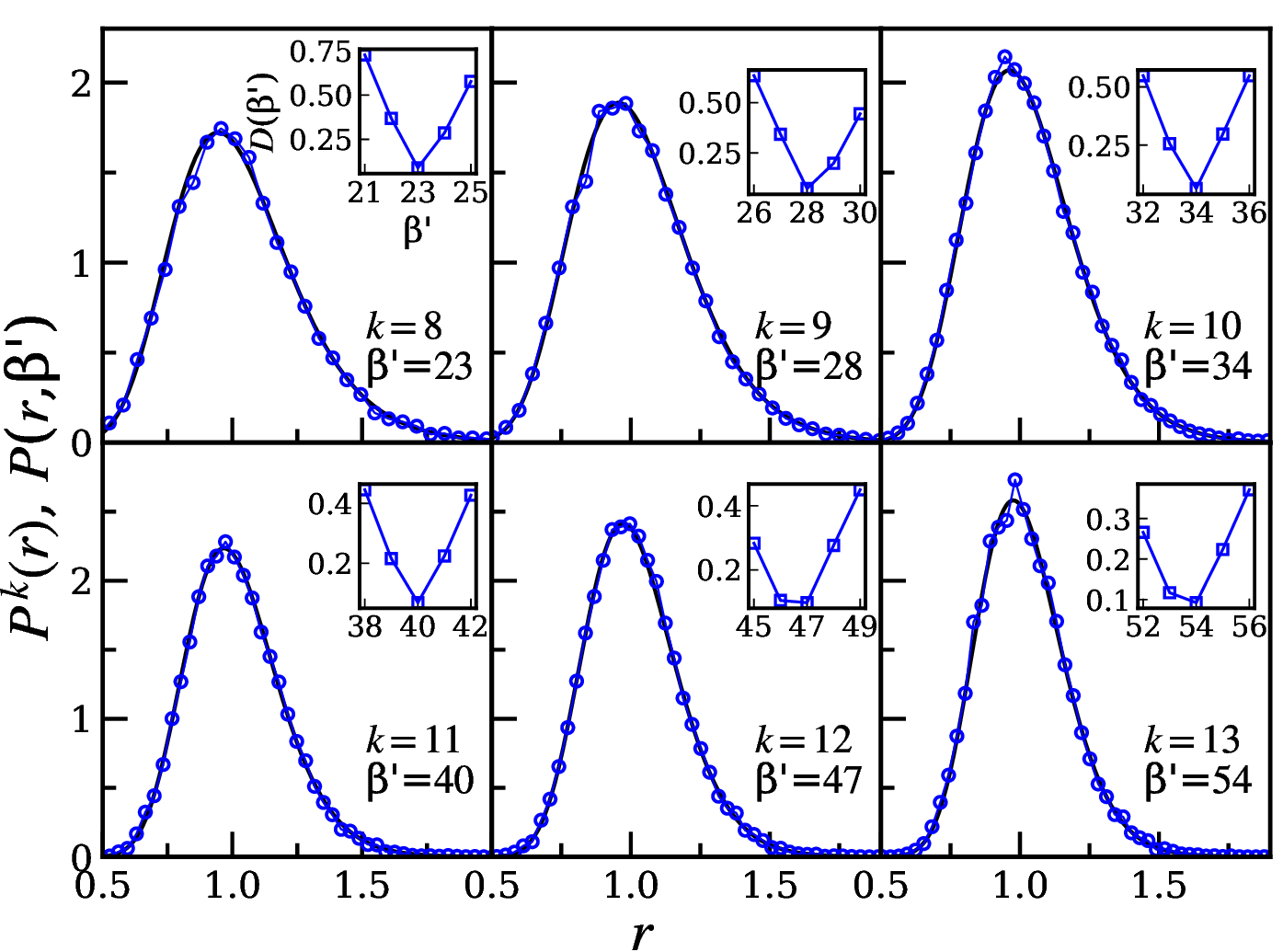}
\caption{\justifying Plot of HOSR distribution $P^{k}(r)$ of eigenangles of QKT for various $k$ (circles). Here, $N= 2001$ ($j= 1000$) and $n= 50$ such that $\tilde{k}= 10$ to $59$. Solid line corresponds to $P(r, \beta')$.}
\label{fig:k_8_to_13_sratios_QKT}
\end{center}
\end{figure}

\begin{figure}[H]
\begin{center}
\includegraphics*[scale=0.45]{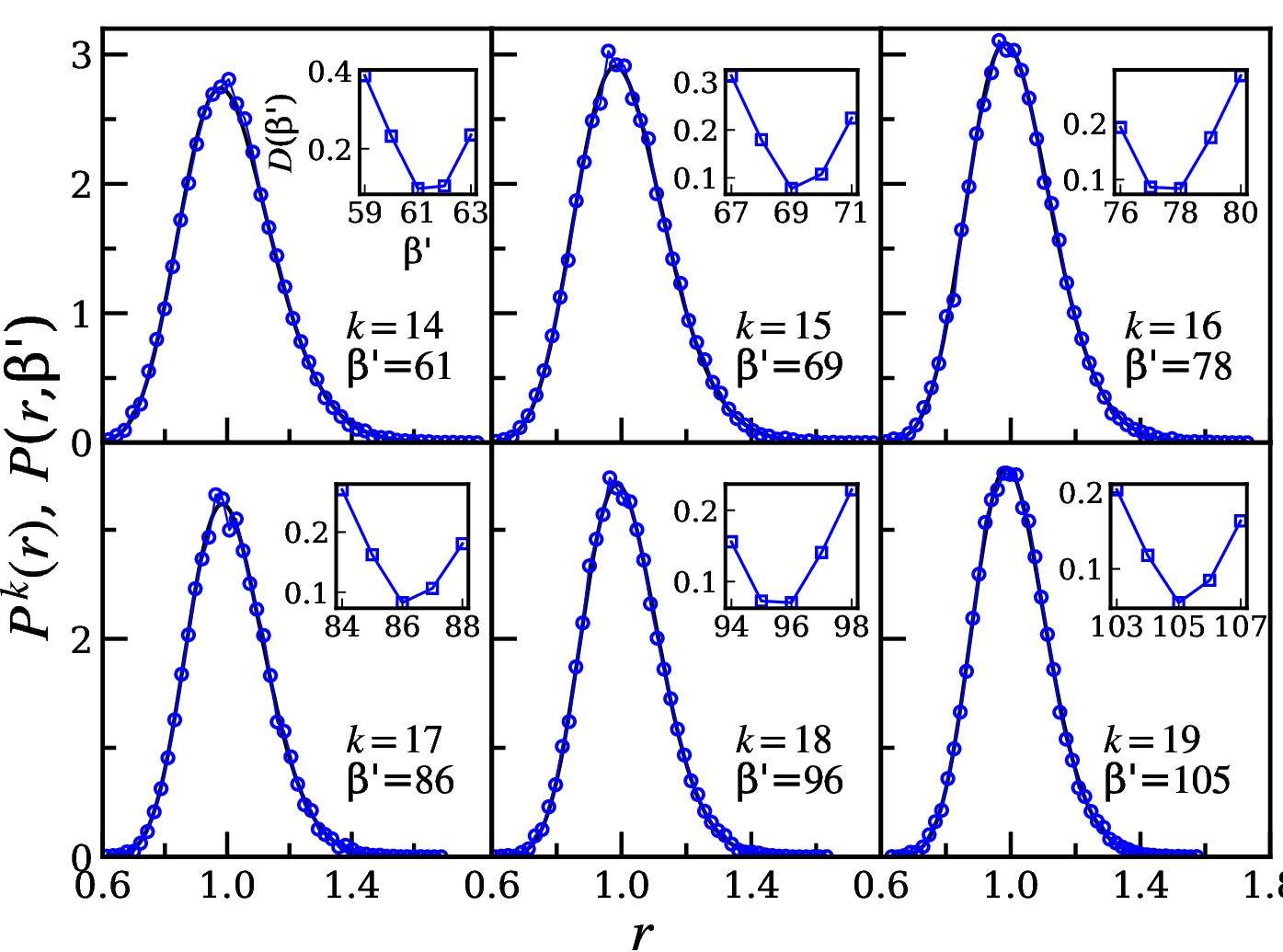}
\caption{\justifying Same as Fig.~\ref{fig:k_8_to_13_sratios_QKT} but for different values of $k$ and $\beta'$.}
\label{fig:k_14_to_19_sratios_QKT}
\end{center}
\end{figure}

\suppsection{Quantum Kicked Top}
\label{sec:more_plots_QKT}
In this section, we have presented few more plots corresponding to spacings in Figs.~\ref{fig:k_4_to_9_spacings_QKT}-\ref{fig:k_15_to_20_spacings_QKT} and ratios in Figs.~\ref{fig:k_8_to_13_sratios_QKT}-\ref{fig:k_14_to_19_sratios_QKT} for quantum chaotic kicked top (QKT).

\begin{figure}[H]
\begin{center}
\includegraphics*[scale=0.5]{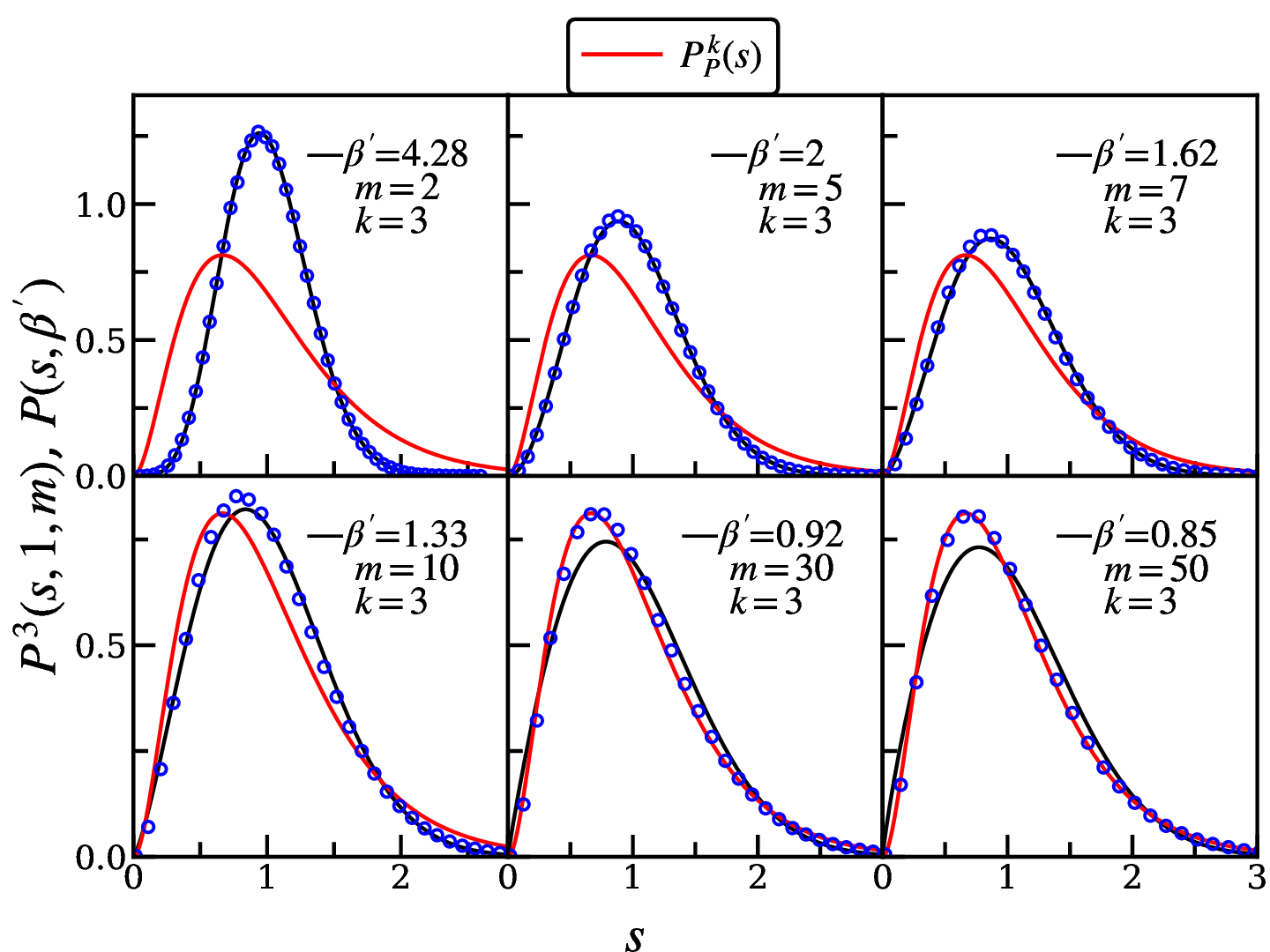}
\caption{\justifying Third order ($k= 3$) spacing distribution $P^{3}(s, 1, m)$ in the $m$ superposed spectra of COEs, denoted by circles. Here, the black solid line corresponds to the generalized Wigner-Dyson distribution corresponding to $\beta'$ for spacings $P(s, \beta')$. Red solid line corresponds to the third-order spacing distribution of the Poisson ensemble $P^{k}_{P}(s)$, see Eq.~ (\ref{Eq: HOS_Poisson}). Here, in each case $N= 5000$ and $n= 600, 1000, 1001, 900, 900$, and $900$, respectively, for $m= 2, 5, 7, 10, 30$, and $50$.}
\label{fig:k_3_COE_m_2_5_7_10_30_50_SPACINGS}
\end{center}
\end{figure}

\begin{figure}[H]
\begin{center}
\includegraphics*[scale=0.37]{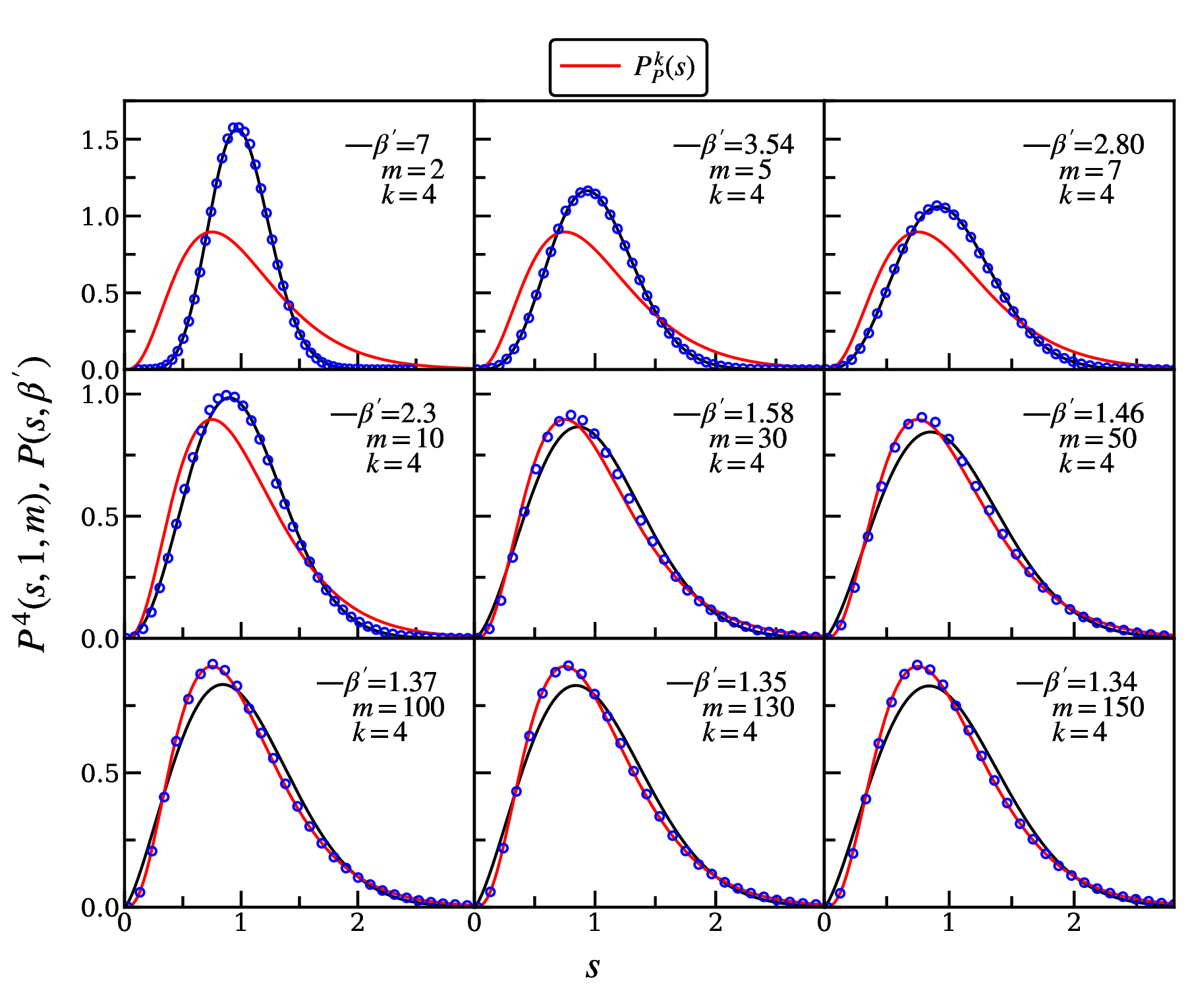}
\caption{\justifying Same as Fig.~\ref{fig:k_3_COE_m_2_5_7_10_30_50_SPACINGS} but for $k= 4$ and higher $m$ values. Here, in each case $N= 5000$ and $n= 600, 1000, 1001, 900, 900, 900, 900, 910$, and $900$, respectively, for $m= 2, 5, 7, 10, 30, 50, 100, 130$, and $150$.}
\label{fig:k_4_COE_m_2_5_7_10_30_50_100_130_150_SPACINGS}
\end{center}
\end{figure}

\begin{figure}[H]
\begin{center}
\includegraphics*[scale=0.37]{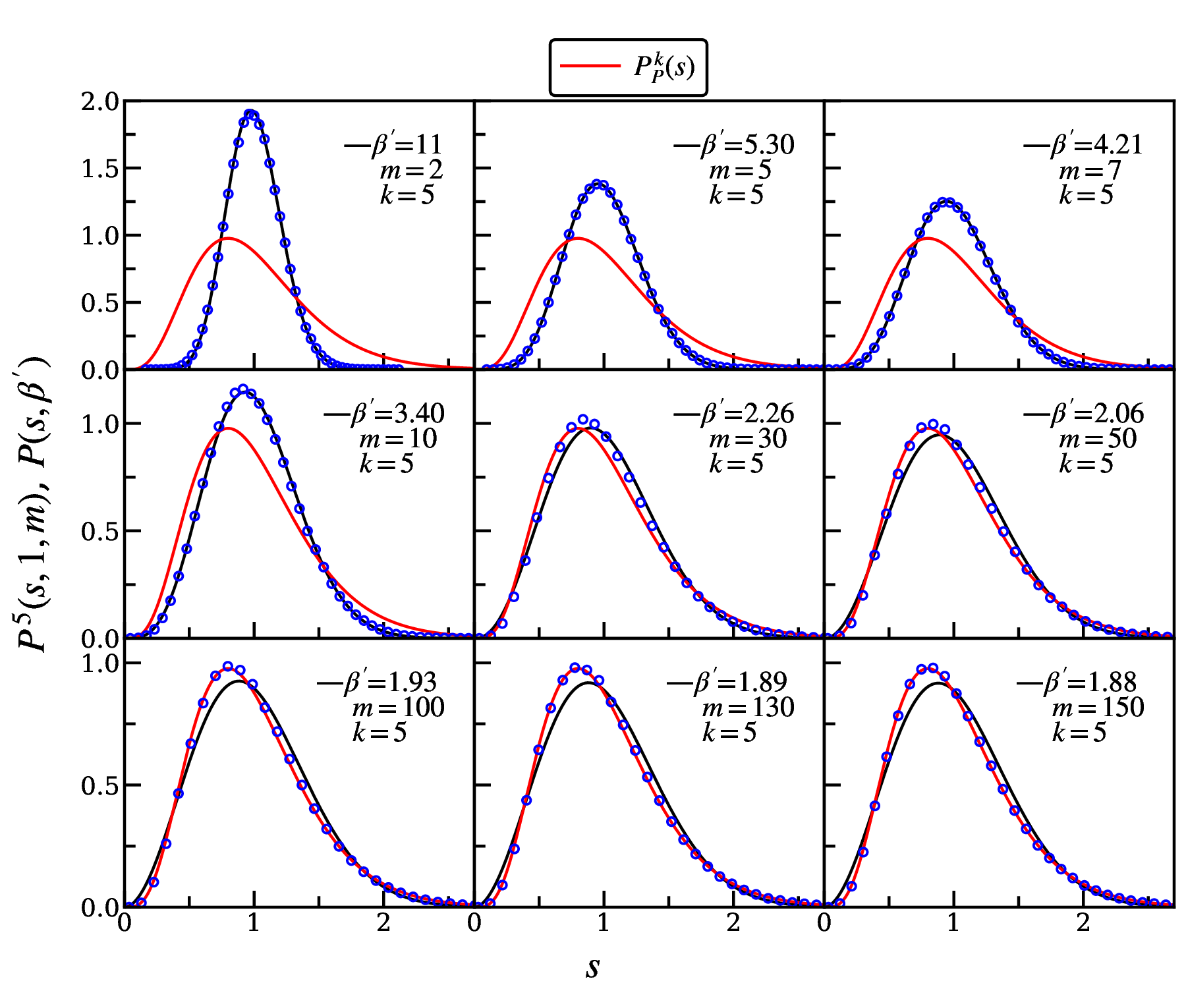}
\caption{\justifying Same as Fig.~\ref{fig:k_4_COE_m_2_5_7_10_30_50_100_130_150_SPACINGS} but for $k= 5$.}
\label{fig:k_5_COE_m_2_5_7_10_30_50_100_130_150_SPACINGS}
\end{center}
\end{figure}

\begin{figure}[H]
\begin{center}
\includegraphics*[scale=0.48]{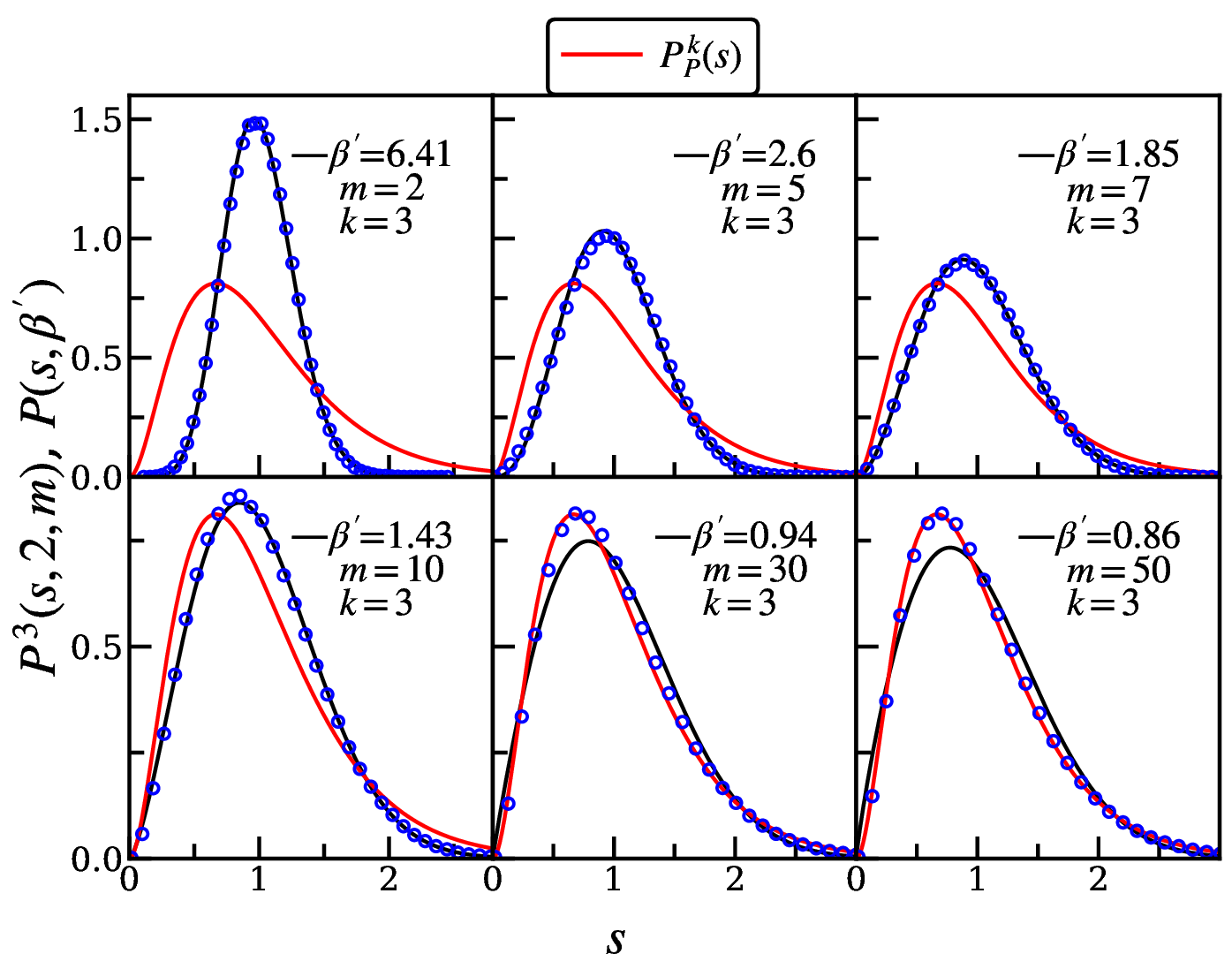}
\caption{\justifying Same as Fig.~\ref{fig:k_3_COE_m_2_5_7_10_30_50_SPACINGS} but for CUE. Here, in each case $N= 5000$ and $n= 600, 1000, 1001, 1000, 990$, and $1000$ respectively for $m= 2, 5, 7, 10, 30$, and $50$.}
\label{fig:k_3_CUE_m_2_5_7_10_30_50_SPACINGS}
\end{center}
\end{figure}

\begin{figure}[H]
\begin{center}
\includegraphics*[scale=0.41]{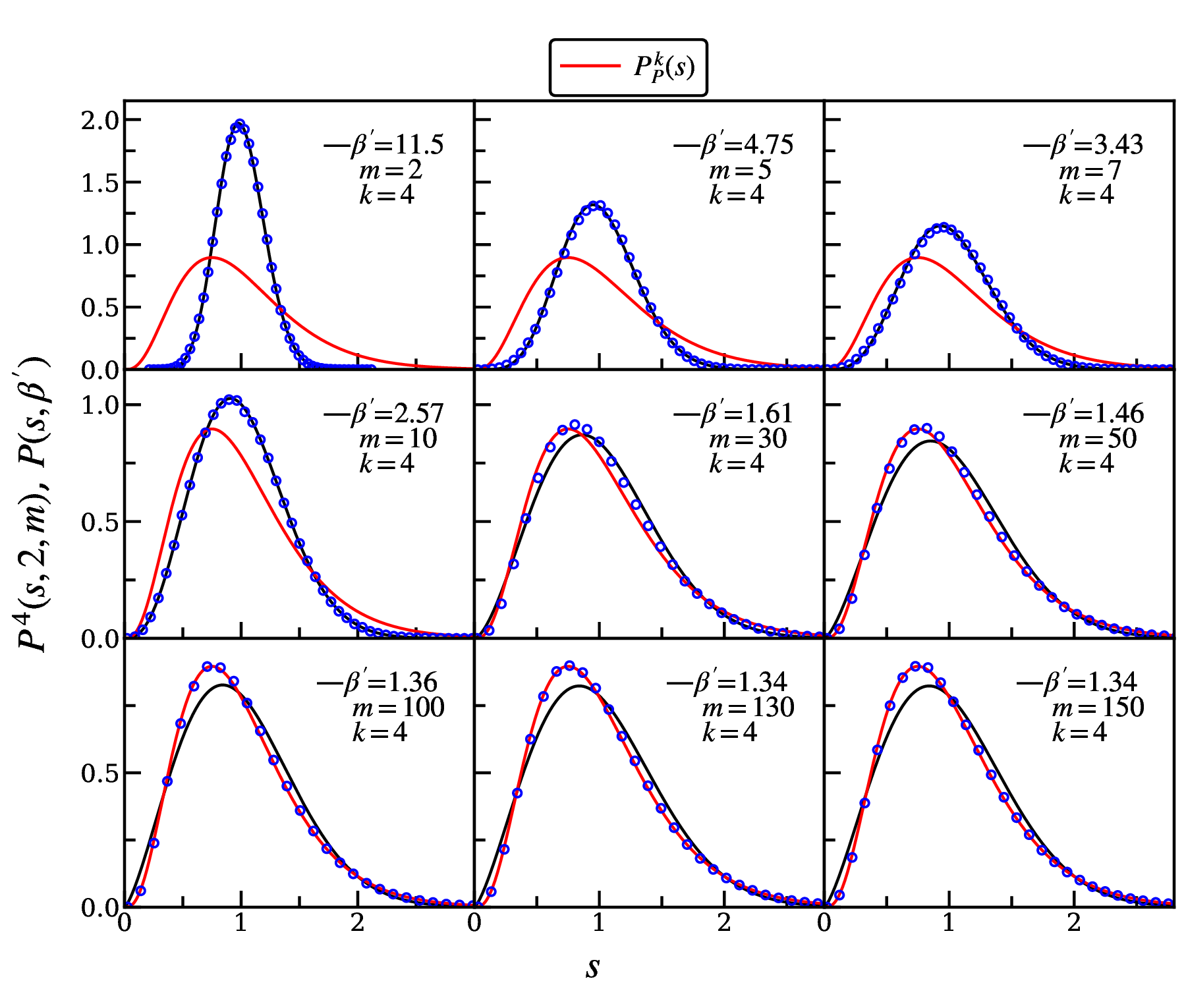}
\caption{\justifying Same as Fig.~\ref{fig:k_3_CUE_m_2_5_7_10_30_50_SPACINGS} but for $k= 4$ and higher $m$ values. Here, in each case $N= 5000$ and $n= 600, 1000, 1001, 1000, 990, 1000, 1000, 910$, and $900$, respectively, for $m= 2, 5, 7, 10, 30, 50, 100, 130$, and $150$.}
\label{fig:k_4_CUE_m_2_5_7_10_30_50_100_130_150_SPACINGS}
\end{center}
\end{figure}

\begin{figure}[H]
\begin{center}
\includegraphics*[scale=0.43]{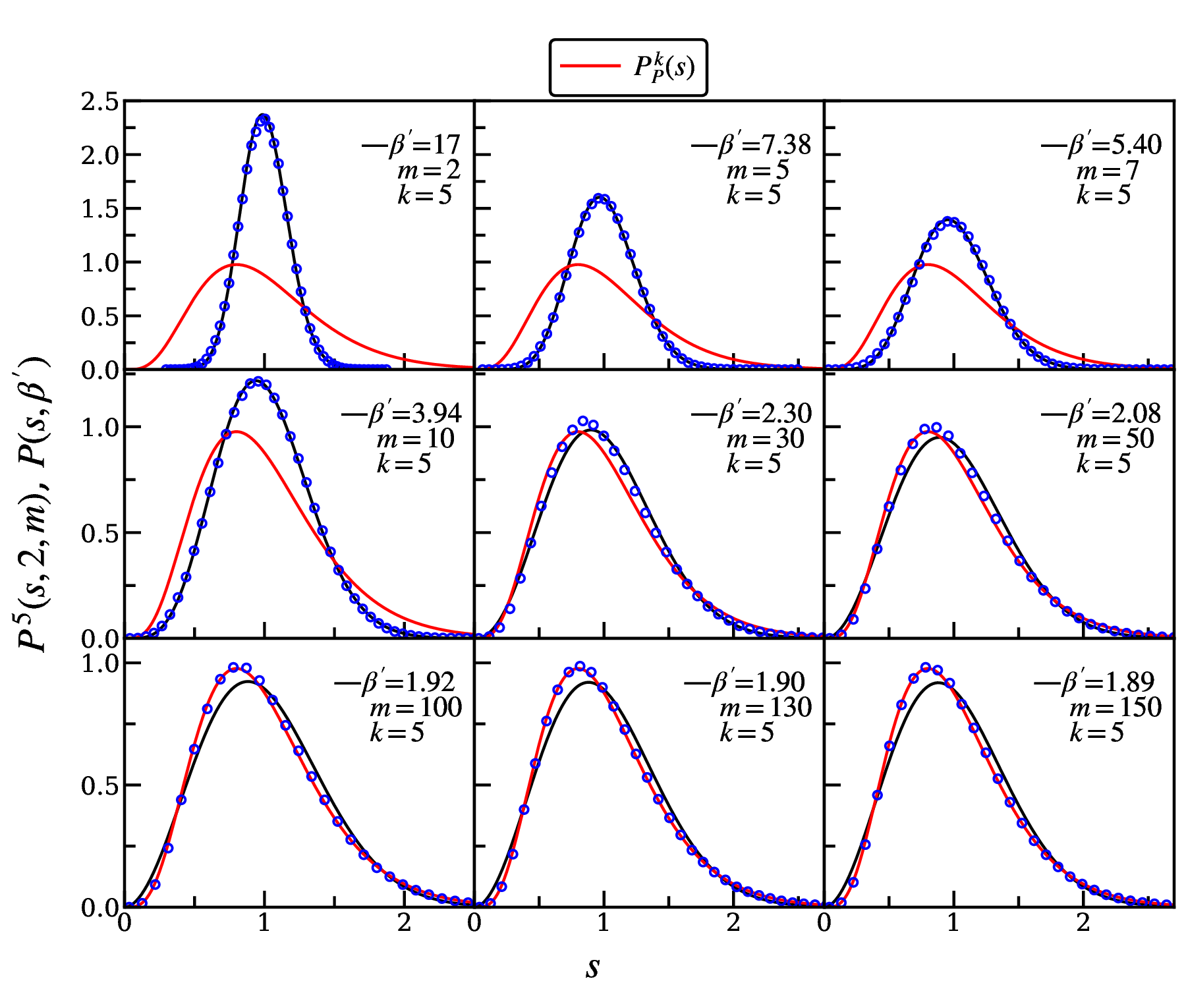}
\caption{\justifying Same as Fig.~\ref{fig:k_4_CUE_m_2_5_7_10_30_50_100_130_150_SPACINGS} but for $k= 5$.}
\label{fig:k_5_CUE_m_2_5_7_10_30_50_100_130_150_SPACINGS}
\end{center}
\end{figure}

\begin{figure}[H]
\begin{center}
\includegraphics*[scale=0.47]{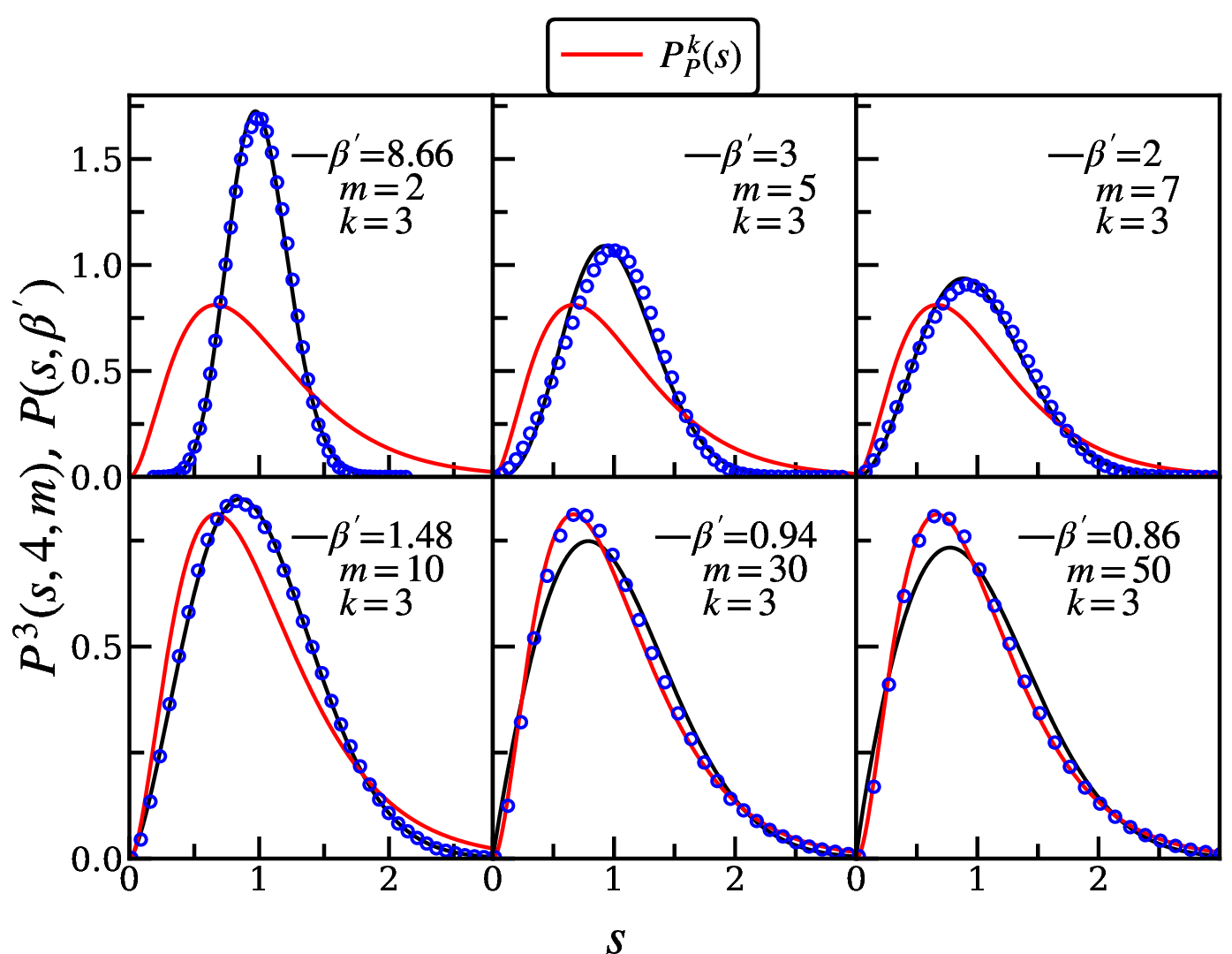}
\caption{\justifying Same as Fig.~\ref{fig:k_3_COE_m_2_5_7_10_30_50_SPACINGS} but for CSE. Here, in each case $N= 5000$ and $n= 600, 1000, 1001, 1000, 990$, and $1000$ respectively for $m= 2, 5, 7, 10, 30$, and $50$.}
\label{fig:k_3_CSE_m_2_5_7_10_30_50_SPACINGS}
\end{center}
\end{figure}

\begin{figure}[H]
\begin{center}
\includegraphics*[scale=0.38]{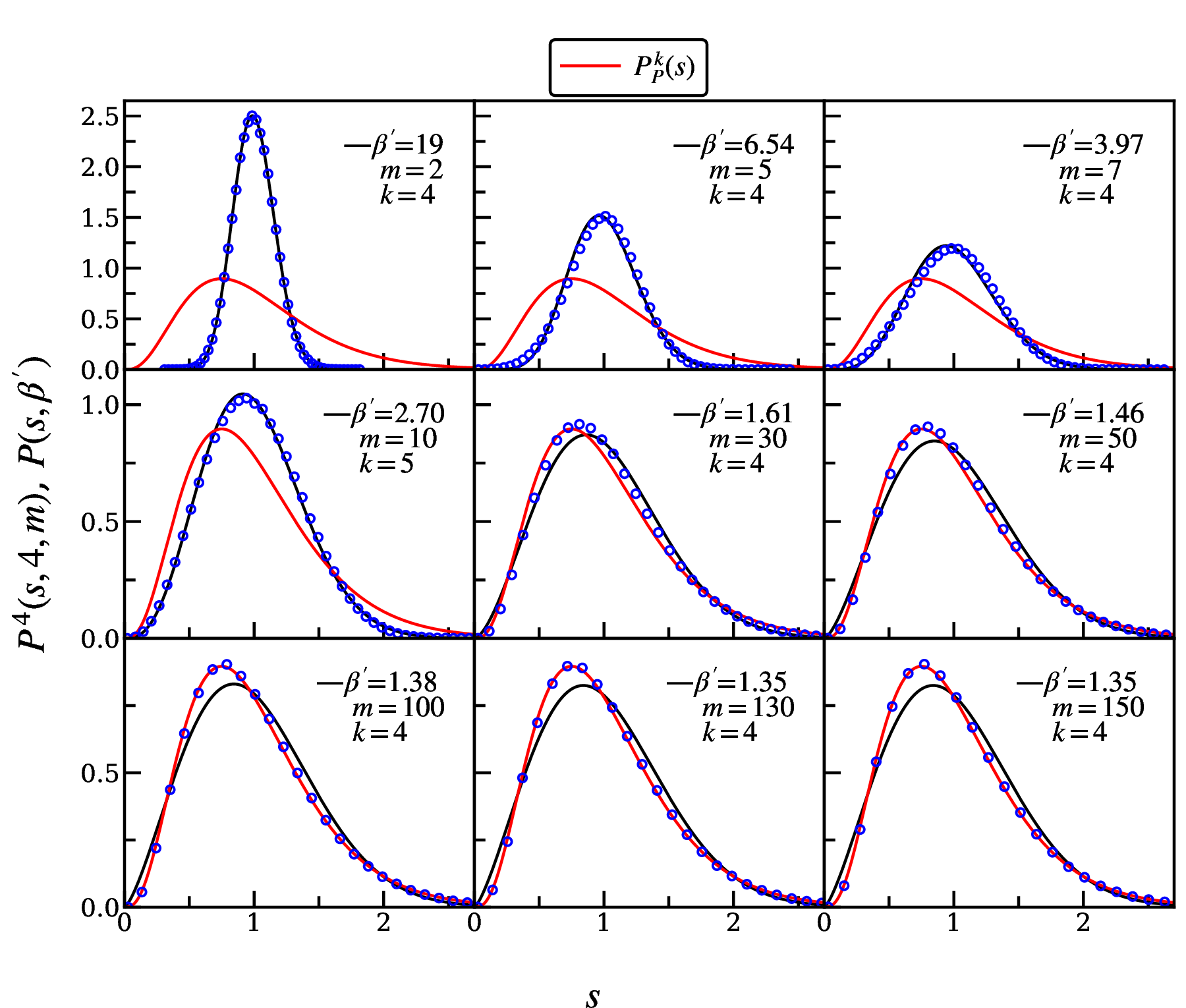}
\caption{\justifying Same as Fig.~\ref{fig:k_3_CSE_m_2_5_7_10_30_50_SPACINGS} but for $k= 4$ and higher $m$ values. Here, in each case $N= 5000$ and $n= 600, 1000, 1001, 1000, 990, 1000, 1000, 910$, and $900$, respectively, for $m= 2, 5, 7, 10, 30, 50, 100, 130$, and $150$.}
\label{fig:k_4_CSE_m_2_5_7_10_30_50_100_130_150_SPACINGS}
\end{center}
\end{figure}

\begin{figure}[H]
\begin{center}
\includegraphics*[scale=0.38]{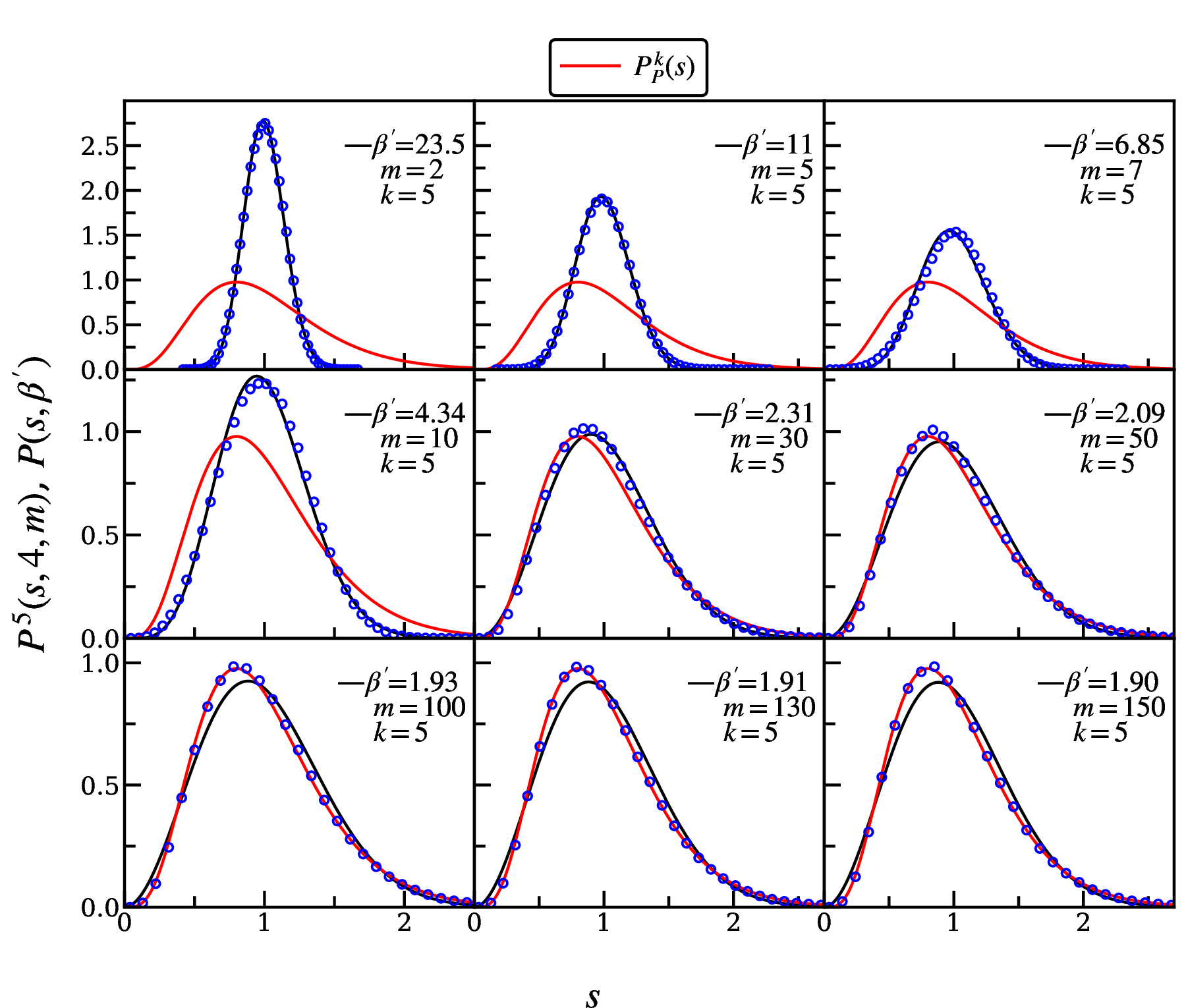}
\caption{\justifying Same as Fig.~\ref{fig:k_4_CSE_m_2_5_7_10_30_50_100_130_150_SPACINGS} but for $k= 5$.}
\label{fig:k_5_CSE_m_2_5_7_10_30_50_100_130_150_SPACINGS}
\end{center}
\end{figure}

\begin{figure}[H]
\begin{center}
\includegraphics*[scale=0.35]{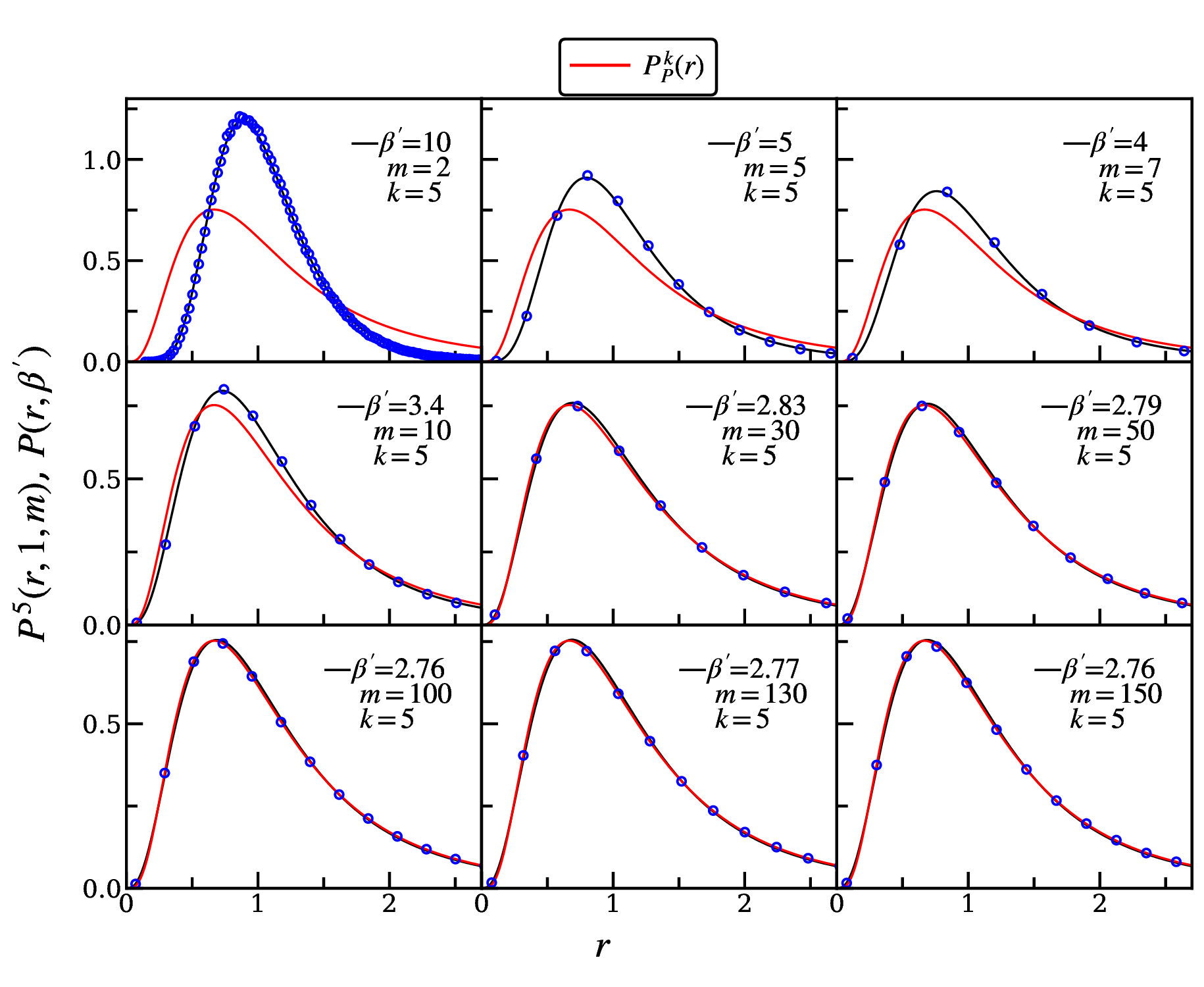}
\caption{\justifying Fifth order ($k= 5$) spacing ratio distribution $P^{5}(r, 1, m)$ in the $m$ superposed spectra of COEs, denoted by circles. Here, the  black solid line corresponds to the generalized Wigner-Dyson distribution corresponding to $\beta'$ for spacing ratios $P(r, \beta')$. The red solid line corresponds to the fifth-order spacing ratio distribution of the Poisson ensemble $P^{k}_{P}(r)$, see Eq.~(\ref{Eq: HOSR_Poisson}). Here, in each case $N= 5000$ and $n= 500, 995, 994, 900, 900, 900, 900, 910$, and $900$ respectively for $m= 2, 5, 7, 10, 30, 50, 100, 130$, and $150$.}
\label{fig:k_5_COE_m_2_5_7_10_30_50_100_130_150_SPACING_RATIOS}
\end{center}
\end{figure}

\begin{figure}[H]
\begin{center}
\includegraphics*[scale=0.35]{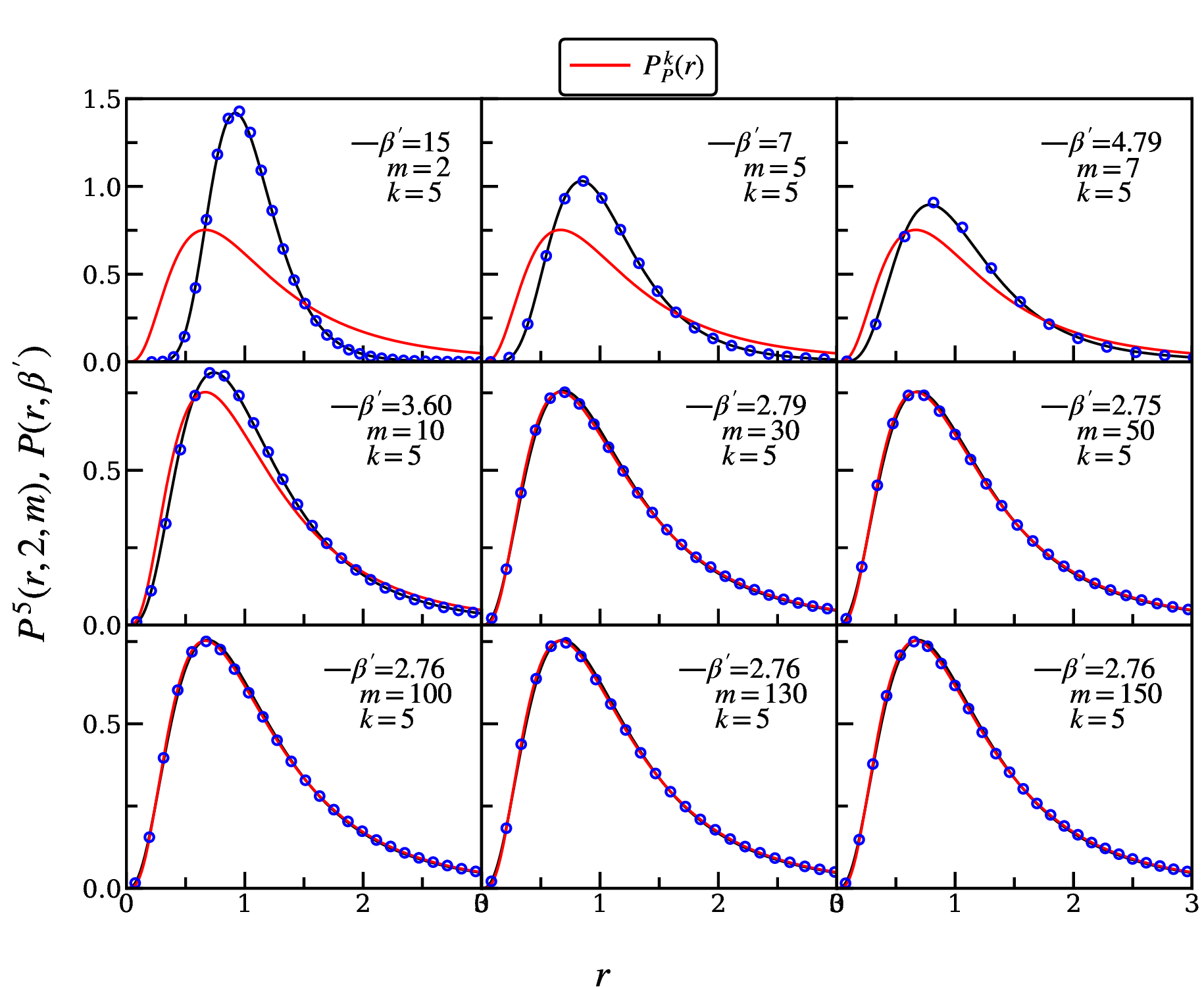}
\caption{\justifying Same as Fig.~\ref{fig:k_5_COE_m_2_5_7_10_30_50_100_130_150_SPACING_RATIOS} but for CUE. Here, in each case $N= 5000$ and $n= 500, 1000, 1001, 1000, 990, 1000, 1000, 910$, and $900$ respectively for $m= 2, 5, 7, 10, 30, 50, 100, 130$, and $150$.}
\label{fig:k_5_CUE_m_2_5_7_10_30_50_100_130_150_SPACING_RATIOS}
\end{center}
\end{figure}

\begin{figure}[H]
\begin{center}
\includegraphics*[scale=0.5]{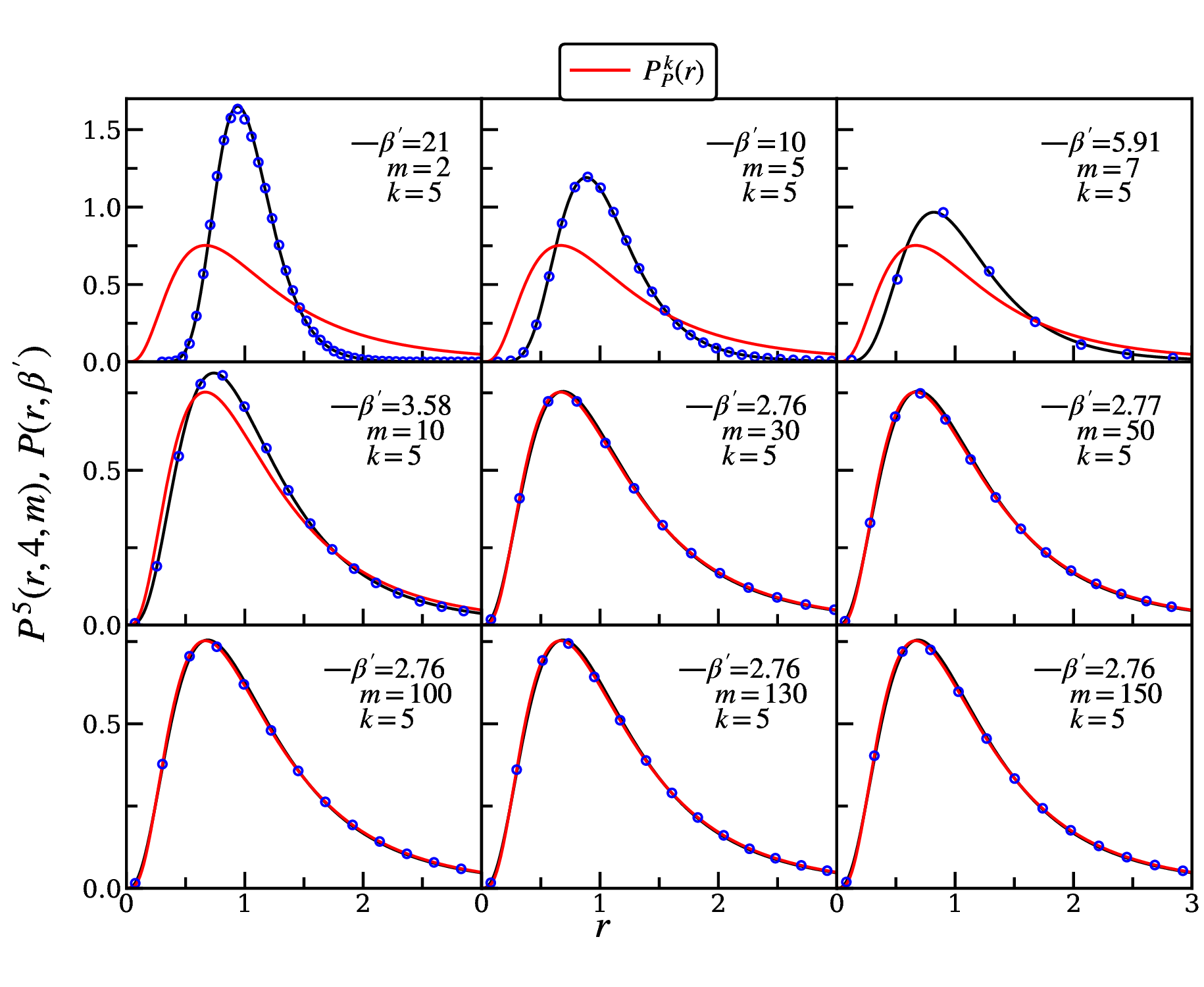}
\caption{\justifying Same as Fig.~\ref{fig:k_5_CUE_m_2_5_7_10_30_50_100_130_150_SPACING_RATIOS} but for CSE.}
\label{fig:k_5_CSE_m_2_5_7_10_30_50_100_130_150_SPACING_RATIOS}
\end{center}
\end{figure}

\suppsection{Convergence to the Poisson distribution}
\label{poisson_convergence}
In this section, we have shown a few more plots corresponding to the $k$-th order spacings and spacing ratios in the $m$ superposed spectra of all three classes of circular ensemble. Particularly, we find that they converge to the corresponding $k$-th order Poisson statistics with an increase in $m$. 
A conjecture based on these results is stated in the main text (see Sec. VIIA therein).
The results are shown in the Figs.~\ref{fig:k_3_COE_m_2_5_7_10_30_50_SPACINGS}-\ref{fig:k_5_CSE_m_2_5_7_10_30_50_100_130_150_SPACINGS} for spacings and in Figs.~\ref{fig:k_5_COE_m_2_5_7_10_30_50_100_130_150_SPACING_RATIOS}-\ref{fig:k_5_CSE_m_2_5_7_10_30_50_100_130_150_SPACING_RATIOS} for spacing ratios. 
The expression for higher-order spacing distribution in the Poisson ensemble is given by \cite{rao2020higher}:
\begin{equation}
P_{P}^{k}(s)=\frac{k^k}{(k-1)!}s^{k-1}e^{-ks}.
\label{Eq: HOS_Poisson}
\end{equation}
Whereas, the expression for higher order spacing ratio distribution in the Poisson ensemble is given by \cite{tekur2020symmetry}:
\begin{equation}
P_{P}^{k}(r)=\frac{(2k-1)!}{[(k-1)!]^2}\frac{r^{k-1}}{(1+r)^{2k}}.
\label{Eq: HOSR_Poisson}
\end{equation}
We use these expressions in the plots of this section.

\suppsection{Kolmogorov-Smirnov test}
\label{sec:KS test} 
Here, we have performed the K-S test \cite{boes1974introduction} using only one realization of superposed spectra of $m$ COEs, CUEs, and CSEs each separately, and each of dimension $N = 5000$. The function $P(s, \beta')$, for which $D(\beta')$ is minimum for the spacing distribution of one realization of superposed spectra, is used as the theoretical distribution function for the K-S test and the corresponding $D_{KS}$ and $p$ values are mentioned in Tables~\ref{Table: K-S Test COE1}-\ref{Table: K-S Test CSE2}. We can observe that the values of $D_{KS}$ are not very large and $p$ are not very small (the $p$ values are greater than the significance level of $0.05$), which indicates that there is no strong discrepancy between the data and the model function. But, for a given $k$ as $m$ becomes large, the distributions start deviating from the Wigner-Dyson and tend towards the corresponding higher-order Poisson statistics. This deviation is reflected in the values of $p$, which are smaller than $0.05$, when the theoretical distribution is taken from the generalized Wigner-Dyson distribution.
\begin{table*}     
\renewcommand{\arraystretch}{1.6} 
\setlength{\tabcolsep}{10pt}  
\begin{center}
\begin{tabular}{|c|ccc|ccc|ccc|}
\hline                                                                                                                                         
\rule{0pt}{12pt}  
Order&\multicolumn{3}{c|}{$m=2$}&\multicolumn{3}{c|}{$m=3$}&\multicolumn{3}{c|}{$m=4$}\\ [1.5ex]
\cline{2-10}  
\rule{0pt}{10pt}  
$k$&$\beta'$&$D_{KS}$&\multicolumn{1}{c|}{$p$}&$\beta'$&$D_{KS}$&\multicolumn{1}{c|}{$p$}&$\beta'$&$D_{KS}$&\multicolumn{1}{c|}{$p$}\\
\hline
2&2&5.674&\multicolumn{1}{c|}{9.025224}&1.27&7.255&\multicolumn{1}{c|}{4.068858}&1&4.667&\multicolumn{1}{c|}{7.745752}\\
3&4.35&9.403&\multicolumn{1}{c|}{3.376031}&3&6.184&\multicolumn{1}{c|}{6.125793}&2.40&5.991&\multicolumn{1}{c|}{4.677218}\\
4&7&8.317&\multicolumn{1}{c|}{4.912329}&5&8.289&\multicolumn{1}{c|}{2.528497}&4&7.289&\multicolumn{1}{c|}{2.373022}\\
5&11&8.589&\multicolumn{1}{c|}{4.496976}&8&6.454&\multicolumn{1}{c|}{5.578437}&6&7.386&\multicolumn{1}{c|}{2.242912}\\
6&15&6.943&\multicolumn{1}{c|}{7.182336}&11&4.986&\multicolumn{1}{c|}{8.481185}&9&10.427&\multicolumn{1}{c|}{0.2569796}\\
7&19&7.579&\multicolumn{1}{c|}{6.115366}&14&4.981&\multicolumn{1}{c|}{8.490280}&11&5.977&\multicolumn{1}{c|}{4.709073}\\
8&24&5.744&\multicolumn{1}{c|}{8.945369}&18&4.950&\multicolumn{1}{c|}{8.542279}&14&5.811&\multicolumn{1}{c|}{5.075510}\\
9&30&6.492&\multicolumn{1}{c|}{7.911891}&22&5.162&\multicolumn{1}{c|}{8.172908}&17&5.269&\multicolumn{1}{c|}{6.335885}\\
10&37&7.936&\multicolumn{1}{c|}{5.524909}&26&5.717&\multicolumn{1}{c|}{7.090673}&21&5.458&\multicolumn{1}{c|}{5.888839}\\
11&43&8.032&\multicolumn{1}{c|}{5.369546}&30&4.951&\multicolumn{1}{c|}{8.542617}&25&6.156&\multicolumn{1}{c|}{4.331986}\\
12&49&9.013&\multicolumn{1}{c|}{3.894269}&35&4.220&\multicolumn{1}{c|}{9.512138}&29&4.955&\multicolumn{1}{c|}{7.082915}\\
13&57&9.623&\multicolumn{1}{c|}{3.113281}&41&4.464&\multicolumn{1}{c|}{9.249343}&33&4.279&\multicolumn{1}{c|}{8.561655}\\
14&64&8.709&\multicolumn{1}{c|}{4.325845}&47&3.754&\multicolumn{1}{c|}{9.836405}&38&3.613&\multicolumn{1}{c|}{9.557343}\\
15&72&15.228&\multicolumn{1}{c|}{0.1929063}&53&4.772&\multicolumn{1}{c|}{8.828238}&42&3.580&\multicolumn{1}{c|}{9.591412}\\
16&81&7.170&\multicolumn{1}{c|}{6.809104}&59&5.833&\multicolumn{1}{c|}{6.855675}&48&6.250&\multicolumn{1}{c|}{4.142448}\\
17&90&6.224&\multicolumn{1}{c|}{8.316773}&66&6.318&\multicolumn{1}{c|}{5.858270}&54&6.330&\multicolumn{1}{c|}{3.983116}\\
18&100&7.472&\multicolumn{1}{c|}{6.302586}&72&5.089&\multicolumn{1}{c|}{8.306147}&59&6.057&\multicolumn{1}{c|}{4.541227}\\
19&111&7.951&\multicolumn{1}{c|}{5.506880}&79&6.528&\multicolumn{1}{c|}{5.436206}&65&4.078&\multicolumn{1}{c|}{8.924599}\\
20&121&6.230&\multicolumn{1}{c|}{8.308914}&86&4.915&\multicolumn{1}{c|}{8.604472}&71&2.965&\multicolumn{1}{c|}{9.944765}\\
\hline
\end{tabular}
\caption{\justifying Tabulation of higher-order indices $\beta'$, $D_{KS}$, and $p$ values for the spacing distributions for various values of $k$ in the superposed spectra of COE. Here, $N = 5000$ for each case. Also, $n=2, 3,$ and $4$ for $m = 2, 3$, and $4$ cases, respectively. Here, $D_{KS}$ and $p$ values are in the units of $10^{-3}$ and $10^{-1}$ , respectively.}
\label{Table: K-S Test COE1}
\end{center}
\end{table*}

\begin{table*}      
\renewcommand{\arraystretch}{1.5} 
\setlength{\tabcolsep}{10pt}  
\begin{center}
\begin{tabular}{|c|ccc|ccc|ccc|}
\hline
\rule{0pt}{12pt}  
Order&\multicolumn{3}{c|}{$m=5$}&\multicolumn{3}{c|}{$m=6$}&\multicolumn{3}{c|}{$m=7$}\\ [1.5ex]  
\cline{2-10}  
\rule{0pt}{10pt}  
$k$&$\beta'$&$D_{KS}$&\multicolumn{1}{c|}{$p$}&$\beta'$&$D_{KS}$&\multicolumn{1}{c|}{$p$}&$\beta'$&$D_{KS}$&\multicolumn{1}{c|}{$p$}\\
\hline
2&0.78&8.024&\multicolumn{1}{c|}{0.7957724}&0.68&11.004&\multicolumn{1}{c|}{0.01388298}&0.62&12.832&\multicolumn{1}{c|}{0.0001955976}\\
5&5.28&5.624&\multicolumn{1}{c|}{4.065013}&4.55&5.772&\multicolumn{1}{c|}{2.693894}&4.19&3.664&\multicolumn{1}{c|}{7.337375}\\
7&9.36&6.475&\multicolumn{1}{c|}{2.444979}&8.28&3.203&\multicolumn{1}{c|}{9.169527}&7.46&3.209&\multicolumn{1}{c|}{8.627344}\\
8&12&6.408&\multicolumn{1}{c|}{2.552406}&10.46&4.191&\multicolumn{1}{c|}{6.663464}&9.42&3.427&\multicolumn{1}{c|}{8.043157}\\
10&18&8.252&\multicolumn{1}{c|}{0.6616439}&15&4.859&\multicolumn{1}{c|}{4.768124}&14&3.888&\multicolumn{1}{c|}{6.638486}\\
13&28&6.326&\multicolumn{1}{c|}{2.688517}&24&5.986&\multicolumn{1}{c|}{2.319697}&21&5.081&\multicolumn{1}{c|}{3.259718}\\
15&36&3.992&\multicolumn{1}{c|}{8.191802}&31&4.675&\multicolumn{1}{c|}{5.271726}&27&3.701&\multicolumn{1}{c|}{7.227033}\\
17&44&3.802&\multicolumn{1}{c|}{8.615694}&39&5.390&\multicolumn{1}{c|}{3.470970}&34&4.046&\multicolumn{1}{c|}{6.143099}\\
18&49&3.222&\multicolumn{1}{c|}{9.569879}&43&5.566&\multicolumn{1}{c|}{3.096404}&37&4.929&\multicolumn{1}{c|}{3.621372}\\
20&59&5.081&\multicolumn{1}{c|}{5.373733}&51&3.646&\multicolumn{1}{c|}{8.189053}&45&2.697&\multicolumn{1}{c|}{9.605002}\\
\hline
\end{tabular}
\caption{\justifying Same as Table~\ref{Table: K-S Test COE1}, but for $n=5, 6,$ and $7$ for $m = 5, 6$, and $7$ cases, respectively.}
\label{Table: K-S Test COE2}
\end{center}
\end{table*}

\begin{table*}      
\renewcommand{\arraystretch}{1.5} 
\setlength{\tabcolsep}{10pt}  
\begin{center}
\begin{tabular}{|c|ccc|ccc|ccc|}
\hline
\rule{0pt}{12pt}  
Order&\multicolumn{3}{c|}{$m=2$}&\multicolumn{3}{c|}{$m=3$}&\multicolumn{3}{c|}{$m=4$}\\ [1.5ex]  
\cline{2-10}  
\rule{0pt}{10pt}  
$k$&$\beta'$&$D_{KS}$&\multicolumn{1}{c|}{$p$}&$\beta'$&$D_{KS}$&\multicolumn{1}{c|}{$p$}&$\beta'$&$D_{KS}$&\multicolumn{1}{c|}{$p$}\\
\hline
2&3.19&9.266&\multicolumn{1}{c|}{3.550465}&1.83&12.388&\multicolumn{1}{c|}{0.1987026}&1.22&10.283&\multicolumn{1}{c|}{0.2892552}\\
5&16&9.965&\multicolumn{1}{c|}{2.721868}&11.37&10.298&\multicolumn{1}{c|}{0.8257023}&9&6.888&\multicolumn{1}{c|}{2.975610}\\
7&30&7.878&\multicolumn{1}{c|}{5.618439}&20&7.060&\multicolumn{1}{c|}{4.416139}&16&6.252&\multicolumn{1}{c|}{4.135523}\\
10&58&8.364&\multicolumn{1}{c|}{4.843238}&39&9.156&\multicolumn{1}{c|}{1.609590}&30&5.076&\multicolumn{1}{c|}{6.795805}\\
13&89&7.893&\multicolumn{1}{c|}{5.597968}&63&4.410&\multicolumn{1}{c|}{9.313387}&48&7.859&\multicolumn{1}{c|}{1.683409}\\
17&146&7.761&\multicolumn{1}{c|}{5.818572}&100&6.922&\multicolumn{1}{c|}{4.674300}&77&6.659&\multicolumn{1}{c|}{3.367250}\\
20&200&6.033&\multicolumn{1}{c|}{8.584305}&133&5.405&\multicolumn{1}{c|}{7.718005}&103&5.153&\multicolumn{1}{c|}{6.616816}\\
\hline
\end{tabular}
\caption{\justifying Tabulation of higher-order indices $\beta'$, $D_{KS}$, and $p$ values for the spacing distributions  for various values of $k$ in the superposed spectra of CUE. Here, $N = 5000$ for each case. Also, $n=2, 3,$ and $4$ for $m = 2, 3$, and $4$ cases, respectively. Here, $D_{KS}$ and $p$ values are in the units of  $10^{-3}$ and $10^{-1}$, respectively.}
\label{Table: K-S Test CUE1}
\end{center}
\end{table*}
\begin{table*}      
\renewcommand{\arraystretch}{1.5} 
\setlength{\tabcolsep}{10pt}  
\begin{center}
\begin{tabular}{|c|ccc|ccc|ccc|}
\hline
\rule{0pt}{12pt}  
Order&\multicolumn{3}{c|}{$m=5$}&\multicolumn{3}{c|}{$m=6$}&\multicolumn{3}{c|}{$m=7$}\\ [1.5ex]  
\cline{2-10}  
\rule{0pt}{10pt}  
$k$&$\beta'$&$D_{KS}$&\multicolumn{1}{c|}{$p$}&$\beta'$&$D_{KS}$&\multicolumn{1}{c|}{$p$}&$\beta'$&$D_{KS}$&\multicolumn{1}{c|}{$p$}\\
\hline
2&0.95&5.213&\multicolumn{1}{c|}{5.035162}&0.80&4.731&\multicolumn{1}{c|}{5.113386}&0.69&7.104&\multicolumn{1}{c|}{0.5815864}\\
5&7.27&6.567&\multicolumn{1}{c|}{2.301912}&6.20&7.483&\multicolumn{1}{c|}{0.6914974}&5.43&9.747&\multicolumn{1}{c|}{0.02571103}\\
7&13&5.281&\multicolumn{1}{c|}{4.869146}&12&7.091&\multicolumn{1}{c|}{0.9749037}&10&7.637&\multicolumn{1}{c|}{0.3359062}\\
10&25&3.910&\multicolumn{1}{c|}{8.379448}&21&6.559&\multicolumn{1}{c|}{1.507753}&19&6.183&\multicolumn{1}{c|}{1.371116}\\
13&39&5.176&\multicolumn{1}{c|}{5.131428}&34&4.047&\multicolumn{1}{c|}{7.083832}&30&3.716&\multicolumn{1}{c|}{7.179475}\\
17&65&5.567&\multicolumn{1}{c|}{4.195399}&54&4.732&\multicolumn{1}{c|}{5.113430}&48&4.833&\multicolumn{1}{c|}{3.859638}\\
20&85&4.515&\multicolumn{1}{c|}{6.868324}&74&4.497&\multicolumn{1}{c|}{5.777395}&64&6.137&\multicolumn{1}{c|}{1.428037}\\
\hline
\end{tabular}
\caption{\justifying Same as Table~\ref{Table: K-S Test CUE1}, but for $n=5, 6,$ and $7$ for $m = 5, 6$, and $7$ cases, respectively.}
\label{Table: K-S Test CUE2}
\end{center}
\end{table*}

\begin{table*}      
\renewcommand{\arraystretch}{1.5} 
\setlength{\tabcolsep}{10pt}  
\begin{center}
\begin{tabular}{|c|ccc|ccc|ccc|}
\hline
\rule{0pt}{12pt}  
Order&\multicolumn{3}{c|}{$m=2$}&\multicolumn{3}{c|}{$m=3$}&\multicolumn{3}{c|}{$m=4$}\\ [1.5ex]  
\cline{2-10}  
\rule{0pt}{10pt}  
$k$&$\beta'$&$D_{KS}$&\multicolumn{1}{c|}{$p$}&$\beta'$&$D_{KS}$&\multicolumn{1}{c|}{$p$}&$\beta'$&$D_{KS}$&\multicolumn{1}{c|}{$p$}\\
\hline
3&8.66&11.438&\multicolumn{1}{c|}{1.450868}&7.17&6.649&\multicolumn{1}{c|}{5.189583}&4.67&24.857&\multicolumn{1}{c|}{\num{3.632607e-10}}\\
7&45&9.259&\multicolumn{1}{c|}{3.562173}&30&4.986&\multicolumn{1}{c|}{8.481145}&23&12.560&\multicolumn{1}{c|}{0.03611715}\\
10&97&7.655&\multicolumn{1}{c|}{5.989631}&57&4.628&\multicolumn{1}{c|}{9.036655}&43&8.452&\multicolumn{1}{c|}{1.143134}\\
11&104&9.010&\multicolumn{1}{c|}{3.897824}&69&7.583&\multicolumn{1}{c|}{3.530272}&53&8.786&\multicolumn{1}{c|}{0.9083464}\\
15&186&7.541&\multicolumn{1}{c|}{6.184944}&128&6.183&\multicolumn{1}{c|}{6.133853}&94&6.601&\multicolumn{1}{c|}{3.470496}\\
18&&&\multicolumn{1}{c|}{}&180&5.597&\multicolumn{1}{c|}{7.337262}&128&6.814&\multicolumn{1}{c|}{3.101069}\\
20&&&\multicolumn{1}{c|}{}&204&6.942&\multicolumn{1}{c|}{4.637339}&163&7.258&\multicolumn{1}{c|}{2.420551}\\
\hline
\end{tabular}
\caption{\justifying Tabulation of higher-order indices $\beta'$, $D_{KS}$, and $p$ values for the spacing distributions  for various values of $k$ in the superposed spectra of CSE. Here, $N = 5000$ for each case. Also, $n=2, 3,$ and $4$ for $m = 2, 3$, and $4$ cases, respectively. Here, $D_{KS}$ and $p$ values are in the units of  $10^{-3}$ and $10^{-1}$, respectively.}
\label{Table: K-S Test CSE1}
\end{center}
\end{table*}
\begin{table*}      
\renewcommand{\arraystretch}{1.5} 
\setlength{\tabcolsep}{7pt}  
\begin{center}
\begin{tabular}{|c|ccc|ccc|ccc|}
\hline
\rule{0pt}{12pt}  
Order&\multicolumn{3}{c|}{$m=5$}&\multicolumn{3}{c|}{$m=6$}&\multicolumn{3}{c|}{$m=7$}\\ [1.5ex]  
\cline{2-10}  
\rule{0pt}{10pt}  
$k$&$\beta'$&$D_{KS}$&\multicolumn{1}{c|}{$p$}&$\beta'$&$D_{KS}$&\multicolumn{1}{c|}{$p$}&$\beta'$&$D_{KS}$&\multicolumn{1}{c|}{$p$}\\
\hline
3&3&26.966&\multicolumn{1}{c|}{\num{3.179622e-15}}&2.39&19.931&\multicolumn{1}{c|}{\num{8.783530e-10}}&2&14.974&\multicolumn{1}{c|}{\num{3.632607e-10}}\\
7&17.63&6.210&\multicolumn{1}{c|}{2.888306}&16&4.532&\multicolumn{1}{c|}{5.672581}&14&6.748&\multicolumn{1}{c|}{0.8220500}\\
10&37&6.222&\multicolumn{1}{c|}{2.869156}&29.25&12.851&\multicolumn{1}{c|}{0.0009893334}&25&6.944&\multicolumn{1}{c|}{0.6818955}\\
11&43&5.875&\multicolumn{1}{c|}{3.529606}&36&10.561&\multicolumn{1}{c|}{0.02468826}&30&10.190&\multicolumn{1}{c|}{0.01386992}\\
15&78&4.207&\multicolumn{1}{c|}{7.666690}&62&7.162&\multicolumn{1}{c|}{0.9179467}&56&7.033&\multicolumn{1}{c|}{0.6249694}\\
18&104&7.342&\multicolumn{1}{c|}{1.346445}&92&6.471&\multicolumn{1}{c|}{1.615989}&75&7.972&\multicolumn{1}{c|}{0.2332682}\\
20&130&3.625&\multicolumn{1}{c|}{8.967848}&106&5.764&\multicolumn{1}{c|}{2.711234}&95&7.062&\multicolumn{1}{c|}{0.6074371}\\
\hline
\end{tabular}
\caption{\justifying Same as Table~\ref{Table: K-S Test CSE1}, but for $n=5, 6,$ and $7$ for $m = 5, 6$, and $7$ cases, respectively.}
\label{Table: K-S Test CSE2}
\end{center}
\end{table*}